\let\KernelLabel\label
\documentclass[letterpaper,aps,rmp,twocolumn,notitlepage,floatfix,nofootinbib,longbibliography,nobalancelastpage,10pt,superscriptaddress]{revtex4-2}
\let\label\KernelLabel

\usepackage{array}[=2016-10-06]

\makeatletter
\def\NAT@sort{0}
\makeatother

\ifdefined\pdfminorversion
\fi
\usepackage{amssymb,amsmath,amsthm}
\usepackage{mathtools}
\usepackage{ifthen}
\usepackage[dvipsnames]{xcolor}
\usepackage{enumitem}
\usepackage{graphicx}
\usepackage[ colorlinks = true,
             linkcolor = blue,
             urlcolor  = blue,
             citecolor = red,
             anchorcolor = green,
]{hyperref}

\usepackage[LGR,T1]{fontenc}
\usepackage[utf8]{inputenc}
\usepackage[greek.polutoniko,english]{babel}
\usepackage{xpatch}
\xapptocmd{\greektext}{\edef~{\string~}}{}{}

\newcommand{\id}{\one}
\newcommand{\idc}{\mathcal{I}} %
\newcommand{\bbC}{\mathbb{C}}  %
\newcommand{\bbR}{\mathbb{R}}  %
\newcommand{\bbN}{\mathbb{N}}  %
\newcommand{\cL}{\mathcal{L}}  %
\newcommand{\cP}{\mathcal{P}}  %
\newcommand{\cS}{\mathcal{S}}  %

\newcommand{\cD}{\mathcal{D}}
\newcommand{\cE}{\mathcal{E}}
\newcommand{\cF}{\mathcal{F}}
\newcommand{\cG}{\mathcal{G}}
\newcommand{\cR}{\mathcal{R}}
\newcommand{\cM}{\mathcal{M}}  %
\newcommand{\cN}{\mathcal{N}}

\DeclareMathOperator{\tr}{Tr}  %
\DeclareMathOperator{\spec}{spec}  %
\DeclareMathOperator{\supp}{supp}  %

\newcommand{\ket}[1]{|#1\rangle}  %
\newcommand{\bra}[1]{\langle #1|}
\newcommand{\braket}[2]{\langle#1|#2\rangle}
\newcommand{\proj}[1]{|#1\rangle\!\langle #1|}

\renewcommand{\dag}{\dagger}  %

\newcommand{\eps}{\varepsilon}
\newcommand{\epsclose}{\approx_\eps}
\newcommand{\D}{\mathbb{D}} %
\DeclareMathOperator{\reg}{reg}
\newcommand{\Dreg}{\D^{\infty}} %
\newcommand{\meas}{{\smash{\mathbb{M}}}} %
\newcommand{\Dmeas}{\D^{\meas}} %
\newcommand{\prep}{\mathbb{P}}
\newcommand{\Dprep}{\D^{\prep}} %
\newcommand{\Dcl}{\D_{\textnormal{cl}}}
\newcommand{\Hcl}{\H_{\textnormal{cl}}}%
\renewcommand{\H}{\mathbb{H}} %
\newcommand{\I}{\mathbb{I}} %

\DeclareMathOperator{\dual}{dual}

\newcommand{\abs}[2][\Biggerthanbig]%
{\ifthenelse{\isundefined{#1}}{\left| #2 \right|}{#1| #2 #1|}}
\newcommand{\norm}[2][\Biggerthanbig]%
{\ifthenelse{\isundefined{#1}}{\left\| #2 \right\|}{#1\| #2 #1\|}}

\usepackage{etoolbox}
\usepackage{refcount}

\usepackage{mathrsfs}
\DeclareFontFamily{U}{rsfs}{\skewchar\font127}
\DeclareFontShape{U}{rsfs}{m}{n}{%
  <-6> rsfs5
  <6-8> rsfs7
  <8-> rsfs10
}{}

\makeatletter

\patchcmd{\NAT@test}{\else \NAT@nm}{\else \NAT@nmfmt{\NAT@nm}}{}{}

\DeclareRobustCommand\poscite
  {\begingroup
   \let\NAT@nmfmt\NAT@posfmt%
   \NAT@swafalse\let\NAT@ctype\z@\NAT@partrue
   \@ifstar{\NAT@fulltrue\NAT@citetp}{\NAT@fullfalse\NAT@citetp}}

\let\NAT@orig@nmfmt\NAT@nmfmt
\def\NAT@posfmt#1{\NAT@orig@nmfmt{#1's}}

\newcommand{\customtextcite}[2]{%
  \hyper@@link[cite]{}{cite.#1}{#2} (\citeyear{#1})}

\newcommand{\horcite}[2]{
  (\hyper@@link[cite]{}{cite.#1}{Horodecki$^{\otimes #2}$}, \citeyear{#1}) }

\newcommand{\hortextcite}[2]{
  \hyper@@link[cite]{}{cite.#1}{Horodecki$^{\otimes #2}$} (\citeyear{#1}) }

\newcommand{\fawzitextcite}[1]{
  \hyper@@link[cite]{}{cite.#1}{Fawzi$^{\otimes 2}$} (\citeyear{#1}) }

\makeatother

\newtheorem*{example}{Example}

\theoremstyle{definition}
\newtheorem*{definition}{Definition}
\newtheorem*{property}{Property}

\newcounter{appcounter}

\newcommand{\apprefcont}[2]{\hyperref[#1]{\theappcounter #2}}

\usepackage{accents}

\usepackage{newtxtext,newtxmath}

\let\ve\eps

\newcommand{\EE}{\mathbb{E}}
\newcommand{\Prob}{\mathbb{P}}
\let\Tr\tr

\renewcommand{\widebar}[1]{\overline{\mkern-1mu#1\mkern-1mu}}

\renewcommand{\d}{\mathrm{d}}
\newcommand{\cX}{\mathcal{X}}

\newcommand{\ketbra}[2]{|#1\rangle\!\langle#2|}

\DeclareMathOperator{\rank}{rank}

\let\wt\widetilde

\newcommand{\infspec}{\underline{D}}
\newcommand{\supspec}{\overline{D}}

\usepackage[most]{tcolorbox}
\usepackage{tabularx}

\tcbset{
  enhanced,
  unbreakable,
  arc=0mm,
  left=1.5mm,
  right=1.5mm,
  boxrule=0.3mm,
}

\usepackage{tikz}
\usetikzlibrary{quantikz2}
\usepackage{graphics}

\usepackage{xparse}

\newcommand{\binent}{h}

\usepackage{epsdice}

\newcommand{\smooth}[2]{{#1,\mspace{2mu}#2}}

\newcommand{\upuparrow}{\uparrow\uparrow}

\newcommand{\ball}{\mathcal{B}}

\newcommand{\dom}{\mathrm{dom}}
\newcommand{\G}{\mathrm{\scriptscriptstyle G}}
\newcommand{\dd}{\mathrm{d}}
\DeclareMathOperator{\ran}{ran}

\usepackage{dsfont}
\newcommand{\one}{\mathds{1}}

\DeclareMathOperator*{\EV}{\mathbb{E}}

\newcounter{application}
\renewcommand{\theapplication}{\arabic{application}}

\makeatletter
\newcommand{\applabel}[2]{%
  \@ifundefined{phantomsection}{}{\phantomsection}%
  \protected@edef\@currentlabel{#1}%
  \label{#2}%
}
\makeatother

\ExplSyntaxOn
\prop_new:N \g_app_total_prev_prop
\prop_new:N \g_app_total_now_prop
\prop_new:N \g_app_base_prop
\prop_new:N \g_app_part_prop
\tl_new:N \l_app_number_tl

\NewDocumentCommand{\appboxtotal}{mm}
  { \prop_gput:Nnn \g_app_total_prev_prop {#1} {#2} }

\cs_new_protected:Npn \appbox_number:nn #1 #2
  {
    \prop_get:NnNTF \g_app_total_now_prop {#1} \l_tmpa_tl
      { \int_set:Nn \l_tmpa_int {\l_tmpa_tl + 1} }
      { \int_set:Nn \l_tmpa_int {1} }
    \prop_gput:Nnx \g_app_total_now_prop {#1} {\int_use:N \l_tmpa_int}

    \prop_if_in:NnF \g_app_base_prop {#1}
      {
        \refstepcounter{application}
        \prop_gput:Nnx \g_app_base_prop {#1} {\theapplication}
        \prop_gput:Nnn \g_app_part_prop {#1} {0}
      }

    \prop_get:NnN \g_app_part_prop {#1} \l_tmpa_tl
    \int_set:Nn \l_tmpa_int {\l_tmpa_tl + 1}
    \prop_gput:Nnx \g_app_part_prop {#1} {\int_use:N \l_tmpa_int}

    \prop_get:NnN \g_app_base_prop {#1} \l_tmpa_tl
    \prop_get:NnNTF \g_app_total_prev_prop {#1} \l_tmpb_tl
      {
        \int_compare:nNnTF {\l_tmpb_tl} > {1}
          { \tl_set:Nx \l_app_number_tl {\l_tmpa_tl(\int_to_alph:n{\l_tmpa_int})} }
          { \tl_set:Nx \l_app_number_tl {\l_tmpa_tl} }
      }
      { \tl_set:Nx \l_app_number_tl {\l_tmpa_tl} }

    \applabel{\tl_use:N \l_app_number_tl}{#2}
  }

\NewDocumentCommand{\appboxnumber}{mm}
  { \appbox_number:nn {#1}{#2} }

\NewDocumentCommand{\currentappnumber}{}
  { \tl_use:N \l_app_number_tl }

\cs_new_protected:Npn \appbox_write_totals:
  {
    \prop_map_inline:Nn \g_app_total_now_prop
      { \iow_now:cx { @auxout } { \string\appboxtotal{##1}{##2} } }
  }

\AtEndDocument{\appbox_write_totals:}
\ExplSyntaxOff

\NewDocumentEnvironment{appbox}{m m m}
{
  \begin{figure}[!htbp]%
  \normalsize
  \appboxnumber{#1}{#3}%
  \begin{tcolorbox}[
    before upper={\justifying\setlength{\parindent}{0pt}},
    halign title=flush left,
    grow to left by=1mm,
    grow to right by=1mm,
    title={\vspace*{1mm}\textbf{Application~\currentappnumber:} #2\vspace*{1mm}}
  ]
}
{
  \end{tcolorbox}
  \end{figure}
}

\newenvironment{inplacefigure}{\begin{figure}[!htbp]\centering}{\end{figure}}

\makeatletter
\patchcmd{\do@startpage}
  {\check@deferlist@stuck\do@startpage}
  {}
  {}
  {\PackageError{review-layout}{REVTeX float scheduler patch failed}{}}
\makeatother

\usepackage{microtype}
\usepackage{ragged2e,xurl,needspace}
\apptocmd{\thebibliography}{\raggedright}{}{}
\DeclareFontFamily{LGR}{ntxtlf}{}
\DeclareFontShape{LGR}{ntxtlf}{m}{n}{<->ssub*cmr/m/n}{}
\DeclareFontShape{LGR}{ntxtlf}{m}{it}{<->ssub*cmr/m/it}{}
\DeclareFontShape{LGR}{ntxtlf}{b}{n}{<->ssub*cmr/bx/n}{}
\DeclareFontShape{LGR}{ntxtlf}{b}{it}{<->ssub*cmr/bx/it}{}
\makeatletter
\input{t1qhv.fd}
\makeatother
\DeclareFontShape{T1}{qhv}{m}{sl}{<->ssub*qhv/m/it}{}

\newcommand{\sneakycomma}{, }

\begin{document}
\raggedbottom

\title{Entropic measures of quantum information}

\author{Roberto Rubboli}
\affiliation{Department of Mathematical Sciences, University of Copenhagen\sneakycomma Universitetsparken 5\sneakycomma 2100 Copenhagen\sneakycomma Denmark}

\author{Bartosz Regula}
\affiliation{Mathematical Quantum Information RIKEN Hakubi Research Team, RIKEN Pioneering Research Institute (PRI) and RIKEN Center for Quantum Computing (RQC), Wako\sneakycomma Saitama 351-0198\sneakycomma Japan}

\author{Mario Berta}
\affiliation{Institute for Quantum Information, RWTH Aachen University\sneakycomma Germany}
\affiliation{Department of Computing, Imperial College London\sneakycomma United Kingdom}

\author{Ludovico Lami}
\affiliation{Scuola Normale Superiore, Piazza dei Cavalieri 7\sneakycomma 56126 Pisa\sneakycomma Italy}

\author{Marco~Tomamichel\vspace*{5pt}}
\email{marco.tomamichel@nus.edu.sg}
\affiliation{Department of Electrical and Computer Engineering, National University of Singapore\sneakycomma Singapore}
\affiliation{Centre for Quantum Technologies, National University of Singapore\sneakycomma Singapore}

\begin{abstract}
Entropy has played a pivotal role in physics ever since its inception by Clausius in the early industrial age. Shannon ushered the concept into the information age by using it to describe fundamental limits of information processing, founding the field of information theory. Today, a fruitful interaction between information theory and physics is pushing our understanding of entropy forward. It has revealed, in particular, that no single notion of entropy suffices to capture all relevant properties of physical systems, making it necessary to consider a variety of entropic measures of quantum information, each with distinct features and uses. This review surveys these recent advances in a structured way, covering a comprehensive array of information measures, including Rényi and smooth entropies, together with their interrelations. Starting with a systematic characterization of entropies that motivates their definitions from natural physical principles, we discuss their applications ranging from quantum communication and cryptography to resource theories, thermodynamics, and many-body physics.
\end{abstract}

\maketitle

\tableofcontents


\section{Introduction}
\label{sec:intro}

\subsection{A short history of entropy: from physics to information theory and back again}

Any account of the history of entropy \cite{MUELLER} must start with %
\poscite%
{CARNOT} realization that the process now known as a Carnot cycle has optimal efficiency, as no ``useless re-establishment of equilibrium in the caloric'' takes place. It is quite remarkable that Carnot could reach such a profound conclusion, clearly hinting at the notion of irreversibility in thermodynamics, by employing the erroneous caloric theory. After this was falsified, and the first law of thermodynamics clearly established, it was \textcite{Clausius1854, Clausius1856, Clausius1865} who arrived at a precise formulation of the notion of entropy, in a sequence of works culminating with the 1865 paper where the name is introduced. There, he proposes~\cite
{Clausius1865} to name the newly introduced quantity\footnote{From the surrounding discussion, where the similarity with the word ``energy'' and the interpretation of the concept as \emph{transformation content} (``Verwandlungsinhalt'') are remarked, it seems likely that Clausius's intended etymology involves also the Greek prefix \textgreek{>en}-, meaning ``in''.}
\begin{quote}
the \emph{entropy} of the body, after the Greek word\\ \textgreek{<h trop'h}, the transformation.
\end{quote}

With \poscite{Boltzmann1877, BOLTZMANN} works, the notion of entropy makes the first of its many interdisciplinary leaps, revealing its profound connections with statistical physics: studying the kinetic theory of gases, he posited that the entropy $S$ is proportional to the logarithm of the number $W$ of microstates associated with a given macrostate. The formula $S = k \log W$, which he however never quite wrote as such, is famously engraved on his tombstone in Vienna.

Yet another leap came with the advent of quantum mechanics. In the first mathematical systematization of the theory, \textcite%
{vonneumann1932} concerns himself with the determination of the entropy of a quantum ensemble of $N$ particles, each represented by a density operator $\rho$, and proves that it is given by $S = - Nk \tr \rho \log \rho$, where $k$ is the same constant appearing in Boltzmann's formula. %
Von Neumann is also credited, perhaps apocryphally, with ushering the %
word `entropy' from physics into information theory by briefing Shannon, who had just stumbled upon the same concept, on its meaning in physics~\cite{tribus1971information}.

In \poscite{shannon1948mathematical} information theory, entropy measures the uncertainty we have about the outcome of a random experiment. Let us describe such an experiment by a random variable, $X$, and denote its Shannon entropy by $H(X)$. The first thing to note is that this makes information-theoretic entropy a fundamentally subjective concept, in contrast to its objective interpretation as a property of a thermodynamic system. This conceptual difference is important to keep in mind --- in the information-theoretic perspective that we will embrace in this review, entropy is subjective to our knowledge. It can be reduced, for example, if we are privy to some side information, that is, a random variable $Y$, that is potentially correlated with $X$. This gives rise to the concept of conditional entropy, $H(X|Y) = H(XY) - H(Y)$, which measures the uncertainty of $X$ from the perspective of an observer with access to side information $Y$. Generally, we observe that $H(X|Y) < H(X)$ but they can be equal if $Y$ does not contain information about $X$.
Indeed, this gap leads to the definition of mutual information, $I(X\!:\!Y) = H(X) - H(X|Y)$, which measures by how much the entropy of $X$ reduces when we take $Y$ into account. Mutual information is thus an entropic measure of correlation between two random variables. %

In his seminal work, \textcite{shannon1948mathematical} established that entropy characterizes the fundamental limit of compression of memoryless sources that output strings of random and independent symbols $X$. Shannon showed that there exist schemes that compress such a source to $H(X)$ bits per symbol and faithfully recover the string from its compression. Moreover, he showed there cannot exist any reliable algorithm that compresses more efficiently than $H(X)$ bits per symbol. In this sense, $H(X)$ demarcates a phase transition for compression: schemes compressing at rates below $H(X)$ will always fail, whereas compression at rates above $H(X)$ is always possible. Beyond compression, he also established the channel capacity (the maximal mutual information between channel inputs and outputs) as a similar phase transition point for reliable communication over memoryless noisy channels. Because of their prominent appearance in many problems, we call Shannon's entropy, conditional entropy, and mutual information standard information measures.

It should not come as a surprise, however, that the breadth of properties and applications of uncertainty in physics cannot be reduced to just a  single number.
To motivate the many information measures that we will concern ourselves with, it is beneficial to understand Shannon's entropy in terms of an underlying random variable: the \emph{surprisal}.
Given a random variable $X$ that takes the value $x$ with probability $p_x$, its surprisal is another random variable that takes the value $S(x) = \log \frac{1}{p_x}$ with probability $p_x$. Thus, the less likely an event, the higher its surprisal, with the logarithmic scaling ensuring that the surprisal of observing independent events equals the sum of the individual surprisals.
The Shannon entropy $H(X) = \sum_x p_x \log \frac{1}{p_x}$ is then the expected surprisal, and emerges in many operational settings when the distribution of the surprisal concentrates around its mean, for example when analyzing sources or channels that are memoryless.
However, understanding the full distribution of the surprisal beyond its average is crucial in many applications, as it enables a more fine-grained quantification of uncertainty.

In cryptography, for example, we are often interested in the minimal surprisal, the min-entropy $H_{\min}(X) = \min_x S(x)$, as it characterizes how hard the value of $X$ is to guess with just a single attempt.  %
This value is closely connected to how much uniform randomness can be extracted from $X$ via the leftover hash lemma~\cite{impagliazzo1989leftover}.
To illustrate this, consider a biased die where the six faces appear with probabilities $\bigl(\frac{1}{2}, \frac{1}{8}, \frac{1}{8}, \frac{1}{8}, \frac{1}{16}, \frac{1}{16}\bigr)$. The min-entropy is $H_{\min}(X) = 1$ bit, determined entirely by the dominant face $\epsdice{1}$. Indeed, since the remaining five faces collectively occur with probability $\frac{1}{2}$, one can extract a uniformly random bit by grouping them against the dominant face.
The Shannon entropy is $H(X) = \frac{17}{8} \approx 2.1$ bits\,---\,more than twice as large\,---\,reflecting the genuine spread of probability across all six faces. Regardless of that spread, however, by simply guessing $\epsdice{1}$ an adversary is correct with probability $\frac{1}{2}$, showing that it is the min-entropy that plays the relevant role here.

The full distribution of the surprisal is characterized by its moments, which can be computed via the moment generating
function, conveniently parametrized by the Rényi entropies $H_{\alpha}(X)$, where $\alpha \geq 0$ is a real parameter~\cite{renyi1961divergence}. As $\alpha$ varies, $H_\alpha$ assigns a different weight to the coefficients of the surprisal distribution, and the resulting function
interpolates between the maximal entropy $H_0$ (the logarithm of the support size), the Shannon entropy $H = H_1$, and the min-entropy $H_{\min} = H_{\infty}$, each capturing a different aspect of the surprisal distribution. The notions of conditional entropy and mutual information can equally be generalized to this fine-grained characterization of correlation. %

R\'{e}nyi information measures arise naturally in a variety of operational settings. In source
coding, for example, \textcite{campbell1965renyi} showed that $H_\alpha$ characterizes the optimal exponentially-weighted codeword length: whereas Shannon entropy governs the
expected length of an optimal prefix code, R\'{e}nyi entropy governs the optimal code when long codewords are penalized exponentially.
In channel coding, \textcite{gallager1965reliability} showed that the reliability
function\,---\,the optimal exponential decay rate of the error probability at
rates below capacity\,---\,is governed by a quantity later identified by
\textcite{csiszar1995cutoff} as the R\'{e}nyi mutual information of
order $\alpha \in (0,1)$ introduced by \textcite{sibson1969radius}. The appearance of Rényi information measures in these settings is not surprising, given the moment generating function's role in governing tail bounds for sums of independent variables in statistics~\cite{cramer1938largedev}.


When moving beyond stationary ergodic sources or channels, one immediately loses many useful concentration properties that underlie the usefulness of entropies based on average asymptotic behavior. It then becomes necessary to study operational problems and the distribution of the surprisal in the `one-shot', unstructured setting, where one considers a single instance of a task rather than many independent repetitions. %
The first systematic approach of this form was given by the information spectrum method, introduced by \textcite{verdue1994infospectrum}
and comprehensively treated in \poscite{han02book} monograph, although the roots of this method in information theory can be traced back to the study of one-shot reliable communication by \textcite{feinstein1954oneshot}.
Conceptually, the technique separates information-theoretic arguments from statistical ones by relating unstructured problems to information measures that can be characterized using probabilistic techniques,  such as using concentration bounds to simplify the resulting expressions for memoryless cases.
The information spectrum method admits also an equivalent formulation in terms of families of one-shot entropic measures.
This is closely connected to the notion of smooth entropies\,---\,an approach based on R\'enyi information measures, formally introduced by~\textcite{renner_2004} in cryptographic contexts, building on the earlier frameworks for randomness extraction due to~\textcite{bennett_1995} and~\textcite{hastad_1999}. By providing a convenient framework for applications to less structured settings, smooth entropies found widespread use also in information theory.

The various notions of entropy that we discussed can be understood as emanating from a single unifying mathematical object, a \emph{divergence} or \emph{relative entropy} between two probability distributions $p$ and $q$, allowing for further extensions and convenient formalization.
The natural concept to study then is the generalization of the surprisal in the form of the \emph{log-likelihood ratio}, given by the random variable $L = \log \frac{p_x}{q_x}$ governed by $p_x$. %
The standard relative entropy of~\textcite{kullback1951divergence} emerges as its expectation, $D(p\|q) = \sum_x p_x \log \frac{p_x}{q_x}$, whereas the~\textcite{renyi1961divergence} divergences are in correspondence with its moment generating function.

All these %
entropic information measures are necessary for a complete understanding of information processing tasks. When the surprisal or log-likelihood ratio concentrates tightly around its expectation\,---\,as it does when one considers many independent repetitions of the same experiment, the so-called independent and identically distributed (i.i.d.)\ setting\,---\,the fundamental phase transition in the achievable rates of operational tasks is captured by standard measures like Shannon entropy or mutual information. When one is interested not only in where this transition occurs, but also in a refined description that includes aspects such as how fast the transition is approached, R\'{e}nyi entropies become the appropriate tool.
Finally, when the problem lacks sufficient structure to guarantee such concentration, one-shot entropies provide tight characterizations without any structural assumptions. Indeed, all such one-shot entropies converge to standard measures in the asymptotic i.i.d.\ setting.

This review is primarily concerned with quantum generalizations of all these concepts,
their interrelations, and their operational applications.
The most natural starting point is the aforementioned von Neumann entropy, $H(A) = -\tr[\rho_A \log \rho_A]$, which replaces the probability distribution $p_X$ with a density matrix $\rho_A$ of some quantum system~$A$~\cite{vonneumann1932}.
Just as Shannon entropy emerges as the expectation of the surprisal, the von Neumann entropy arises as the expected surprisal of the eigenvalues of $\rho$, and characterizes the fundamental limits of quantum data compression~\cite{schumacher1995coding}. While entropies can thus be lifted to the quantum setting quite naturally by evaluating them on the spectrum, the situation already becomes much more intriguing once we move to conditional entropies.

The conditional von Neumann entropy can be defined by direct analogy with the classical case, $H(A|B) = H(AB) - H(B)$, yet this formal similarity conceals a striking departure: unlike its classical counterpart, the quantum conditional entropy can be negative, a phenomenon with no classical interpretation in terms of residual uncertainty. This negativity, and the question of which structural inequalities survive the passage to the quantum setting, attracted considerable attention from mathematical physicists. The central result is strong subadditivity (SSA), first conjectured by \textcite{lanford1968} and proved by \textcite{lieb1973proof} via the deep concavity theorem now bearing \poscite{lieb73c} name. Despite the potential negativity of the conditional entropy, SSA recovers a crucial piece of classical intuition: conditioning on more information cannot increase entropy, $H(A|BC) \leq H(A|B)$, which admits the clean operational reading that access to an additional system $C$ can only improve, or leave unchanged, one's ability to predict $A$ given $B$. Strong subadditivity underpins a vast range of subsequent results, from channel capacity theorems to entanglement theory and quantum thermodynamics.

Going beyond conditional entropy, and in particular into the realm of more general divergences, leads to even more difficulty.
Historically, it was not evident how to properly generalize classical entropic measures and divergences to the quantum setting\,---\,the log-likelihood ratio does not have a natural quantum analogue, and non-commutativity offers several potential candidates, notable variants of which were studied by~\textcite{umegaki1962conditional}, \textcite{belavkin1982c}, and~\textcite{donald1986relent}.
The operational viewpoint offers one way out of this conundrum: if a classical entropic measure is the answer to a question about classical physics, information theory, or statistics, then surely its proper quantum generalization should be the answer to the corresponding quantum question.
A prominent role is then played by the question of distinguishability of quantum states by measurements, reflecting the natural role of divergences in quantifying the difference between two distributions.
This can be formalized in the setting of quantum hypothesis testing, generalizing a central problem of classical statistics. Indeed, just as the Chernoff--Stein lemma~\cite{chernoff_1956} gives an interpretation to the %
Kullback--Leibler divergence as the exponential rate of decay of error in binary hypothesis testing, the study of its quantum generalization led to the first operational justification that singles out one variant of relative entropy:~\textcite{hiai1991proper} established that it is \poscite{umegaki1962conditional} relative entropy that mirrors the role of Kullback--Leibler divergence in this task. The subsequent applications of this quantity in a number of other problems solidified its position as the quantum relative entropy\,---\,although other variants of quantum relative entropy also find uses. We will discuss these standard information measures, together with their main applications, in Section~\ref{sec:standard}.

The landscape becomes more complicated still when considering quantum generalizations beyond the standard information measures.  %
Even operationally, no single notion of a quantum R\'enyi divergence suffices to describe all relevant asymptotic properties of quantum states, and multiple quantum Rényi divergences meaningfully generalize the classical definition to the non-commutative setting: most prominently the Petz--Rényi divergence~\cite{petz1986quasi} as well as the sandwiched Rényi divergence~\cite{lennert13renyi,wilde13sandwich}, both of which govern  exponential error rates in quantum hypothesis testing and other quantum information processing tasks such as channel coding.
Quantum %
R\'enyi divergences and related information measures are covered in Section~\ref{sec:additive}.

The applications of one-shot information measures have also played important roles in the study of quantum information.
An extension of the information spectrum method to the quantum
setting~\cite{hayashi2003capacity} provided the first one-shot and
non-i.i.d.\ characterizations of quantum information-processing tasks without
any assumption on the structure of the underlying quantum states or channels.
An alternative approach to one-shot information theory is offered by
\poscite{renner2008security} framework of smooth entropies, which notably has found widespread applications in quantum cryptography through a quantum generalization of min-entropy and its role in key distribution.
The one-shot entropic approaches to quantum information are discussed in Section~\ref{sec:one-shot}.

Together, these entropic measures form the backbone of modern quantum information theory and will be the central objects of this review. Their study is an interdisciplinary effort using advanced methods from mathematics and information theory, motivated by operational problems in physics and quantum information processing. And, indeed, results and techniques developed in the theoretical study of quantum entropies have
not only shed light on the quantification of properties of quantum systems in information-theoretic contexts, but they have also
found their way back to physics, where they have enabled the understanding of problems such as the behavior of correlations in many-body quantum systems~\cite{wolf08} or the black hole information paradox~\cite{hayden07,Almheiri2020}.
The remarkable progress achieved in these areas over the last ten years, however, has left the literature fractured. The purpose of this review is to bring some order into the zoo of measures that are currently used in the literature, survey the most important quantities and results, and explain how they are used in practice.

\subsection{Philosophy and structure of this review}

In this review, we want to explore the full breadth of entropic measures of quantum information and show how they are used in different contexts. This departs from the approach of earlier reviews of quantum entropy by \textcite{wehrl1978review} and \textcite{vedral02review}, which focused on von Neumann entropy and closely related quantities\,---\,the standard measures.

Our approach is twofold: axiomatic as well as operational. In the first part of the review we take an axiomatic angle and categorize different types of entropic measures, quantifying the distinguishability of quantum states (relative entropies), uncertainty about quantum states (entropies), uncertainty from the perspective of an observer with side information (conditional entropies), and correlation between subsystems (mutual information).

Once we have decided, for example, that conditional entropies are supposed to measure uncertainty about a target system from the perspective of an observer with access to side information, we can then list desiderata for these measures to make such an interpretation meaningful. For example, if an observer were to lose access to part of the side information, or if the side information became corrupted by noise, then we expect that the uncertainty about the target system increases in general.
Hence, it is natural for us to postulate this property as a formal mathematical axiom for all conditional entropies. In Section~\ref{sec:class} we follow exactly this approach and introduce physically motivated properties for the different categories of entropic measures.

Having settled on some such axioms, we can then explore their implications. We can show, for example, that all relative entropies satisfying certain axioms have some more intricate properties that non-trivially follow from them (and in many cases turn out to be useful in applications). This approach helps us in the later parts of the review where we discuss specific instances of entropic measures. We now only have to verify that a quantity satisfies the axioms to welcome it into the respective category of measures and thereby inherit all their features.

We will see that from any relative entropy satisfying the axioms, we can define well-behaved entropies, conditional entropies, and mutual information measures. We also explore how classical entropic measures can be lifted to quantum measures systematically, and the difficulties of this approach. The main hurdle here is the non-commutativity of quantum states, which in many cases leads to a whole collection of mathematically meaningful (meaning that they satisfy our axioms) quantum generalizations of a single classical quantity.
On the one hand, we have chosen our axioms carefully so that they are indeed necessarily satisfied by entropic measures with operational meaning. On the other hand, not all measures satisfying the axioms have found applications.

In this spirit, we recall \poscite{shannon1948mathematical} own perspective, expressed after introducing an axiomatic characterization of his entropy:
\begin{quote}
This theorem, and the assumptions required for its proof, are in no way necessary for the present theory. It is given chiefly to lend a certain plausibility to some of our later definitions. The real justification of these definitions, however, will reside in their implications.
\end{quote}

In the second part of the review, starting with Section~\ref{sec:standard}, we thus take a more operational approach and discuss specific examples of entropic measures of quantum information that have proven useful in quantum information and physics. We ask: can we find a physically meaningful and mathematically precise question for which a specific information measure gives the answer? Some of the main applications of these measures are discussed within the text, directly where the quantities themselves are discussed. For quick reference, we summarize these applications in Table~\ref{tab:qinfo-apps}.

\begin{table}[tb]
    \centering
    \small
    \renewcommand{\arraystretch}{1.1}
    \begin{tabularx}{\columnwidth}{@{}>{\raggedright\arraybackslash}X@{\hspace{1em}}>{\raggedright\arraybackslash}p{0.32\columnwidth}@{}}
        \hline
        Topic & Box(es) \\
        \hline
        Entanglement measures and transformations
        & \ref{app:axiom-a}, \ref{app:axiom-b},  \ref{app:entanglement}, \ref{app:entanglement-testing}, \ref{app:max-entanglement}, \ref{app:singletfraction}, \ref{app:axiom-c} \\

        Entropic uncertainty relations
        & \ref{app:uncertainty-a}, \ref{app:uncertainty-b},
          \ref{app:uncertainty-c}, \ref{app:uncertainty-smooth} \\

        Thermodynamics
        & %
        \ref{app:thermo-b},
          \ref{app:free-energy}, \ref{app:Second laws} \\

        Holevo's bound
        & \ref{app:holevo} \\

        Source compression and state merging
        & \ref{app:data-compression},
        \ref{app:state-merging},
        \ref{app:state-merging-exp},
          \ref{app:one-shot-compression} \\

        Quantum state discrimination
        & \ref{app:HT},
          \ref{app:HT-d}, \ref{app:HT-b}, \ref{app:HT-e},
          \ref{app:hockey-stick-guessing},
          \ref{app:quantum-illumination}, \ref{app:HT-f} \\

        Contraction coefficients
        & \ref{app:relaxation},
          \ref{app:contraction-coefficient} \\

        Quantum channel capacities
        & \ref{app:CC}, \ref{app:one-shot-coding},
          \ref{app:gaussian-capacity} \\

        Thermal correlations
        & \ref{app:thermal-mutual}, \ref{app:SSSAI-a} \\

        Strong subadditivity
        & \ref{app:SSSAI-b} \\

        Tensor networks and replica methods
        & \ref{app:renyi-mps}, \ref{app:purity} \\

        Guessing probability
        & \ref{app:guessing-a} \\

        Quantum key distribution
        & \ref{app:entropy-accumulation},
          \ref{app:qkd} \\

        Randomness extraction
        & \ref{app:privacy} \\

        Quantum channel simulation
        & \ref{app:channel-simulation} \\

        \hline
    \end{tabularx}
    \caption{\textbf{Guide to applications.} Related discussions are grouped in a single row; letters link to the individual discussions.}
    \label{tab:qinfo-apps}
\end{table}

\subsection{Scope}

The landscape of concepts with the label ``entropy'' attached is vast, and we do not attempt an exhaustive treatment. Our focus is on notions of entropy that have found operational significance in quantum information processing.

For example, \textcite{tsallis2009introduction} entropies, which have been studied in the context of nonextensive statistical mechanics, will not be covered in detail; we simply note that they are in one-to-one correspondence with R\'enyi entropies in Section~\ref{sec:further_divergences}. Bregman divergences are important in optimisation theory but only briefly mentioned within this review, a fate they share with Csisz\'ar $f$-divergences as well as computational entropies.

Although we discuss communication through quantum channels, our entropic quantities are functions of states. Extensions to divergences and entropies of channels, including their role in adaptive channel discrimination, form a separate theory beyond the scope of this review~\cite{Gour21,wilde20-2}.

\subsection{Preliminaries and notation}
\label{sec:prelim}

We use standard notation and conventions in quantum information theory that we will quickly summarize here. For a more comprehensive modern introduction to the quantum formalism in the context of quantum information we refer to the textbooks by \textcite{hayashi2006book}, \textcite{holevo2012book}, \textcite{wilde2013book}, \textcite{gour2025book} or \textcite{tomamichel2016book}.

We denote the logarithm by $\log$ and its inverse by $\exp$. Unless otherwise stated, our results are independent of the base of the logarithm; however, in numerical examples we will take the binary logarithm, reflecting information content being measured in \emph{bits}, and if the natural logarithm is convenient we explicitly say that a quantity is measured in \emph{nats}.

\subsubsection{Hilbert space formalism}

We use letters $A, B, \ldots, E$ to denote \emph{quantum systems} with their respective \emph{Hilbert spaces}.
For simplicity of exposition, we will for now assume these spaces to be finite-dimensional, and we will address continuous-variable systems separately later.
We use the Dirac bra-ket notation for vectors. The letters $X, Y, Z$ are reserved for \emph{classical systems}, that is, systems with a designated orthonormal basis $\{ \ket{i} \}_{i=1}^d$, where $d$ is the \emph{dimension} --- this is called the \emph{computational basis}. For qubits ($d = 2$) we also use the conjugate basis, given by $\ket{\pm} = \frac{1}{\sqrt{2}} \big( \ket{0} \pm \ket{1} \big)$.

Our main mathematical objects of interest are \emph{linear operators} acting on Hilbert spaces. We denote them by letters $K, L, \ldots, W$. The \emph{adjoint} (Hermitian conjugate) of $L$ with regard to the inner product is denoted $L^\dag$, and
$\tr[\cdot]$ denotes the \emph{trace}. An operator $L$ is \emph{Hermitian} or self-adjoint if $L^\dag = L$. An operator $U$ is an \emph{isometry} if $U^\dag U = \id$, where $\id$ is the \emph{identity operator}, and \emph{unitary}
if further $UU^{\dag} = \id$. We say that $L$ and $K$ \emph{commute} if $[L, K] = LK - KL = 0$.

To denote \emph{positive (semi-definite) operators} we often simply write $R \geq 0$. This induces the \emph{Loewner partial order}
$R \geq S \iff R - S \geq 0$, and analogously $R > S \iff R - S > 0$ in terms of positive definite (strictly positive) operators.
We will distinguish \emph{projectors}, which are positive operators $\Pi$ satisfying $\Pi^2 = \Pi$. We denote by $\Pi_{R}$ the projector onto the \emph{support} of $R$, which is the subspace spanned by the eigenvectors corresponding to the non-zero eigenvalues of $R$.

\subsubsection{Quantum states and channels}

\emph{Quantum states} are positive operators with unit trace. We refer to them using lowercase Greek letters $\rho, \sigma, \tau$, etc. The fully mixed state is $\pi = \id/d$ and its \emph{purifications}, commonly denoted by $\phi$, are called maximally entangled. Pure states are projectors and can be written as $\psi = \proj{\psi}$ using the bra-ket notation. States that are not pure are called \emph{mixed}. \emph{Classical states} are diagonal in the computational basis,
and are in one-to-one correspondence with probability mass functions, typically denoted as $p$ or $q$.

\emph{Quantum channels} in the Schrödinger picture are completely positive trace-preserving (cptp) maps on linear operators and are denoted by calligraphic capital letters $\cE$, $\cF$ or $\cG$. The identity channel is denoted by $\idc$. Following \poscite{stinespring1954isometry} dilation theorem, channels can be decomposed into isometries followed by a partial trace. In particular,  for any isometry $U$, we denote by $\mathcal{U}$ the channel that maps $\rho \mapsto U \rho U^{\dag}$. In the Heisenberg picture we are interested in the adjoint $\cE^{\dag}$ of a quantum channel $\cE$, which is defined via the relation $\tr[\cE(\rho) A] = \tr[\rho \cE^{\dag}(A)]$. Adjoint channels are completely positive and unital (cpu) maps and for clarity we always decorate them with~$\dag$.

\emph{Unitary channels} have the form $\rho \mapsto U \rho U^{\dag}$ for some unitary $U$. A \emph{mixing channel} $\cE$ is a convex combination (or mixture) of unitary channels. For a Hermitian $L$, we denote by $\cP_L$ the \emph{pinching channel} associated with $L$, defined by its action  $\cP_L(K)
    = \sum_{\lambda}
    \Pi_{\lambda}\, K \, \Pi_{\lambda}$,
where $\Pi_{\lambda}$ is the orthogonal projector onto the eigenspace of $L$ corresponding to the eigenvalue $\lambda$ of $L$. These are examples of mixing channels.
A \emph{measurement} is a channel with a classical output and a preparation is a channel with a classical input. An \emph{instrument} is a collection of completely positive trace non-increasing maps $\{ \cG_x \}_x$ such that $\sum_x \cG_x$ is a quantum channel. Here we can think of $x$ as a classical output. A \emph{conditional mixing channel} is a bipartite channel $\cE = \sum_x \cF_x \otimes \cG_x$ where $\cF_x$ are mixing channels and $\{ \cG_x \}_x$ is an instrument.

\subsubsection{Composite systems}

The composition of two quantum systems $A$ and $B$ is described by the \emph{tensor product} $AB = A \otimes B$. When multiple subsystems are present, mathematical objects are equipped with a subscript indicating the space they belong to. This allows us to omit identities and use the shorthand $K_A \ket{\psi}_{AB} = (K_A \otimes \id_B) \ket{\psi}_{AB}$ where $K_A$ acts on $A$ and $\id_B$ is the identity operator on $B$. Similarly, $\cE_{A \to C}(\rho_{AB}) = (\cE_{A \to C} \otimes \idc_B)(\rho_{AB})$. For a state $\rho_{AB}$ on a joint system $AB$, we implicitly define its marginals $\rho_A$ and $\rho_B$, defined via the partial trace $\rho_A = \tr_B [\rho_{AB}]$ and $\rho_B = \tr_A[\rho_{AB}]$. Mixed states $\rho_A$ have a \emph{purification}, a pure state $\ket{\rho}_{AA'}$ where $A'$ is isometric to $A$ that has $\rho_A$ as its marginal. We can write such pure states in their Schmidt decomposition $\ket{\rho}_{AA'} = \sum_i \sqrt{\lambda_i} \ket{\theta_i}_A \otimes \ket{\theta_i}_{A'}$ where $\lambda_i$ and $\ket{\theta_i}$ are the eigenvalues and eigenvectors of $\rho_A$, respectively. In particular, this implies that the marginals $\rho_A$ and $\rho_{A'}$ have the same spectrum.

A psd operator $R_{AB}$ is called \emph{separable} if it can be decomposed into tensor products of positive local operators as $R_{AB} = \sum_x R_A^x \otimes R_B^x$ where $\{ R_A^x \}_x$ are psd operators on $A$ and $\{ R_B^x \}$ are psd operators on $B$. \emph{Classical-quantum states} $\rho_{XB} = \sum_{x} \rho_X(x)\, \proj{x} \otimes \rho_B^x$ for some probability mass function $\{ \rho_X(x) \}_x$ and states $\rho_B^x$ are examples of separable states. A quantum state is called \emph{entangled} if it is not separable.

\subsubsection{Standard norms and metrics}\label{subsec:standard_metrics}

We use two distance measures on quantum states in this review, trace distance and fidelity in the form of purified distance. The \emph{trace distance} is a metric and has a direct statistical interpretation~\cite{helstrom1967detection}. It is defined as $T(\sigma,\tau) = \frac12 \| \sigma - \tau\|_1$, where $\|L\|_p$ denotes the Schatten $p$-norm throughout. The \emph{fidelity} can be expressed as $F(\sigma, \tau) = \| \sqrt{\sigma}\sqrt{\tau}\|_1^2 = \max_{\ket{\sigma}, \ket{\tau}} |\braket{\sigma}{\tau}|^2$, where the equality is due to \poscite{uhlmann_1976} theorem and the maximization is over all purifications $\tau$ and $\sigma$, respectively. Uhlmann's theorem offers us closure in the sense that if two states are close in terms of fidelity we can always find extensions and even purifications of them that are equally close.

We have the \customtextcite{fuchsvandegraaf1999}{Fuchs--van de Graaf} inequalities relating fidelity and trace distance as
\begin{align}
    1 - \sqrt{F(\tau, \sigma)} \leq T(\tau, \sigma) \leq \sqrt{1-F(\tau, \sigma)},
\end{align}
where the second bound is an equality for pure states. This leads us to define the \emph{purified distance}~\cite{rastegin2006sinedistance,tomamichel2010duality} as
\begin{align}
    P(\tau, \sigma) = \min_{\ket{\tau}, \ket{\sigma}} T(\proj{\tau}, \proj{\sigma}) = \sqrt{1 - F(\tau, \sigma)} .
\end{align}
The purified distance inherits operational significance and metric properties from the trace distance\footnote{It can be interpreted as a minimal extension of the trace distance from pure to mixed states~\cite{gour2020extensions}.} and closure under extensions from the fidelity, making it particularly convenient.


\section{Classification of entropic information measures}
\label{sec:class}

\label{sec: classification}

The purpose of this section is to classify existing entropic information measures and discuss their universal properties, allowing us to later refer to them when we discuss specific instances of such measures.

The approach we will follow has a long history, dating back to~\poscite{shannon1948mathematical} and \poscite{renyi1961divergence} axiomatic introduction of entropy.\footnote{See also~\cite{Faddeev1956concept,khintchine1957mathematical,aczel1975measures,kullback1997information,jizba2004world,chakrabarti2005shannon,baez_2014} for similar systematic approaches to classical information measures.}
For quantum systems,
a parallel viewpoint was developed by \textcite{petz_1992} and \textcite{matsumoto2010relative}, while, in the context of thermodynamics, similar efforts have been directed towards the derivation of the free energy from first principles~\cite{GILES,Lieb-Yngvason,Wilming2017,boes_2019}.
We will discuss this further in Section~\ref{sec:standard}.

The idea is to start with mathematical properties of entropic measures that reflect desirable physical principles, rather than being mere mathematical assumptions as in previous approaches. As these properties are common to all entropic measures that characterize meaningful operational tasks, our presentation aims to be deeply rooted in physics.
We will also see that many more advanced properties of information measures already follow from simple information-theoretic axioms. The exposition here loosely follows~\textcite{gour2020extensions, gour2021axiomatic,gour2025book}.

\subsection{Entropy}
\label{subsec: Entropy}

Entropies are the simplest information measures under consideration here, quantifying the uncertainty of the state of a quantum system. As such, entropies are very suitable to quantify uncertainty relations (see Application~\ref{app:uncertainty-a}).
Throughout this review we use the letter `H' to refer to entropies, following the conventions in information theory, rather than the letter `S', which is more commonly used in other areas of physics.

A generic entropy is a functional on states of the form
\begin{align}
	\H: \bigcup_{A} \cS(A) \to \mathbb{R}, \quad \rho \mapsto \H(\rho) \,
\end{align}
where $A$ goes over all finite-dimensional quantum systems and $\cS(A)$ denotes quantum states on $A$.
For a state $\rho_A$ on $A$, we will sometimes abuse notation slightly and write $\H(A)_{\rho} = \H(\rho_A)$. This shorthand is common in the literature and allows us to concisely describe entropies of parts of a multi-partite system. We often also drop the subscript $\rho$ when the state is obvious from the context.

Entropies are measures of uncertainty or mixedness of a state and thus only depend on its spectrum, i.e., on the eigenvalues of the state. The first property we require is invariance under isometries (or embeddings) since they preserve the non-zero spectrum.

\begin{property}[Isometric invariance] We say that $\H$ satisfies \emph{isometric invariance} if for all states $\rho$ and isometries $U$, it holds that
\begin{align}
    \H(\rho) = \H( U \rho U^{\dag}) \,.
\end{align}
\end{property}

Moreover, entropies are expected to be monotone non-decreasing under channels that add randomness to the system. These are quantum channels that are convex combinations (mixtures) of unitary channels. Examples of such channels are dephasing (or pinching) channels $\rho \mapsto \sum_i P_i \rho P_i$ for some projective measurement $\{P_i\}_i$ that removes off-diagonal elements from $\rho$. We call them mixing channels. Mixing channels $\cE_{A \to A}$ are unital, i.e., they satisfy $\cE(\id_A) = \id_{A}$ and thus leave the fully mixed state invariant. However, not all unital channels are mixing channels in the above sense~\cite{landau_1993}.
\begin{example}
The prototypical mixing channel is the depolarizing channel, which acts as $\rho \mapsto (1-p)\rho + p \frac{\id}{d}$. It increases entropy as it replaces the input state $\rho$ with
the fully mixed state $\frac{\id}{d}$ with probability $p$. An example of a channel that is not mixing is the amplitude damping channel $\rho \mapsto M_p^{\vphantom{\dag}} \rho M_p^\dag + p \ketbra{0}{1} \rho \ketbra{1}{0}$ on a two-level system (qubit), with $M_p = \ketbra{0}{0} + \sqrt{1-p} \ketbra{1}{1}$. This models the coherent process by which excitations (state $\ket{1}$) are lost to the environment; it reduces entropy as it pushes the input state towards the ground state.
\end{example}

Physically, we can think of mixing channels as describing uncertainty about the exact time evolution of a system. Since entropy is a measure of uncertainty, we expect that this additional uncertainty will generally increase entropy.

\begin{property}[Data-processing] We say that $\H$ satisfies \emph{data-processing} if for all mixing channels $\cE$ and any state $\rho$, it holds that
\begin{align}
    \H(\rho) \leq \H(\cE(\rho)) \,.
\end{align}
\end{property}

These two requirements, invariance under isometries and data-processing, can be collected into data-processing for a larger class of channels, namely channels that are sub-unital, i.e., channels $\cE_{A \to A'}$ that satisfy $\cE(\mathds{1}_A) \leq \mathds{1}_{A'}$. In some literature data-processing for sub-unital channels is postulated, and indeed we will see that most entropies satisfy this stronger relation.
However, we focus here on mixing channels as those are physically well motivated.

\begin{appbox}{Axiomatic}{Data-processing in pure-state entanglement transformation}{app:axiom-a}
Consider the transformation of bipartite pure states, where two parties are allowed to perform local operations and communicate classically (LOCC). \poscite{nielsen1999} theorem states that a bipartite pure state $\ket{\psi}$ can be transformed into another bipartite pure state $\ket{\phi}$ by LOCC if and only if the marginals $\rho$ of $\psi$ and $\tau$ of $\phi$ satisfy $\rho \prec \tau$. This relation is invariant under local isometries---reflecting the fact that we can do invertible local operations for free. The relation is also transitive and, by \poscite{birkhoff1946tres} theorem, any pre-processing with an operation that acts as a mixing channel on the marginal $\rho$ will leave the relation intact.

\smallskip

Isometric invariance and the data-processing property thus single out monotones for such transformations: if an LOCC transformation is possible then we have $\H(\rho) \geq \H(\tau)$ for any quantity satisfying these two properties.
\end{appbox}

For probability vectors of equal length, $p\prec q$ means that the partial sums of the decreasingly ordered components of $p$ do not exceed the corresponding sums for $q$. The Hardy--Littlewood--P\'olya theorem states that this holds if and only if $p=Dq$ for some doubly stochastic matrix $D$~\cite{hardy1952inequalities}. Birkhoff's theorem, in turn, expresses every such matrix as a convex combination of permutation matrices~\cite{birkhoff1946tres}. Vectors of different lengths are compared by first padding the shorter one with zeros. For quantum states, $\rho\prec\sigma$ denotes majorization of their spectra. Working in their respective eigenbases, the permutations above lift to unitary transformations, whose mixture maps $\sigma$ to $\rho$. When the dimensions differ, zero-padding the spectra corresponds to isometrically embedding the states into a common space. Isometric invariance and data processing under mixing channels therefore imply
\begin{align}
    \H(\rho) \geq \H(\sigma)
    \qquad \textnormal{if} \qquad
    \rho \prec \sigma .
\end{align}
In particular, $\H$ attains its minimum on every pure state.

The final desirable property is additivity under tensor products. This distinguishes quantities that are extensive, namely they grow linearly with the number of subsystems if these subsystems are uncorrelated.

\begin{property}[Additivity] We say that $\H$ is \emph{additive} if for any states $\rho$ and $\sigma$, we have
\begin{align}
    \H(\rho \otimes \sigma) = \H(\rho) + \H(\sigma) \,.
\end{align}
\end{property}

In particular, additivity ensures that for independent subsystems $A_1A_2 \cdots A_n$ with identical state $\rho_A$, we have
\begin{align}
    \H(A_1A_2\cdots A_n) = n \H(A)
\end{align}
where $\rho_{A_1 A_2 \cdots A_n} = \rho_A^{\!\otimes n}$.
A related property is sub-additivity, which states that $\H(\rho_{AB}) \leq \H(\rho_A) + \H(\rho_B)$ for a general bipartite state $\rho_{AB}$. This property is however not shared universally and we discuss it in Section~\ref{sec:standard}.

\begin{appbox}{Axiomatic}{Additivity and rescaling in asymptotic pure-state entanglement transformation}{app:axiom-b}
    Additivity naturally arises in problems where one considers the limit of many copies, and the task becomes invariant under rescaling the number of copies.

    Going back to LOCC transformations of pure states, we now consider asymptotic transformations where $n$ copies of $\psi$ must be transformed to $n$ copies of $\phi$ with asymptotically (in $n$) vanishing error. Such an LOCC transformation is possible if $H(\rho)>H(\tau)$ and impossible if $H(\rho)<H(\tau)$, with $H$ the von Neumann entropy~\cite{bennett_1996-2,bennett2000exact}. When $H(\rho)=H(\tau)$, the entropy comparison alone does not guarantee vanishing error at these fixed copy numbers; this boundary case requires a finer analysis, as discussed in Application~\ref{app:axiom-c}.

    Grouping copies into blocks of any positive integer size $m$ replaces $\psi$ and $\phi$ by $\psi^{\otimes m}$ and $\phi^{\otimes m}$. By additivity, $H(\rho^{\otimes m})=mH(\rho)$ and $H(\tau^{\otimes m})=mH(\tau)$, so both strict entropy inequalities are preserved under this common rescaling. Thus, additivity makes the entropy comparison compatible with grouping copies into blocks.

\end{appbox}

We sometimes also require normalization, namely that $\H(\pi) = \log 2$ for the qubit fully mixed state $\pi$.

Entropies need to satisfy all of these properties, but in analogy to relative entropy we also introduce the concept of one-shot entropies that do not need to be additive.

\begin{definition}[Entropy]
    A functional $\H$ is called an \emph{entropy} if it is normalized, isometrically invariant, additive and satisfies data-processing. %
\end{definition}

It is noteworthy that entropies, since they only depend on the spectrum of the state, are in one-to-one correspondence with classical entropies of the probability vector with the state's eigenvalues. Since normalized classical entropies are well understood and can always be expressed as convex combinations of R\'enyi entropies~\cite{gour2021axiomatic,mu2021renyi}, the same holds for quantum entropies. One consequence of this, which will be further elaborated in Section~\ref{sec:additive/entropy}, is that
\begin{align}
    H_{\infty}(\rho) \leq \H(\rho) \leq H_{0}(\rho) \label{eq:entropy-minmax}
\end{align}
holds for any entropy $\H$ and any state $\rho$, where the indices $0$ and $\infty$ identify the two extreme elements of the spectrum of R\'enyi entropies:
\begin{align}
    H_{\infty}(\rho) &= -\log \lambda_{\max}(\rho), \label{eq:min_entropy} \\
    H_{0}(\rho) &= \log \tr [\Pi_\rho] = \log \rank(\rho)\, , \label{eq:max_entropy}
\end{align}
with $\lambda_{\max}(\cdot)$ denoting the maximal eigenvalue, and $\Pi_\rho$ the projector onto the support of $\rho$.
The smallest entropy $H_{\infty}~=~H_{\min}$ is called the min-entropy, while $H_0$ is often termed the max-entropy; we will adopt the former name but will refrain from using the latter, since it leads to ambiguities when defining conditional entropies of quantum states (Section~\ref{subsec:renyi_conditional}).
The two entropies characterize the peak and width of the spectrum of $\rho$, respectively. The von Neumann entropy,
\begin{align}
    H(\rho) = -\tr[\rho \log \rho],
\end{align}
the most widely used entropy, lies in between them and characterizes the average asymptotic behavior of entropies in a sense that we will make precise later.

\begin{appbox}{uncertainty}{Entropic uncertainty relations}{app:uncertainty-a}
Entropic uncertainty relations~\cite{coles2017review} quantify the incompatibility of measurements in terms of the entropy of the outcomes. Heisenberg's uncertainty principle prohibits preparation of a state that has deterministic outcomes for two non-commuting measurements, and indeed, the sum of the entropies of the measurement outcomes satisfies a lower bound.

Specifically, consider any qubit $\rho$ and measurements in the computational and diagonal bases. This results in binary random variables $Z$ and $X$, respectively, with
\begin{align*}
    &\Pr[Z = 0] = \bra{0}\rho\ket{0}, \ &\Pr[Z = 1] &= \bra{1}\rho\ket{1} \notag \\
    &\Pr[X = 0] = \bra{+}\rho\ket{+}, \ &\Pr[X = 1] &= \bra{-}\rho\ket{-} \notag\,.
\end{align*}
\citet{maassen1988uncertainty} showed that the classical entropies of these random variables always satisfy
\begin{align*}
    H(X) + H(Z) \geq \log{2}, \quad \textnormal{and} \notag\\
    H_{\infty}(X) + H_{0}(Z) \geq \log{2} \notag,
\end{align*}
independently of the state. Equality is achieved, for example, if $\rho = \proj{0}$, so that $H(Z) = 0$. In contrast, if the state is mixed, the former relation can be strengthened to
\begin{align*}
    H(X) + H(Z) \geq \log{2} + H(\rho)
\end{align*}
as shown by~\citet{berta2010uncertainty}. This relates the Shannon entropies of the measurement results to the von Neumann entropy of the state prior to measurement.
\end{appbox}

Additivity and isometric invariance have some immediate implications via~\eqref{eq:entropy-minmax}. For example, since $H_{\min}(\rho)
\geq 0$, with equality if and only if $\rho = \proj{\phi}$ is pure, we get that $\H(\rho) \geq 0$ for any entropy. Moreover, $\H(\proj{\phi}) = 0$ for pure states since taking a tensor product with $\proj{\phi}$ is equivalent to an isometric embedding. Thus, we have
\begin{align}
    \H(\rho) \geq 0
\end{align}
with equality if and only if $\rho$ is pure. Similarly, we find
\begin{align}
    \H(\rho) \leq \log d
\end{align}
from the definition of $H_{0}$. This maximum is achieved by fully mixed states for any $\H$ (due to normalization), but, for $H_{0}$ specifically, any full rank state will do.

\subsection{Divergence and relative entropy}
\label{sec: relative entropy}

Divergences and relative entropies are measures of distinguishability between pairs of quantum states, or positive operators more generally. We next focus on them because they can be seen as parent quantities for conditional entropies and mutual information.

\subsubsection{Some remarks about terminology}

Traditionally, in statistics and information geometry, the term ``divergence'' is used to denote (usually asymmetric) measures quantifying the distinctness of one distribution from another. As such, divergences are required to be positive and faithful, by which we mean that they vanish if and only if the two distributions are the same (see, e.g., \cite{bregman1967divergence} and \cite{amari2000infogeometry} for this use of the term). From an information-theoretic perspective, monotonicity under the application of channels takes a more central role, because it allows us to interpret divergences as measures of distinguishability. This is because we expect that, according to any such measure, processing of the states can never render them easier to distinguish. Thus, in this review, the term ``divergence'' is reserved for monotone measures and positivity and faithfulness are not required. Many classical divergences are in fact positive, faithful and monotone, for example~\textcite{renyi1961divergence} divergences or \textcite{csiszar1967divergence} $f$-divergences, and are thus divergences in either sense.

In classical information theory, the term ``relative entropy'' is mainly used as a synonym for~\customtextcite{kullback1951divergence}{Kullback--Leibler} divergence. In contrast and somewhat frustratingly, in quantum information the terms divergence and relative entropy are often used interchangeably. Here we advocate using ``relative entropy'' only for divergences that are extensive, i.e., additive under tensor products of states. Under this convention \customtextcite{kullback1951divergence}{Kullback--Leibler} and \textcite{renyi1961divergence} divergences as well as their additive quantum generalizations are relative entropies, while most \textcite{bregman1967divergence} and \textcite{csiszar1967divergence} $f$-divergences are not.
We will also encounter various notions of ``one-shot relative entropies'' that one can informally regard as divergences that regularize to a relative entropy.

We hope that this terminology clarifies the relationship between measures while at the same time not being significantly in tension with the use of the terms in the existing literature.

\subsubsection{Classification}
\label{subsec: classification}

To make these notions precise, let us first consider very general functionals $\D$ of the form
\begin{align}
    \D : \ \bigcup_{A} \, \bigl( \cP(A) \times \cP(A) \bigr) \to \bbR \cup \{+ \infty\}  \label{eq:D}
\end{align}
that map any pair of positive semidefinite operators on the same (finite-dimensional) Hilbert space $A$ to the extended real line. While we often define these functionals for general positive operators we mostly use them for pairs of states or at least when the first argument is a state.

\begin{property}[Faithfulness]
    We say that $\D$ is \emph{faithful} if $\D(\cdot\|\cdot)$ admits a minimum over all pairs of states $(\rho, \sigma)$ that is achieved iff $\rho = \sigma$.
\end{property}

We say that $\D$ is \emph{positive} if, for any state pair $(\rho, \sigma)$, $\D(\rho\|\sigma) \geq 0$ with equality if $\rho = \sigma$. We note that any faithful functional $\D$ is \emph{bounded from below}, i.e., there exists a constant $c \in \mathbb{R}$ such that $\D + c$ is positive. Positive and faithful functionals quantify the distinctness of a pair of states as a positive real number that vanishes if and only if the two states are equal (but might diverge to $+\infty$ for some pairs).
Trace and purified distance as well as any other metrics are examples of positive faithful functionals and many of the entropic information measures we consider share these two properties. On the other hand, some commonly encountered one-shot entropies are faithful and bounded from below but not positive. %

If we are interested in measures of intrinsic information-theoretic distinguishability, we need to require that pairs of states cannot become more distinguishable when they are exposed to noise or any other physical process. Indeed, this is a defining property of almost every relevant measure in quantum information theory.

\begin{inplacefigure}
    \centering
{\includegraphics[width=0.85\linewidth]{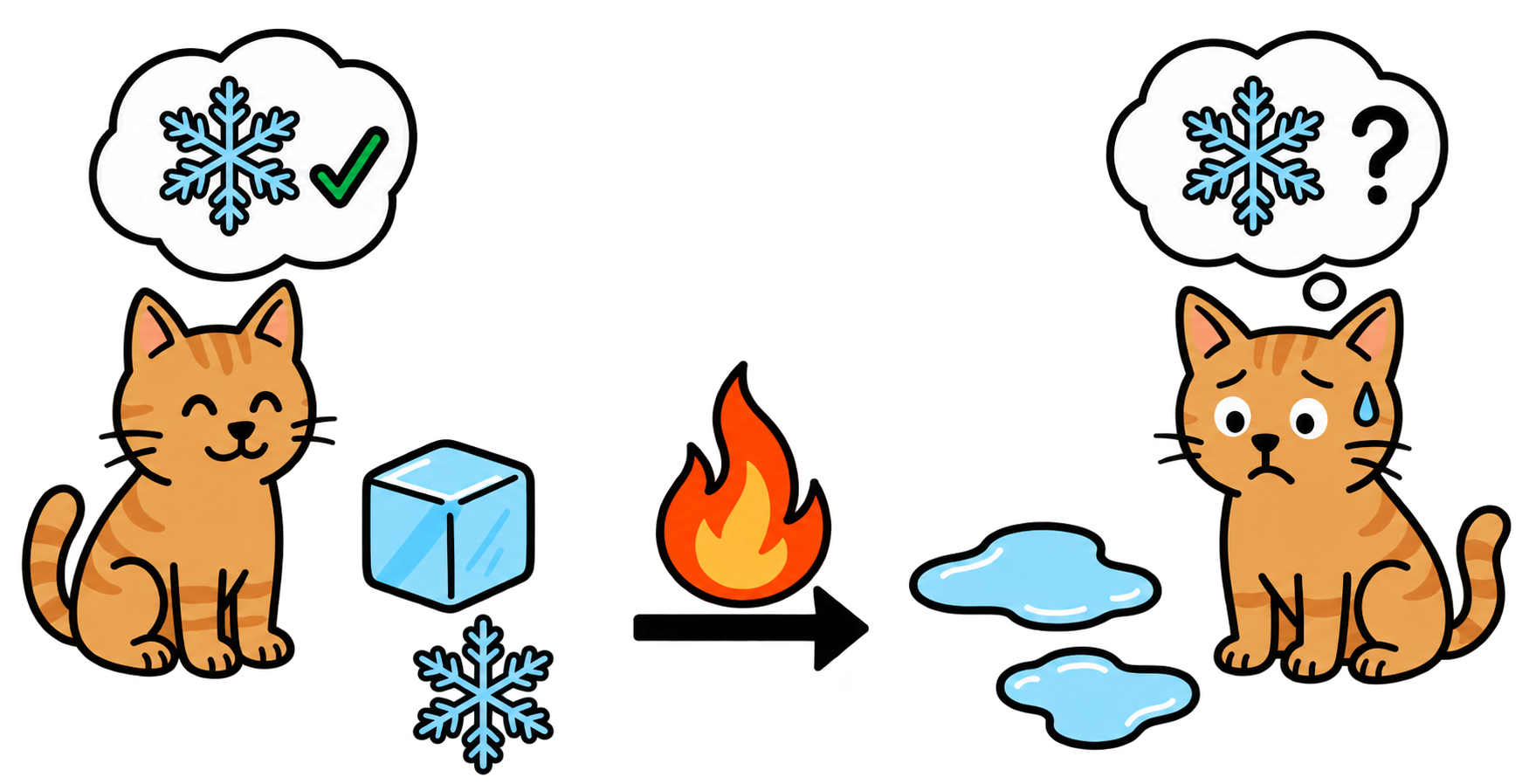}}
    \caption{\textbf{Monotonicity.} Any measure of distinguishability should satisfy a data-processing inequality, which ensures that a pair of quantum systems (represented here by ice and snow crystals) can only become harder to distinguish after we apply a channel (heat).}
    \label{fig:DPI}
\end{inplacefigure}

Formally, the quantity needs to be non-increasing when the same quantum channel is applied to both arguments.
\begin{property}[Monotonicity]
    We say that $\D$ is \emph{monotone} if it satisfies the \emph{data-processing inequality (DPI)}. Namely, for any pair of states $(\rho, \sigma)$ and any channel $\cE$, we have
    \begin{align}
        \D(\rho\|\sigma) \geq \D(\cE(\rho)\|\cE(\sigma)) \,.
    \end{align}
\end{property}
In particular, the DPI for measurement channels implies that distinguishability can never increase when we have only access to measurement outcomes instead of the original quantum states. Depending on the choice of measurement, this loss of distinguishability can be smaller or greater.

\begin{example}
Consider two pure qubit states $\rho = \proj{0}$ and $\sigma = \proj{1}$. These states are perfectly distinguishable, but a measurement in the conjugate basis $\{\ket{+},\ket{-}\}$, with $\ket{\pm} = \tfrac{1}{\sqrt2} (\ket{0} \pm \ket{1})$, leads to uniformly random outputs for both states\,---\,two indistinguishable distributions. %
\end{example}

An immediate consequence of monotonicity is invariance under isometries, namely,
\begin{align}
    \D(\rho\|\sigma) = \D(U \rho U^{\dag} \|U \sigma U^{\dag})
\end{align}
for isometries $U$. Conversely, due to \poscite{stinespring1954isometry} dilation theorem, invariance under isometries together with data-processing under the partial trace is sufficient to show monotonicity for general channels.

We may further ask for the data-processing inequality to be satisfied for any pair of positive operators. Many quantities of interest are monotone not only under channels but under more general classes of positive trace-preserving maps.

\begin{definition}[Divergence]
    A functional $\D$ is called a \emph{divergence} if it is monotone.
\end{definition}

In fact, for any triple $(\rho, \sigma, \tau)$, the relation $\D(\rho\|\sigma) \geq \D(1\|1) = \D(\tau\|\tau)$ follows from monotonicity under trace and a channel that prepares $\tau$. This ensures that divergences are bounded from below. %
The monotonicity property makes divergences suitable as resource monotones (see Application~\ref{app:thermo-b}).

Closely related to monotonicity are properties such as \emph{joint convexity}, which demands that
\begin{align}
   \D\left( \sum_x p(x) \rho_x \,\middle\|\, \sum_x p(x) \sigma_x \right) \leq \sum_{x} p(x)\, \D(\rho_x \| \sigma_x)
\label{eq:joint_convexity}
\end{align}
for all finite collections of states $(\rho_x)_x$ and $(\sigma_x)_x$ and all associated probability distributions $p(x)$, and \emph{joint quasi-convexity}, which only requires that
\begin{align}
   \D\left( \sum_x p(x) \rho_x \,\middle\|\, \sum_x p(x) \sigma_x \right) \leq \max_x \D(\rho_x \| \sigma_x)\, .
\label{eq:joint_quasiconvexity}
\end{align}
Joint concavity and quasi-concavity are defined analogously, with the inequalities in~\eqref{eq:joint_convexity} and~\eqref{eq:joint_quasiconvexity} reversed and $\min_x$ replacing $\max_x$ on the right-hand side of~\eqref{eq:joint_quasiconvexity}. Note that joint (quasi-)convexity of $\D(\cdot\|\cdot)$ entails (quasi-)convexity in each of the two arguments when the other is kept fixed.


Many quantities of interest in physics (e.g., internal and free energy, volume and statistical entropy) are extensive for weakly interacting systems close to thermal equilibrium. This means that these quantities typically scale linearly in the system size. Specifically, for statistical entropy, we see that the number of microstates simply multiplies for two non-interacting systems, and the equilibrium Gibbs state factorizes into a tensor product. Mathematically, extensivity thus requires that measures %
be additive for such tensor product states. Many relevant entropic measures reflect this property.

\begin{property}[Additivity]
	We say that $\D$ is \emph{additive} if, for any two pairs of states $(\rho, \sigma)$ and $(\tau, \omega)$, we have
	\begin{align}
	   \D(\rho \otimes \tau \| \sigma \otimes \omega) = \D(\rho\|\sigma) + \D(\tau\|\omega) \,. \label{eq:class-add}
	\end{align}
    Moreover, it is called \emph{weakly additive} if%
    \begin{align}
\D(\rho^{\otimes n} \| \sigma^{\otimes n}) = n\,  \D(\rho \| \sigma)
    \end{align}
    for all pairs of states $(\rho, \sigma)$ and $n \in \bbN$.
\end{property}

\begin{appbox}{thermo}{Constraints on thermal processes}{app:thermo-b}
Divergences can be used to give necessary conditions for thermodynamic processes to exist\,---\,microscopic second laws of thermodynamics for single quantum systems. Consider a system with Hamiltonian $X$ and Gibbs state $\gamma = \frac{1}{Z}\exp(-\beta X (\log e))$ with partition function $Z$. Thermal operations model all processes that can be realized without an external source of work and preserve the Gibbs state. Thus, if a state $\rho$ on this system can be transformed to $\sigma$ by a thermal operation $\cE$, we must have
\begin{align*}
    \D(\rho\|\gamma) \geq \D(\cE(\rho)\|\cE(\gamma)) = \D(\sigma\|\gamma) \notag
\end{align*}
for any divergence $\D$. In other words, the quantity $\D( \cdot \|\gamma)$ is a monotone for thermal operations. Hence, if monotonicity is violated for any divergence $\D$, then such a thermal operation cannot exist. Moreover, if $\D$ is faithful and positive then it measures athermality, i.e., how distinguishable $\rho$ is from the thermal equilibrium. See~\cite{janzing2000thermodynamic, brandao2013resource, horodecki2013fundamental}.

\smallskip
In the macroscopic regime of thermodynamic processes involving a large number of particles, conversion is often studied as being given access to $n$ non-interacting systems, each prepared in state $\rho$ and aim to convert this to $n$ copies of a state $\sigma$. That is, we want to convert $\rho^{\otimes n}$ to $\sigma^{\otimes n}$ for large enough $n$.
    From the monotones discussed earlier, the weakly additive ones survive. Namely, for any weakly additive divergence $\D$ and any thermal operation $\cE_n$ acting on $n$ systems, we have
    \begin{align*}
        \D(\rho\|\gamma) &= \frac{1}{n} \D\big(\rho^{\otimes n} \big\| \gamma^{\otimes n}\big) \geq \frac{1}{n} \D\big( \cE_n(\rho^{\otimes n}) \big\| \gamma^{\otimes n} \big) \notag\\
        &= \frac{1}{n} \D\big(\sigma^{\otimes n} \big\| \gamma^{\otimes n}\big) = \D(\sigma\|\gamma)\,, \notag
    \end{align*}
    where we used the fact that $\gamma^{\otimes n}$ is the Gibbs state of $n$ non-interacting systems with local Hamiltonian $H$.

    For catalytic conversion we instead allow access to a catalyst system that can have a Hamiltonian and state of our choice; however, our thermal operations must preserve this state exactly. An argument similar to the one above then shows that every additive divergence $\D$ yields a monotone $\D(\cdot\|\gamma)$. See, e.g., \cite{brandao_2015-2}.
\end{appbox}

It is often convenient to fix a normalization for divergences. For this, we require that for one (and thus all) qubit pure states $\phi$, it holds that $\D(\phi\|\pi) = \log 2$ for the qubit fully mixed state $\pi$. Note that this seemingly innocuous requirement breaks the symmetry between the two arguments of the relative entropy.

Throughout this review, we reserve the term \emph{relative entropy} for functionals that are both monotone and additive for states.
\begin{definition}[Relative entropy]
    A functional $\D$ is called a \emph{relative entropy} if it is a normalized additive divergence.
\end{definition}

Relative entropies have been fully characterized in the classical setting by~\textcite{mu2021renyi} and they correspond to~\textcite{renyi1961divergence} divergences. Quantum relative entropies are thus quantum generalizations of the same; however, there are many possible generalizations due to non-commutativity\,---\,they will be explored in Section~\ref{sec:additive}.

\begin{example}
Prominent examples of relative entropies include \poscite{umegaki1962conditional}, often simply referred to as relative entropy, as well as the min- and max-relative entropies~\cite{renner2008security,datta2009min}. They are defined (under some regularity conditions, cf.\ Sections~\ref{sec:standard} and~\ref{sec:additive} for a formal treatment) as
\begin{align}
    D(\rho\|\sigma) &= \tr [\rho (\log \rho - \log \sigma)] , \label{eq:intro-D} \\
    D_{\min}(\rho\|\sigma) &= -\log \tr[ \Pi_\rho \sigma ], \label{eq:intro-Dmin}   \\
    D_{\max}(\rho\|\sigma) &= \inf \{ \lambda \in \bbR : \rho \leq \exp(\lambda) \sigma \} \label{eq:intro-Dmax} \\
    &= \log \big\| \sigma^{-1/2} \rho \sigma^{-1/2} \big\|_\infty.
\end{align}
\end{example}

The min- and max-relative entropies are the smallest and largest relative entropies, respectively~\cite{gour2020extensions}. %
Specifically, for any quantum relative entropy $\D$ and any pair of states $\rho$ and $\sigma$,
\begin{align}
D_{\min}(\rho\|\sigma) \le \D(\rho\|\sigma) \le D_{\max}(\rho\|\sigma) ,
\end{align}
justifying their names.\footnote{We note that in the literature the label $D_{\min}$ is sometimes used for a different relative entropy related to the fidelity; this is however not the smallest relative entropy. We will return to this issue in Sec.~\ref{subsec:renyi_conditional}. }

So far we only discussed relative entropies for pairs of states. The additivity requirement offers a natural way to extend the definition to general positive operators; namely, we will simply set
\begin{align}
    \D( r \rho \| s \sigma) = \D(\rho\|\sigma) + \log \frac{r}{s}
\label{eq:rescaling}
\end{align}
for any pair of states $(\rho, \sigma)$ and any $r, s \geq 0$. We will often find this extension useful, e.g., when defining conditional entropies via relative entropies with a second argument that is not normalized. Note that if the first argument is allowed to be general, the closed-form expressions in~\eqref{eq:intro-D} and~\eqref{eq:intro-Dmin} need to be adjusted, i.e., $D(R\|S) = \frac{1}{\tr[R]} \tr [R(\log R - \log S)]$.

Relative entropies satisfy several additional properties that follow directly from their definition. We list these properties in Appendix~\ref{app:universal}.

\subsubsection{Additivity and regularization}\label{sec:class/reg}


Although additivity of relative entropies is a natural property to expect, some %
divergences may satisfy only relaxed notions of additivity.
A functional $\D$ is called \emph{sub-additive} if it satisfies $\D(\rho \otimes \tau \| \sigma \otimes \omega) \leq \D(\rho\|\sigma) + \D(\tau\|\omega)$ for all states, and \emph{super-additive} if the inequality is reversed. A \emph{weakly sub-additive} functional is then one that satisfies $\D(\rho^{\otimes n} \| \sigma^{\otimes n}) \leq n \D(\rho \| \sigma)$, and analogously for weakly super-additive ones.

A natural way to recover some form of additivity is through the procedure of \emph{regularization}, which can be understood as extracting the extensive component of a (possibly non-extensive) functional.
\begin{definition}[Regularization]
    For any $\D$ we define its \emph{regularization}, $\reg[\D]$, as the pointwise limit\footnote{The existence of the limit can be guaranteed for all weakly sub- or super-additive divergences by \poscite{fekete1923} lemma.}
\begin{align}
    \reg[\D](\rho\|\sigma) = \lim_{n \to \infty} \frac{1}{n} \D(\rho^{\otimes n} \| \sigma^{\otimes n}) \,,
\end{align}
    if this limit exists for all $(\rho,\sigma)$.
\end{definition}
 Weak additivity of the regularized quantity $\reg[\D]$ is now guaranteed by construction, %
and monotonicity as well as positivity are inherited if satisfied by the underlying~$\D$. In the literature, a superscript is commonly used to denote regularized quantities, e.g.\ $\Dreg = \reg[\D]$, which we avoid so as not to confuse the order in which various operations are applied to the given divergence.

\begin{example}
    Faithfulness is not always preserved. The trace distance is faithful and bounded. The latter property means that its regularization is always zero, and thus not faithful.
\end{example}

We will encounter regularization in two contexts.
First, quantities such as \emph{one-shot relative entropies}, which aim to measure distinguishability between unstructured states in non-asymptotic scenarios, are typically not additive by design. In such cases, we expect some form of a law of large numbers, i.e.\ asymptotic equipartition, to recover an average quantity such as the Umegaki relative entropy $D$ by evaluating one-shot relative entropies for many i.i.d.\ copies. In other words, for any one-shot relative entropy $\D$, we expect that $\reg[\D] = D$.
Representative examples of such quantities are smooth entropies, which we discuss in Section~\ref{sec:one-shot}.
Furthermore, regularization emerges also as a way to naturally restore additivity to quantities that, though constructed from axiomatic principles, may not satisfy it. We will now discuss an important case of this.

From a physical perspective, the above regularization procedure turns a one-shot, ``microscopic'' quantity into a (weakly) additive, ``macroscopic'' one via the many-copy limit. It may be thought of as the analogue, in quantum information theory, of the thermodynamic limit in statistical physics, which extracts the bulk, extensive component of a state function while washing away boundary effects.


\subsubsection{Lifting classical divergences}
\label{sec:class/lifting}

Taking classical divergences as a starting point, it is conceptually appealing to define quantum ones by systematically extending their classical counterparts.
Perhaps the most intuitive such approach, first formalized in \poscite{donald1986relent} study of the relative entropy, is to use measurements: since divergences aim to quantify the distinguishability of quantum states, and the only way to access the statistics of quantum systems in practice is by measuring them, it is natural to define quantum divergences in terms of the distributions of the post-measurement outcomes. One alternative to this approach, popularized by \textcite{matsumoto2010relative,matsumoto2018maxdivergence} is to look at quantum states from an opposite angle, in terms of a channel that prepares them from different classical distributions. Although these two extensions are not the only ways to extend an underlying classical divergence to quantum states, they play important roles due to forming the smallest and largest extensions.

We start with a classical divergence $\Dcl$ that is defined on pairs of probability distributions and is monotone for stochastic maps on probability distributions. We can then formally define the two extensions as follows.

\begin{definition}[Minimal and maximal extension]
    Let $\Dcl$ be a classical divergence. Its \emph{minimal (measured) extension} is defined as
    \begin{align}
        \Dmeas(\rho\|\sigma) = \sup_{\cM} \Dcl (\cM(\rho) \| \cM(\sigma) )\,,
    \end{align}
    where the maximum is taken over all measurement channels.
    We further define its \emph{maximal (preparation) extension} by
    \begin{align}
        \Dprep(\rho\|\sigma) = \inf_{p, q, \cF} \Dcl(p\|q)\, ,
    \end{align}
    where the minimum is taken over probability distributions $p, q$ and preparations $\cF$ so that $\cF(p) = \rho$ and $\cF(q) = \sigma$.
\end{definition}

Both $\Dmeas$ and $\Dprep$ are divergences and reduce to $\Dcl$ on classical states.
The names ``minimal'' and ``maximal'' are justified by the important property that all divergences obtained from $\Dcl$ are bounded by these two extremal extensions.
\begin{enumerate}[label=\Roman*.]
    \item Any divergence $\D$ that reduces to $\Dcl$ on classical states  satisfies $\Dmeas \leq \D \leq \Dprep$.
\end{enumerate}
This follows as a consequence of the monotonicity under measurement and preparation maps.
Functionals defined through $\Dmeas$ are often called measured divergences, encountered especially in the study of R\'enyi divergences~\cite{fuchs_1996,matsumoto_2014} and often used as a proof tool to reduce quantum problems to the classical domain.

Sometimes the two extensions coincide, which leads to a unique quantum extension of the classical divergence. For instance, we have that $D_{\max}^{\meas} = D_{\max}^{\prep} = D_{\max}$.
This is however not always the case: the most prominent relative entropy, the \customtextcite{kullback1951divergence}{Kullback--Leibler} divergence, does not have a unique quantum extension (see Sections~\ref{sec:standard} and~\ref{sec:additive}).

\begin{example}
The commonly used distance metrics in quantum information are indeed the minimal extensions of their classical counterparts. The trace distance naturally extends by measurements the total variation distance between probability distributions~\cite{helstrom1967detection,holevo_1973}. Analogously, the purified distance, $\sqrt{1-F(\rho,\sigma)}$, is the minimal extension of its classical variant, where the fidelity $F$ is replaced by the squared Bhattacharyya coefficient~\cite{fuchs_1995}.
\end{example}

In addition to the data processing inequality, the two extensions inherit some desirable properties from their underlying  divergences: for example, $\Dmeas$ and $\Dprep$ are faithful iff $\Dcl$ is faithful on classical distributions. However, some properties are not guaranteed. Importantly, $\Dmeas$ and $\Dprep$ need not be additive even if $\Dcl$ is, which means that they are not guaranteed to be bona fide relative entropies.
While this could hint at a potential drawback of the quantum extensions, this motivates precisely a systematic way to construct (weakly) additive quantum divergences:
the procedure of regularization. %
\begin{enumerate}[resume,label=\Roman*.]
\item If $\Dcl$ is additive for classical distributions, then $\Dmeas$ is super-additive and $\Dprep$ is sub-additive, while $\reg[\Dmeas]$ and $\reg[\Dprep]$ are weakly additive divergences.
\item Any weakly additive divergence $\D$ that reduces to $\Dcl$ on classical distributions satisfies $\reg[\Dmeas] \leq \D \leq \reg[\Dprep]$.
\end{enumerate}

The combined procedure of lifting classical relative entropies to quantum ones and then regularizing them thus gives a natural way to obtain quantum divergences. In practice, quantities obtained from regularization are typically fully additive and thus are relative entropies; this is, however, not guaranteed.

\begin{example}
The measured relative entropy $D^{\meas}(\rho\|\sigma)$, defined through the Kullback--Leibler divergence $D(p\|q)$, is itself not additive. However, its regularization is exactly the Umegaki relative entropy $D(\rho\|\sigma)$, a fact famously shown by \textcite{hiai1991proper} and used to motivate the latter quantity as the most suitable definition of a quantum relative entropy.
\end{example}


\subsection{Conditional entropy and mutual information}
\label{subsec: CE and MI}

The purpose of this subsection is to discuss properties we expect conditional entropies and mutual information to satisfy, and discuss some general ways in which we can construct entropies and mutual information from relative entropies. Explicit examples of entropies and conditional entropies will be discussed in the later sections.

\subsubsection{Classification of conditional entropy}
\label{sec: classification of conditional entropy}

Conditional entropies are measures of uncertainty from the perspective of an observer with side information. They are defined on bipartite states, and are generically functionals of the form
\begin{align}
	\H: \bigcup_{A, B} \cS(AB) \to \mathbb{R} \,,
\end{align}
where $A$ and $B$ go over all finite-dimensional quantum systems. We call $A$ the target system and $B$ the side information.
Entropy is just a special case of conditional entropy with trivial side information, i.e., a one-dimensional system $B$. For a bipartite state, say $\rho_{AB}$, we use the notation $\H(A|B)_{\rho}$ to denote the conditional entropy and further drop the subscript $\rho$ if the state is unmistakable from context.

We are clearly still interested in monotonicity under mixing channels on $A$ as for unconditional entropy. However, for a conditional entropy to be meaningful, we further require that any processing of the side information system $B$ cannot decrease uncertainty about the system. Indeed, any such local operation cannot increase correlations with $A$, by the non-signalling principle, and this should be reflected in conditional entropy measures.

Here, we consider an even larger class of operations. Consider an observer that, depending on the result of a measurement on the side information $B$, can control the time evolution of the target system $A$. This ability does not reveal any new knowledge about the target system to the observer, and thus we expect that conditional entropy cannot decrease even in this case. Such channels are collected as conditional mixing channels, bipartite channels $\cE_{AB \to AB'} = \sum_x \cF_{A\to A}^x \otimes \cG_{B\to B'}^x$ where $\cF^x$ are mixing channels and $\{ \cG^x \}_x$ is an instrument.
The definition of conditional mixing implicitly allows side-information to leak into the target system. However, there can be no signalling in the other direction, as this could evidently reduce uncertainty.

\begin{property}[Data-processing and isometric invariance]
    We say that $\H$ satisfies \emph{data-processing} if, for all conditional mixing channels $\cE_{AB \to AB'}$ and any state $\rho_{AB}$,
    \begin{align}
        \H(A|B)_{\rho} \leq \H(A|B')_{\sigma}\,,
    \end{align}
    where $\sigma_{AB'} = \cE_{AB \to AB'}(\rho_{AB})$.
    Moreover, it satisfies \emph{isometric invariance} if, for any isometries $U$ from $A$ to $A'$ and $V$ from $B$ to $B'$, we have
    \begin{align}
        \H(A|B)_{\rho} = \H(A'|B')_{\sigma}\,,
    \end{align}
    where $\sigma_{A'B'} = (U \otimes V) \rho_{AB} (U^{\dag} \otimes V^{\dag} )$.
\end{property}

Note that data-processing already implies isometric invariance of the side information, but not of the first system, because isometries are not mixing channels.

Conditional mixing channels have the property that, for every $\sigma_B$, there exists a $\tau_{B'}$ such that
\begin{align}
    \cE_{AB \to AB'}( \id_A \otimes \sigma_B) = \id_A \otimes \tau_{B'} \,. \label{eq:cond-dpi-enhanced}
\end{align}
Data-processing under larger classes of channels, namely all those satisfying~\eqref{eq:cond-dpi-enhanced}, is investigated by \textcite{vempati2022conditional} and, with some additional non-signalling constraints, by \textcite{gour2024conditional}. We restrict our attention to our notion of conditional mixing channels since it is more easily motivated physically.

The data-processing inequality we require has strong implications as conditional mixing channels allow for information transmission from $B$ to $A$. Consider, for example, classical-quantum states $\rho_{XX'AB}$ where $X$ and $X'$ are two copies of the same classical information. Such states can be written in the form
\begin{align} \label{eq:axiomatic-cqcqc}
    \rho_{XX'AB} = \sum_{x} p_x \proj{x}_X \otimes \proj{x}_{X'} \otimes \rho_{AB}^x \,,
\end{align}
where $p_x$ is a probability distribution and $\{ \rho_{AB}^x \}_x$ are bipartite quantum states.
For any such state, data-processing and isometric invariance imply the relations\footnote{The equality is seen as follows. First, isometric invariance allows for a tensor product with $\proj{0}$ on the target system. The state $\ket{0}$ can then be rotated to $\ket{x}$ with knowledge of $x$ using a conditional unitary channel, an operation that can also be reversed. The inequality follows from data-processing on the side information.}
\begin{align}
    \H(A|BX') = \H(AX|BX') \leq \H(AX|B) \,,
\end{align}
which states that moving classical information from side information to the target system can at most increase the measure. A stronger property is often desirable, namely that classical registers always increase entropy.
\begin{property}
    We say that $\H$ satisfies \emph{monotonicity under coarse-graining} if, for any states $\rho_{XAB}$ where $X$ is classical and any function $f: \mathcal{X} \to \mathcal{Z}$, we have
    \begin{align}
        \H(AX|B) \geq \H(A f(X)|B) \,.
    \end{align}
\end{property}
The special case of dropping the classical register altogether occurs when $f$ is constant. We note that if $f$ is injective the above holds with equality as a consequence of isometric invariance and data-processing, but generally this natural property is not implied by these properties. Indeed, now using the full power of coarse-graining and conditional mixing channels, we find that, for any function $g: \mathcal{X} \times \mathcal{Y} \to \mathcal{Z}$, we have
\begin{equation}
\begin{aligned}
    \H(XA|YB) &= \H(g(X,Y)XA|YB) \\
    &\geq \H(g(X,Y)A|YB)
\end{aligned}
\end{equation}
for any state $\rho_{XYAB}$ that is classical on $X$ and $Y$. This makes explicit that conditional mixing channels may communicate information from the side information system to the target system without increasing entropy.

The final desideratum we have is again additivity, which ensures that the measure is extensive for independent systems.

\begin{property}[Additivity] We say that $\H$ is additive if for any states $\rho_{AB}$ and $\tau_{A'B'}$, we have
\begin{align}
    \H(AA'|BB')_{\rho\otimes\tau} = \H(A|B)_{\rho} + \H(A'|B')_{\tau} \,.
\end{align}
\end{property}

Additionally, we often require that $\H$ is normalized, which is equivalent to demanding that it is normalized on the fully mixed qubit state $\pi$ with trivial side information.

Again, we reserve the term conditional entropy for quantities that satisfy all the above properties but do not require additivity for one-shot entropies.
\begin{definition}[Conditional entropy]
    A functional $\H$ is called a \emph{conditional entropy} if it is normalized, isometrically invariant, additive, and satisfies data-processing.
\end{definition}

Note that we do not require the coarse-graining inequality in general --- imposing it would restrict the class of conditional entropies~\cite{goodarzi2026sharp}, while we will in fact see operationally relevant entropies that do not satisfy it. Conditional entropies have been fully characterised in the classical setting by~\textcite{rubboli2026complete}, but are much less understood in the quantum case.

The most prominent conditional entropy is the conditional von Neumann entropy, $H(A|B)_{\rho} = \log d_A - D(\rho_{AB} \| \pi_A \otimes \rho_B)$. We can also define a minimal and a maximal conditional entropy:
\begin{align}
H^{\smash\downarrow}_{\min}(A|B)_{\rho} &= \log d_A - D_{\max}(\rho_{AB}\|\pi_A \otimes \rho_B) \,, \label{eq:conditional_min_entropy} \\
H^{\uparrow}_{\max}(A|B)_{\rho} &= \log d_A - \inf_{\sigma_B} D_{\min}(\rho_{AB}\|\pi_A \otimes \sigma_B) \,. \label{eq:conditional_max_entropy}
\end{align}
Here, $D$, $D_{\min}$ and $D_{\max}$ denote the Umegaki, min- and max-relative entropies introduced in~\eqref{eq:intro-D}--\eqref{eq:intro-Dmax}. The structure of these definitions and the arrow notation will be motivated shortly in Sec.~\ref{sec:class/canonical}.
The  inequalities
\begin{align}
\label{eq: bound Hmin and Hmax}
H^{\smash\downarrow}_{\min}(A|B) \leq \H(A|B) \leq H^{\uparrow}_{\max}(A|B)\,,
\end{align}
hold for any conditional entropy $\H$ that satisfies data-processing for all channels satisfying~\eqref{eq:cond-dpi-enhanced} and non-signalling constraints~\cite{gour2024conditional, ji2025fundamental}. More precisely, while the upper bound holds for any conditional entropy, the lower bound can only be shown using the additional structure connected to~\eqref{eq:cond-dpi-enhanced}; however, we conjecture it to hold for any conditional entropy.

The above result directly implies that~\cite{tomamichel2016book}
\begin{align}
-\log{\min\{d_A, d_B\}} \leq \H(A|B)_{\rho} \leq \log d_A \,. \label{eq:condentropy-dimensionbounds}
\end{align}
Due to isometric invariance, these dimensions may always be replaced by the ranks of the respective marginals of $\rho$.

\begin{example}
    For any entropy $\H$, the maximum is achieved by any tensor product state $\rho_{AB} = \pi_A \otimes \rho_B$ that is fully mixed on $A$. The minimum is achieved by maximally entangled states.
\end{example}

It is worth pausing here to note that the upper bound implies that any  conditional entropy necessarily becomes negative on maximally entangled states, as first noted by~\textcite{gour2024conditional}. This reflects a fundamental departure from the classical setting, where conditional entropies are always non-negative.

\begin{appbox}{Axiomatic}{Entanglement indicators}{app:entanglement}
 The lower bound in~\eqref{eq: bound Hmin and Hmax}, expressed in terms of the min-entropy, implies that any conditional entropy is non-negative on separable states. This is because, for any separable state, necessarily of the form $\rho_{AB} = \sum_{x} p_{x} \rho_A^{x} \otimes \rho_B^{x}$, we have $\rho_{AB} \leq \id_A \otimes \rho_B$ since $\rho_A^x \leq \id_A$ holds for all $x$ \horcite{Horodecki1999}{2}. Consequently, we can bound
 \begin{align*}
     H_{\min}^{\downarrow}(A|B) &= -\log \Big\| \big(\id_A \otimes \rho_B^{-1/2} \big) \rho_{AB} \big(\id_A \otimes \rho_B^{-1/2}\big) \Big\|_{\infty} \notag\\
     &\geq - \log \| \id_A \|_{\infty} = 0  \notag\,.
 \end{align*}
 As a result, all conditional entropies satisfying~\eqref{eq: bound Hmin and Hmax} are non-negative on separable states. They serve as entanglement indicators (or nonlinear entanglement witnesses):
 a negative value certifies that the underlying state is entangled.

\smallskip

 The converse is certainly not true. For the entangled state $\rho_{AA'B} = \pi_{A} \otimes \proj{\phi}_{A'B}$ with $\phi$ maximally entangled and $B$ isomorphic to $A'$, we find
 \begin{align*}
     \H(AA'|B) = \H(A) + \H(A'|B) = \log d_A - \log d_{A'}\,, \notag
 \end{align*}
 which can be either positive or negative depending on the ratio of the two dimensions.
\end{appbox}

Similar to divergences, \emph{one-shot conditional entropies} have the property that their regularization is the von Neumann entropy, i.e., they satisfy the relation
\begin{align}
   \reg[\H](\rho_{AB}) = \lim_{n \to \infty} \frac{1}{n}
\H(A^n|B^n)_{\rho^{\otimes n}} = H(A|B)_{\rho} \,.
\end{align}
One-shot conditional entropies have found many applications in quantum information theory and cryptography, and are the topic of Section~\ref{sec:one-shot}.

\subsubsection{Entropic duality}
\label{sec:class/duality}

As a consequence of the Schmidt decomposition, for any pure state on $AB$, we have $\H(A) = \H(B)$. This leads to one of the most intriguing and truly quantum properties of entropy: the entropic duality relation. As discussed in more detail in  Section~\ref{sec:standard/entropy_cond}, for a tripartite quantum system $ABC$ in a pure state the conditional von Neumann entropy satisfies
\begin{equation}
\begin{aligned}
    H(A|B) &= H(AB) - H(B) \\
    &= H(C) - H(AC) = -H(A|C) \,.
\end{aligned}
\end{equation}
This directly relates the uncertainty of $A$ for an observer with access to $B$ to the same uncertainty for an observer with access to $C$. This relation has deep implications, e.g., for quantum cryptography: namely, it entails that we can estimate the uncertainty of an adversary ($C$, in this example) by just investigating the marginal $\rho_{AB}$ shared between two honest parties ($A$ and $B$). This relation does not have any classical analogue\,---\,the purification offers a notion of closure that simply does not exist for classical information.

Inspired by this observation, for any entropy $\H$, we can define a dual entropy through the following construction:

\begin{definition}[Entropic dual]\label{def:entropic-dual}
    Assume that $\H$ satisfies isometric invariance. We define its dual
    \begin{align}
        \dual[\H] : \rho_{AB} \mapsto -\H(A|C)_{\rho} ,
    \end{align}
    where $\rho_{ABC}$ is any purification of $\rho_{AB}$.
\end{definition}

This is well defined (i.e., independent of the choice of purification) due to Uhlmann's theorem relating all purifications and the assumed isometric invariance of $\H$. Moreover, the dual inherits data-processing and additivity from $\H$. This implies, in particular, that if $\H$ is a (one-shot) entropy, then so is $\dual[\H]$. An alternative implicit definition, for the case when $\H$ satisfies data-processing, is given through the relation
\begin{align}
    \H(A|B)_{\rho} + \dual[\H](A|C)_{\rho} \geq 0\,,
\end{align}
with equality if $\rho_{ABC}$ is pure.

\begin{example}
\label{ex: duality min and max conditional entropy}
Important examples of duality relations include
\begin{align}
    \dual[H] = H \quad \textrm{and} \quad \dual\big[H_{\min}^{\downarrow}\big] = H_{\max}^{\uparrow}
\end{align}
Indeed, the von Neumann entropy is self-dual (see Section~\ref{sec:standard/entropy_cond}).
\end{example}

\subsubsection{Classification of mutual information}
\label{sec:class/mutual-info}

Mutual informations are measures of correlation between two systems. They are defined on bipartite states, and are of the form
\begin{align}
	\I: \bigcup_{A, B} \cS(AB) \to \mathbb{R} \,,
\end{align}
where $A$ and $B$ are arbitrary finite-dimensional quantum systems.
We use the notation $\I(A:B)_{\rho}$ to denote the mutual information of a bipartite state $\rho_{AB}$.

We require monotonicity under general local quantum channels on subsystems $A$ and $B$. This reflects the fact that local processing cannot increase correlations between subsystems, as it involves no interaction between them. %

\begin{property}[Data processing]
    We say that $\I$ satisfies \emph{data-processing} if, for all channels $\cF_{A\to A'}$ and $\cG_{B\to B'}$ and any state $\rho_{AB}$ and $\sigma_{A' B'} = \cF_{A\to A'}\otimes \cG_{B\to B'}(\rho_{AB})$, we have
    \begin{align}
        \I(A \!:\! B)_{\rho} \geq \I(A' \!:\! B')_{\sigma} \,.
    \end{align}
\end{property}

The second property we require is again additivity, which ensures that the measure is extensive for independent systems.

\begin{property}[Additivity] We say that $\I$ is additive if for any states $\rho_{AB}$ and $\tau_{A'B'}$, we have
\begin{align}
    \I(AA'\!:\!BB')_{\rho\otimes\tau} = \I(A\!:\!B)_{\rho} + \I(A'\!:\!B')_{\tau} \,.
\end{align}
\end{property}

One may further impose a normalization condition on $\I$ requiring that it takes the value $\log 2$ on a perfectly correlated classical two-bit state.
We use the term mutual information for quantities that satisfy all of the above properties.

\begin{definition}[Mutual information]
    A functional $\I$ is called a \emph{mutual information} if it is normalized, additive, and satisfies data-processing.
\end{definition}

The most common mutual information is based on the von Neumann entropy, $I(A\!:\!B) = D(\rho_{AB} \| \rho_A \otimes \rho_B)$, as discussed in Section~\ref{sec:standard}. This definition is symmetric under the exchange of $A$ and $B$; however, generally we do not require mutual information to be symmetric and we will consider such examples in Section~\ref{sec:additive}.

\begin{appbox}{holevo}{Holevo's bound}{app:holevo}
  A prominent use of the data-processing inequality for mutual information is the \textcite{holevo_1973-1} bound. Depending on a classical input $X$, a sender sends a quantum signal $\rho_B^x$ to a receiver. The resulting joint state, $\rho_{XB} = \sum_x p_x \proj{x}_X \otimes \rho_B^x$ describes this situation. The receiver now measures $B$ to obtain a classical outcome $Y$; by data-processing, we have
  \begin{align*}
      \I(X\!:\!Y) \leq \I(X\!:\!B) \notag\,,
  \end{align*}
  and specifically, for standard mutual information,
  \begin{align*}
      I(X\!:\!Y) \leq I(X\!:\!B) = H(\rho_B) - \sum_x p_x H(\rho_B^x) = \chi \notag
  \end{align*}

  A celebrated consequence of the Holevo bound is that a qubit can at most store a single bit of classical information. This is because $H(\rho_B) \leq \log{2}$ for qubits, and thus
  \begin{align*}
      I(X \!:\! Y) \leq \log{2} \notag
  \end{align*}
  for any $Y$ that can be attained by measuring $B$. This bound can be achieved if $X = Y$ are both bits.
\end{appbox}

Mutual information is much less well understood axiomatically compared to entropy and conditional entropy; nonetheless, one property we often ask for but that is not implied by the previous properties and is not universally shared by mutual information quantities is ``no information locking''. It states that a classical register $X$ can increase mutual information by at most $\log d_X$ in the chosen entropy units.

Here $X$ plays the role of a classical key. The term ``locking'' refers to the possibility that revealing a small key unlocks a much larger amount of correlation, as measured by the quantity in question. The bound below rules out such an amplification.

\begin{property}[No information locking]
    We say that a mutual information satisfies no information locking if
    \begin{align}
        \I(A:BX) \leq \I(A:B) + \log d_X
    \end{align}
    for any state $\rho_{ABX}$ that is classical on $X$.
\end{property}

\begin{definition}[One-shot mutual information]
A functional $\I$ is called a \emph{one-shot mutual information} if it satisfies data-processing and if $\reg[\I]$, defined by the pointwise limits
\begin{align}
   \reg[\I]: \rho_{AB} \mapsto \lim_{n \to \infty} \frac{1}{n}
\I(A^n:B^n)_{\rho^{\otimes n}}\, ,
\end{align}
exists and is a mutual information.
\end{definition}

\subsubsection{Canonical constructions from relative entropy}
\label{sec:class/canonical}

The useful chain rule $H(A|B) = H(AB) - H(B)$ obeyed by the von Neumann conditional entropy is sometimes taken as the definition of this quantity. This, however, may be misleading, as the construction $\H(AB) - \H(B)$ does not always lead to a valid notion of conditional entropy in the sense studied here\,---\,affecting, for instance, the commonly encountered R\'enyi conditional entropies~\cite{fehr14renyi}. This point will be discussed further in  Section~\ref{subsec:renyi_conditional}.

We thus need an alternative systematic approach to the construction of conditional entropies. A %
general idea is to employ divergences. In this perspective, a divergence quantifies a notion of distinguishability between states; we are thus implicitly adopting a geometric viewpoint: entropic quantities can be interpreted as a distance between a given state and a suitably chosen set of reference states, quantifying the deviation of the state from a set representing maximal uncertainty or lack of correlations.
This provides a direct way to construct entropic measures with the desired properties, which are inherited from the underlying divergence.
Generalizing~\poscite{renner2008security} early definitions of conditional min- and max-entropies, a general version of this geometric approach was put forward by~\textcite{datta2009min}, where divergences were used to define entanglement monotones.
We call these \emph{canonical} constructions and will use them often.

Let us first consider conditional entropies. Mimicking~\eqref{eq:conditional_min_entropy}--\eqref{eq:conditional_max_entropy}, for any divergence $\D$ we set
\begin{align}
	\H^{\downarrow}(A|B)_{\rho} &= \log d_A - \D(\rho_{AB} \| \pi_A \otimes \rho_B), \\
	\H^{\uparrow}(A|B)_{\rho} &= \log d_A - \min_{\sigma_B} \D(\rho_{AB} \| \pi_A \otimes \sigma_B) \,,
\end{align}
where the minimization is over states $\sigma_B$ on $B$. The arrows remind us of the trivial relation $\H^{\downarrow} \leq \H^{\uparrow}$.

For any relative entropy $\D$, the quantities $\H^{\downarrow}$ and $\H^{\uparrow}$ defined above
satisfy normalization, isometric invariance, and data-processing. These properties are inherited from $\D$ \cite{gour2024conditional}. The additivity property is, however, more subtle. While $\H^{\downarrow}$ can be shown to be additive for any relative entropy $\D$, it remains unclear whether $\H^{\uparrow}$ is additive in general, with current proofs for specific relative entropies in the literature relying on their particular structure.

\begin{example}
When $D$ is the Umegaki relative entropy, the two definitions coincide, recovering the von Neumann conditional entropy $H$
discussed in Section~\ref{sec:standard}. Generally the two constructions differ; examples are discussed in Section~\ref{sec:additive}.
\end{example}

In the case of mutual information, we are interested in how far a state is from being in a product form, as these correspond to uncorrelated states. For any divergence $\D$, we define
\begin{align}
\I^{\uparrow\uparrow}(A\!:\!B)_\rho
&= \D\!\left(\rho_{AB}\,\middle\|\, \rho_A \otimes \rho_B\right) \\
\I^{\downarrow\downarrow}(A\!:\!B)_\rho
&= \inf_{\sigma_A,\sigma_B} \D\!\left(\rho_{AB}\,\middle\|\, \sigma_A \otimes \sigma_B\right) ,
\end{align}
where the minimization is taken over states on $A$ and $B$, respectively.
Both these definitions are symmetric under exchange of $A$ and $B$. But this is not always necessary or desirable. Indeed, the following, asymmetric, construction has found many applications in both classical~\cite{arimoto1977information} and quantum~\cite{mosonyi2011quantum} communication theory:
\begin{align}
\I^{\uparrow\downarrow}(A\!:\!B)_\rho
&= \inf_{\sigma_B} \D\!\left(\rho_{AB}\,\middle\|\, \rho_A \otimes \sigma_B\right)
\end{align}
The arrows are meant as a reminder of which marginals are optimized over.

All these definitions satisfy normalization and the data-processing inequality, while additivity is more subtle and generally not known to hold. For the Umegaki relative entropy, these definitions all coincide, as discussed in Section~\ref{sec:standard}; for more general divergences, however, they differ, as examined in Section~\ref{sec:additive}.


\section{Standard information measures}
\label{sec:standard}

The purpose of this section is to introduce the most prevalent entropic measures, based on \textcite{vonneumann1932} entropy, which we will see as emanating from \poscite{umegaki1962conditional} quantum relative entropy. As we will see, these entropic quantities are distinguished by possessing a remarkably rich set of strengthened properties, not shared by any other family of information measures. This, in turn, makes these standard or von Neumann–type information measures omnipresent in applications across both information theory and physics. We refer to \poscite{Carlen25} recent textbook for an in-depth presentation of mathematical derivations.


\subsection{Entropy}
\label{sec:standard/entropy}

The standard measure of entropy of quantum states is the \textcite{vonneumann1932} entropy, defined for a state $\rho$ on a $d$-dimensional system as
\begin{align}
    H(\rho) = -\tr[\rho\log\rho]\,, %
    \label{eq:von_Neumann}
\end{align}
The logarithm is understood to be taken only on the support of $\rho$ in the above.

The von Neumann entropy can alternatively be understood as the \textcite{shannon1948mathematical} entropy of the eigenvalue distribution (or spectrum) of the quantum state.
Von Neumann's original definition was motivated in particular by the information that an observer can extract from a quantum state through measurements --- in the sense considered in Section~\ref{sec:class/lifting}, this can be understood as a maximal extension of the classical entropy. However, an important property of the entropy is that any quantum extension yields the same function, and the von Neumann entropy is the unique extension of the Shannon entropy to quantum states; this will be discussed further in Section~\ref{sec:additive/entropy}.

The entropy is isometrically invariant, satisfies data processing, is normalized with
\begin{align}
    0\leq H(\rho)\leq\log d\, ,  \label{eq:vN-entropy-bounds}
\end{align}
and it is additive under tensor products\,---\,thus satisfies all the properties we require of an entropy. Moreover, $H$ is concave in its argument, that is,
\begin{align}
    \sum_i p_i H(\rho_i) \leq H\left( \sum_i p_i \rho_i \right)
\end{align}
for any probability distribution $\{p_i\}_i$ and state ensemble~$\{ \rho_i \}_i$.

\begin{example}
    The upper bound in~\eqref{eq:vN-entropy-bounds} can be seen as a consequence of concavity by choosing a unitary design $p_i, \{U_i\}_i$ so that the ensemble given by $\rho_i = U_i^{\vphantom{\dag}} \rho U_i^{\dag}$ satisfies $\sum_i p_i \rho_i = \pi$. The upper bound is then a consequence of isometric invariance and concavity of the entropy.
\end{example}

For multipartite quantum states $\rho_{ABC}$, we write $H(A)_\rho =  H(\rho_A)$ for the von Neumann entropy of its marginals. Due to the Schmidt decomposition of pure states, we get that for any pure state $\ket{\psi}_{AB}$, the relation $H(A)_{\psi} = H(B)_{\psi}$ holds. Two fundamental properties of von Neumann entropy, which are not universally shared by other entropies, are sub-additivity and strong sub-additivity. Sub-additivity~\cite{araki70inequalities} refers to the fact that
\begin{align}
    H(AB)_{\rho} \leq H(A)_{\rho} + H(B)_{\rho} \label{eq:sub-additivity-1}
\end{align}
for any bipartite state $\rho_{AB}$ (with equality achieved by product states, due to additivity). This can be strengthened to strong sub-additivity (SSA), which states that~\cite{lieb1973proof}
\begin{align}
     H(ABC)_{\rho} + H(B)_{\rho} \leq H(AB)_{\rho} + H(BC)_{\rho} \,, \label{eq:strong-sub-additivity-1}
\end{align}
for any tripartite state $\rho_{ABC}$. The seminal proof of this relation borrows from \textcite{lieb73c} and settled an important conjecture by \textcite{robinson1967conjecture} and \textcite{lanford1968}.
These two fundamental properties can be expressed in various forms, most of which we will discuss later when we are concerned with conditional entropy and mutual information. An important equivalent formulation (via entropy duality) of sub-additivity is the \textcite{araki70inequalities} triangle inequality:
\begin{align}
\label{eq:arakilieb}
    H(AB)_{\rho} \geq \left| H(A)_{\rho} - H(B)_{\rho} \right| .
\end{align}
The inequality $H(XB)_{\rho} \geq H(B)_{\rho}$, which holds for classical registers $X$, does not extend to the fully quantum case. Indeed, for general states, \eqref{eq:arakilieb} is the strongest bound we can hope for.

In gapped two-dimensional ground states, suitable geometric combinations of region entropies cancel boundary contributions and isolate a universal topological contribution, providing a diagnostic of topological order~\cite{Kitavev06,LevinWen2006}.
\begin{example}
    Consider a pure state $\ket{\psi}_{AB}$. While $H(AB)_{\psi} = 0$, the marginal entropy $H(A)_{\psi}$ can reach $\log \min\{ d_A, d_B \}$, a quantum violation of the classical bound $H(XY) \geq H(X)$.
\end{example}

\begin{appbox}{compression}{Quantum source compression}{app:data-compression}
One fundamental operational interpretation of the von Neumann entropy is in terms of quantum source compression. In the classical setting, \poscite{shannon1948mathematical} source coding theorem establishes that an information source can be compressed with vanishing loss to an asymptotic rate given by its Shannon entropy. The quantum analogue considers a memoryless source emitting quantum states according to an ensemble $\{p_x,\proj{\psi_x}\}$, with corresponding average state $\rho=\sum_xp_x\proj{\psi_x}$. \textcite{schumacher1995coding} then shows that asymptotically faithful compression is possible at a rate equal to $H(\rho)$. As such, the von Neumann entropy quantifies the minimal number of qubits required to faithfully represent the typical outputs of a quantum information source.
\end{appbox}

The von Neumann entropy also has a variational form, given by~\cite{petz88}
\begin{align}
    H(\rho)=\inf_{X=X^\dagger}\Big\{\log\tr[\exp X]-\tr[\rho X]\Big\} .
\label{eq:variational_H}
\end{align}

\begin{example}
   The above variational form implies the \textcite{gibbs1902principles} variational principle for quantum entropy,
\begin{align}
   - H(\rho) + \log Z(\beta) + \beta \tr[\rho X] \geq 0
\end{align}
    that relates the energy with the partition function $Z(\beta) = \tr[\exp(-\beta X)]$ and the entropy.
\end{example}

Generally von Neumann entropy is expensive to compute for large systems since it requires a full diagonalization of the state, which takes time exponential in the number of qubits involved. This is particularly undesirable if the state is available in a compressed form, e.g., as a general matrix product state~\cite{fannes1992mps}. In this case proxies for von Neumann entropy like the R\'enyi-$2$ entropy from Section~\ref{sec:additive/entropy} become desirable~\cite{hastings2010renyi}. Alternatively, Eq.~\eqref{eq:variational_H} can be used to construct linear upper bounds.

Assume now that we have sample access to independent copies of a system of dimension $d$ in state $\rho$ and want to estimate the von Neumann entropy up to additive error $\varepsilon$. \textcite{acharya2020entropy}, \textcite{gao2026breaking}, and \textcite{wang2026unified} show that this can be done using
$\widetilde{\Theta}(d^2/\varepsilon+1/\varepsilon^2)$ samples, where the
tilde suppresses polylogarithmic factors in $d$ and
$1/\varepsilon$. Estimating von Neumann entropy is therefore expensive and requires nearly
quadratically many copies in the dimension, and hence
exponentially many in the number of qubits\,---\,similar to full tomography of the state.

The continuity of von Neumann entropy was studied by \textcite{fannes73}, \textcite{audenaert07}, and \textcite{petzbook08}, who credits Csiszár for the classical special case. They find
\begin{align}
|H(\rho)-H(\tilde\rho)| \leq
 \eps\log(d-1)+\binent(\eps)
\label{eq:Fannes_Audenaert}
\end{align}
for $\eps = \frac12 \norm{\rho - \tilde\rho}_1$ and the binary entropy $\binent(\eps)= -\eps\log\eps-(1\!-\!\eps)\log(1\!-\!\eps)$. Moreover, for every $\eps$ and $d$, there exists a pair of states that achieves this bound. If we take more information about the states into account, the bound can be tightened further, see, e.g.,~\cite{jabbour24continuity,berta25}.


\subsection{Relative entropy}
\label{sec:relative-entropy}

Unlike the entropy, a unique definition of a standard relative entropy for quantum states has historically been significantly more difficult to justify.
There is nevertheless a distinguished information measure that found prominent use, and one that we will see to typically emerge as the most appropriate extension of classical relative entropy to the quantum realm.
Its definition is due to \textcite{umegaki1962conditional}, and it is commonly simply termed the quantum relative entropy.
For a state $\rho$ relative to a positive operator~$\sigma$ it is defined as
\begin{align}
    D(\rho\|\sigma) =  \tr [\rho (\log \rho - \log \sigma)]
    \label{eq:Umegaki}
\end{align}
whenever the support of $\rho$ is contained in the support of $\sigma$, and set to $+\infty$ otherwise.\footnote{Here, the logarithm is to be taken only on the support of the respective operator. Alternatively, one may define $D$ for strictly positive operators first and then take limits to cover the rank-deficient case to reach the same definition.} For classical states, it simplifies to the \customtextcite{kullback1951divergence}{Kullback--Leibler} divergence. %

\begin{example}
    For two classical-quantum states of the form $\rho = \sum_x p_x \proj{x} \otimes \rho_x$ and $\sigma = \sum_x q_x \proj{x} \otimes \sigma_x$, we have
    \begin{align}
        D(\rho\|\sigma) = \sum_x p_x D(\rho_x\|\sigma_x) + D(p\|q) \,,
    \end{align}
    where $D(p\|q)$ is the Kullback--Leibler divergence between the probability distributions $\{p_x\}_x$ and $\{q_x\}_x$.
\end{example}

The von Neumann entropy can be recovered from the relative entropy via the relation
\begin{align}
H(\rho) = \log d - D(\rho\|\pi)\, ,
\label{eq:entropy_from_relative_entropy}
\end{align}
with $\pi = \id/d$ the fully mixed state of dimension $d$.

While the additivity of the quantum relative entropy simply follows from the tensor rule formula of the logarithm, its monotonicity is mathematically more involved and was first proven by \textcite{lindblad1974expectations,lindblad1975completely}. It states that
\begin{align}
    \label{eq:dpi-umegaki}
    D(\rho\|\sigma) \geq D(\cE(\rho)\|\cE(\sigma))\,.
\end{align}
for any state $\rho$, positive operators $\sigma$ and quantum channels $\mathcal{E}$. This is arguably the most fundamental inequality in quantum information theory, and equivalent to strong sub-additivity; see, e.g., \textcite{ruskai2002ssa} and~\textcite{nielsen2005ssa} for discussions of this relationship and modern proofs. Particularly noteworthy is that~\eqref{eq:dpi-umegaki} not only holds for quantum channels but also for general positive trace-preserving maps~\cite{reeb17}. Equality in~\eqref{eq:dpi-umegaki} holds if and only if $\mathcal{R}_\sigma(\mathcal{E}(\rho)) = \rho$, where
\begin{align}
    \mathcal{R}_\sigma(\cdot) = \sigma^{1/2}\, \mathcal{E}^\dagger \Big(
  \mathcal{E}(\sigma)^{-1/2} (\cdot) \, \mathcal{E}(\sigma)^{-1/2}
\Big) \sigma^{1/2} \label{eq:petz-recovery-map}
\end{align}
is the so-called~\textcite{petz1986sufficiency, petz1988sufficiency} recovery map. In this case, both $\rho$ and $\sigma$ can be recovered exactly from their respective images under $\mathcal{E}$. More generally, when the data-processing inequality is only approximately saturated, a suitable recovery map constructed from rotated Petz maps still recovers $\sigma$ exactly and $\rho$ approximately. Quantitatively, the DPI gap is lower bounded by the measured relative entropy between $\rho$ and its recovered state~\cite{sutter17}.

As a particularly important example, the quantum relative entropy is monotone under measurement channels.
A consequence of this is that $D(\rho\|\sigma) \geq 0$ for all pairs of states since the classical quantity is positive. Positivity can also be derived from \poscite{klein1931quantum} inequality, preceding the full proof of monotonicity.
Moreover, the relative entropy is faithful, meaning that $D(\rho\|\sigma) = 0$ if and only if $\rho = \sigma$.

\begin{appbox}{thermo}{Free energy from relative entropy and state conversion}{app:free-energy}
The relative entropy to a Gibbs state has a direct thermodynamic interpretation. Using natural logarithms in this box, let $X=X^\dag$ be the Hamiltonian, $\beta=1/(kT)>0$ the inverse temperature, and\vspace*{-.5\baselineskip}
\begin{align*}
\gamma=\frac{e^{-\beta X}}{Z},\qquad Z=\tr[e^{-\beta X}]
\end{align*}
the Gibbs state and partition function. The thermodynamic free energy introduced by~\textcite{Helmholtz1882} is $F(\rho)=\tr[\rho X]-\beta^{-1}H(\rho)$. Then
\begin{align*}
D(\rho\|\gamma)
&=-H(\rho)+\beta\tr[\rho X]+\ln Z\\
&=\beta\big(F(\rho)-F(\gamma)\big),
\end{align*}
where $F(\gamma)=-\beta^{-1}\ln Z$. Thus, the free-energy excess above thermal equilibrium is quantified by the relative entropy to the Gibbs state.

The Umegaki relative entropy provides criteria for state transformations
under channels that preserve the Gibbs state. In particular,
$D(\rho\|\gamma)>D(\sigma\|\gamma)$ guarantees the existence of a sequence
of Gibbs-preserving channels transforming $\rho^{\otimes n}$ into
$\sigma^{\otimes n}$ with vanishing error as
$n\to\infty$~\cite{matsumoto2010relative,buscemi2019information}.

This many-copy result also yields a necessary and sufficient relative-entropy criterion for correlated catalysis, in which a catalyst assists the transformation while its marginal state is returned exactly, although it may become correlated with the system~\cite{shiraishi_2021}. Analogous standard entropy criteria for correlated catalysis govern bipartite pure-state transformations under LOCC~\cite{kondra_2021,lipka2024catalysisreview} and state transformations under global unitaries~\cite{wilming2021entropy,wilming_2022,boes_2019}.
\end{appbox}

We can infer this from the quantum Pinsker's inequality, which follows by~\eqref{eq:dpi-umegaki} from the classical \poscite{pinsker1960inequality} inequality (in the form of \textcite{kullback1967pinsker}) by choosing a measurement that achieves the trace distance:
\begin{align}
    \label{eq:pinsker}
    D(\rho\|\sigma) \geq \frac{\log e}{2} \| \rho - \sigma \|_1^2 \,.
\end{align}
Upper bounds on the relative entropy in terms of the trace distance\,---\,so-called reverse Pinsker inequalities\,---\,have been investigated by \textcite{audenaert2005upper,audenaert2011upper} and \textcite{hirche2024quantum}, for example. The latter yields
\begin{align}
    D(\rho\|\sigma) \leq \frac{\max \{ D_{\max}(\rho\|\sigma), D_{\max}(\sigma\|\rho) \} }{2} \| \rho - \sigma\|_1 \,.
\end{align}
A lower bound in terms of the fidelity follows by choosing a measurement that achieves the fidelity, giving: %
\begin{align}
    D(\rho\|\sigma) \geq - \log F(\rho, \sigma) \,.
\end{align}

Indeed, the local behavior of $D$ is quadratic when we vary around a fixed state $\tau$~\cite{lesniewski1999metrics}, i.e.,
\begin{align}
   D((1-\eps)\tau + \eps \rho \| \tau )
   = \frac{\eps^2 \log e}{2}\, \chi^2(\rho\|\tau) + O(\eps^3) ,
\end{align}
where the quantum $\chi^2$-divergence,\footnote{More precisely, it is the $\chi^2$-divergence corresponding to the \textcite{bogoliubov1970fisher}--\textcite{kubo1957fisher}--\textcite{mori1958fisher} (BKM) metric. It is by no means the unique quantum generalization of $\chi^2$ divergence~\cite{petz1996metrics,lesniewski1999metrics}. Others will be discussed in Section~\ref{sec:chi2}.} or quantum Fisher information~\cite{petz1996metrics}, is given by
\begin{align}
\label{eq: chi square Umegaki}
    \chi^2(\rho\|\sigma) = \int_{0}^{\infty} \tr \left[ \big( (\rho - \sigma) (t\id + \sigma)^{-1} \big)^2 \right] \mathrm{d}t\,.
\end{align}
This is a type of $f$-divergence and will be discussed further in Section~\ref{sec:integral-f}. %
Moreover, we have
\begin{align}
    D(\rho\|\sigma) \leq \log \left(1 + \chi^2(\rho\|\sigma) \right) \leq (\log e)\, \chi^2(\rho\|\sigma)
\end{align}
for all pairs of states. This is by no means an exhaustive list of relevant bounds on the relative entropy.

Equivalent to monotonicity is the fact that the map $(\rho, \sigma) \mapsto D(\rho\|\sigma)$
is jointly convex \cite{Uhlmann73}, meaning that
\begin{align}
   \sum_{i} p_i D(\rho_i \| \sigma_i) \geq D\left( \sum_i p_i \rho_i \middle\| \sum_i p_i \sigma_i \right)
\end{align}
for any probability distribution $\{ p_i \}_i$ and ensembles $\{ \rho_i \}_i$ and $\{ \sigma_i \}_i$ of states and positive operators, respectively. This also implies convexity in each of the arguments.

An alternative but conceptually appealing way to define a relative entropy of quantum states is to simply optimize the right-hand side of~\eqref{eq:dpi-umegaki} over all measurement channels, reflecting the best classical relative entropy accessible by measurements. This leads to the minimal extension of the classical \customtextcite{kullback1951divergence}{Kullback--Leibler} divergence, termed the measured relative entropy
\begin{align}\label{eq:measured-entropy}
    D^{\mathbb{M}}(\rho\|\sigma) =  \max_{\cM} D(\cM(\rho) \|\cM(\sigma) )\,,
\end{align}
which was first formally considered by \textcite{donald1986relent}.\footnote{Without loss of generality it is sufficient to consider only projective measurements \cite{berta2017variational}.} It is worth emphasizing that this quantity is not equal to the Umegaki relative entropy in general: we have that \cite{petz1986sufficiency}
\begin{align}
    D^{\mathbb{M}}(\rho\|\sigma) = D(\rho\|\sigma)\quad \iff \quad[\rho,\sigma]=0\, . \label{eq:gt-equality}
\end{align}
What this entails is that for non-commuting quantum states, any measurement channel acting on them will contract the relative entropy and thus lead to a loss of distinguishability. However, the connection between the measured relative entropy and the Umegaki relative entropy is restored by considering the regularization (see Section~\ref{sec:class/reg}) of the former, giving~\cite{hiai1991proper}
\begin{align}
    \reg\left[D^{\mathbb{M}}\right](\rho\|\sigma)=D(\rho\|\sigma)\,.
\end{align}
Hence, the quantum relative entropy is the minimal additive quantum generalization of the Kullback--Leibler divergence in the sense of Section~\ref{sec:class/lifting}. As shown by \textcite{matsumoto2018maxdivergence}, the maximal additive quantum extension equals the \textcite{belavkin1982c} relative entropy, given by
\begin{align}
    D^{\prep}(\rho\|\sigma) = \tr\left[\rho \log{\left(\rho^{1/2}\sigma^{-1}\rho^{{1/2}}\right)}\right] .
\end{align}
This expression also satisfies all the properties of a relative entropy (and, in particular, no regularization is needed). It will be discussed further in Section~\ref{sec:renyi/maximal}.

Since the quantum extension of the classical \customtextcite{kullback1951divergence}{Kullback--Leibler} is thus not unique, an axiomatic derivation singling out the quantum relative entropy among all quantum extensions is of interest. Such a derivation was, e.g., given by \textcite{matsumoto2010relative}. He showed that every weakly additive relative entropy extension $D(\rho\|\sigma)$ of the Kullback--Leibler divergence, which is also asymptotically continuous in the first argument, i.e., it satisfies
\begin{align}
    &%
\lim_{n\to\infty} \norm{ \rho_n - \rho^{\otimes n}}_1=0\nonumber\\
    &\ \implies \quad\lim_{n\to\infty}\frac{1}{n}\Big(D\left(\rho_n\middle\|\sigma^{\otimes n}\right)-D\left(\rho^{\otimes n}\middle\|\sigma^{\otimes n}\right)\Big)=0\,,
\end{align}
is equal to the quantum relative entropy. %
Another axiomatic derivation of the quantum relative entropy among all faithful relative entropies continuous in the first argument, is via the super-additivity property,
\begin{align}\label{eq:super-additivity}
    D(\rho_{AB}\|\sigma_A\otimes\omega_B)\geq D(\rho_A\|\sigma_A)+D(\rho_B\|\omega_B)\, ,
\end{align}
to be obeyed for all states $\rho_{AB}$, $\sigma_A$ and $\omega_B$ on bipartite systems~\cite{Wilming2017}. More precisely, the super-additivity property~\eqref{eq:super-additivity} is satisfied by both $D$ and $D_{\min}$, but the latter is not faithful.
We refer to \textcite{capel18} for a generalization of super-additivity when the second argument is not a product state. Further, it is also possible to derive the quantum relative entropy from the form of the von Neumann entropy, reconstructing~\eqref{eq:Umegaki} from~\eqref{eq:von_Neumann}. Namely, we can interpret $D(\rho\|\sigma)$ as the excess entropy that derives from randomizing the position of a single copy of $\rho$, placed into a chain of identical copies of $\sigma$: setting $\Omega_n = \frac1n \left( \rho \otimes \sigma^{\otimes (n-1)} + \sigma \otimes \rho \otimes \sigma^{\otimes (n-2)} + \ldots + \sigma^{\otimes (n-1)} \otimes \rho \right)$, we have \cite{Diosi2006, Csiszar2007}
\begin{align}
D(\rho\|\sigma) = \lim_{n\to\infty} \Big\{ H(\Omega_n) - H(\rho) - (n-1) H(\sigma) \Big\}\,.
\end{align}
Last but not least, \textcite{petz_1992} axiomatically derives quantum relative entropy via the so-called conditional expectation property, which is related to the quantum relative entropy satisfying plentiful chain rules. We further discuss these in Sections~\ref{sec:standard/mutual_cond} -- \ref{sec:standard/inequalities}.

The distinguished character of the quantum relative entropy among all relative entropies was historically further solidified with the discovery of its information-theoretic operational meaning in quantum hypothesis testing \cite{hiai1991proper}\,---\,discussed in Application~\ref{app:HT}%
. We continue by presenting its mathematical properties.

\begin{appbox}{HT}{%
Quantum state discrimination}{app:HT}
One of the most fundamental interpretations of the Umegaki relative entropy is as a measure of distinguishability of two quantum states.

\smallskip
Consider a device that outputs either $n$ copies of a state $\rho_0$, or $n$ copies of a state $\rho_1$; our aim is then to distinguish between these two scenarios by means of a global measurement on the $n$ systems.
This setting is illustrated in Fig.~\ref{fig:HT}.
In asymmetric hypothesis testing, the goal is to minimize the probability of incorrectly concluding that the state is $\rho_0$ (a so-called ``type II error''), while keeping the probability of incorrectly concluding that the state is $\rho_1$ (a ``type I error'') below a fixed threshold $\varepsilon$.
This error decreases exponentially with the number of copies $n$, and the Stein exponent quantifies the corresponding rate of decay. The quantum Stein’s lemma~\cite{hiai1991proper,ogawa2000strong} shows that this exponent is given by the Umegaki relative entropy between the two states, $D(\rho_0 \| \rho_1)$.

\smallskip
This asymmetric treatment of the two errors is natural in many contexts. In MRI, for example, the state of hydrogen spin ensembles in tissue is genuinely a quantum state\,---\,healthy and malignant tissue correspond to different density matrices $\rho_0$ and $\rho_1$. A missed  diagnosis of an illness can be catastrophically worse than a false alarm, so we naturally want to fix the false positive rate and ask how fast we can drive false negatives to zero, which is exactly asymmetric testing.
\end{appbox}

\begin{inplacefigure}
\centering
\begin{tikzpicture}

    \node at (0,0) {\includegraphics[width=0.32\textwidth]{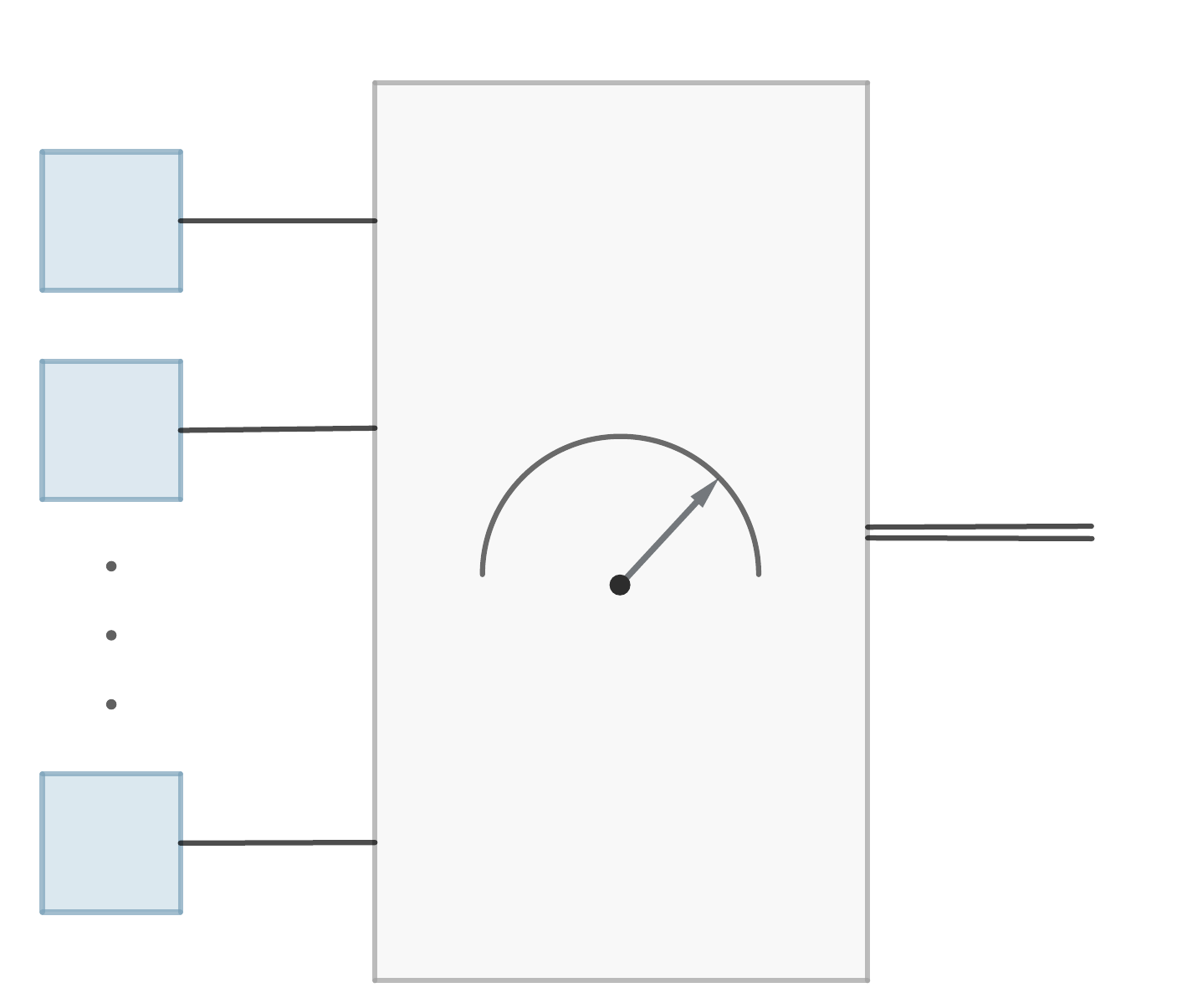}};

      \node at (2.68,-0.07) {\large$\hat{x}$};
      \node at (-2.3,1.38) {\large$\rho_x$};
       \node at (-2.3,0.35) {\large$\rho_x$};
       \node at (-2.3,-1.65) {\large$\rho_x$};

\end{tikzpicture}

\caption{\textbf{The binary quantum hypothesis testing problem.} Given two hypotheses $\rho_0$ or $\rho_1$, and $n$ copies of one of the two states, the goal is to determine the true state by performing a joint measurement. The measurement outputs a classical bit $\hat{x}$, which aims to match the true label $x$ of the input state $\rho_x$. Two types of errors may occur: either $\hat{x}=1$ while the state is $\rho_0$ (type I error), or $\hat{x}=0$ while the state is $\rho_1$ (type II error).}
\label{fig:HT}
\end{inplacefigure}

Generalizing classical results for the \customtextcite{kullback1951divergence}{Kullback--Leibler} divergence, and building on previous developments by \textcite{Araki1973}, \textcite{petz88} gives the following variational formula for the relative entropy, which appears through the Legendre transformation:
\begin{align}
    D(\rho\|\sigma)
    &=
\sup_{W>0}\Big\{
\tr[\rho\log W]
\notag\\
&\;\;+
(\log e)\,
\big(1-\tr[\exp(\log\sigma+\log W)]\big)
\Big\}\, . \label{eq:qreltentropy-variational}
\end{align}
In contrast, the measured relative entropy can be written as \cite{donald1986relent,berta2017variational}
\begin{align}
    D^{\mathbb{M}}(\rho\|\sigma)
    =\sup_{W>0}\Big\{\tr [\rho\log W]+(\log{e})(1-\tr [\sigma W])\Big\}\,.\label{eq:qreltentropy-variational4}
\end{align}
In this representation, monotonicity of the relative entropy under measurements and the equivalence condition~\eqref{eq:gt-equality} are thus direct consequences of the Golden--Thompson~\cite{golden1965inequality,thompson1965inequality} inequality.
The optimization in \eqref{eq:qreltentropy-variational} is concave in $\log W$, while \eqref{eq:qreltentropy-variational4} is concave in $W$, and particularly useful as the dependence on $\rho$ and $\sigma$ in the objective functions is separated.

Another variational expression for the quantum relative entropy in terms of an integral is due to \textcite{kosaki86} (see also \textcite%
{petz93-book}), with further variants discussed by \textcite{donald1986relent}. It has strong applications in device-independent quantum cryptography \cite{brown24} and takes the form
\begin{align}
&D(\rho\|\sigma)= \sup_{\left(n,X,Y\right)} \bigg\{ \log n - (\log{e}) \int_{\frac1n}^\infty \!\bigg( \tr\big[\rho Y(t)^\dag Y(t)\big] \nonumber \\
&\hspace{20.7ex} + \frac1t \tr\big[\sigma X(t) X(t)^\dag\big] \bigg) \frac{\mathrm{d}t}{t} \bigg\} ,
\end{align}
where $n$ runs over all positive integers, and $X,Y:\big(\tfrac1n,\infty) \to \cL(A)$, with $A$ being the underlying Hilbert space, are two arbitrary step functions taking on finitely many values and such that $X(t) + Y(t) = I$ for all $t$. In spite of the lengthy form, Kosaki's integral formula makes several properties of the relative entropy immediately apparent. For instance, $D(\rho\|\sigma)$ is easily seen to be jointly convex, as it is expressed as a point-wise supremum of linear functions in both $\rho$ and $\sigma$.

More recently, and generalizing classical works by \textcite{sason16}, a different integral representation of the quantum relative entropy has been given by \textcite{frenkel2023integral}. Subsequently refined by \textcite{jencova24} and \textcite{hirche2024quantum}, its so-called hockey stick form is
\begin{align}
 &\frac{1}{\log e} \, D(\rho\|\sigma) \notag\\
 &\quad=\int_1^\infty\!\!\left(\frac{1}{\gamma}\tr[\rho\!-\!\gamma\sigma]_+ + \frac{1}{\gamma^2}\tr[\sigma\!-\!\gamma\rho]_+\right)\mathrm{d}\gamma\,,
\label{eq:Frenkel}
\end{align}
where everything is measured in nats, and, for a Hermitian operator $X$ with spectral decomposition $X = \sum_i x_i \ketbra{i}{i}$, we have denoted its positive part as $[X]_+ = \sum_i \max\{x_i,0\} \ketbra{i}{i}$. The above expression is particularly useful as a plethora of mathematical properties can directly be lifted from the Schatten 1-norm to the quantum relative entropy, and because it enables strong applications in quantum information theory \cite{cheng25} and device-independent quantum cryptography \cite{kossmann24}. Further %
mathematical extensions are given by \textcite{friedland26}.

The quantum relative entropy is lower semi-continuous \cite{lindblad1974expectations}, and, more quantitatively, one has the following tight semi-continuity relation in the first argument~\cite{audenaert25, berta25}:
\begin{align}
    &\frac12 \norm{\rho-\tilde\rho}_1\leq\eps \quad \Longrightarrow \nonumber \\
    &D(\rho\|\sigma)-D(\tilde\rho\|\sigma)\leq \begin{cases} \eps\log(d_{\rm eff}\!-\!1)+\binent(\eps) & \eps \leq 1 \!-\! \frac{1}{d_{\rm eff}}\, , \\
    \log d_{\rm eff} & \eps > 1 \!-\! \frac{1}{d_{\rm eff}}\, , \end{cases}
    \label{eq:qrelentropy-cont}
\end{align}
featuring $d_{\rm eff} = \exp (D_{\max}(\rho\|\sigma))$. Continuity in the second argument is also well studied, but leads to expressions that diverge when the minimal eigenvalue is small~\cite{bluhm23, audenaert25}.

Another useful property of the quantum relative entropy is its operator anti-monotonicity in the second argument,
\begin{align}
  \sigma_1\geq\sigma_2\quad \implies \quad D(\rho\|\sigma_1)\leq D(\rho\|\sigma_2)\,,
\end{align}
which follows from the operator monotonicity of the logarithm and has various applications in quantum information.

The monotonicity of the quantum relative entropy can sometimes be strengthened to  so-called strong data processing inequalities, featuring for a channel $\cE$ and a state $\sigma$ the contraction coefficient \cite{lesniewski1999metrics,hiai15}
\begin{align}
\eta(\cE,\sigma)= \sup_{\rho\neq\sigma}\frac{D(\cE(\rho)\|\cE(\sigma))}{D(\rho\|\sigma)},\quad\eta(\cE)= \sup_{\sigma}\eta(\cE,\sigma)\,.
\end{align}
Contraction coefficients have applications for bounding mixing times \cite{temme10,muller-hermes26}, e.g., for quantifying the limitations of noisy quantum algorithms \cite{stilck-franca21}, or in quantum machine learning around differential privacy \cite{angrisani25}.

\begin{example}
    A simple example of a contraction coefficient in closed form is available for the depolarizing channel, $\cD_p(\cdot)=(1-p)(\cdot)+p\frac{\id}{2}$; \textcite{hirche22} find $\eta(\cD_p)=(1-p)^2$.
    For \poscite{leditzky18-2} dephrasure channel
    \begin{align}
        \cD_{\eps,p}(\cdot)= (1-\eps)\Big((1-p)(\cdot)+pZ(\cdot)Z\Big)+\eps\ketbra{e}{e}
    \end{align}
    with the $Z$ the Pauli operator, \textcite{hirche22} find $\eta(\cD_{\eps,p})=1-\eps$, independent of $p\in[0,1]$.
\end{example}

\begin{appbox}{Contraction}{Relaxation to equilibrium}{app:relaxation}
Dissipative dynamics induced by coupling to an environment, such as a thermal bath, are often modeled by a time-independent master equation $\dot{\rho}_t=\mathcal L(\rho_t)$,
whose generator $\mathcal L$, also called a Lindbladian, has the GKSL form~\cite{Gorini1976,Lindblad1976,AlickiLendi2007}. This provides an accurate effective description of many open quantum systems when the system--environment coupling is sufficiently weak and their characteristic timescales are well separated~\cite{BreuerPetruccione2002,GardinerZoller2004,RivasHuelga2012,Vacchini2013}.

In the simplest setting, the state converges to a unique full-rank Gibbs state $\gamma$, and a central question is how quickly this equilibrium is reached. A standard approach is to show that entropy production satisfies the modified log-Sobolev inequality
$-\frac{\dd}{\dd t}D(\rho_t\|\gamma)\geq2\kappa D(\rho_t\|\gamma)$
for some $\kappa>0$, uniformly over initial states $\rho_0$. Integrating this inequality and applying Pinsker's inequality~\eqref{eq:pinsker} yields
\begin{align*}
D(\rho_t\|\gamma)&\leq e^{-2\kappa t}D(\rho_0\|\gamma),\\
\|\rho_t-\gamma\|_1&\leq\sqrt{\frac{2D(\rho_0\|\gamma)}{\log e}}\,e^{-\kappa t}.
\end{align*}
The largest such $\kappa$ is the modified log-Sobolev constant in this convention. Since $D(\rho_0\|\gamma)\leq\log(1/\gamma_{\min})$, these estimates bound the thermalization time uniformly over initial states~\cite{kastoryano2013quantum}. Application~\ref{app:contraction-coefficient} discusses local methods for estimating contraction rates.
\end{appbox}

Approximate preservation of relative entropy also supports entanglement-wedge reconstruction in holography: universal recovery channels reconstruct bulk information from the corresponding boundary region~\cite{Cotler19}.

As thoroughly discussed for general relative entropies in Section~\ref{sec:class}, the quantum relative entropy then acts as a parent quantity for the von Neumann entropy, the quantum conditional entropy (Section~\ref{sec:standard/entropy_cond}), the quantum mutual information as well as the quantum conditional mutual information (Section~\ref{sec:standard/mutual_cond}).


\subsection{Conditional entropy}
\label{sec:standard/entropy_cond}

The quantum conditional entropy of $A$ given $B$ for bipartite states $\rho_{AB}$ is generated from the quantum relative entropy as \cite{cerf1997negative}
\begin{align}
    H(A|B)_\rho &= \log d_A-D(\rho_{AB}\|\pi_A\otimes\rho_B) \label{eq:cond_entropy_2} \\
    &=\log d_A-\min_{\sigma_B}D(\rho_{AB}\|\pi_A\otimes\sigma_B)\,, \label{eq:cond_entropy_3}
\end{align}
i.e., both canonical constructions of the conditional entropy are equivalent.
Data-processing, isometric invariance, and additivity are then inherited from the quantum relative entropy. As such, the quantum conditional entropy is a measure for how much uncertainty one has about $A$ given access to $B$.

The quantum conditional entropy can further be conveniently decomposed as
\begin{align}
    H(A|B)_\rho
    &=H(AB)_\rho-H(B)_\rho\, , \label{eq:cond_entropy_1}
\end{align}
which is also called the chain rule for quantum conditional entropy, and more generally can be formulated as the rule
\begin{align}
    H(AB|C)_{\rho} = H(A|BC)_{\rho} + H(B|C)_{\rho} .
\end{align}
for any tripartite state $\rho_{ABC}$.
 However, this is where the parallels end as one can easily see that entangled states $\rho_{AB}$ can have negative quantum conditional entropy, i.e.,
\begin{align}\label{eq:cond-dim}
    -\log{\min\{d_A, d_B\}} \leq H(A|B)_\rho \leq \log d_A \,.
\end{align}
This demonstrates that it is possible to know more about the full system than about its parts in quantum mechanics \cite{horodecki05,delrio11negative}. The information-theoretic formalization of this underlies the task of quantum state merging \cite{Horodecki2007}.

\begin{example}\label{ex:MES}
    For a maximally entangled state $\phi_{AB}=\ketbra{\phi}{\phi}_{AB}$ in dimension $d_A = d_B = d$ one indeed finds that $H(AB)_\phi=0$ and $H(B)_\phi=\log d$, so that
    \begin{align}\label{eq:MES-entropy}
        H(A|B)_\phi=-\log d\, .
    \end{align}
    In contrast, for a separable (i.e., unentangled) quantum state $\sigma_{AB}$, the reduction criterion $\sigma_{AB} \leq \mathds{1}_A \otimes \sigma_B$ \horcite{Horodecki1999}{2} and the operator monotonicity of the logarithm imply
    \begin{align}
    H(AB)_\sigma &= - \tr \big[\sigma_{AB} \log \sigma_{AB}\big] \\
    &\geq - \tr \big[\sigma_{AB} \log \big(\mathds{1}_A \otimes \sigma_{B}\big)\big] = H(B)_\sigma\, .
    \end{align}
    That is, the quantum conditional entropy remains positive: $H(A|B)_\sigma \geq 0$, further see \textcite{Plenio2000} for an extension of this inequality to general states $\sigma_{AB}$.
\end{example}

\begin{appbox}{compression}{Quantum state merging}{app:state-merging}
Quantum state merging is a fundamental task in quantum information theory that generalizes source compression, with broad applications to quantum communication and entanglement distillation. It also gives a precise meaning to the entropic measures of quantum correlation.

\smallskip
Quantum conditional entropy $H(A|B)$ quantifies the cost of transferring quantum information to a receiver with side information. Suppose Alice holds $A$, Bob holds $B$, and their state $\rho_{AB}$ has a purification $\psi_{ABR}$, with $R$ an inaccessible reference. Using local operations, classical communication, and shared entanglement, they aim to transfer $A$ to Bob while preserving the full purified system $ABR$. For many independent copies and asymptotically vanishing error, the optimal net entanglement cost is $H(A|B)_\rho$ ebits per copy~\cite{horodecki05}. A positive conditional entropy therefore specifies how many shared Bell pairs must be consumed, whereas a negative value means that merging generates $-H(A|B)_\rho$ Bell pairs per copy.

\smallskip
Meanwhile, mutual information characterizes the classical communication required in the same task. The minimum rate is $I(A:R)_\psi$ bits per copy from Alice to Bob, and this rate can be achieved simultaneously with the optimal entanglement cost~\cite{Horodecki2007}. Thus, conditional entropy and mutual information quantify two distinct resources needed for merging. Comparing the entanglement cost with the $H(A)_\rho$ ebits required to compress and teleport Alice’s share without exploiting Bob’s side information gives a further interpretation: $I(A:B)_\rho=H(A)_\rho-H(A|B)_\rho$ measures the saving in net entanglement afforded by access to $B$.
\end{appbox}

For classical side information $X$ and any state $\rho_{ABX} = \sum_x p_x \ketbra{x}{x}_X \otimes \rho_{AB}^x$, we have the decomposition
\begin{align}
    H(A|BX)_{\rho} = \sum_x p_x H(A|B)_{\rho^x} ,
\end{align}
mimicking the behavior of Shannon entropy. Moreover, conditional entropy satisfies monotonicity under coarse graining. In particular, as a consequence of the chain rule and positivity for separable states, we get
\begin{align}
    H(XA|B)_{\rho} = H(X|AB)_{\rho} + H(A|B)_{\rho} \geq H(A|B)_{\rho} \, ,
\end{align}
showing that classical registers always increase entropy.

Even though it can get negative, the conditional entropy still has many useful intuitive properties.
For bipartite quantum states $\rho_{AB}$, conditioning on the $B$-system reduces the entropy, i.e., one has the sub-additivity relation \cite{araki70inequalities}
\begin{align}\label{eq:sa}
    H(A)_\rho\geq H(A|B)_\rho\, ,
\end{align}
which can be seen as equivalent to~\eqref{eq:sub-additivity-1} by the chain rule. In the same spirit, \poscite{lieb1973proof} strong sub-additivity of~\eqref{eq:strong-sub-additivity-1} is equivalent to the relation
\begin{align}
\label{eq:ssa}
    H(A|B)_{\rho} \geq H(A|BC)_{\rho}
\end{align}
for any tripartite state $\rho_{ABC}$. Operationally, this implies that removing the system $C$ cannot decrease the uncertainty an observer has on $A$. By Stinespring dilation, strong sub-additivity further implies that any data-processing on the side information cannot decrease uncertainty.

Another form of the strong sub-additivity inequality, achieved by taking the entropy dual of the second entropy in~\eqref{eq:ssa}, is known as weak monotonicity of quantum entropy~\cite{ruskai2002ssa}. It states that
\begin{align}\label{eq:weak-mono}
    H(A|B)_\rho + H(A|C)_\rho\geq0\,,
\end{align}
which fundamentally quantifies a form of monogamy of entanglement. That is, the more entangled parties $A$ and $B$ are, as measured by the negative conditional entropy, the less entangled can $A$ be with the third party $C$. As an extreme example, if $A$ and $B$ are maximally entangled, $A$ has to be product with~$C$.

\begin{example}
    It is easily verified that for pure tripartite states $\psi_{ABC}=\ket{\psi}\bra{\psi}_{ABC}$ Eq.~\eqref{eq:weak-mono} becomes an equality. Thus, the conditional entropy is self-dual in the sense of Definition~\ref{def:entropic-dual}, i.e., we have $\dual[H] = H$.
\end{example}

\begin{appbox}{uncertainty}{Entropic uncertainty relations}{app:uncertainty-b}
Entropic uncertainty relations can be extended to incorporate quantum correlations between the observed object and its environment~\cite{berta2010uncertainty}. For example, for two-qubit states $\rho_{AB}$ and measurements on system $A$ in the conjugate basis $X$ and computational basis $Z$, one finds
\begin{align*}
    H(X|B)_\rho+H(Z|B)_\rho \geq \log{2}+H(A|B)_\rho \notag
\end{align*}
with the joint classical-quantum post-measurement states $\rho_{XB}$ and $\rho_{ZB}$, respectively. This pleasantly generalises the strengthened \textcite{maassen1988uncertainty} relation from Application~\ref{app:uncertainty-a}. Here, $H(A|B)_\rho$ denotes the quantum conditional entropy before the measurement is applied. It can become negative for entangled quantum states, effectively lowering the uncertainty (relative to $B$) all the way down to zero for maximally entangled qubits\,---\,which have $H(A|B)_\rho=-\log{2}$ as in Eq.~\eqref{eq:MES-entropy}. And indeed, for the maximally entangled state, whatever measurement $X$ or $Z$ was performed on $A$, the corresponding statistics can be read off by performing the same measurement on $B$. Nevertheless, for general states the equation above gives a trade-off relation between the entanglement present and the uncertainty available \cite{coles2017review}.
\end{appbox}

The quantum conditional entropy is a concave function of the quantum state, a statement that is in fact equivalent to weak monotonicity \cite{ruskai2002ssa}; it is moreover continuous and, for sufficiently small $\eps$, satisfies the tight bound~\cite{berta2026sharpcontinuityquantumconditional}
\begin{align}
& \frac12 \left\| \rho_{AB} - \tilde{\rho}_{AB} \right\|_1 \leq\eps\quad \implies \nonumber \\
&\qquad \big|H(A|B)_\rho-H(A|B)_{\tilde\rho}\big| \leq\eps\log\big(d_A^2-1\big) + \binent(\eps)\, ,
\label{eq:continuity_conditional_entropy}
\end{align}
Strikingly, the bound is independent of the dimension of the $B$-system. The relation~\eqref{eq:continuity_conditional_entropy} improves on the earlier works of \textcite{alicki04CMI,synak-radtke06,mosonyi2011quantum,winter16continuity,alhejji20,wilde20,audenaert25,berta25}. For separable quantum states and sufficiently small $\eps$, the right-hand side can be improved to the again tight $\eps\log(d_A-1)+\binent(\eps)$. We refer to \textcite{berta2026sharpcontinuityquantumconditional} for the precise validity ranges of these bounds.


\subsection{Mutual information}
\label{sec:standard/mutual}

The quantum mutual information (QMI) of a bipartite quantum state $\rho_{AB}$ is constructed from the relative entropy as
\begin{align}
    I(A\!:\!B)_\rho &= D(\rho_{AB}\|\rho_A\otimes\rho_B)\\
    &= \min_{\omega_B} D(\rho_{AB}\|\rho_A\otimes\omega_B) \label{eq:mutual-opt} \\
    &= \min_{\sigma_A,\,\omega_B}D(\rho_{AB}\|\sigma_A\otimes\omega_B) \label{eq:mutual-opt2} \,.
\end{align}
where the second and third equalities can be deduced from the super-additivity in~\eqref{eq:super-additivity}.
Thus, all canonical constructions of Section~\ref{sec:class/canonical} coincide. The mutual information is zero if and only if $\rho_{AB}$ is a product state.
Indeed, QMI is a mutual information in the sense of Section~\ref{sec:class/mutual-info}, and, as such, a natural measure for the total correlations between $A$ and $B$ \cite{groisman05}. In particular, it is additive and satisfies data-processing for any local operations on $A$ or $B$, i.e., the mutual information cannot increase under such operations. It can be decomposed as
\begin{align}
    I(A\!:\!B)_{\rho} &= H(A)_\rho+H(B)_\rho-H(AB)_\rho\,,\label{eq:mutual-chain} \\
    &= H(A)_{\rho} - H(A|B)_{\rho} \\
    &= H(B)_{\rho} - H(B|A)_{\rho} \,.
\end{align}
These decompositions are generally not shared by other measures of mutual information and distinguish the standard quantum mutual information in many applications.

\begin{appbox}{thermal correlations}{An area law for thermal correlations}{app:thermal-mutual}
Consider a Gibbs state $\gamma_\beta=e^{-\beta X}/Z$ of a local Hamiltonian $X=X_A+X_B+X_\partial$, where $X_\partial$ contains the interactions crossing the cut between $A$ and $B$. Although each region can have an extensive entropy, their mutual information satisfies~\cite{wolf08}
\begin{align*}
I(A:B)_{\gamma_\beta}\leq2\beta(\log e)\,\|X_\partial\|_\infty.
\end{align*}
For bounded, finite-range interactions, the right-hand side grows at most with the boundary area. Thus, mutual information obeys an area law at any fixed positive temperature, without assuming a spectral gap. It bounds the total classical and quantum correlations across the cut.
\end{appbox}

From the decompositions it also becomes clear that positivity of the mutual information is equivalent to sub-additivity of entropy, and furthermore, that the QMI is bounded as
\begin{align}\label{eq:mutual-dim}
    0\leq I(A:B)_\rho\leq2\min\{H(A)_\rho,H(B)_\rho\}\,,
\end{align}
where the factor of two is not needed for separable quantum states (cf.~the previous Example \ref{ex:MES}). Moreover, for tripartite states $\rho_{ABC}$, we have the relations
\begin{align}
    \label{eq:mutual-nolocking}
    I(A\!:\!B)_{\rho} + 2\log d_C \geq I(A\!:\!BC)_{\rho} \geq I(A\!:\!B)_{\rho} \,,
\end{align}
where the factor two can be dropped if the system $C$ is classical. Quantum mutual information thus satisfies the ``no information locking'' property. The lower bound is an instance of data-processing for a channel that takes a partial trace on~$C$.

\begin{example}
    To verify~\eqref{eq:mutual-nolocking}, we may expand $I(A\!:\!BC) - I(A\!:\!B) = H(C|B) - H(C|AB)$,
    and the lower bound is then equivalent to strong sub-additivity. The upper bound follows by the dimension bounds on conditional entropy in~\eqref{eq:cond-dim} and the fact that $H(C|AB)$ is positive when $C$ is classical.
\end{example}

QMI is neither a convex nor a concave function of the quantum state; it is, however, continuous, a fact that is captured by the relation \cite{shirokov17}
\begin{align}
\big|I(A\!:\!B)_\rho-I(A\!:\!B)_{\tilde\rho}\big| \leq 2 \eps\log \min\{d_A,d_B\} + 2g(\eps)
\label{eq:mutual-cont}
\end{align}
for $\frac12 \| \rho_{AB} - \tilde{\rho}_{AB} \|_1 \leq \eps$, and $g(\eps)=(1+\eps)\binent\!\left(\frac{\eps}{1+\eps}\right)$.

\begin{appbox}{capacity}{Quantum channel capacities}{app:CC}
As illustrated in Fig.~\ref{fig:Capacity}, the goal of quantum channel coding is to transmit a message $M$ over noisy channels $\mathcal{N}$ connecting Alice to Bob by designing appropriate maps $\mathcal{E}$ (encoder) and $\mathcal{D}$ (decoder). For classical information transmission one might use channels $\mathcal{N}_{X\to B}$ with classical inputs and quantum outputs, e.g., mapping a classical bit to non-orthogonal qubit states $\ket{0}$ and $\ket{+}$ with probabilities $p$ and $1-p$, respectively. The optimal rate of information transmission in the asymptotic limit of many channel uses $\mathcal{N}^{\otimes n}$ with vanishing error probability defines the (classical) channel capacity. It is quantified by an entropic mutual information formula called the Holevo information \cite{Schumacher97,holevo1998capacity}
\begin{align*}
    C(\mathcal{N}_{X\to B})=\max_{(p_x)_x}I(X:B)_\omega\,,
\end{align*}
where $\omega_{XB}=\sum_x p_x\ketbra{x}{x}_X\otimes\mathcal{N}_{X\to B}(\ketbra{x}{x})$ is the joint state of the classical input label $X$ and the channel output $B$, and the maximum is over all probability distributions $(p_x)_x$ on the input alphabet. This generalizes \poscite{shannon1948mathematical} classical formula.

\smallskip
When sending quantum information over fully quantum channels $\mathcal{N}_{A\to B}$, such as e.g.\ the qubit amplitude damping channel,
the corresponding notion of quantum capacity $Q(\mathcal{N})$ is defined analogously. However, due to the involved optimization over many channel uses, in general no computable entropic formula for $Q(\mathcal{N})$ is known \cite{Cubitt15}. For the special class of degradable channels~\cite{devetak_2005-1} or the more general less noisy channels \cite{Watanabe12}\,---\,for which more information is transmitted than leaked to the environment\,---\,one finds the capacity is given by the coherent information \cite{Lloyd97,Devetak05}
\begin{align*}
    Q(\mathcal{N}_{A\to B})=\max_{\rho_A}\big[-H(A'|B)_\omega\big]\,,
\end{align*}
where $\omega_{A'B}=(\mathcal{I}_{A'}\otimes\mathcal{N}_{A\to B})(\psi_{A'A})$ and $\psi_{A'A}$ is a purification of the input state $\rho_A$.
Simplifications for general channels can be obtained by exploiting additional shared resources on top of the noisy channel connection, such as entanglement between the communicating parties. This leads to different capacity notions with their own entropic formulas, and in particular to the equality between the quantum communication capacity and one-half of a channel's mutual information \cite{Bennett02}.
\end{appbox}

\begin{inplacefigure}
\centering
\begin{tikzpicture}

    \node at (0,0) {\includegraphics[width=0.45\textwidth]{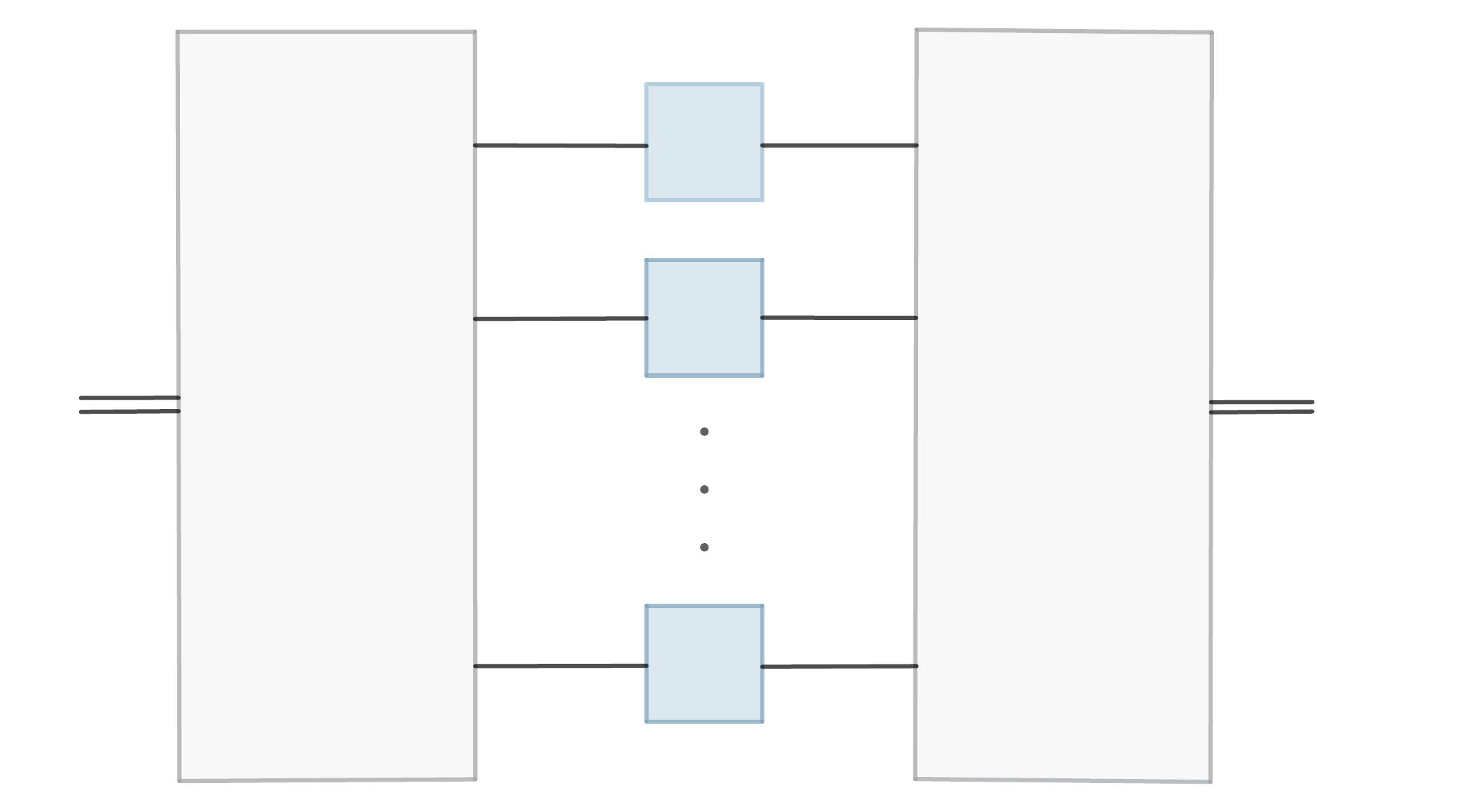}};

    \node at (-3.9,0.04) {\large$M$};
     \node at (-2.25,0.04) {\large$\mathcal{E}$};
     \node at (1.85,0.04) {\large$\mathcal{D}$};
      \node at (3.55,0.04) {\large$\hat{M}$};
       \node at (-0.15,0.5) {\large$\mathcal{N}$};
       \node at (-0.15,1.45) {\large$\mathcal{N}$};
       \node at (-0.15,-1.42) {\large$\mathcal{N}$};

\end{tikzpicture}

\caption{\textbf{Quantum channel coding.} Using a noisy channel $\mathcal{N}$ between two parties, Alice (using an encoder $\mathcal{E}$) and Bob (using a decoder $\mathcal{D}$), the goal is to transmit a message $M$ from Alice to Bob, such that Bob's estimate $\hat{M}$ agrees with high probability with $M$. The capacity of $\mathcal{N}$ is the largest achievable transmission rate in the limit $n\to\infty$, under the requirement that the error probability vanishes asymptotically. Depending on the communication task and the available resources, one obtains different notions of capacity, including capacities for transmitting classical, quantum, or private information, as well as assisted capacities in the presence of additional resources shared between Alice and Bob, such as entanglement.}
\label{fig:Capacity}
\end{inplacefigure}

\subsection{Conditional mutual information}
\label{sec:standard/mutual_cond}

The QMI is extended to the conditional quantum mutual information (CQMI) for tripartite quantum states as
\begin{align}
    & I(A\!:\!B|C)_\rho \nonumber \\
    &\quad = H(AC)_\rho+H(BC)_\rho-H(ABC)_\rho-H(C)_\rho\label{eq:cqmi-linear}\\
    &\quad = H(A|C)_\rho-H(A|BC)_\rho\label{eq:cqmi-cond}\\
    &\quad = I(A\!:\!BC)_{\rho} - I(A\!:\!C)_\rho\,.\label{eq:cqmi_mutual}
\end{align}
The strong sub-additivity of the quantum conditional entropy in~\eqref{eq:ssa} is equivalently cast as the non-negativity of the CQMI. The conditional mutual information is a measure for the total correlations between $A$ and $B$ when having access to $C$. In particular, if the conditioning system is classical, i.e., for states $\rho_{ABX} = \sum_x p_x \ketbra{x}{x}_X \otimes \rho_{AB}^x$, the CQMI decomposes as the average QMI as follows:
\begin{align}
    I(A\!:\!B|X)_{\rho} = \sum_x p_x \, I(A\!:\!B)_{\rho^x} \,.
\end{align}
Such a decomposition and corresponding interpretation are no longer available when the conditioning system is quantum.

The CQMI is monotone under local operations on the $A$ and $B$ systems. However, importantly, there is no monotonicity on the $C$ system. In particular, when removing information from the conditioning system, one can neither say that the CQMI decreases nor increases. In fact, this statement equally holds for the conditional mutual information of multipartite probability distributions, as can be seen by the following %
example.

\begin{example}
    Following \textcite{Renner02}, let $X=Y=Z$ be three uniformly distributed binary random variables. Then, conditioning on $Z$ decreases the mutual information between $X$ and $Y$, i.e., for this probability distribution $p_{XYZ}$ we have $I(X\!:\!Y)_p\geq I(X\!:\!Y|Z)_p$. On the other hand, if $X$ and $Y$ are independent, conditioning on $Z=X\oplus Y$ increases their mutual information, i.e., for that probability distribution $q_{XYZ}$ we have $I(X\!:\!Y)_q\leq I(X\!:\!Y|Z)_q$.
\end{example}

State independent upper bounds on the CQMI are directly available from~\eqref{eq:cqmi-cond} and \eqref{eq:cqmi_mutual} and the corresponding bounds from~\eqref{eq:cond-dim} and~\eqref{eq:mutual-dim}. By choosing the conditioning system $C$ to be trivial, one immediately has that the CQMI is neither convex nor concave as a function of the quantum state. The continuity bound from~\eqref{eq:mutual-cont} for the QMI also holds for the CQMI \cite{shirokov17}.

The information-theoretic operational meaning of the CQMI can be formalized through the task of quantum state redistribution~\cite{devetak_2008-1,abeyesinghe2009mother}.

The CQMI has applications to mixed-state entanglement and the structure of correlations in many-body states~\cite{christandl_2004,kato2019quantum}. As the CQMI is not directly generated from one quantum relative entropy term in the usual way, the conditions when it vanishes are highly non-trivial. For the classical case and tripartite probability distributions $p_{XYZ}$, one has the structural identity
\begin{align}
I(X\!:\!Y|Z)_p
&=
D\left(p_{XYZ}\middle\|p_{Y|Z}p_{XZ}\right).
\label{eq:CMI-recover}
\end{align}
This says that the CMI can be viewed as the relative-entropy distance to
$p_{Y|Z}p_{XZ}$, namely the distribution obtained by locally recovering the
global distribution on $XYZ$ from the marginal $XZ$, by acting only with a
channel on $Z$.

Thanks to more refined notions of monogamy of entanglement, weaker analogues of Eq.~\eqref{eq:CMI-recover} can still be obtained in the quantum setting.
Namely, a series of works, starting from \poscite{fawzi15CMI} breakthrough result, culminated in the following (incomparable) strengthenings of the strong sub-additivity of Eq.~\eqref{eq:strong-sub-additivity-1}
\begin{align}
    &I(A\!:\!B|C)_\rho\nonumber\\
    &\geq
    \begin{cases}
     -\int\beta_0(t)\log F\left(\rho_{ABC},\cR^{[t]}_{C\to BC}(\rho_{AC})\right)\mathrm{d}t \\[1.5ex]
    D^{\mathbb{M}}\!\left(\rho_{ABC}\middle\|\int\beta_0(t)\,\cR^{[t]}_{C\to BC}(\rho_{AC})\,\mathrm{d}t\right)\,,
   \end{cases}\label{eq:ssa-stronger}
\end{align}
with the probability density $\beta_0(t)=\frac{\pi}{2}(\cosh(\pi t)+1)^{-1}$ and the local rotated Petz recovery maps
\begin{align}
    \cR^{[t]}_{C\to BC}(\cdot)
    = \rho_{BC}^{\frac{1+it}{2}}
      \rho_C^{-\frac{1+it}{2}}(\cdot)
      \rho_C^{-\frac{1-it}{2}}
      \rho_{BC}^{\frac{1-it}{2}}\,,
    \label{eq:petz-recovery-map-rotated}
\end{align}
These are rotated versions of the Petz recovery map in Eq.~\eqref{eq:petz-recovery-map}, associated with the partial trace $\mathcal{E}=\Tr_B$ and reference state $\sigma=\rho_{BC}$; the ordinary Petz map is recovered at $t=0$.
An earlier fidelity-based recovery bound, optimized over the rotation parameter, was established by \textcite{wilde_2015} using complex interpolation.
The top lower bound in Eq.~\eqref{eq:ssa-stronger} was derived by \textcite{junge18}, and the bottom one by \textcite{sutter17}.
 While the two bounds are not comparable in general, the latter reduces to~\eqref{eq:CMI-recover} for classical distributions and is thus tight for these. These bounds also show that vanishing conditional mutual information is equivalent to exact recoverability: if $I(A\!:\!B|C)_\rho=0$, then there exists a recovery channel that reconstructs $\rho_{ABC}$ exactly from $\rho_{AC}$. %
 It is an open question if weaker lower bounds on the CQMI in terms of a distance to quantum Markov chains (see below) are possible \cite{ibinson08,salzmann24}, and we discuss related lower bounds in terms of separability in Application \ref{app:SSSAI-b}.

\begin{appbox}{thermal correlations}{Local recovery of thermal states}{app:SSSAI-a}
Let $\rho_{\beta,ABC}$ be a Gibbs state, with regions $A,B,C$ partitioning the lattice and $B$ separating $A$ from $C$. For classical local Hamiltonians, conditioning on $B$ makes $A$ and $C$ independent~\cite{clifford1971markov}, so $I(A:C|B)_{\rho_\beta}=0$. This conditional independence of regions given their separating boundary has applications, for instance, in learning the Hamiltonian from copies of the Gibbs state (see, e.g.,~\cite{haah2022optimal}). Quantum Gibbs states generally lack this exact property, but in several regimes their QCMI can be shown to decay with the distance between $A$ and $C$.

For one-dimensional finite-range Gibbs states, QCMI decay at arbitrary finite temperature was established by Kato and Brandão~\cite{kato2019quantum}. Kuwahara~\cite{Kuwahara2024} obtained more general bounds, independent of region size and with improved polynomial dependence on inverse temperature in one dimension. These bounds imply local recoverability: there exists a channel $\mathcal R_{B\to BC}$ such that~\cite{fawzi15CMI}
\begin{align*}
F\!\left(\rho_{\beta,ABC},
(\idc_A\otimes\mathcal R)(\rho_{\beta,AB})\right)
\geq \exp\!\left[-I(A:C|B)_{\rho_\beta}\right].
\end{align*}
\end{appbox}

Conversely, approximate recoverability forces the conditional mutual information
to be small. That is, if a channel on $C$ approximately reconstructs
$\rho_{ABC}$ from $\rho_{AC}$, then $I(A\!:\!B|C)_\rho$ must be
correspondingly small~\cite{berta14CMI,fawzi15CMI}. From this it follows that the CQMI vanishes if and only if $\rho_{ABC}$ forms a quantum Markov chain in the order $A-C-B$, namely, if there exists a channel $\cR_{C\to BC}$ such that\footnote{Indeed, looking back at Eq.~\eqref{eq:cqmi-cond}, vanishing CQMI can be seen as equality for data-processing under the partial trace on $B$. The recovery channel $\cR$ can thus be chosen as a Petz recovery map of Eq.~\eqref{eq:petz-recovery-map} or any of its rotated variants from Eq.~\eqref{eq:petz-recovery-map-rotated}.}
\begin{align}
    \rho_{ABC} = \cR_{C\to BC}(\rho_{AC})\,.
\end{align}
In words, this means that $\rho_{ABC}$ can be reconstructed from its marginal $\rho_{AC}$ by a channel acting only on the conditioning system $C$. This pleasantly generalises the classical notion of conditional independence to the fully quantum regime, with plentiful applications in physics around the (efficient) description of quantum many body systems. We detail one example in Application~\ref{app:SSSAI-a}. Quantum Markov chains admit a structural characterisation in terms of a direct-sum decomposition~\cite{hayden04CMI}:
\begin{align}\label{eq:direct-sum}
\rho_{ABC}=\bigoplus_k p_k\, \rho^k_{AC_k^L}\otimes\rho^k_{C_k^RB}\,, \end{align}
with respect to some induced decomposition $C=\bigoplus_kC_k^L\otimes C_k^R$ and some probability distribution $\{p_k\}$.

The notion of Markov states naturally motivates considering the relative-entropy distance to the set of such states.
 In the classical case, one has
\begin{align}
I(X\!:\!Y|Z)_p &=\inf_{\text{$q$ Markov}} D(p_{XYZ}\|q_{X-Z-Y})\,,\label{eq:CMI-Markov}
\end{align}
and hence the CMI neatly quantifies the relative entropy distance to the set of Markov chains $X-Z-Y$.
For a general quantum state $\rho_{ABC}$, quantum correlations pose a substantial obstruction to direct extensions of Eq.~\eqref{eq:CMI-Markov}. In particular, the analogue of Eq.~\eqref{eq:CMI-Markov} holds, in general, only as the one-sided bound~\cite{ibinson08}
\begin{align} I(A\!:\!B|C)_\rho \leq \inf_{\sigma \text{ Markov}}D(\rho_{ABC}\|\sigma_{A-C-B})\,.
\end{align}

\begin{appbox}{SSA}{Stronger strong sub-additivity}{app:SSSAI-b}
Following \textcite{brandao11}, \textcite{li14}, and \textcite{berta24}, a first step in the direction of proving lower bounds on the CQMI in terms of a distance to quantum Markov chains is another lower bound on the CQMI in terms of the distance to separable states $\sigma_{A:B}$. Namely, it holds that
\begin{align*}
    I(A:\!B|C)_\rho&\geq\min_{\sigma_{A:B}}D^{\text{LOCC}(B\to A)}(\rho_{AB}\|\sigma_{A:B})\notag\\
    &\geq\frac{c}{d}\cdot\min_{\sigma_{A:B}} \frac12\norm{\rho_{AB} -\sigma_{A:B}}_1^2\notag
\end{align*}
for some constant $c>0$, $d= \min\{d_A,d_B\}$, and $D^{\text{LOCC}(B\to A)}$ denoting a measured relative entropy as in Eq.~\eqref{eq:measured-entropy}, but with the maximization restricted to local measurements that use a single round of classical communication from $B$ to $A$. Note that when tracing out the $C$-system of quantum Markov states in Eq.~\eqref{eq:direct-sum}, the resulting state is indeed separable in the cut $A:B$. Unfortunately, the dimensional factor in the bound is necessary due to the infamous anti-symmetric state \cite{christandl12}. %
\end{appbox}

The CQMI also defines squashed entanglement, $E_{\rm sq}(\rho_{AB})=\frac12\inf_{\rho_{ABE}}I(A:B|E)_\rho$, with the infimum over extensions of $\rho_{AB}$~\cite{tucci1999,christandl_2004}. This additive entanglement measure upper-bounds distillable entanglement. In many-body physics, QCMI bounds for Gibbs states of local Hamiltonians can thus be used to establish that entanglement between distant lattice regions is small (see Application~\ref{app:SSSAI-a}). However, since QCMI also captures classical correlations, such bounds cannot detect phenomena such as the sudden death of entanglement above a constant temperature~\cite{bakshi2024high}.


\subsection{Further entropic inequalities}
\label{sec:standard/inequalities}

For multipartite states $\rho_{A_1A_2BC}$ the CQMI enjoys\,---\,by inspection\,---\,the chain rule
\begin{align}\label{eq:cqmi-chain}
    I(A_1A_2\!:\!B|C)_\rho=I(A_1\!:\!B|C)_\rho+I(A_2\!:\!B|CA_1)_\rho\,,
\end{align}
and similarly for more parties via induction. This iteratively builds up the quantum correlations, which consequently allows us to quantify the monogamy of entanglement. For example, one might use Eq.~\eqref{eq:cqmi-chain} for proving quantum de Finetti theorems with quantum side information, directly in relative entropy distance \cite{brandao17,berta22,berta24-2}. Finally, multipartite generalizations of the CQMI \cite{avis08,yang09} as well as the related conditional entanglement of mutual information (CEMI) \cite{yang08} are also available.

As discussed in Sec.~\ref{sec:relative-entropy}, the quantum relative entropy is singled out by the conditional expectation property \cite{petz_1992}, and we saw many of the resulting chain rules in Section \ref{sec:standard/mutual_cond}. Now, starting with strong sub-additivity from Eq.~\eqref{eq:ssa}, weak monotonicity from Eq.~\eqref{eq:weak-mono}, and their equivalence, \textcite{pippenger03} asked more generally if any linear inequality for the von Neumann entropy of reduced states of multipartite quantum systems is effectively of that form.

That is, for which real coefficients $\{c_S\}_{S\subseteq[n]}$, indexed by subsets $S$ of $[n]$, does one have for all states $\rho_{A_1\cdots A_n}$ that
\begin{align}
    \sum_{S\subseteq[n]} c_S H(\rho_S) \geq 0 \,\, ?
    \label{eq:general_entropy_inequality}
\end{align}
 This so-called entropy-cone problem remains open for $n\geq4$, whereas for $n\leq3$ strong subadditivity is effectively the only inequality~\cite{pippenger03}. Nevertheless, starting from classical non-standard information inequalities \cite{zhang97,zhang98}\,---\,also termed non-Shannon type\,---\,infinitely many constrained quantum entropy inequalities independent of strong sub-additivity have been derived by \textcite{linden2005,cadney2012}. The crucial proof tool is thereby \poscite{hayden04CMI} direct sum decomposition of quantum Markov chains given in Eq.~\eqref{eq:direct-sum}. The simplest non-standard inequality is as follows.

\begin{example}
    \textcite{majenz_thesis} finds that for quantum states $\rho_{ABCD}$ that satisfy $I(A:C|B)_\rho=0=I(B:C|A)_\rho$ one also has
    \begin{align}
        I(C\!:\!AB)_\rho\leq I(C\!:\!D)_\rho+I(A\!:\!B|D)_\rho\,,
    \end{align}
    which is then independent of strong sub-additivity.
\end{example}

The above general entropy-cone problem asks for inequalities that apply to all quantum states on $n$-partite systems. If one restricts attention, for example, to stabilizer states, then \textcite{Linden2013} and \textcite{Gross2013} have proved that additional non-Shannon entropy inequalities do hold. An explicit example is the Ingleton inequality,
\begin{align}
    I(A\!:\!B|C)_\rho+I(A\!:\!B|D)_\rho+I(C\!:\!D)_\rho\geq I(A\!:\!B)_\rho\,.
\end{align}
This inequality does not hold for general quantum states. Another example are Gaussian quantum states, for which all balanced inequalities \eqref{eq:general_entropy_inequality} with $\sum_{S:\,j\in S} c_S=0$ hold, if they hold classically (cf.~Section \ref{sec:gaussian} for details). This then includes, e.g., \poscite{zhang97,zhang98} non-standard information inequalities. The validity of these inequalities for general quantum states remains open.

Lastly, holographic states obey further constraints: when entropies are given by the \customtextcite{Ryu06}{Ryu--Takayanagi} prescription, mutual information is monogamous, $I(A:BC)\geq I(A:B)+I(A:C)$, a relation that fails for general quantum states~\cite{hayden13,bao15-2}.

The variational expression for the quantum relative entropy from Eq.~\eqref{eq:qreltentropy-variational} allows us to state certain types of entropy inequalities in a dual form in terms of Young-type inequalities for Schatten norms. In its simple finite-dimensional quantum form, this so-called \customtextcite{brascamp76}{Brascamp--Lieb} duality states that, for channels $(\mathcal{E}_1,\ldots,\mathcal{E}_n)$, probability vectors $(q_1,\ldots,q_n)$, and some constant $C$ \cite{Carlen04},
\begin{align}\label{eq:BL-duality}
    &H(\rho)\leq\sum_kq_kH(\mathcal{E}_k(\rho))+C\quad\forall\rho\quad\Longleftrightarrow\\
    &\tr\left[\exp\left(\sum_k\mathcal{E}_k^{\dagger}\!\left(\log\omega_k\right)\right)\right]\leq \exp(C)\cdot\prod_k\lVert\omega_k\rVert_{1/q_k}\quad\forall\omega_k\,,
\end{align}
where the $\mathcal{E}_k^{\dagger}$ denote the corresponding adjoint maps (in Hilbert--Schmidt inner product). This duality often offers additional insights into proof strategies for quantum entropy inequalities. As an example, we derive in Application~\ref{app:uncertainty-c} \poscite{maassen1988uncertainty} entropic uncertainty relation from Application \ref{app:uncertainty-a}. Such proofs via duality were pioneered by \textcite{frank13}, and more general dualities for quantum relative entropy inequalities are presented by \textcite{berta23}.

\begin{appbox}{uncertainty}{Derivation of uncertainty relations}{app:uncertainty-c}
The strengthened \textcite{maassen1988uncertainty} inequality for a single-qubit state $\rho$, measured either in the computational or diagonal basis, is
\begin{align*}
    H(X)+H(Z)\geq \log{2}+H(\rho)\quad\forall\rho\,,\notag
\end{align*}
By the duality of Eq.~\eqref{eq:BL-duality}, the claim then becomes equivalent to
\begin{align*}
    \tr\left[\exp\left(\mathcal{M}_X\left(\log\omega_1\right)+\mathcal{M}_Z\left(\log\omega_2\right)\right)\right]\leq\frac{1}{2}\quad\forall\omega_1,\omega_2\notag
\end{align*}
with the measurement maps $\mathcal{M}_X(\cdot)= \sum_x\bra{x}\cdot\ket{x}\ket{x}\bra{x}$ and $\mathcal{M}_Z(\cdot)$ similarly. This, however, follows straightforwardly from the operator Jensen inequality \cite{hansen03} to take the measurement map inside the logarithm, followed by the Golden--Thompson inequality to split up the exponential and let the trace act on the orthogonal computational and diagonal basis states.
\end{appbox}


\subsection{Relative entropy optimization}

We saw that the standard information measures including conditional entropy and mutual information emerge from an optimization of the quantum relative entropy over a set of states. This basic concept underlies many other quantities in quantum information, such as evaluating channel capacities (cf.~Application \ref{app:CC}) or channel entropies~\cite{wilde20-2,gour_2021,Gour21} by optimizing over input states of the channel, computing key rates in quantum cryptography by making use of observed test statistics~\cite{fawzi18,fawzi18-2}, or estimating ground-state energy in many-body systems~\cite{fawzi_2024}.
Various numerical methods have been developed to handle such optimization problems in practice. Different algorithms exploit quantum entropy inequalities \cite{ramakrishnan21}, approximations of the operator logarithm function \cite{fawzi18,fawzi18-2}, integral representations as in Eq.~\eqref{eq:Frenkel} \cite{kossmann26}, and other methods rooted in optimization theory \cite{girard14,sutter16}.

Optimizing relative entropy over a set of reference states can also be used to define distance measures that often find use in quantifying quantum resources~\cite{horodecki_2013-1,chitambar_2019}.
A representative example is the relative entropy of entanglement (REE), defined by \textcite{vedral_1997} as
\begin{align}\label{eq:ree}
E_R(\rho_{AB})=\min_{\sigma_{AB}\in\mathrm{SEP}_{A:B}} D(\rho_{AB}\|\sigma_{AB})
\end{align}
where $\mathrm{SEP}_{A:B}$ denotes the set of states separable across the bipartition $A\!:\!B$.
For a pure state $\rho_{AB} = \psi_{AB}$, this reduces to the entropy of entanglement $H(\psi_A)$.
Induced distance measures of the form~\eqref{eq:ree} are not entropies in general, but they obey a related form of monotonicity under channels that preserve the relevant set of reference states\,---\,in particular, in the case of $E_R$, under local operations and classical communication.
Some of the operational roles played by the entropy and relative entropy itself can be generalized to such induced entropic distances: mirroring the interpretation of relative entropy in state discrimination, the generalized quantum Stein's lemma~\cite{hayashi_2024,lami_2025-1} establishes the regularization of the REE as the figure of merit in entanglement testing (see Application~\ref{app:entanglement-testing}), and similar to how thermodynamic state transformations can be constrained by relative entropy, the REE plays a role in bounding tasks such as entanglement distillation~\cite{vedral_1998,rains_2001,hayashi2006book}.
However, in general such measures are harder to motivate from first principles: while an axiomatization of the entropy of entanglement of pure states is possible~\cite{donald_2002}, the REE does not play the role of a unique entropic entanglement measure on similar terms as thermodynamic or entropic axiomatizations~\cite{vidal_2001,lami_2023}.

\begin{appbox}{Axiomatic}{Entanglement testing}{app:entanglement-testing}
The relative entropy of entanglement is not additive in general~\cite{vollbrecht_2001}, which means that in many-copy, asymptotic settings we need to make use of its regularization,
\begin{align*}
    E_R^\infty(\rho_{AB}) = \lim_{n \to \infty} \frac{1}{n}\, E_R(\rho_{AB}^{\otimes n}) ,
\end{align*}
where the entanglement of each $n$-copy system is evaluated with regards to the bipartition $A^n\!:\!B^n$. This quantity attains an operational interpretation in the context of entanglement testing, namely, distinguishing many copies of a given entangled state $\rho_{AB}$ from all separable states. Specifically, consider a joint hypothesis test that accepts $\rho_{AB}^{\otimes n}$ with probability at least $1-\ve$, and let $\beta_n^\ve$ be the smallest worst-case probability of accepting a state separable across $A^n:B^n$ instead (type II error). The generalized quantum Stein's lemma~\cite{hayashi_2024,lami_2025-1} then shows that the optimal error exponent is given precisely by the regularized relative entropy of entanglement, namely
\begin{align*}
\liminf_{n\to\infty} \left\{-\frac1n\log\beta_n^\ve\right\} = E_R^\infty(\rho_{AB})
\end{align*}
for all $\ve \in (0,1)$. This is a significant generalization of the quantum Stein's lemma discussed in Application~\ref{app:HT}, as the separable alternative is not of an i.i.d.\ form, and may instead contain correlations across the many copies.

\smallskip
The generalized quantum Stein's lemma finds direct use in asymptotic transformations of quantum resources. In particular, it underlies \poscite{brandao_2010-1} observation that, under the axiomatic class of asymptotically non-entangling operations, the regularized relative entropy of entanglement $E_R^\infty(\rho_{AB})$ becomes a unique entropic measure of  transformations in the resource theory of quantum entanglement. This, however, requires a departure from more practical sets of operations such local operations and classical communication~\cite{vidal_2001}, and even small restrictions on the allowed transformations to bring them more in line with the assumptions used in standard entropic or thermodynamic settings make the existence of a unique asymptotic entropy of entanglement impossible~\cite{lami_2023}.
\end{appbox}


\section{R\'enyi information measures}
\label{sec:additive}

In this section, we discuss entropic measures based on R\'enyi entropies, including relative entropies, conditional entropies, and mutual information--type quantities.
They extend the standard measures of information like the von Neumann entropy by incorporating information about higher moments,  allowing for a much more comprehensive understanding of the spectral properties of quantum states.
The applicability of the R\'enyi measures stems from the fact that they typically satisfy both the data-processing inequality under relevant classes of channels and additivity under tensor products, making them ideally suited for systems that we expect to be extensive and the study of operational tasks in many-copy settings.

\subsection{Rényi entropies}
\label{sec:additive/entropy}

 Quantum Rényi entropies extend the classical Rényi entropies~\cite{renyi1961divergence} from probability distributions to quantum states.
For a quantum state, the Rényi entropy is defined for $\alpha \in (0,1) \cup (1,\infty)$ as
\begin{align} \label{eq:def_Renyi_entropy}
    H_\alpha(\rho) = \frac{1}{1-\alpha}\log{\tr\big[\rho_A^\alpha\big]}\,.
\end{align}
This coincides with the classical Rényi entropy of the eigenvalue spectrum, reflecting the fact that any quantum state gives rise to a probability distribution via its spectral decomposition.

At the boundary points $\alpha=0,1,\infty$, the Rényi entropies are defined by taking the corresponding pointwise limits. In particular, the von Neumann entropy is recovered at $\alpha=1$, a fact that follows from an application of l’Hôpital’s rule, while the limits $\alpha\to\infty$ and $\alpha\to0$ yield the extremal entropies that we encountered in~\eqref{eq:min_entropy} and~\eqref{eq:max_entropy}. Another commonly encountered case is $\alpha=2$, which corresponds to the collision entropy; it owes its name to the fact that $H_{2}(X)_p = - \log \Pr[X =X']$ for two independent random variables distributed as $X,X'\sim p$. Their explicit forms are
\begin{align}
    H_{0}(\rho) &= \log \tr[\Pi_\rho]
    = \log \rank(\rho)\,,\\
    H(\rho) &= -\tr[\rho \log \rho]\,,\\
    H_{2}(\rho) &= -\log \tr[\rho^2]\,,\\
    H_{\infty}(\rho) &= -\log \lambda_{\max}(\rho)\,.
\end{align}

Rényi entropies are monotonically decreasing in the parameter $\alpha$, making $H_\infty$ the minimal and $H_0$ the maximal entropy. Moreover, they are concave for $\alpha \leq 1$ and quasi-concave for $\alpha > 1$ as functions of the input state. This also makes them continuous in the interior of the state space; in fact, except for $H_{0}$ (which jumps when the support changes), they are continuous everywhere.

R\'enyi entropies can behave very differently for different values of $\alpha$, since they are more sensitive to large eigenvalues ($\alpha > 1$) and small eigenvalues ($\alpha < 1$), respectively. For example, $H_2$ directly measures the purity, $\tr[\rho^2]$, of the state and $H_{\infty}$ ignores everything but the largest eigenvalue. In contrast, $H_{0}$ is a function of the support size only.
\begin{example}
Let $\rho_d$ be a quantum state with eigenvalue spectrum
\begin{align}
    \Bigg\{1-\frac1d,\, \frac{1}{d(d-1)},\, \dots,\, \frac{1}{d(d-1)}\Bigg\}
\end{align}
Then, a simple calculation reveals that, for large values of the dimension $d$,
$H_{\min}(\rho_d) \sim \log{e}/d$, whereas $H(\rho_d) \sim (2\log{d})/d$ and $H_{\max}(\rho_d) = \log d$. Thus, the three quantities exhibit distinct asymptotic scalings, $H_{\min}(\rho_d) \ll H(\rho_d) \ll H_{\max}(\rho_d)$, with dimension-dependent ratios.
\end{example}

\begin{appbox}{tensor networks}{R\'enyi entropies and tensor-network approximation}{app:renyi-mps}
Cut a pure spin chain into two parts, $A$ and $B$.
The eigenvalues $p_1\geq p_2\geq\cdots$ of the reduced
state $\rho_A$ are the squared Schmidt coefficients:
they give the weights of the terms in the Schmidt
decomposition. Retaining only the largest $\chi$ terms
discards total weight $\epsilon_\chi=\sum_{j>\chi}p_j$.
For $0<\alpha<1$, this tail satisfies
\begin{align*}
\epsilon_\chi\leq
\exp\!\left[\frac{1-\alpha}{\alpha}
\big(H_\alpha(\rho_A)-\log\chi\big)\right].
\end{align*}
A matrix product state (MPS) encodes the wavefunction
as a chain of local tensors. Its bond dimension is the
maximum dimension of the indices connecting these
tensors, and bounds the number of Schmidt terms across
each cut. The inequality therefore helps determine how
large a bond dimension is sufficient to retain most of
the state's weight. Controlling the discarded weights
across all cuts gives bounds on the accuracy of an MPS
approximation to the full chain~\cite{eisert10}.
\end{appbox}

A different application uses the classical R\'enyi entropy of squared Pauli expectation values. For a pure $n$-qubit state, the distribution $\Xi_P=2^{-n}|\bra{\psi}P\ket{\psi}|^2$ over phase-free Pauli strings defines the stabilizer entropy $M_\alpha(\psi)=H_\alpha(\Xi)-n\log2$~\cite{leone_2022}; it quantifies non-stabilizerness and has an operational interpretation in quantum property testing~\cite{bittel2026operational}.

For critical one-dimensional systems described by conformal field theory, the logarithmic growth of interval entanglement entropies also reveals the central charge~\cite{calabrese2004entanglement}.

\begin{appbox}{tensor networks}
{Replicas and collision entropy}
{app:purity}
The collision entropy, $H_2(\rho)=-\log\tr[\rho^2]$, can be evaluated using the expression
\begin{align*}
\tr\big[\rho^2\big]=\tr[(\rho\otimes\rho)F]
\end{align*}
for the purity of $\rho$, where $F$ is the swap operator. This is the two-replica form of the {replica trick}, which expresses moments of a state through permutations of multiple copies. It underlies calculations of entanglement in quantum Monte Carlo simulations~\cite{hastings2010renyi} and quantum field theory, where path integrals for the replicas are joined along the region of interest~\cite{calabrese2004entanglement}.

\smallskip
A matrix product state (MPS) represents a many-body wavefunction as a chain of local tensors connected by auxiliary indices whose maximum size is the bond dimension. The replica trick is particularly useful when computing marginals of disconnected subsystems of an MPS. Here $\tr\big[\rho^2\big]$ can be evaluated by contracting two replicas, swapping the physical indices at the retained sites and tracing out the others. This contraction proceeds along the chain with cost polynomial in its length and bond dimension, avoiding the potentially exponential cost of explicitly computing the reduced-state spectrum~\cite{rolandi2020extensive} and von Neumann entropy.

\smallskip
Experimentally, the same identity gives a controlled swap test using two copies and an ancilla~\cite{ekert2002direct}:
\begin{center}
\begin{quantikz}[row sep=0.3cm, column sep=0.4cm]
    \lstick{$\ket{0}$}
        & \gate{H} & \ctrl{1} & \gate{H} & \meter{} \\
    \lstick{$\rho$}
        &           & \swap{1} &           &          \\
    \lstick{$\rho$}
        &           & \targX{} &           &
\end{quantikz}
\end{center}
Here $H$ denotes a Hadamard gate. Measuring the ancilla in the computational basis gives
\begin{align*}
    \Pr(0) =\frac{1+\tr\big[\rho^2\big]}{2}, \quad
    \Pr(1) =\frac{1-\tr\big[\rho^2\big]}{2}. \notag
\end{align*}
Thus, smaller $\tr[\rho^2]$ makes the outcomes closer to a fair coin, requiring more repetitions to resolve their bias, which gives a worst-case bound of $O(d^2/\varepsilon^2)$ samples to estimate purity up to additive error $\varepsilon$. Joint measurements on more samples can achieve a $\Theta(d)$ scaling~\cite{acharya2020entropy}.
\end{appbox}

An important aspect of quantum R\'enyi entropies is that there is no ambiguity in their definition: they constitute the unique quantum extension of classical Rényi entropies. To make this precise, in close analogy with the construction discussed in Section~\ref{sec: classification} for relative entropies, one can introduce minimal and maximal quantum extensions of a classical entropy. Let $\Hcl$ be an entropy defined on probability distributions and
\begin{align}
\H^{\meas}(\rho) = \min_{\mathcal{M}} \quad \Hcl\big(\mathcal{M}(\rho)\big), \;
\H^{\prep}(\rho) = \max_{p,\,\mathcal{F}} \, \Hcl(p) \,.
\end{align}
Here the minimum is taken over all rank-$1$ projective measurement channels $\mathcal{M}$, where the channel output is the classical register indicating the measurement outcome; or, equivalently, over all mixing channels with diagonal output.
The maximum is taken over all probability distributions $p$ and mixing channels $\mathcal{F}$ such that $\mathcal{F}(p) = \rho$.
It turns out that these constructions in fact coincide. In particular, any quantum entropy $\H$ that reduces to $\Hcl$ on classical states must be given by the classical entropy of the eigenvalue distribution. %

Combined with the fact that Rényi entropies constitute the most general class of entropies, as established in~\textcite{gour2021axiomatic} and~\textcite{mu2021renyi},\footnote{Neither \cite{mu2021renyi} nor \cite{gour2021axiomatic} proves this claim explicitly, but it follows by combining their results.} this implies that any entropy $\H$ satisfying the axioms of Section~\ref{sec: classification} admits a representation as a convex combination of quantum Rényi entropies: namely,
\begin{align}\label{eq:Entropy barycentre}
    \H(\rho) = \int_{[0,\infty]} \d\mu(\alpha) \, H_{\alpha}(\rho)
\end{align}
for some probability measure $\mu$ on the extended positive reals.

As in the discussion of von Neumann entropy in Section~\ref{sec:standard/entropy}, consider estimating an entropy from independent copies of an unknown $d$-dimensional state, allowing arbitrary collective measurements. For R\'enyi entropies, the sample complexity depends sharply on the order $\alpha$~\cite{acharya2020entropy,chen2026nearly,wang2026unified}. Particularly intriguing is that for $\alpha > 1$ the sample complexity scales as $\tilde{\Theta}(d^2)$ except at integer points $\alpha \in \mathbb{N}$ where only $\Theta(d^{2 -2 /\alpha})$ samples are required. Application~\ref{app:purity} covers the case $\alpha=2$.

\subsection{Rényi divergences}

Additive relative entropies are generalizations of classical Rényi divergences.
We already saw that there is in general no unique way to extend such classical measures to quantum states, hindered in particular by the lack of a unique notion of log-likelihood ratio between quantum states. However, in the case of the standard relative entropy (Section~\ref{sec:standard}), this is to a large extent mitigated by the fact that both applications and axiomatic formalizations point to a single, distinguished notion of quantum relative entropy.
That is not the case for quantum R\'enyi divergences: neither axiomatic nor operational considerations appear to single out a unique privileged notion of quantum R\'enyi divergence, and multiple such quantities need to be considered to cover the full scope of applications.

We thus begin our discussion with the classical case, leading us to its natural quantum extensions and their relations.

\subsubsection{Classical R\'enyi relative entropies}

The Rényi entropies $\alpha \mapsto H_\alpha$ are directly related to the moment generating function of the surprisal, which allows for its precise characterization in regimes where tail events are important. A more general approach can be obtained by studying the moment generating function of the log-likelihood ratio $L$, defined as
\begin{align}
\Lambda_t[L] =  \EV \nolimits_p\! \big[\exp(t L)\big], \quad L(X) = \log \frac{p(X)}{q(X)}
\end{align}
for probability distributions $p$ and $q$ on $\cX$ and a random variable $X$, whose distribution on $\cX$ is governed by $p$.
Rényi relative entropies form a one-parameter family indexed by $\alpha \geq 0$ and are related to the moment generating function of the log-likelihood ratio via
\begin{align}
D_{\alpha}(p\|q) =&\ \frac{1}{\alpha-1}\, \log\Lambda_{\alpha-1}[L] \\
=&\ \frac{1}{\alpha-1}\log\!\left(\sum_x p(x)^\alpha q(x)^{1-\alpha}\right). \label{eq:classical_Renyi_divergence_2}
\end{align}
If the probability distributions do not have full support, the above expression is taken to be $+\infty$ when $p$ and $q$ are orthogonal for $\alpha \in [0,1)$, or when the support of $q$ does not contain that of $p$ for $\alpha \in [1,\infty]$.
Furthermore, the Rényi relative entropies at the boundary points $\alpha = 0, 1, \infty$ are defined through their respective pointwise limits.

For $\alpha=1$, one recovers the \customtextcite{kullback1951divergence}{Kullback--Leibler} divergence.
The case $\alpha=1/2$ is  the \textcite{bhattacharyya1946measure} distance (up to a factor of $2$):
\begin{equation}
D_{1/2}(p\|q) = -2\, \log \sum_{x \in \mathcal{X}} \sqrt{p(x) q(x)}\, ,
\end{equation}
The term inside the logarithm is known as the Bhattacharyya coefficient. The case $\alpha=\infty$ is the max-relative entropy~\cite{sibson1969radius}, defined as
\begin{equation}
D_{\max}(p\|q) = \max_{x \in \mathcal{X}} \log \frac{p(x)}{q(x)}.
\end{equation}

Rényi relative entropies constitute the most general form of an additive divergence. Indeed, \textcite{mu2021renyi} show that any additive divergence $\D$ that is finite on pairs of probability distributions with coinciding supports can be represented, for any such pair $p$ and $q$, as a positive linear combination of classical Rényi relative entropies. Namely,
\begin{align}\label{eq:Divergence barycentre}
    \D(p\|q) = \int_{[\frac{1}{2},\infty]} \d\mu(\alpha) \, D_{\alpha}(p\|q) + \int_{[\frac{1}{2},\infty]} \d\nu(\alpha) \, D_{\alpha}(q\|p)
\end{align}
for positive measures $\mu$ and $\nu$ on the extended positive reals.\footnote{The characterization above does not explicitly include $D_{\min}$, as this divergence is trivial on pairs of states with coinciding supports. Pairs with non-coinciding supports require a more refined treatment. One possible approach is to extend the analysis directly to non-full-rank states, along the lines of~\cite{verhagen2025matrix}; another is to consider additional continuity assumptions, as in~\cite{gour2026quantum}. Finally, we note that the functionals $(p,q)\mapsto D_{\alpha}(q\|p)$
are not quantum relative entropies in the sense of Section~\ref{sec: classification}, since they fail to satisfy the normalization axiom.
}
Rényi relative entropies are monotonically increasing in the parameter $\alpha$, which motivates the name of the max-relative entropy as detailed in Section~\ref{sec: relative entropy}. Moreover, as a well-known consequence of the data-processing inequality, they are jointly convex in their arguments for $\alpha \leq 1$, and jointly quasi-convex for $\alpha > 1$. The quantity inside the logarithm, instead, is jointly concave for $\alpha \leq 1$ and jointly convex for $\alpha > 1$ (for a definition of these properties, see~\eqref{eq:joint_convexity} and~\eqref{eq:joint_quasiconvexity}). These R\'enyi divergences are widely used in classical statistics, cryptography as well as learning and information theory, motivating us to take a comprehensive look at their quantum generalizations.

\begin{appbox}{thermo}{Catalytic transformation of states under thermal operations}{app:Second laws}
Classical Rényi divergences are useful for characterizing the transformation of states that are diagonal in the energy eigenbasis under thermal operations with the assistance of a catalyst that is returned uncorrelated at the end of the transformation.

Specifically, let $\rho$ and $\sigma$ be states that are diagonal in the energy eigenbasis, with corresponding eigenvalue vectors $r$ and $s$. Let $\gamma$ denote the Gibbs state with eigenvalue vector $g$. Then, it was shown in~\cite{brandao_2015-2, verhagen2025matrix} that $\rho$ can be transformed into $\sigma$ under catalytic thermal operations if and only if, for all $\alpha \ge 1/2$,
\begin{align*}
    &D_\alpha(r\|g) \ge D_\alpha(s\|g),  \;\,\text{and} \;\,
    D_\alpha(g\|r) \ge D_\alpha(g\|s)\,.\notag
\end{align*}

R\'enyi relative entropies thus govern uncorrelated catalysis, whereas
correlated catalysis is characterized by the Umegaki relative entropy
($\alpha=1$), as discussed in Application~\ref{app:free-energy}.
As in the correlated setting, this characterization is closely linked
to many-copy conversion. R\'enyi relative entropies provide criteria
for \textit{exact} transformations from $\rho^{\otimes n}$ to $\sigma^{\otimes n}$
for all sufficiently large $n$~\cite{farooq2024matrix,verhagen2025matrix},
and the construction of~\textcite{duan_2005-1} turns such conversions
into uncorrelated catalytic ones. Analogous results for catalytic
pure-state entanglement transformations under LOCC were obtained
by~\textcite{klimesh_2007,turgut_2007}.

\end{appbox}

\subsubsection{Max-relative entropy}
\label{subsec:max-relative}

Unlike generic R\'enyi orders, the classical max-relative
entropy admits a unique quantum extension satisfying the
data-processing inequality: the measured and prepared
constructions introduced in Section~\ref{sec:class/lifting} both recover $D_{\max}$%
~\cite{mosonyi2015quantum}. We recall from~\eqref{eq:intro-Dmax} that it is defined as
\begin{equation} \label{eq:Dmax}
\begin{aligned}
D_{\max}(\rho\|\sigma)
&= \log \inf\{\lambda>0:\rho\leq\lambda\sigma\} \\
&= \log\norm{\sigma^{-1/2}\rho\sigma^{-1/2}}_\infty,
\end{aligned}
\end{equation}
where the second expression uses an inverse on the support of $\sigma$ and is only valid when the first expression is finite.
Operator domination bounds of this form already appeared in the quantum substate theorem~\cite{jain2002privacy} and underlie
\poscite{renner2008security}'s min-entropy.
Their systematic study as relative entropies was developed
by~\textcite{datta2009min}.

The operator inequality $\rho\leq\lambda\sigma$ is equivalent to
$\tr[E\rho]\leq\lambda\tr[E\sigma]$ for every measurement effect
$0\leq E\leq\id$.
Thus, $D_{\max}$ quantifies the largest log-likelihood ratio
attainable by a measurement outcome. This worst-case statistical interpretation leads to
applications in privacy: bounds on $D_{\max}$ between
channel outputs corresponding to neighboring inputs
characterize quantum differential
privacy~\cite{hirche_2023}.

An early application is the max-relative entropy of
entanglement~\cite{datta2009min}, whose relation to
generalized robustness is explained in
Application~\ref{app:max-entanglement}.
Extensions of this construction have found uses as measures of more general quantum resources, whose meaning is again connected to statistical distinguishability~\cite{napoli_2016,takagi_2019-1}.
The max-relative entropy is commonly encountered in a \emph{smoothed} form, which we will return to in Section~\ref{sec:one-shot}.
Through such extensions, $D_{\max}$ is used to characterize operational tasks related to resource dilution~\cite{liu_2019} such as entanglement cost~\cite{brandao_2011} or the work of formation in quantum thermodynamics~\cite{horodecki2013fundamental}.
Building on the reverse-testing framework
of~\textcite{matsumoto2010relative},
\textcite{wang2019resource} also gave an interpretation of $D_{\max}$ as the one-shot cost
of asymmetric distinguishability dilution.
\begin{appbox}{Axiomatic}
{Max-relative entropy and robustness of entanglement}
{app:max-entanglement}
\textcite{datta2009min} introduced the max-relative entropy
of entanglement of a bipartite state $\rho_{AB}$ as
\begin{align*}
E_{\max}(A:B)_\rho = \min_{\sigma_{AB}\in\mathrm{SEP}_{A:B}}
D_{\max}(\rho_{AB}\|\sigma_{AB}),
\end{align*}
where $\mathrm{SEP}_{A:B}$ is the set of
separable (unentangled) states. Earlier, building on work by~\textcite{VidalTarrach}, \textcite{harrow_2003} and independently \textcite{Steiner2003} had introduced the generalized robustness of entanglement, which quantifies the noise needed to make $\rho_{AB}$ separable:
\begin{align*}
R_g(\rho_{AB}) = \min_{\mu\geq0,\,\tau_{AB}}
\left\{ \mu:\frac{\rho_{AB}+\mu\,\tau_{AB}}{1+\mu}
\in\mathrm{SEP}_{A:B}
\right\},
\end{align*}
where $\tau_{AB}$ is any normalized state, possibly
entangled.
A suitable noise state $\tau_{AB}$ exists exactly when $\rho_{AB}\leq(1+\mu)\sigma_{AB}$ for some separable $\sigma_{AB}$, which means that the quantities are in fact equivalent in the sense that $E_{\max}(A:B)_\rho= \log\bigl(1+R_g(\rho_{AB})\bigr)$~\cite{datta2009robustness}.

\smallskip
The robustness of entanglement exactly measures the advantages that an entangled state can provide over unentangled ones in quantum channel discrimination: precisely, if $p_{\rm guess}(\{\cN_i\}, \rho)$ denotes the best average probability of distinguishing a set of quantum channels $\{\cN_i\}$ by inputting the state $\rho$, then~\cite{takagi_2019-1,bae_2019}
\begin{align*}
\max_{\{\cN_i\}} \frac{p_{\rm guess}(\{\cN_i\}, \rho_{AB})}{\displaystyle \max_{\sigma_{AB}\in\mathrm{SEP}_{A:B}} p_{\rm guess}(\{\cN_i\}, \sigma_{AB})} = 1 + R_g(\rho_{AB}),
\end{align*}
with the maximum over all channel discrimination games.
Variants of $E_{\max}$ also find use in characterizing quantum entanglement dilution~\cite{brandao_2011}
and in bounding quantum channel capacities~\cite{christandl_2017,berta_2018-1}.
\end{appbox}

Two symmetrizations of $D_{\max}$, namely the Hilbert
projective metric and the Thompson metric, defined as
\begin{align}
D_{\Omega}(\rho_0\|\rho_1) &= D_{\max}(\rho_0\|\rho_1) + D_{\max}(\rho_1\|\rho_0),\\
D_{\Xi}(\rho_0\|\rho_1) &= \max\bigl\{ D_{\max}(\rho_0\|\rho_1),\, D_{\max}(\rho_1\|\rho_0) \bigr\},
\end{align}
characterize the exact asymptotic error exponents in
asymmetric and symmetric postselected quantum hypothesis
testing, respectively~\cite{regula2024postselected}.
Unlike conventional state discrimination (Application~\ref{app:HT}), in this postselected setting inconclusive outcomes are allowed and error
probabilities are conditioned on obtaining a conclusive result, without imposing a bound on the probability
of it occurring.

\subsubsection{Measured R\'enyi relative entropy}
\label{subsec: measured}
As we already saw in several important examples, the fundamental role of measurements in quantum theory\,---\,and in particular in quantum state discrimination\,---\,points to a natural way to define a quantum counterpart to any classical relative entropy, in the form of the largest divergence that can be achieved through a physical measurement process.
The measured R\'enyi relative entropy is thus
\begin{align}
    D^{\mathbb{M}}_\alpha(\rho\|\sigma) = \max_{\cM} D_\alpha (\cM(\rho) \| \cM(\sigma) ),
\end{align}
where the maximization is taken over all measurements $\cM$, and $D_\alpha$ is the classical R\'enyi relative entropy.
As discussed in Section~\ref{sec:class/lifting}, this is the minimal extension of the classical R\'enyi divergence that obeys data processing.
The earliest definitions of such quantities were considered by~\textcite{fuchs_1996}, with a complete formalization due to \textcite{berta2017variational}.
The latter showed that the maximization can equivalently be restricted to projective measurements.

A motivating example of the measured R\'enyi relative entropy is the case $\alpha = \frac12$, in which case it holds that $D^{\mathbb{M}}_{1/2}(\rho\|\sigma) = - \log F(\rho,\sigma)$, reflecting the fundamental fact that the fidelity of quantum states can be achieved by a measurement~\cite{fuchs_1995}.
Other special cases of the measured R\'enyi divergences are $D^{\mathbb{M}}_0 = D_{\min}$ and $D^{\mathbb{M}}_\infty = D_{\max}$~\cite{mosonyi2015quantum}. As mentioned in Section~\ref{sec:relative-entropy}, the limiting case of $\alpha\to1$ does not equal the Umegaki relative entropy in general, being strictly smaller for all non-commuting states.

For general values of $\alpha$, the measured Rényi relative entropies do not admit a closed-form formula. They nevertheless admit a useful variational expression, namely,~\cite{berta2017variational}
\begin{align}
    Q_\alpha^{\mathbb{M}}(\rho \| \sigma) &= \inf_{W > 0} \left\{ \alpha  \Tr\big[\rho  W^{\frac{\alpha-1}{\alpha}}\big] + (1 - \alpha)  \Tr\big[\sigma  W\big] \right\} \nonumber\\
    &=\inf_{W > 0} \left\{  \Big(\Tr\big[\rho  W^{\frac{\alpha-1}{\alpha}}\big]\Big)^\alpha \Big(\Tr\big[\sigma  W\big]\Big)^{1-\alpha} \right\}
\end{align}
for all $\alpha \in (0,1)$,
where $Q_\alpha^{\mathbb{M}}(\rho \| \sigma) = \exp{\big((\alpha-1)D_\alpha^{\mathbb{M}}(\rho \| \sigma)\big )}$. For $\alpha>1$, the same variational expression holds, with the infimum replaced by a supremum.

These variational formulas also permit numerical evaluation. For rational
$\alpha\in(0,1)\cup(1,\infty)$, $Q_\alpha^{\mathbb{M}}$ admits an exact
semidefinite programming (SDP) formulation~\cite{huang2026measured}.
For real orders in the same range, accelerated projected-gradient methods
provide an alternative, with convergence guarantees established for positive
definite $\rho$ and $\sigma$~\cite{huang2025accelerated}.

The measured R\'enyi divergences are superadditive and, in general, non-additive, hindering their direct application in the asymptotic analysis of quantum information tasks.
They still find a number of uses in the one-shot setting, which we will return to in Section~\ref{sec:one-shot}.
Their usefulness in asymptotic quantum information is restored by regularization, which, in this case, has a closed-form solution (see~\eqref{eq:regularized_measured_Rényi_above_1/2} and~\eqref{eq: Reverse sandwiched from regularization} below). Indeed, as discussed in Section~\ref{sec:class/reg}, regularization is a natural and minimal way to obtain additive quantities from non-additive ones. This insight leads to important families of quantum R\'enyi relative entropies.

\subsubsection{Sandwiched R\'enyi relative entropy}
\label{sec: sandwiched Renyi relative entropy}

The sandwiched Rényi relative entropy was introduced in~\cite{lennert13renyi} and~\cite{wilde13sandwich}.\footnote{The main definitions and some of the results in~\cite{lennert13renyi} were already publicly presented in~\cite{tomamichel12smooth_tutorial}, and later published in~\cite{dupuis13noisystorage,lennert13renyi_thesis}. \textcite{wilde13sandwich} coined the term ``sandwiched'' with reference to the structure of the trace term and provided the first application of the divergence.} It takes the form
\begin{align}\label{eq:sandwiched-renyi}
    \widetilde{D}_{\alpha}(\rho\|\sigma) = \frac{1}{\alpha - 1} \log \tr \left[ \left( \sigma^{\frac{1-\alpha}{2\alpha}} \rho \sigma^{\frac{1-\alpha}{2\alpha}} \right)^{\alpha} \right] \,,
\end{align}
provided that the natural support conditions between the two states $\rho$ and $\sigma$ are satisfied: specifically, the quantity evaluates to $ \infty $ if $ \rho $ and $ \sigma $ are orthogonal for $ \alpha < 1 $, or if the support of $ \sigma $ fails to contain that of $ \rho $ for $\alpha>1$.
In the limit $\alpha \to 1$, the sandwiched R\'enyi relative entropy reduces to the Umegaki relative entropy $D(\rho\|\sigma)$, while, for
$\alpha=1/2$ and $\alpha=\infty$, it equals the corresponding measured R\'enyi relative entropy, entailing in particular that $\widetilde{D}_{1/2} = - \log F$ and $\widetilde{D}_{\infty} = D_{\max}$.

The sandwiched R\'enyi relative entropies found early use in bounds
revealing the exponential decay of the success probability in quantum
information tasks in the so-called strong converse
setting~\cite{wilde13sandwich}.
Their role goes beyond providing bounds: Mosonyi and
Ogawa~\cite{mosonyi2015quantum} showed that they characterize the
exact strong converse exponent for binary quantum hypothesis testing
(see Application~\ref{app:HT-d}).
They also characterize the exact strong converse exponent for
classical--quantum channel coding~\cite{mosonyi2017strong}.
For entanglement-assisted classical communication, Gupta and
Wilde~\cite{gupta_2015} derived a lower bound on the strong converse
exponent in terms of these quantities, which was subsequently shown
to be tight by Li and Yao~\cite{li2023strong}.
For quantum communication, Tomamichel, Wilde, and
Winter~\cite{tomamichel_2017} used sandwiched R\'enyi generalizations
of the Rains information to show that a channel's Rains information
is a strong converse rate, with the transmission fidelity decaying
exponentially at higher rates.
Beyond the strong converse setting, the sandwiched R\'enyi relative
entropies characterize the optimal exponential decay of the smoothing
error for the max-relative entropy, with smoothing measured in
purified distance~\cite{li_2023}.
They also define generalized free energies that give necessary
conditions for catalytic thermal state
transformations~\cite{brandao_2015-2}.
Conditional entropies and mutual information derived from
$\widetilde{D}_\alpha$ also yield exponential strong converse bounds
for state redistribution and the classical communication cost of
measurement compression with quantum side
information~\cite{leditzky_2016}.
We will return to these applications in
Sections~\ref{subsec:renyi_conditional}--\ref{subsec:renyi_mutual}.

\begin{appbox}{HT}{Strong converse exponents of state discrimination}{app:HT-d}
Consider discriminating $\rho_0^{\otimes n}$ from $\rho_1^{\otimes n}$, with $\rho_0$ as the null hypothesis. Requiring the type~II error probability (i.e., the probability of incorrectly concluding that the state is $\rho_0$) to decay at an exponential rate $r$ that is strictly larger than the Stein exponent $D(\rho_0\|\rho_1)$ forces, owing to the discussion in Application~\ref{app:HT}, the type~I error probability away from zero\,---\,in fact, it will converge exponentially to one. The strong converse exponent $\mathrm{sc}(r)$ captures how fast this happens: it is defined as the smallest achievable exponential decay rate of one minus the type~I error probability under the constraint that $r > D(\rho_0\|\rho_1)$.

\smallskip

\textcite{hayashi2006book} outlined a limiting characterization of this exponent using spectral pinching:
\begin{align*}
\mathrm{sc}(r)
&= \sup_{\alpha>1}
   \bigg\{
   \frac{\alpha-1}{\alpha}
   \bigg( r
   -\lim_{n\to\infty}\frac{1}{n}
   D_\alpha\big(
   \cP_{\rho_1^{\otimes n}}(\rho_0^{\otimes n})
   \big\| \rho_1^{\otimes n}
   \big)
   \bigg)
   \bigg\}.
\end{align*}
Here, $\smash{\cP_{\rho_1^{\otimes n}}}$ denotes spectral pinching with respect to $\rho_1^{\otimes n}$. Since its output commutes with $\rho_1^{\otimes n}$, $D_\alpha$ is simply the classical R\'enyi relative entropy evaluated in a common eigenbasis. \textcite{mosonyi2015quantum} identified the limit in this expression with $\widetilde{D}_\alpha(\rho_0\|\rho_1)$ for
$\alpha>1$, as per~\eqref{eq: sandwiched through pinching}, and established the resulting single-letter formula
\begin{align*}
\mathrm{sc}(r) = \sup_{\alpha>1} \left\{ \frac{\alpha-1}{\alpha} \bigl(r-\widetilde{D}_\alpha(\rho_0\|\rho_1)\bigr) \right\}.
\end{align*}

The proof of the strong-converse result reveals why the sandwiched R\'enyi relative entropy plays such a prominent operational role across several tasks. The key property is its minimality: indeed, for any additive quantum extension $\D_\alpha$ of the classical R\'enyi relative entropy satisfying data processing, the binary test $\{T_n,I-T_n\}$ yields
\begin{align*}
  \D_\alpha(\rho^{\otimes n}\|\sigma^{\otimes n})
  &\geq
  D_\alpha\!\left(
    (1-\alpha_n,\alpha_n)
    \,\big\|\,
    (\beta_n,1-\beta_n)
  \right),
\end{align*}
where $\alpha_n=\operatorname{Tr}[(I-T_n)\rho^{\otimes n}]$ and $\beta_n=\operatorname{Tr}[T_n\sigma^{\otimes n}]$. Since the right-hand side is classical, it is independent of the chosen quantum extension. For $\alpha>1$, the sandwiched divergence is the smallest such extension and therefore yields the sharpest bound obtainable through this argument. %
\end{appbox}

Importantly for many applications, for $\alpha \in [1/2,\infty]$, the sandwiched R\'enyi relative entropies equal the regularization of the measured R\'enyi relative entropies,
\begin{align} \label{eq:regularized_measured_Rényi_above_1/2}
     \widetilde{D}_\alpha(\rho\|\sigma) = \reg\big[D_{\alpha}^{\meas}\big](\rho\|\sigma) = \lim_{n \rightarrow \infty }\frac{1}{n}D^{\mathbb{M}}_\alpha(\rho^{\otimes n}\|\sigma^{\otimes n})\,,
\end{align}
making them the minimal additive extensions of the classical R\'enyi divergences.
This result was first proved for $\alpha=1$ by \textcite{hiai1991proper}, and subsequently by \textcite{mosonyi2015quantum} and \textcite{hayashi_2016-1} for $\alpha \in (1,+\infty)$ and $\alpha \in (1/2,1)$, respectively.
This property is key to many applications of the sandwiched R\'enyi relative entropies, as it allows one to translate many properties of the classical Rényi divergences directly into the quantum setting, naturally yielding many operational achievability results.
An important consequence of the minimality of the sandwiched R\'enyi relative entropies is the optimality of the bounds built from them, which notably includes their role in hypothesis testing as discussed in Application~\ref{app:HT-d}.

Another convenient regularized form of the sandwiched R\'enyi relative entropies, valid for all $\alpha\geq 0$, is
\begin{align}
\label{eq: sandwiched through pinching}
\widetilde{D}_\alpha(\rho\|\sigma) = \lim_{n \to \infty} \frac{1}{n}D_\alpha\big(\cP_{\sigma^{\otimes n}}(\rho^{\otimes n}) \| \sigma^{\otimes n}\big),
\end{align}
where $\cP_{\sigma^{\otimes n}}$ denotes the spectral pinching map with respect to $\sigma^{\otimes n}$. This result for all $\alpha \geq 0$ was established by~\textcite{hayashi_2016-1}, extending earlier results by~\textcite{mosonyi2015quantum} as well as the case $\alpha = 1$ treated by~\textcite{hiai1991proper}.
The result in equation~\eqref{eq: sandwiched through pinching} tells us how to construct, in principle, a measurement that achieves~\eqref{eq:regularized_measured_Rényi_above_1/2}: it is a projective (von Neumann) measurement, with corresponding basis obtained by joining the eigenbases of the restrictions of $\rho^{\otimes n}$ to each eigenspace of $\sigma^{\otimes n}$.

The mathematical structure of the sandwiched R\'enyi divergences allows for connections with non-commutative $L_p$ spaces~\cite{beigi_2013,berta18}. This helped establish a number of their properties, including in particular the data-processing inequality (DPI):
$\widetilde{D}_\alpha$ indeed satisfies the DPI for $\alpha \geq 1/2$. This was shown first for the regime $\alpha \in (1,2]$ by~\textcite{lennert13renyi_thesis} and~\textcite{wilde13sandwich}, then for the full parameter range by~\textcite{frank_2013} and independently by \textcite{beigi_2013} for $\alpha \geq 1$. \textcite{lennert13renyi} observed via numerical counterexamples that the DPI fails for $\alpha<1/2$, while a fully analytical proof of this fact was later provided by~\textcite{berta2017variational}. By standard arguments, DPI is equivalent to joint convexity for $\alpha>1$ or joint concavity for $\alpha\in [1/2,1)$ of the function $(\rho,\sigma)\mapsto \exp\big((\alpha-1)\widetilde{D}_\alpha(\rho\|\sigma)\big)$.
Consequently, in the range $\alpha \geq 1/2$, the sandwiched Rényi relative entropy defines a relative entropy and satisfies a variety of fundamental properties shared by all relative entropies as discussed in Appendix~\ref{app:universal}. Finally, the family is monotonically increasing in the parameter $\alpha$~\cite{beigi_2013,lennert13renyi}.

A variational formula for $\widetilde{D}_\alpha$ was established by \textcite{frank_2013}, who proved that
\begin{align}
    \widetilde{Q}_{\alpha}(\rho \| \sigma)
    &= \inf_{W > 0}
    \Big\{
        \alpha \Tr\big[\rho W
        \big]  \notag \\
    &\qquad
        + (1 - \alpha) \Tr\big[
        (
            W^{-\frac{1}{2}}
            \sigma^{\frac{1-\alpha}{\alpha}}
            W^{-\frac{1}{2}}
        )^{\frac{\alpha}{1-\alpha}}
        \big]
    \Big\}\label{eq: sandwiched variational form}\\
    &= \inf_{W > 0}
    \Big\{
        \Big( \Tr\big[
        \rho W
        \big]\Big)^\alpha \notag\\
        &\qquad \times
    \Big( \Tr\big[
        (
            W^{-\frac{1}{2}}
            \sigma^{\frac{1-\alpha}{\alpha}}
            W^{-\frac{1}{2}}
        )^{\frac{\alpha}{1-\alpha}}
        \big]\Big)^{1 - \alpha}
    \Big\},
\end{align}
where $\widetilde{Q}_{\alpha}(\rho \| \sigma) = \exp{\big((\alpha-1)\widetilde{D}_{\alpha}(\rho \| \sigma)\big)}$. For $\alpha>1$, the same variational expression holds, with the infimum replaced by a supremum. This variational representation has been used to prove the data-processing inequality and also to characterize the equality cases~\cite{leditzky2017equality}.
An infinite-dimensional extension of this variational
representation was used by~\textcite{mosonyi2023strong}
to establish finite-dimensional approximation properties
underlying the exact strong converse exponent of quantum
state discrimination.

One distinguishing property of the sandwiched R\'enyi  divergences is that, unlike their classical counterparts, they do not satisfy the symmetry relation $D_{\alpha}(p\|q) = \frac{\alpha}{1-\alpha} D_{1-\alpha}(q\|p)$.
Because of this, for $\alpha \in [0, 1/2]$, a regularization of the measured Rényi relative entropy leads to a notion of so-called \emph{reverse} sandwiched Rényi relative entropy. Explicitly, we have %
\begin{align}
\label{eq: Reverse sandwiched from regularization}
     \lim_{n \rightarrow \infty} \frac{1}{n}D^{\mathbb{M}}_\alpha(\rho^{\otimes n}\|\sigma^{\otimes n}) = \frac{\alpha}{1-\alpha}\, \widetilde{D}_{1-\alpha}(\sigma\|\rho)
\end{align}
as first reported by \textcite{mosonyi2024geometric}.
The importance of this follows from the fact that regularization always yields the minimal (weakly) additive extensions of divergences --- specifically,
if $\D_\alpha$ is any quantum relative entropy that coincides with the classical Rényi relative entropy $D_\alpha$ on commuting states and satisfies data processing, then
\begin{equation}
    \D_\alpha(\rho\|\sigma) \geq  \begin{cases}
    \frac{\alpha}{1-\alpha}\, \widetilde{D}_{1-\alpha}(\sigma\|\rho), & \alpha \in [0,1/2]\\[1ex]
         \widetilde{D}_{\alpha}(\rho\|\sigma), & \alpha \geq 1/2\,.
    \end{cases}
\end{equation}

Although they naturally emerge in the regime of $\alpha \leq 1/2$, the reverse sandwiched relative entropies, $\frac{\alpha}{1-\alpha}\, \widetilde{D}_{1-\alpha}(\sigma\|\rho)$, can be studied for general $\alpha$. In the limit $ \alpha\to 1$,  they converge to \poscite{keyl2006quantum} rate function $D^\star$~\cite{audenaert2015alpha}. \textcite{keyl2006quantum} originally introduced this quantity as the large-deviation rate function arising in quantum state estimation. The underlying representation-theoretic framework was subsequently extended to quantum measurement settings~\cite{noetzel2015permutation}, as well as to moment-map estimation~\cite{franks2020minimal,botero2021large}. It has since found applications including transformations of pairs of quantum states~\cite{lipka-bartosik_2024}, quantum hypothesis testing~\cite{lipka-bartosik_2024,ji2025beyond,beigi2025additivity,hayashi2026operational}, and quantum thermodynamics~\cite{watanabe2026reliability}.
$D^\star$ coincides with the KL divergence on classical states. However, it may be strictly smaller than the Umegaki relative entropy and hence does not satisfy the data-processing inequality in general.

For optimization problems involving the sandwiched R\'enyi relative entropy, recent advances have established self-concordant barriers for the associated trace functional for $\alpha\in[1/2,1)\cup(1,2]$, enabling efficient minimization of the divergence under semidefinite constraints via interior-point methods~\cite{he2026operator}.

\subsubsection{Petz--R\'enyi relative entropy}
\label{sec: Petz Renyi relative entropy}

A different generalization of the classical Rényi relative entropy was studied by~\textcite{petz1986quasi}:
\begin{align}\label{eq:Petz-Renyi}
    \widebar{D}_\alpha(\rho\|\sigma) = \frac{1}{\alpha-1}\log{\tr\left[\rho^\alpha\sigma^{1-\alpha}\right]}\,,
\end{align}
under the standard support conditions.
This definition is typically motivated from a mathematical perspective, as it corresponds to a generalization of the classical $D_\alpha$ where the role of the likelihood ratio between classical probability distributions is replaced by its canonical extension to the non-commutative setting of operator algebras, namely, the relative modular operator $\Delta_{\rho|\sigma} = \cL_\rho \cR_\sigma^{-1}$ with $\cL_\rho$ and $\cR_\sigma$ denoting the left- and right-multiplication super-operators. A more direct justification is that~\eqref{eq:Petz-Renyi} is the most immediate quantum extension of the classical definition~\eqref{eq:classical_Renyi_divergence_2}, obtained by replacing $p$ with $\rho$, $q$ with $\sigma$, and the sum over symbols with the trace.

On top of its mathematical motivation, the Petz--R\'enyi relative entropy has important operational and physical applications.
In quantum hypothesis testing, for example, it determines the quantum Hoeffding and Chernoff exponents\,---\,see Application~\ref{app:HT-b}.
These results underpin quantitative demonstrations of quantum advantages in target detection~\cite{tan2008illumination} and optical-memory readout~\cite{pirandola2011reading}. It also characterizes fluctuations of dissipated work in driven quantum systems through the Petz divergence between the final state and the corresponding equilibrium state~\cite{wei2017relations}.

In the limit $\alpha \to 1$, the Petz--Rényi relative entropy reduces to the standard (Umegaki) relative entropy, while for $\alpha=0$, it reduces to the min-relative entropy~\cite{datta2009min}.
The Petz--Rényi relative entropies satisfy the data-processing inequality for $\alpha \in [0,2]$~\cite{petz1986quasi}.
This can also be seen as a consequence of Ando’s convexity theorem~\cite{ando1979concavity}, which characterizes the conditions under which the trace term is jointly concave or convex, together with the well-known fact that joint convexity is equivalent to the DPI. Moreover, they violate the DPI for $\alpha>2$~(\citealp{hiai2013concavity}; see also~\citealp{mosonyi2015quantum}). A well-known consequence of the DPI is that the function $(\rho,\sigma)\mapsto \exp((\alpha-1)\widebar{D}_\alpha(\rho\|\sigma))$ is jointly concave for $\alpha \in [0,1)$ and jointly convex for $\alpha \in (1,2]$, entailing that the Petz--Rényi relative entropy is jointly convex in the arguments for $\alpha \in [0,1]$ and jointly quasi-convex for $\alpha \in (1,2]$. Consequently, in the range $\alpha \in [0,2]$, the Petz--Rényi relative entropy defines a relative entropy and satisfies the properties in Appendix~\ref{app:universal}.

An important representation introduced by~\textcite{nussbaum2009chernoff} allows one to express the Petz--Rényi relative entropies between two quantum states in terms of the classical R\'enyi relative entropies between two particular probability distributions constructed from the states, known as the Nussbaum--Szkoła distributions. This representation also allows for lifting properties of classical R\'enyi relative entropies.
Explicitly, writing the states in their spectral decompositions as $\rho=\sum_x \lambda_x\ketbra{e_x}{e_x}$ and $\sigma=\sum_x \mu_x\ketbra{f_x}{f_x}$, define
\begin{align}
    p^{\rho,\sigma}(x,y) = \lambda_x|\braket{e_x}{f_y}|^2, \, \ q^{\rho,\sigma}(x,y) = \mu_y|\braket{e_x}{f_y}|^2 \,.
\end{align}
Then, for all $\alpha$ it holds that
\begin{align}
    \widebar{D}_\alpha(\rho\|\sigma) = D_\alpha(p^{\rho,\sigma}\| q^{\rho,\sigma})\,.
\end{align}
This representation shows that the Petz--R\'enyi relative entropy is monotone in $\alpha$, a fact discussed, e.g., by \textcite{mosonyi2011quantum}.
The Nussbaum--Szkoła distributions find key use in proving the optimality of the exponential bounds obtained from $\widebar{D}_\alpha$~\cite{Nagaoka2006converse,nussbaum2009chernoff}. On the other hand, the  fact that these bounds are achievable in practice often relies on the celebrated inequality of Audenaert et al.~\cite{audenaert2007discriminating} of the form
\begin{equation}\begin{aligned}
  1 - \frac12 \norm{\rho - \sigma}_1 \leq \exp\big((\alpha-1)\widebar{D}_\alpha(\rho \| \sigma)\big ),
\end{aligned}\end{equation}
which shows that the asymptotic behavior of the quantum trace distance can be tightly bounded by the Petz--R\'enyi divergences.

A variational expression for $\widebar{D}_\alpha$ can be obtained as a special case of a general variational form for the $\alpha$-$z$ Rényi relative entropy, which we will discuss in Section~\ref{subsec: other generalizations}.

\bigskip

\begin{appbox}{HT}{Error exponents of state discrimination}{app:HT-b}
In the setting of asymmetric hypothesis testing, we may be interested in fixing the rate at which the error probability of incorrectly concluding that the state is $\rho_0$ (type II error) decays, by requiring it to scale as $\sim \exp(-rn)$ in terms of the number of copies $n$,
where the exponent $r$ is smaller than the Umegaki relative entropy between $\rho_0$ and $\rho_1$. %
In this setting, the Petz--Rényi relative entropy characterizes the achievable decay rate of the probability of the other (type I) error, which consists in incorrectly concluding that the state is $\rho_1$. This probability scales as $\sim \exp(-e(r)\,n)$, where $e(r)$ is called the direct exponent, or also the error exponent. The quantum Hoeffding bound~\cite{Nagaoka2006converse, Hayashi2007error} states that this is given by
\begin{align*}
e(r) = \sup_{\alpha \in (0,1)}\left\{\frac{1-\alpha}{\alpha}\big(\widebar{D}_\alpha(\rho_0\| \rho_1) - r\big)\right\}\,.\notag
\end{align*}
\smallskip

An analogous question can be asked in the setting of symmetric hypothesis testing, where we require that both errors decay equally fast --- equivalently, we aim to maximize the rate of decay of their sum.
The quantum Chernoff bound states that this sum decays with the exponent given by~\cite{audenaert2007discriminating, nussbaum2009chernoff}
\begin{align*}
C(\rho_0,\rho_1) &= \sup_{\alpha\in (0,1)} \big\{ (1-\alpha)\;\widebar{D}_\alpha(\rho_0\| \rho_1)\big\} , \\
&= -\log \inf_{\alpha\in (0,1)} \Tr\big[ \rho_0^\alpha \rho_1^{1-\alpha}\big]\, ,
\end{align*}
also called Chernoff exponent.

\end{appbox}

\begin{inplacefigure}
    \centering
{\includegraphics[width=\linewidth]{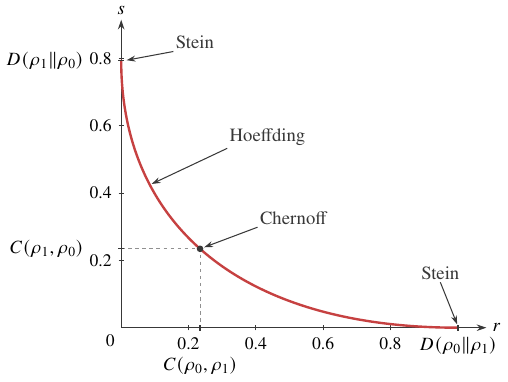}}
    \caption{Error exponents for the example states
\vskip-5pt
\begin{equation*}
 \rho_0 = \tfrac{1}{32}
 \left(\begin{smallmatrix}
 13 & -7 & 0 \\[1pt]
 -7 & 13 & 0 \\[1pt]
 0 & 0 & 6
 \end{smallmatrix}\right), \quad
 \rho_1 = \tfrac{1}{64}
 \left(\begin{smallmatrix}
 20 & 12 & 1 \\[1pt]
 12 & 20 & 0 \\[1pt]
 1 & 0 & 24
 \end{smallmatrix}\right).
\end{equation*}%
The horizontal axis represents the error exponent $r$ for the Type II error, defined by the asymptotic behavior $\beta_n \approx 2^{-r n}$, where $\beta_n$ is the probability of incorrectly deciding $\rho_0$. The vertical axis represents the error exponent $s$ for the Type I error, defined by $\alpha_n \approx 2^{-s n}$, where $\alpha_n$ is the probability of incorrectly deciding $\rho_1$. The red curve represents the Hoeffding exponent in Application~\ref{app:HT-b}, plotted as a function of $r$. The Chernoff exponent
in Application~\ref{app:HT-b} is obtained when the two error probabilities decay at the same exponential rate, i.e., $r=s$. The Hoeffding curve vanishes when $r$ reaches the Umegaki relative entropy $D(\rho_0\|\rho_1)$, which corresponds to Stein's exponent described in Application~\ref{app:HT} and thus to the largest Type II error exponent for which the Type I error still vanishes asymptotically. Finally, the curve meets the vertical axis at $D(\rho_1\|\rho_0)$, by the same reasoning with the roles of the two states exchanged. In particular, in this regime, the Type I error converges exponentially to one, with a rate given by the strong converse exponent, as discussed in Application~\ref{app:HT-d}.}
    \label{fig:error exp}
\end{inplacefigure}


For optimization problems involving the Petz--R\'enyi relative entropy, SDP formulations of the associated trace functional enable minimization of the divergence under semidefinite constraints for rational $\alpha\in(0,1)\cup(1,2]$~\cite{fawzi2017lieb}, while interior-point methods also support real orders in the same range~\cite{fawzi2023optimal}.

\subsubsection{Relating sandwiched and Petz--R\'enyi divergences}\label{sec:relating_sandwich_Petz}

The additive quantum extensions of R\'enyi relative entropies that we discussed so far form the backbone of many information-theoretic tasks in quantum theory.
They generally find applications in complementary regimes:
the study of error exponents of so-called packing-type problems in information theory (such as hypothesis testing, source compression, or channel coding)
typically relies on R\'enyi divergences in the range $\alpha \in (0,1)$, and it is common to encounter the Petz--R\'enyi divergences $\widebar{D}_\alpha$ in such contexts; on the other hand, the asymptotics of so-called covering-type problems (such as randomness extraction, decoupling, or simulation of quantum channels) typically depend on the orders $\alpha > 1$, and there the sandwiched R\'enyi divergences $\widetilde{D}_\alpha$ often find use.
This disjuncture can be related to the fact that the corresponding divergences satisfy the data-processing inequality in these complementary regimes of the parameters~$\alpha$.
Note, however, that this is not a definitive rule~\cite{li2024operational,rubboli2026strong}, and furthermore that the roles of the different entropies are often reversed in the study of strong converse exponents, as we saw in Applications~\ref{app:HT-d}--\ref{app:HT-b}.

The minimality of the sandwiched R\'enyi divergences ensures that $\widebar{D}_\alpha(\rho\|\sigma) \geq \widetilde{D}_\alpha(\rho\|\sigma)$ for all $\alpha \geq \frac12$; more directly,
the inequality
\begin{align}
\widebar{D}_\alpha(\rho\|\sigma) &\geq \widetilde{D}_\alpha(\rho\|\sigma)\qquad \forall\ \alpha\geq 0
\label{eq:Petz_sandwiched_inequality}
\end{align}
is a consequence of the Araki--Lieb--Thirring trace inequalities~\cite{datta_2014,wilde13sandwich}. The inequality is, in general, strict: for $\alpha \neq 1$, $\widetilde{D}_\alpha(\rho\|\sigma) = \widebar{D}_\alpha(\rho\|\sigma)$ holds if and only if $[\rho,\sigma]=0$~\cite{hiai_1994}.  In fact, the gap can be arbitrarily large for any $\alpha>1$, as the following example demonstrates.

\begin{example}
Fix $\alpha>1$ and let $\sigma_d = \big(1-\frac{1}{d}\big)\proj{0} + \frac{1}{d}\proj{1}$ and $\ket{\psi_{d}} = \big(1-d^{-\frac{\alpha-1}{\alpha}}\big)^{1/2}\ket{0} + \big(d^{-\frac{\alpha-1}{\alpha}}\big)^{1/2}\ket{1}$. Then, for large $d$,
\begin{align}
&\widebar{D}_\alpha\big(\proj{\psi_d} \big\|\sigma_d\big) \sim ((\alpha-1)/\alpha)\log d\,,\;\;\\
&\widetilde {D}_\alpha\big(\proj{\psi_d} \big\|\sigma_d\big)   \sim (\alpha/(\alpha-1)) \log{2} \,,
\end{align}
demonstrating that $\widebar{D}_\alpha\big(\proj{\psi_d} \big\|\sigma_d\big) \gg \widetilde{D}_\alpha\big(\proj{\psi_d} \big\|\sigma_d\big)$ for large $d$. The gap in~\eqref{eq:Petz_sandwiched_inequality}, therefore, can be arbitrarily large for any $\alpha>1$.
\end{example}

Reverse inequalities have also been studied by~\textcite{iten_2017} and~\textcite{wilde_2018-1}, who showed that
\begin{align}
    &\widetilde{D}_{\frac{1}{2-\alpha}}(\rho\|\sigma) \geq \widebar{D}_\alpha(\rho\|\sigma) \quad \alpha\in[0,2] \label{eq:reverse_inequality_Petz_sandwiched_1} \\
    &\alpha \widebar{D}_\alpha(\rho\|\sigma) \leq \widetilde{D}_{\alpha}(\rho\|\sigma) \quad \alpha\in[0,1] \label{eq:reverse_inequality_Petz_sandwiched_2}
\end{align}
for $\alpha\in[0,2]$.

\begin{example}
In an attempt to generalize~\eqref{eq:reverse_inequality_Petz_sandwiched_1}, one may wonder whether, for $\alpha>2$, it is possible to upper bound $\widebar{D}_{\alpha}$ in terms of a state-independent multiple of some other $\widetilde{D}_\beta$. However, this turns out not to be possible. Indeed, defining $\phi_\eps = \sqrt{1-\eps}\ket{0} + \sqrt{\eps} \ket{1}$ and $\omega_\eps = (1-\eps) \proj{0} + \eps \proj{1}$, we have, for $\alpha>2$ and $\eps\to 0^+$,
\begin{align}
\widebar{D}_{\alpha}(\phi_\eps\|\omega_\eps) &\sim \frac{\alpha-2}{\alpha-1} \log\frac1\eps \to +\infty\, ,\\
\widetilde{D}_\beta(\phi_\eps\|\omega_\eps) &\leq D_{\max}(\phi_\eps\|\omega_\eps) = \log 2\, ,
\end{align}
showing that, for $\alpha>2$, no inequality of the form $\widebar{D}_{\alpha}(\rho\|\sigma) \leq C_{\alpha,\beta}\, \widetilde{D}_\beta(\rho\|\sigma)$ can hold in general.
\end{example}

Nevertheless, surprising and close connections between the two types of divergences emerge from entropic duality relations, which we will discuss in Section~\ref{subsec:renyi_conditional}.

A unifying view of the sandwiched and Petz--R\'enyi relative entropies can be obtained through the class of $\alpha$–$z$ Rényi relative entropies, which encompasses the two families. It was first investigated by~\textcite{jaksic2011entropic} in the context of quantum dynamical systems and entropic fluctuations in non-equilibrium statistical mechanics, and was later studied from an information-theoretic perspective in~\cite{audenaert2015alpha}. The $\alpha$–$z$ Rényi relative entropy is defined as
\begin{align}
\label{def: alpha-z relative entropies}
    D_{\alpha,z}(\rho \| \sigma)=
\frac{1}{\alpha-1}\log{\tr \left[\left(\rho^\frac{\alpha}{2z}\sigma^\frac{1-\alpha}{z}\rho^\frac{\alpha}{2z}\right)^z\right]}\,.
\end{align}
For non–full-rank states, the standard support conditions apply.
The range of parameter values for which it satisfies the data-processing inequality was explored by \textcite{hiai2013concavity, carlen2018inequalities}, and fully characterized by \textcite{zhang2020wigner}. In the limit $\alpha \to 1$ ($z\neq 0$), it reduces to the Umegaki relative entropy; when $z=1$ and $z=\alpha$, it reduces to the Petz and sandwiched Rényi relative entropies, respectively; for $z=1-\alpha$, it reduces
to the reverse sandwiched relative entropies, and for $z\to +\infty$ it gives the so-called log-Euclidean relative entropies, which we will discuss shortly. %

The $\alpha\text{-}z$ Rényi relative entropies satisfy several monotonicity properties in the parameters $\alpha$ and $z$.
An application of the Araki--Lieb--Thirring inequality shows that, as a function of $z$, the $\alpha\text{-}z$ Rényi relative entropies are monotonically increasing for $\alpha < 1$ and monotonically decreasing for $\alpha > 1$ \cite{lin2015investigating}. The monotonicity in $\alpha$ has been established by \textcite{rubboli2024mixed} for $z = |\alpha - 1|$ and for general fixed $z$ by \textcite{hiai2024log}.
They admit a variational representation extending that of the sandwiched relative entropies in~\eqref{eq: sandwiched variational form}, as established by \textcite{zhang2020wigner}, \textcite{hiai2022}, and \textcite{mosonyi2023strong}.
Although these quantities do not yet admit a direct operational interpretation in quantum hypothesis testing, they have nevertheless proven useful in several contexts%
, such as in providing conditions for transformations between specific states~\cite{verhagen2025conditions} or as dual quantities to a class of conditional entropies that characterize the strong converse exponent for randomness extraction under purified distance~\cite{rubboli2026strong}.

\subsubsection{Maximal (prepared) R\'enyi relative entropy}
\label{sec:renyi/maximal}

As discussed in Section~\ref{sec:class/lifting}, quantum relative entropies can also be defined from classical ones in a fashion conceptually reversed from measurements, that is, by minimizing the classical relative entropy over all probability distributions that can be transformed into the corresponding quantum states.
The prepared Rényi relative entropy for $\alpha \in [0,\infty]$ is
\begin{align}
    \label{eq: prepared Renyi relative entropy}D^{\mathbb{P}}_\alpha(\rho\|\sigma) = \min_{p, q, \cF} D_\alpha(p\|q)\,.
\end{align}
where the minimization is over all probability distributions $p$ and $q$ of arbitrary (finite) dimension for which there exists a preparation channel $\cF$ such that $\cF(p) = \rho$ and $\cF(q) = \sigma$. Here, $D_\alpha$ is the classical R\'enyi relative entropy.

The above optimization problem admits a closed-form solution for $\alpha = \infty$, where $D^{\mathbb{P}}_{\infty} = D_{\max}$, and for $\alpha \in [0,2]$, where it reduces to the geometric Rényi relative entropy, defined as
\begin{align}
    \widehat{D}_\alpha(\rho\|\sigma) = \frac{1}{\alpha-1}\log{\tr\left[\sigma\left(\sigma^{-1/2}\rho\sigma^{-1/2}\right)^\alpha\right]} \,.
\end{align}
For states that do not have full support, the standard conventions apply for $\alpha \geq 1$; for $\alpha \in [0,1)$, instead, $\rho$ must be replaced by its absolutely continuous part with respect to $\sigma$~\cite{mosonyi2024geometric}. Namely, decomposing $\sigma = \sigma_{00} \oplus 0$ and writing $\rho = \left(\begin{smallmatrix} \rho_{00} & \rho_{01} \\ \rho_{10} & \rho_{11} \end{smallmatrix}\right)$
with respect to this decomposition,
we have,
\begin{equation}
\widehat{D}_\alpha(\rho\|\sigma) = \frac{1}{\alpha-1}\log{\tr\left[\sigma_{00}\left(\sigma_{00}^{-1/2}(\rho/\rho_{11})\sigma_{00}^{-1/2}\right)^\alpha\right]}\, ,
\end{equation}
where $\rho/\rho_{11} = \rho_{00} - \rho_{01} \rho_{11}^{-1} \rho_{10}$, with the inverse of $\rho_{11}$ calculated on its own support.

For $\alpha\rightarrow 1$, the geometric R\'enyi relative entropy converges to the Belavkin--Staszewski relative entropy~\cite{belavkin1982c}
\begin{align}
    \widehat{D}(\rho\|\sigma) = \tr\left[\rho \log{\left(\rho^{1/2}\sigma^{-1}\rho^{{1/2}}\right)}\right] .
\end{align}
The optimal probability distribution and preparation map are known explicitly, and were derived by \textcite{matsumoto2018maxdivergence}.
The geometric relative entropy contains the trace of the~\poscite{kubo1980means} $\gamma$-weighted geometric mean, defined by $A \#_{\gamma} B = A^\frac{1}{2}\big(A^{-\frac{1}{2}}B A^{-\frac{1}{2}}\big)^\gamma A^\frac{1}{2}$, from which it inherits several useful properties. Notably, these quantities are fully additive, even without regularization. The geometric Rényi relative entropies are monotonically increasing in $\alpha$, as they arise from the prepared relative entropy in~\eqref{eq: prepared Renyi relative entropy}, and the classical Rényi relative entropies are themselves increasing in $\alpha$. From the above considerations, it follows that they satisfy the DPI for $\alpha \in [0,2]$ and satisfy joint convexity properties in the arguments for the same range of the parameter~$\alpha$.

Since the geometric Rényi relative entropy is additive, it constitutes the maximal quantum generalization of the Rényi relative entropy for $\alpha\in[0,2]$.
For $\alpha>2$, the prepared Rényi relative entropies have no known closed-form expression and are subadditive but not additive. After regularization, however, they coincide with the $\alpha$-$z$ Rényi relative entropy at $z=\alpha-1$~\cite{rubboli2026maximal}, defined in~\eqref{def: alpha-z relative entropies}, i.e.,
\begin{align}
\label{eq:regularized_prepared_Rényi_above_2}
   \reg\big[D_{\alpha}^{\mathbb{P}}\big](\rho\|\sigma)
    = D_{\alpha,\alpha-1}(\rho\|\sigma)\,.
\end{align}
Since this quantity is additive under tensor products, it represents the maximal Rényi relative entropy for $\alpha>2$.

The above results together imply that, if $\D_\alpha$ denotes a quantum relative entropy that reduces to the classical Rényi relative entropy $D_\alpha$ for commuting states, then
\begin{equation}
    \D_\alpha(\rho\|\sigma)
    \leq
    \begin{cases}
        \widehat{D}_{\alpha}(\rho\|\sigma), & \alpha\in[0,2],\\[1ex]
        D_{\alpha,\alpha-1}(\rho\|\sigma), & \alpha\geq2\,.
    \end{cases}
\end{equation}

In Fig.~\ref{fig:important_entropies}, we illustrate the relationships among the most relevant entropic measures discussed in this work, while Table~\ref{tab:renyi-divergence-properties} summarizes the parameter ranges for which the Rényi information measures satisfy the data-processing inequality and indicates whether they are additive.

\begin{inplacefigure}
\centering
\begin{tikzpicture}

    \node at (0,0) {\includegraphics[width=0.45\textwidth]{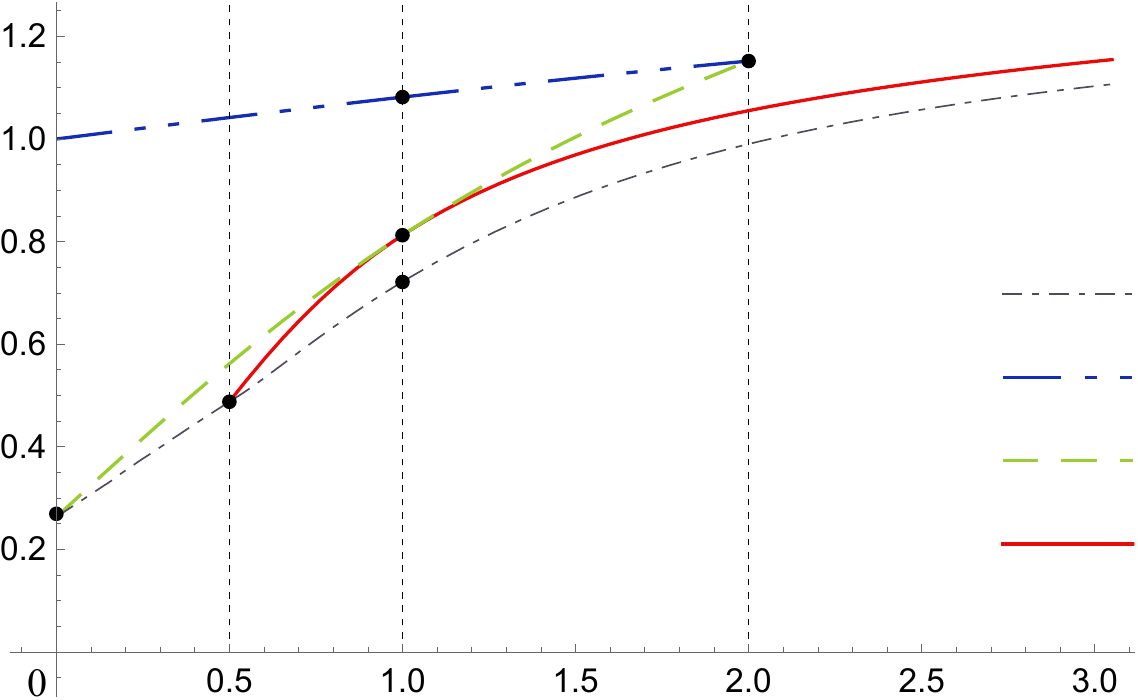}};

    \node at (-3.17,-1.3) {\small$D_{\min}$};
    \node at (-2,-0.58) {\small$-\log{F}$};
    \node at (-1.34,0.96) {\small$D$};
    \node at (-0.83,0.40) {\small $D^{\mathbb{M}}$};
    \node at (-0.98,2.08) {\small$\widehat{D}$};
    \node at (1.05,2.3) {\small$\widebar{D}_2$};
    \node at (2.8,2.18) {\small $D_{\max}$};
    \draw[->] (3.2,2.18) -- (3.5,2.20)
    node[midway, right, sloped] {\tiny $\;+\infty$};
    \node at (2.7,-1.33) {\small$\widetilde{D}_{\alpha}$};
    \node at (2.7,-0.755) {\small$\widebar{D}_{\alpha}$};
    \node at (2.7,-0.13) {\small$\widehat{D}_{\alpha}$};
    \node at (2.7,0.46) {\small$D^{\mathbb{M}}_\alpha$};
    \node at (4.2,-2.2) {\large$\alpha$};

\end{tikzpicture}

\caption{Prominent divergences evaluated for the states defined in Figure~\ref{fig:all_entropies}.
The smallest and largest quantities are given by the min- and max-relative entropies, $D_{\min}$ and $D_{\max}$, respectively. For fixed $\alpha$, the smallest relative entropy is given by the sandwiched R\'enyi relative entropy for $\alpha \geq 1/2$. While the measured relative entropy is smaller than the sandwiched relative entropy, it is however not additive.
Within the range $\alpha \in [0,2]$, the geometric relative entropy is the maximal relative entropy.}
\label{fig:important_entropies}
\end{inplacefigure}

\begin{table*}[t]
\setlength{\tabcolsep}{5pt}
\renewcommand{\arraystretch}{1.25}
\begin{tabular*}{\textwidth}{@{\extracolsep{\fill}}lllllll@{}}
Name & Symbol
& \shortstack{DPI\\$\alpha<1$}
& \shortstack{Limit\\$\alpha\to1$}
& \shortstack{DPI\\$\alpha>1$}
& Additive & Regularization \\
\hline%
\rule{0pt}{3ex}%
Measured & $D_\alpha^{\mathbb M}$ & $[0,1)$ & $D^{\mathbb M}$ & $(1,\infty]$ & $\times$ & \text{$D_{\alpha,1-\alpha}$ for $\alpha\in [0,\frac12]$; $\widetilde{D}_\alpha$ for $\alpha\in [\tfrac12,\infty]$} \\
Sandwiched & $\widetilde D_\alpha$ & $[\tfrac12,1)$ & $D$ & $(1,\infty]$ & $\checkmark$ & \\
Reverse sandwiched & $D_{\alpha,1-\alpha}$ & $[0,\tfrac12]$ & $D^\star$ & $\varnothing$ & $\checkmark$ & \\
Petz & $\widebar{D}_\alpha$ & $[0,1)$ & $D$ & $(1,2]$ & $\checkmark$ & \\
$\alpha$--$z$ & $D_{\alpha,z}$ & $[1-z,z]$ & $D$ & $[z,\min\{2z,z+1\}]$ & $\checkmark$ & \\
Geometric & $\widehat D_\alpha$ & $[0,1)$ & $\widehat D$ & $(1,2]$ & $\checkmark$ & \\
Prepared & $D_\alpha^{\mathbb P}$ & $[0,1)$ & $\widehat D$ & $(1,\infty]$ & $\times^{\mathrm b}$ & \text{$\widehat{D}_{\alpha}$ for $\alpha\in [0,2]$; $D_{\alpha,\alpha-1}$ for $\alpha\in [2,\infty]$}  \\
Log-Euclidean & $D_\alpha^\flat$ & $[0,1)$ & $D$ & $\varnothing$ & $\checkmark$ & \\
Sharp & $D_\alpha^\#$ & $\varnothing$ & $\widehat D$ & $(1,\infty]$ & $\times^{\mathrm c}$ & \text{$\widetilde{D}_\alpha$ for $\alpha\in (1,\infty]$} \\
Hockey-stick & $D_\alpha^{\rm HT}$ & $[0,1)$ & $D$ & $(1,\infty]$ & $\times^{\mathrm d}$ & \text{$\widebar{D}_\alpha$ for $\alpha \in [0,1)$, $\widetilde{D}_\alpha$ for $\alpha\in (1,\infty]$} \\
\end{tabular*}
\caption{
\textbf{Quantum R\'enyi divergences.}
The quantum generalizations of R\'enyi relative entropies are listed together with the parameter ranges in which they satisfy the data processing inequality (DPI), their limits $\alpha \to 1$, and an indication of their additivity. For non-additive divergences, their additive approximations are obtained by regularization. For $D_{\alpha,z}$, the parameter $z>0$ is held fixed when taking limits. For measures marked $\times$, additivity holds generally only at:
$^{\mathrm a}$ $\alpha=0,1/2,\infty$;
$^{\mathrm b}$ $\alpha\in[0,2]\cup\{\infty\}$; $^{\mathrm c}$ $\alpha=1,\infty$; $^{\mathrm d}$ $\alpha=0,1,\infty$.
}
\label{tab:renyi-divergence-properties}
\end{table*}

\subsubsection{Other quantum generalizations}
\label{subsec: other generalizations}
Several other quantum divergences have been introduced in the literature that aim to extend the properties of classical R\'enyi relative entropies.

Among them, the log-Euclidean Rényi relative entropy has played a central technical role in deriving error exponents in quantum information theory~\cite{mosonyi2017strong,mosonyi2020divergence,li2023strong,li2024operational}.
It is defined as~\cite{hiai1993golden,audenaert2015alpha,mosonyi2015quantum}
\begin{align}
    D_\alpha^\flat(\rho\|\sigma) = \frac{1}{\alpha-1}\log{\tr[\exp{(\alpha \log{\rho} +(1-\alpha) \log{\sigma})}]}\, .
\end{align}
For states without full support, replace $\log\rho$ and $\log\sigma$ by
$\Pi(\log\rho)\Pi$ and $\Pi(\log\sigma)\Pi$, where $\Pi$ projects onto
$\supp(\rho)\cap\supp(\sigma)$. Each logarithm is taken on its original
support, and the exponential and trace are evaluated on the intersection~\cite{mosonyi2024geometric}.
For $\alpha>1$, the divergence is $+\infty$ unless
$\supp(\rho)\subseteq\supp(\sigma)$.
The definition is motivated by a variational form of classical R\'enyi divergences, which can be expressed as an optimization of the KL divergence over a suitable domain~\cite{csiszar2003information}.
Replacing the classical with the Umegaki relative entropy then gives precisely the log-Euclidean R\'enyi divergence as
\begin{align}
\label{eq:variational_form_log_Euclidean}
    D_\alpha^\flat(\rho\|\sigma) = \frac{1}{1-\alpha} \min_{\omega}\{\alpha D(\omega\|\rho)+(1-\alpha)D(\omega\|\sigma)\}\, ,
\end{align}
where the minimization is over all normalized quantum states $\omega$.\footnote{This is immediate to establish once one observes that the objective function on the right-hand side of~\eqref{eq:variational_form_log_Euclidean} can be rewritten as $D(\omega\|\exp(\alpha \log \rho + (1-\alpha)\log \sigma))$, and this is minimized when $\omega$ is proportional to the operator in the second argument.}
The log-Euclidean R\'enyi relative entropy satisfies the data-processing inequality for $\alpha < 1$, as can be inferred from the variational representation combined with the DPI of the Umegaki relative entropy. It was shown by \textcite{mosonyi2017strong} that the DPI fails for $\alpha>1$.

\smallskip

Another type of R\'enyi divergence that found important technical applications is $D_\alpha^{\rm HT}$, based on \poscite{hirche2024quantum} hockey-stick integral representations. As part of the broader family of $f$-divergences, it will be discussed in Section~\ref{sec:integral-f}.

\smallskip

Yet another extension is the sharp Rényi divergence, introduced by \textcite{fawzi2021defining}. For $\alpha>1$, it is defined through the convex optimization problem
\begin{align}
D_\alpha^\#(\rho\|\sigma)
&=\frac{1}{\alpha-1}\log \inf_{A\geq0}
\left\{\tr[A] : \rho \leq
\sigma\mathbin{\#}_{1/\alpha}A\right\},
\end{align}
where $A \#_{\gamma} B$ is~\poscite{kubo1980means} $\gamma$-weighted geometric mean. The value is set to $+\infty$ unless $\supp(\rho)\subseteq\supp(\sigma)$. The joint concavity of the geometric mean makes the optimization convex and, for rational $\alpha$, it admits a semidefinite-program representation.
The sharp Rényi divergence agrees with the classical Rényi divergence on commuting states, satisfies the data-processing inequality, and is subadditive under tensor products; it fails, however, to be additive. At the single-copy level, it lies between the sandwiched and geometric R\'enyi divergences,
\begin{align}
\widetilde{D}_\alpha(\rho\|\sigma)
\leq D_\alpha^\#(\rho\|\sigma)
\leq \widehat{D}_\alpha(\rho\|\sigma)\, ,
\end{align}
while its regularization coincides with the sandwiched R\'enyi divergence: $\reg[D_\alpha^\#](\rho\|\sigma)
=\widetilde{D}_\alpha(\rho\|\sigma)$. \textcite{Bergh2021} proved that the $\alpha\to 1^+$ limit equals the Belavkin--Staszewski relative entropy:
$\lim_{\alpha\to 1^+} D_\alpha^\#(\rho\|\sigma) = \widehat{D}(\rho\|\sigma)$. In particular, remarkably, the regularization fails to commute with the limit $\alpha\to 1^+$.
The extension of $D_\alpha^\#$ to quantum channels satisfies a chain-rule inequality
that can be used to characterize strong converse exponents for
channel discrimination and to derive upper bounds on quantum
channel capacities assisted by two-way classical
communication~\cite{fawzi2021defining}.

\textcite{mosonyi_2023} studied
test-measured R\'enyi divergences, obtained by restricting the
optimization defining measured R\'enyi divergences to two-outcome
measurements. For $\alpha>1$, their regularizations coincide with
the sandwiched R\'enyi divergences, whereas for $0<\alpha<1$ they
can be strictly smaller than the classical R\'enyi divergence
even for commuting states.

Another construction, introduced by \textcite{mosonyi2024geometric},
is given by $\gamma$-weighted geometric relative entropies. These
are obtained by replacing the second argument of a general quantum
relative entropy with the $\gamma$-weighted geometric
mean~\cite{kubo1980means} of the two states and applying a suitable
normalization. Choosing the Umegaki relative entropy yields a
divergence that continuously interpolates between the Umegaki
and the Belavkin--Staszewski relative entropies.
The same authors also investigated barycentric R\'enyi divergences,
obtained by replacing the Umegaki relative entropy with a general
quantum relative entropy in the variational
representation~\eqref{eq:variational_form_log_Euclidean}.
When the underlying relative entropy satisfies the data-processing
inequality, so do these divergences for $\alpha\leq 1$; however,
it remains unknown whether they are generally additive.

\subsection{Rényi conditional entropies}
\label{subsec:renyi_conditional}

Several alternative notions of conditional entropy, distinct from the von Neumann conditional entropy, have been considered in the literature. The latter admits the representation $H(A|B) = H(AB) - H(B)$, which may suggest defining R\'enyi conditional entropies as $H_{\alpha}(A|B) = H_\alpha(AB) - H_\alpha(B)$, where $H_\alpha$ denotes a R\'enyi entropy. Despite appearing in the literature, this definition does not satisfy the data-processing inequality for conditional entropies, as the following example using classical distributions shows.

\begin{example}
Let $X$ and $Y$ be binary random variables with joint distribution given by
$p(0,0)=0$, $p(0,1)=p(1,0)=1/4$, and $p(1,1)=1/2$.
Consider the definition
\begin{align}
    H_2(X|Y) = H_2(XY) - H_2(Y) \,.
\end{align}
We have $H_2(X|Y) = \log (5/3) > \log (8/5) = H_2(X)$.
Thus, discarding the side information $Y$ decreases the conditional entropy, contrary to the expected behavior.
\end{example}

Such definitions via a chain rule are thus not consistent with our intuition that side information can only decrease entropy, and consequently find few applications.
Alternatively, the von Neumann conditional entropy can be expressed in the canonical forms
\begin{align}
    H(A|B)_\rho &= -D(\rho_{AB} \| \id_A \otimes \rho_B) \\
    &= \sup_{\sigma_B} -D(\rho_{AB} \| \id_A \otimes \sigma_B)\, ,
\end{align}
as discussed in Section~\ref{sec:class/canonical}. Replacing the Umegaki relative entropy with a R\'enyi divergence in the above often yields valid conditional entropies satisfying the axioms of Section~\ref{subsec: CE and MI}. Though not all conditional entropies arise in this way, most of the quantities discussed below are obtained through this construction.

\subsubsection{Conditional min- and max-entropy, and guessing}

Of particular relevance is the conditional min-entropy, built from R\'enyi divergence of order $\infty$. It admits an operational interpretation in terms of optimal guessing probabilities, as discussed in Application~\ref{app:guessing-a}. For classical distributions, two variants are commonly distinguished: an optimized (average) and a non-optimized (worst-case) version,
\begin{align}
H^\uparrow_{\min}(X|Y)_{p} &= \sup_{q_Y} \left\{ -D_{\max}(p_{XY}\| \one_X \otimes q_Y)\right\} \,,\\
 H^\downarrow_{\min}(X|Y)_{p} &= -D_{\max}(p_{XY}\| \one_X \otimes p_Y) \,.
\end{align}
Here and throughout, the upward and downward arrows indicate whether we maximize the second argument over distributions on $Y$ or instead simply take the marginal of $p$. Due to the first definition's ubiquity in cryptography and information theory, and consistently with most of the existing literature, in the following we omit the up-arrow symbol and simply write $H_{\min} $ for $H^{\uparrow}_{\min}$.

\begin{appbox}{guessing}{Guessing probability}{app:guessing-a}
Conditional entropies can be used to describe how well one can guess the outcome of a random variable $X$ given access to a potentially correlated system $Y$. In particular, the optimized and non-optimized min-entropies quantify the average and worst-case optimal guessing probabilities. To see this, note that upon observing the outcome $Y=y$, the optimal guessing strategy is clearly to guess the value $
x^*(y) = \arg\max_x p_{X|Y}(x|y) $.
The resulting average success probability is therefore~\cite{dodis2004fuzzy}
\begin{align*}
    \sum_y p_Y(y) \max_x p_{X|Y}(x|y) = \exp\!\big(-H^\uparrow_{\min}(X|Y)\big)
\end{align*}
The worst-case success probability (over realizations of the side information) is instead~\cite{renner2005simple}
\begin{align*}
    \max_y \max_x p_{X|Y}(x|y) = \exp\!\big(-H^\downarrow_{\min}(X|Y)\big)\, .
\end{align*}
Here, worst-case is understood from the perspective of secrecy, i.e., it gives the smallest upper bound on the optimal guessing
probability that holds for every $y$ with $p_Y(y)>0$.
Thus, the two min-entropies capture different operational criteria:
average vs.\ worst-case guessing performance.
\end{appbox}

Quantum generalizations of conditional min-entropy are widely used, particularly due to the development of quantum key distribution and the information-theoretic approach to security~\cite{renner2008security}. The optimized quantum conditional min-entropy is
\begin{equation}\begin{aligned}
H_{\min}(A|B)_{\rho} &= H^{\uparrow}_{\min}(A|B)_\rho \\
&= \sup_{\sigma_B \in \mathcal{S}(B)} \left\{-D_{\max}(\rho_{AB}\| \one _A \otimes \sigma_B)\right\} .
\end{aligned}\end{equation}
The conditional min-entropy admits a
semidefinite-program representation, given here in its primal and dual forms:
\begin{align}
  &\exp\!\big(-\!H_{\min}(A|B)_\rho\big) \nonumber\\
  &= \min \big\{ \tr[\sigma_B] : \id_A\otimes\sigma_B \geq \rho_{AB},\, \sigma_B \geq 0 \big\}\\
  &= \max \big\{ \tr[\rho_{AB}X_{AB}] : \tr_A[X_{AB}] \leq \id_B,\,  X_{AB} \geq 0 \big\} . \label{eq:min-entropy-sdp-dual}
\end{align}

For classical-quantum states, $\rho_{XB} = \sum_x p_X(x) \proj{x} \otimes \rho_B^x$, the interpretation as a guessing probability extends. Namely, we may consider the scenario in which one aims to guess the outcome of $X$ given access to the potentially correlated quantum system $B$, on which arbitrary quantum operations can be performed.
Specializing~\eqref{eq:min-entropy-sdp-dual}, we see that the optimal probability that the agent correctly identifies the label $x$ is related to the min-entropy by~\cite{konig2009operational}
\begin{align}
    \exp\!\big(-H_{\min}(X|B)_{\rho} \big) = \max_{\{E_x\}_x} \sum_x p(x) \tr\big(E_x \rho_B^x\big) \, ,
\end{align}
where the optimization is over all quantum measurements $\{E_x\}_x$ on the system $B$.
The fully quantum conditional min-entropy further admits operational interpretations in terms of the maximal fidelity with the maximally entangled state achievable by local operations, as discussed in Application~\ref{app:singletfraction}.

The complementary max-entropy $H_{\max}$ then emerges from the notion of entropic duality.
Given a tripartite pure state $\rho_{ABC}$, \textcite{konig2009operational} established the duality relation
\begin{align}
\dual[H_{\min}] = H_{\max}
\end{align}
where the max-entropy is defined in terms of the square fidelity as
\begin{align}\label{eq:rightafter}
H_{\max}(A|B)_{\rho} = \max_{\sigma_B}\log F(\rho_{AB},\one_A \otimes \sigma_B) \,.
\end{align}

While $H_{\min}$ and $H_{\max}$ are referred to as the min- and max-entropy, respectively, and their relation is commonly termed min–max entropy duality, these quantities do not correspond to the minimal and maximal conditional entropies in the sense adopted in Section~\ref{sec: classification of conditional entropy}; these roles are played by $H^{\downarrow}_{\min}$ and $H_{\max}^{\uparrow}$, respectively, defined as
\begin{align}
H^{\downarrow}_{\min}(A|B)_{\rho} &= \log d_A - D_{\max}(\rho_{AB}\|\pi_A \otimes \rho_B) \,, \ \  \\
H^{\uparrow}_{\max}(A|B)_{\rho} &= \log d_A - \inf_{\sigma_B} D_{\min}(\rho_{AB}\|\pi_A \otimes \sigma_B) \,.
\end{align}
Here, $D$, $D_{\min}$ and $D_{\max}$ denote the Umegaki, min- and max-relative entropies introduced in~\eqref{eq:intro-D}--\eqref{eq:intro-Dmax}.
In particular, $H^{\downarrow}_{\min} \neq H_{\min} = H^\uparrow_{\min}$ and $H_{\max}^{\uparrow} \neq H_{\max} \neq H^\downarrow_{\max}$.
These quantities are also connected via a duality relation
\begin{align}
\dual\big[H_{\min}^\downarrow\big] = H_{\max}^\uparrow .
\end{align}
While the pair $H_{\min}^\downarrow$ and $H_{\max}^\uparrow$ is extremal among all conditional entropies, the more standard choices of $H_{\min}$ and $H_{\max}$ can be understood as extremal within a more constrained family of entropic measures: namely, if we further require that both entropies in the dual pair be monotonic under coarse-graining, then $H_{\min}$ and $H_{\max}$ emerge as, respectively, the smallest and largest among a broad class of quantum R\'enyi conditional entropies~\cite{goodarzi2026sharp}. Since coarse-graining is indeed an expected property in many operational contexts, it provides a stronger justification for the naming convention of $H_{\min}$ and $H_{\max}$.

Here it is worth noting that $H_{\max}$
is equivalent to $\smash{\widetilde{H}_{1/2}^\uparrow}$, a conditional entropy built from $\smash{\widetilde{D}_{1/2}}$ that will be discussed in the next subsection. It is not directly related to $D_{\min}$ and does not reduce to the entropy $H_0$ when the system $B$ is trivial, contrary to what one might expect. The max-entropy also admits an operational interpretation as the decoupling accuracy, which quantifies how close a bipartite state $\rho_{AB}$ is to a product state of the form $\pi_A \otimes \sigma_B$, where $\pi_A$ denotes the maximally mixed state and $\sigma_B$ is an arbitrary state~\cite{konig2009operational}.

The pair $H_{\min}$ and $H_{\max}$ forms the foundation of the smooth entropy framework that will be discussed in Section~\ref{sec:one-shot}. The min-entropy was introduced by \textcite{renner2008security} to analyze security of quantum key distribution, but its interpretations extend far beyond that realm. For example, the min- and max-entropies
provide an entropic formulation of wave–particle duality, in the sense that the tradeoff between path distinguishability and interference visibility can be expressed using them~\cite{coles2014equivalence}.
Separately, families of conditional min-entropies characterize quantum majorization and yield complete entropic criteria for certain symmetry-constrained and thermodynamic state transformations~\cite{gour2018quantum}.

\textcite{renner2008security} established that the min-entropy shares many properties with the von Neumann entropy. For instance, it is monotone under coarse-graining, which ensures that additional classical registers can only increase entropy. It also satisfies crucial chain rules:
\begin{align}
    H_{\min}(A|BX) &\geq H_{\min}(A|B) - \log d_X\,, \label{eq:min-chain-classical} \\
    H_{\min}(A|BC) &\geq H_{\min}(A|B) - 2\log d_C\,. \label{eq:min-chain-quantum}
\end{align}
These imply that conditioning on an additional classical register $X$ can reduce the min-entropy at most by $\log d_X$, while conditioning on a quantum register $C$ can reduce min-entropy by up to $2\log d_C$. The same inequalities also hold for max-entropy~\cite{tomamichel2016book}. The above relations are crucial in the analysis of quantum key distribution and other cryptographic protocols, as they are used to bound the loss in min-entropy when information is leaked to the eavesdropper.

\begin{example}
    Equation~\eqref{eq:min-chain-classical} is satisfied with equality if $\rho_{AX}$ is a classical maximally correlated state, while Equation~\eqref{eq:min-chain-quantum} is saturated if $\rho_{AC}$ is maximally entangled; neither can thus be improved in general.
\end{example}

Both quantities satisfy the data-processing inequality under conditionally mixing channels, a property inherited from the DPI of the underlying relative entropy. Furthermore, the max-entropy is a concave function and the min-entropy a quasi-concave function of the state.

\begin{appbox}{Axiomatic}{Singlet fraction via local operations}{app:singletfraction}
The quantum conditional min-entropy captures how close one can bring a bipartite state to a maximally entangled one using only local operations on one subsystem. For a bipartite state $\rho_{AB}$, it holds that~\cite{konig2009operational}
\begin{align*}
    &d \, \max_{\mathcal{E}} F\big((\one_A \! \otimes\! \mathcal{E})(\rho_{AB}), \proj{\phi}_{AB}\big) \notag\\
    &\quad=\exp\!\big(-H_{\min}(A|B) \big) \,,\notag
\end{align*}
where the optimization is performed over all quantum operations $\mathcal{E}$ on the subsystem $B$.
\end{appbox}

\begin{example}
    Beyond coarse-graining, the relation~\eqref{eq:min-chain-classical} is also quite delicate and only known to hold for the family of optimized sandwiched Rényi conditional entropies that $H_{\min}$ and $H_{\max}$ are part of. For example, consider a uniform random variable $X$ on $[d_X]$ and $Y$ a binary indicator for $X=1$. Then
    \begin{align}
        H_{\min}^{\downarrow}(X|Y) = 0 \quad \textrm{but} \quad H_{\min}(X) = \log d_X ,
    \end{align}
    which violates~\eqref{eq:min-chain-classical} when $d_X > 2$.
\end{example}

Finally, the two min-entropies satisfy the following chain rule~\cite{renner2008security}:
\begin{align}
H_{\min}^{\uparrow}(AB|C)
\geq
H_{\min}^{\downarrow}(A|BC) + H_{\min}^{\uparrow}(B|C) \,.
\end{align}

\subsubsection{Rényi conditional entropies with classical side information}

When the conditioning system is classical, conditional R\'enyi
entropies admit explicit expressions as exponential averages of unconditional entropies. Consider a quantum--classical
state $\rho_{AY} = \sum_y p(y)\,\rho_A^y \otimes \ketbra{y}{y}_Y$. For
$\alpha\in(0,1)\cup(1,\infty)$, two prominent definitions are
\begin{align}
\label{Hayashi conditional entropy}
H_\alpha^\downarrow(A|Y)_\rho
&= \frac{1}{1-\alpha}
\log\bigg(\sum_y p(y)\,
\tr\big[(\rho_A^y)^\alpha\big]\bigg),\\
\label{Arimoto conditional entropy}
H_\alpha^\uparrow(A|Y)_\rho
&= \frac{\alpha}{1-\alpha}
\log\bigg(\sum_y p(y)\,
\Big(\tr\big[(\rho_A^y)^\alpha\big]\Big)^{1/\alpha}\bigg).
\end{align}
Both extend to $\alpha=1$ by continuity, yielding $H(A|Y)$.
The arrows indicate that these are simply the fixed-marginal and optimized constructions discussed in Section~\ref{sec: classification}.

When $A$ is also classical, $\smash{H_\alpha^\downarrow}$ reduces to the conditional R\'enyi entropy introduced by
\textcite{hayashi2011exponential} and \textcite{vskoric2011sharp},
while $\smash{H_\alpha^\uparrow}$ reduces to the definition introduced
by \textcite{arimoto1977information}, commonly called the Arimoto
conditional entropy.
In this classical setting, $\smash{H_\alpha^\uparrow}$ plays a central
role in guessing problems with side
information~\cite{arikan2002inequality,sundaresan2006guessing,sason2018improved}
and is also connected to hypothesis testing~\cite{sason2017arimoto}.

Several other classical definitions of conditional entropy have appeared in the
literature~\cite{cachin1997entropy,renner2005simple,hayashi2016equivocations,tan2018conditional}.
More generally, \textcite{rubboli2026complete} fully characterized the
classical conditional entropies satisfying the axioms in
Section~\ref{sec: classification}.

\subsubsection{Quantum Rényi conditional entropies}

In the fully quantum setting, Rényi conditional entropies become even more intricate. Quantum conditional Rényi entropies are typically defined using the sandwiched and Petz--Rényi relative entropies, with the distinction of whether or not to optimize over the marginal of the second subsystem indicated by the arrow notation. For the sandwiched case, this leads to the following definitions~\cite{lennert13renyi}
\begin{align}
&\widetilde{H}_{\alpha}^\uparrow(A|B)_\rho = \sup_{\sigma_B} \left\{ - \widetilde{D}_{\alpha}(\rho_{AB}\|\one_A \otimes \sigma_B) \right\} \label{eq:sandwiched-renyi-cond-optimised},\\
&\widetilde{H}_{\alpha}^\downarrow(A|B)_\rho = -\widetilde{D}_{\alpha}(\rho_{AB}\|\one_A \otimes \rho_B) \,.
\end{align}
These quantities inherit several properties from the underlying R\'enyi relative entropy. For $\alpha \geq 1/2$, they are conditional entropies satisfying all the properties discussed in Section~\ref{sec: classification of conditional entropy}. In particular, they are monotone under conditionally mixing channels and, as a consequence, are concave in the state for $\alpha<1$ and quasi-concave for $\alpha>1$. They also satisfy monotonicity under coarse-graining. Moreover, both families are monotonically decreasing in $\alpha$ and converge to the von Neumann conditional entropy when $\alpha \to 1$.

The Petz conditional entropies are defined analogously using the Petz relative entropy. They satisfy all the required properties discussed in Section~\ref{sec: classification of conditional entropy} for $\alpha\in[0,2]$, the range in which the underlying relative entropy obeys the data-processing inequality. However, they are not generally monotone under coarse-graining~\cite{goodarzi2026sharp}. Unlike the sandwiched case, the optimized Petz conditional entropy admits a closed-form expression~\cite{sharma2012strong,tomamichel13duality}.
Although a closed-form expression is generally unavailable in the sandwiched case, the optimizer admits a fixed-point characterization that has proved useful in establishing additivity~\cite{hayashi_2016-1,cheng2025tight,rubboli2024quantum,burri2024doubly}, and more recently continuity bounds~\cite{cheng2026sharp} and error exponents~\cite{cheng2026informationtransmissionquantumchannels}.
This fixed-point characterization arises from the vanishing-gradient condition at the optimizer and is therefore naturally connected to numerical methods for computing these conditional entropies, including mirror descent~\cite{you2022minimizing} and Riemannian gradient descent with unit step size, which takes the form of a fixed-point iteration~\cite{brahmachari2025fixed,afham2022quantum}.

The optimized sandwiched Rényi conditional entropy in~\eqref{eq:sandwiched-renyi-cond-optimised} is most versatile as it satisfies various chain rules. Most importantly, the relation~\cite{tomamichel2016book}
\begin{align}
    \widetilde{H}_{\alpha}^{\uparrow}(A|BC) &\geq \widetilde{H}_{\alpha}^{\uparrow}(AC|B) - \log d_C
\end{align}
holds for any tripartite state. Together with coarse-graining this yields two chain rules as for the min- and max-entropy:
\begin{align}
    \widetilde{H}_{\alpha}^{\uparrow}(A|BX) &\geq \widetilde{H}_{\alpha}^{\uparrow}(A|B) - \log d_X\,, \\
    \widetilde{H}_{\alpha}^{\uparrow}(A|BC)
    &\geq \widetilde{H}_{\alpha}^{\uparrow}(A|B) - 2\log d_C\,.
\end{align}
This is not surprising since this family contains the min- and max-entropy as its minimal and maximal elements (still satisfying data-processing):
\begin{align}
    H_{\min} = \widetilde{H}_{\infty}^{\uparrow} \quad \textrm{and} \quad
    H_{\max} = \widetilde{H}_{1/2}^{\uparrow} \,.
\end{align}

We emphasize here again a notational clash permeating the literature: The conditional max-entropy is naturally defined as above, but when taking the side-information to be trivial it reduces to $H_{1/2}$ and not $H_0$. The latter quantity is indeed the largest entropy. %

\medskip

Analogous definitions are obtained by replacing the sandwiched R\'enyi relative entropy with the Petz--R\'enyi relative entropy~\cite{petz1986quasi,tomamichel08_aep}. These yield the two families $\smash{\widebar{H}^\uparrow_{\alpha}}$ and $\smash{\widebar{H}_{\alpha}^{\downarrow}}$. For $\alpha\in[0,2]$, these quantities are conditional entropies satisfying all the properties discussed in Section~\ref{sec: classification of conditional entropy}. Both families are anti-monotone in $\alpha$ and converge to the von Neumann conditional entropy for $\alpha \to 1$.

In recent years, R\'enyi conditional entropies have played an increasingly important role in quantum cryptography, in particular in modern security proofs for quantum key distribution based on Rényi entropy accumulation~\cite{arqand2025generalized}. Beyond security proofs, R\'enyi conditional entropies yield achievable
error exponents for privacy amplification and
decoupling~\cite{hayashi_2015,dupuis2023privacy}, and characterize the exact strong
converse exponent of classical data compression with quantum side
information~\cite{cheng2021nonasymptotic}.

These entropies satisfy duality relations that extend those of the von Neumann conditional entropy as well as the min- and max-entropies.
In particular,
\begin{align}
 	\label{relation 1}
    \dual[\widetilde{H}^\uparrow_{\alpha}] = \widetilde{H}^\uparrow_{\beta} &, \quad \alpha^{-1} + \beta^{-1} = 2;\\
 	\label{relation 2}
 	\dual[\widebar{H}^\downarrow_{\alpha}] = \widebar{H}^\downarrow_{\beta} &, \quad \alpha + \beta = 2;\\
 	\label{relation 3}
 	\dual[\widetilde{H}^\downarrow_{\alpha}]= \widebar{H}^\uparrow_{\beta} &, \quad \alpha \beta = 1.
\end{align}
The first two relations~\cite{tomamichel08_aep,lennert13renyi} demonstrate that the families of optimized sandwiched R\'enyi conditional entropies as well as the Petz--R\'enyi conditional entropies with fixed marginals are closed under duality. In contrast, the third relation, proved in~\cite{tomamichel13duality}, connects a sandwiched and a Petz family.

Recall the chain rule $H(AB|C)=H(A|BC)+H(B|C)$
for the von Neumann entropy, discussed around
Eq.~\eqref{eq:cond_entropy_1}. For Rényi conditional
entropies and $\alpha \neq 1$, this identity generally fails even for classical
states, as the following example illustrates.

\begin{example}
Let $C$ be trivial and consider classical registers $A$ and
$B$ with joint distribution
\begin{align}
p_{AB}(a,b)=
\begin{cases}
    2/3, & (a,b)=(0,0),\\
    1/24, & a\in\{1,\ldots,8\},\ b=1,\\
    0, & \text{otherwise}.
\end{cases}
\end{align}
Thus, $p_B=(2/3,1/3)$, while $A$ is deterministic
conditional on $B=0$ and uniform on eight outcomes
conditional on $B=1$. Using the classical expression
in~\eqref{Arimoto conditional entropy}, we obtain
\begin{align}
H_3^\uparrow(A|B) =\frac32\log\frac43,\quad
H_3(B) =\frac12\log 3.
\end{align}
On the other hand, $H_3(AB)=\log\frac{8}{\sqrt{19}}
<\log\frac83
= H_3^\uparrow(A|B)+H_3(B)$,
so the chain rule fails at order $3$.
\end{example}

This implies that we can in general not expect a chain rule with equality. However, if we allow for a change of Rényi order, inequalities that hold for all states are still possible.
Choose parameters $(\alpha, \beta, \gamma)$ in $[\frac12, 1) \cup (1, \infty]$ such that
\begin{align}
    \frac{\alpha}{1-\alpha} = \frac{\beta}{1-\beta} + \frac{\gamma}{1-\gamma} \,. \label{eq:dupuis-chain-params}
\end{align}
An elegant chain rule was derived by~\textcite{dupuis2015chain}, who showed that
\begin{align}
&\widetilde{H}^\uparrow_{\alpha}(AB|C)\geq	\widetilde{H}^\uparrow_{\beta}(A|BC)+	\widetilde{H}^\uparrow_{\gamma}(B|C) \label{eq:dupuis-chain}
\end{align}
whenever $(\alpha-1)(\beta-1)(\gamma-1) > 0$, with the inequality reversed when $(\alpha-1)(\beta-1)(\gamma-1) < 0$. Certain relaxations of the inequality~\eqref{eq:dupuis-chain} follow directly from the monotonicity of R\'enyi conditional entropy in the order parameter.
However, the boundary given by the parameter relations in~\eqref{eq:dupuis-chain-params} is tight in general, as the above example already shows.

\begin{example}
Since this distribution is nonuniform on its support,
$H_\alpha(AB)$ is strictly decreasing in $\alpha$.
Consequently,
\begin{align}
H_\alpha(AB)
<H_{3/2}(AB)
=\log\frac83
=H_3^\uparrow(A|B)+H_3(B)
\end{align}
for every $\alpha>3/2$. Thus, the triples $(\alpha, 3,3)$ admit~\eqref{eq:dupuis-chain} if and only if $\alpha \leq 3/2$, saturating the parameter relation~\eqref{eq:dupuis-chain-params}.
\end{example}

We call these inequalities ``chain rules'' since they generalize the von Neumann entropy chain rule in the following concrete sense. Setting
$\beta=\gamma=t$ and $\alpha(t) =2t/(1+t)$ satisfies the
parameter relation and gives
\begin{align}
\widetilde{H}^\uparrow_{\alpha(t)}(AB|C)
&\begin{cases}
\geq \widetilde{H}^\uparrow_t(A|BC)
    +\widetilde{H}^\uparrow_t(B|C), & t>1,\\
\leq \widetilde{H}^\uparrow_t(A|BC)
    +\widetilde{H}^\uparrow_t(B|C), & 2/3<t<1.
\end{cases}
\end{align}
By continuity at order one, the limits $t \to 1$ yield the  two opposite inequalities,
and hence the identity $H(AB|C)=H(A|BC)+H(B|C)$.

Roughly speaking, these chain rules imitate the one for von Neumann entropy but we lose something in the Rényi parameter every time we apply one, which makes them unsuitable for many applications. If we vary the state between the different entropic quantities, introducing so-called tilted states, we can get better Rényi parameters, as shown in, e.g.,~\textcite{dupuis_2020,hirche2023chain}.
More recently, \textcite{rubboli2024quantum} established a corresponding set of chain rules for the Petz conditional entropies, as well as mixed chain rules involving both the Petz and sandwiched entropies.

\begin{appbox}{qkdapp}{Entropy accumulation}{app:entropy-accumulation}
The analysis of quantum key distribution (QKD) relies on quantifying how much secret key, in the form of nearly uniform randomness that is guaranteed to be secure from adversaries, remains after public communication~\cite{renner2008security}. Entropy accumulation aims to provide accurate bounds on how much uncertainty successive rounds of a protocol generate, accounting in particular for the memory retained between rounds, estimated by adding the contributions to the entropy from individual rounds~\cite{dupuis_2020,metger2024generalised}.  Sandwiched R\'enyi conditional entropies are useful here because their chain rules combine these contributions without a deterioration in R\'nyi order that grows with the number of rounds $n$.

\smallskip
Precisely, consider sequential channels
$\mathcal M_i:R_{i-1}E_{i-1}\to A_iR_iE_i$,
where $R_i$ carries memory between rounds and $E_i$
represents side information that an adversary may possess. Here one only assumes non-signalling from
$R_{i-1}$ to $E_i$: the reduced output on $E_i$
can be generated by a channel acting only on
$E_{i-1}$. For the resulting state $\rho$ and
$1<\alpha<2$, entropy accumulates as~\cite{metger2024generalised}
\begin{align*}
\widetilde H_\alpha^\downarrow(A_1\cdots A_n|E_n)_\rho
&\geq \sum_{i=1}^n\, \inf_{\omega}
\widetilde H_{1/(2-\alpha)}^\downarrow
(A_i|E_iQ_i)_{\mathcal M_i(\omega)}.
\end{align*}
Here each infimum ranges over all input states
$\omega_{R_{i-1}E_{i-1}Q_i}$, with $Q_i$ a reference
large enough to purify the input. Thus, a
worst-case entropy contribution from each round
bounds the entropy of the entire correlated sequence.

\smallskip
Subsequent generalized R\'enyi formulations incorporated the analysis of observed test statistics directly at the level of R\'enyi conditional entropies, leading to accessible and numerically computable security bounds for a broad range of finite-size QKD protocols~\cite{arqand2025generalized}.
\end{appbox}

\medskip

The canonical constructions introduced in Section~\ref{sec:class/canonical} offer a binary choice between using the marginal state and optimizing over all states. In the classical case, however, the conditional entropy introduced by~\textcite{hayashi2016equivocations} interpolates between these two choices. This raises the question of whether a more general quantum extension of this entropy could provide a unified treatment. Further motivation comes from the duality relation in Eq.~\ref{relation 3}, which connects the Petz and sandwiched conditional entropies and points to a common underlying structure. This question was addressed by~\textcite{rubboli2024quantum}, who introduced a three-parameter family of conditional entropies in which a parameter $\lambda$ interpolates between the marginal and optimized constructions. This family is closed under duality, with the three seemingly distinct duality relations arising as special cases of a single, more general relation. For $\lambda\geq0$, these quantities are defined as
\begin{align}
&H_{\alpha,z}^{\lambda}(A|E)_\rho
=\sup_{\sigma_E}\frac{1}{1-\alpha}\notag\\
&\quad\times\log\Tr\left[
\left(
\rho_{AE}^{\frac{\alpha}{2z}}
\rho_E^{\frac{(1-\lambda)(1-\alpha)}{2z}}
\sigma_E^{\frac{\lambda(1-\alpha)}{z}}
\rho_E^{\frac{(1-\lambda)(1-\alpha)}{2z}}
\rho_{AE}^{\frac{\alpha}{2z}}
\right)^z
\right]\,.
\end{align}
For $\lambda<0$, the supremum is replaced by an infimum. These quantities satisfy all the properties required of a conditional entropy in Section~\ref{sec: classification of conditional entropy}, provided that $(\alpha,z)$ lies in the DPI region of the $\alpha$-$z$ Rényi relative entropies and $1-|z/(1-\alpha)|\leq\lambda\leq1$. The case $z=\alpha$ defines the club-sandwiched conditional entropies, which have found applications in randomness extraction~\cite{rubboli2026strong} and quantum state merging~\cite{berta2026strong}. This family, which includes the sandwiched conditional entropies, satisfies the coarse-graining monotonicity property introduced in Section~\ref{sec: classification of conditional entropy}. This property has proved essential in applications and singles out the club-sandwiched conditional entropies from several other quantum generalizations~\cite{goodarzi2026sharp}.

\begin{appbox}{compression}{Exponents of quantum state merging}{app:state-merging-exp}
Direct applications of different variants of conditional R\'enyi entropy emerge in a refined analysis of quantum state merging. (Application~\ref{app:state-merging}).
Assume that Alice and Bob share many copies of their joint state $\rho_{AB}^{\otimes n}$ with purification $\psi_{ABR}^{\otimes n}$, and the net entanglement cost per copy is fixed as $E$, meaning that $\frac1n (k_n-\ell_n) \to E$, where $k_n$ (respectively, $\ell_n$) is the number of ebits shared by Alice and Bob at the beginning (end) of the protocol. One may then wish to characterize the optimal asymptotic behavior of the achievable error $\ve_n$,  as measured by the purified distance between the output of the protocol and the ideal desired state $\psi_{A'BR}^{\otimes n}$ together with $\ell_n$ ebits.
The optimal error exponent is then~\cite{berta2026tightanyshotquantumdecoupling}
\begin{align*}
    e_{\rm merg}(A|B,E)
    =\sup_{\alpha\in(1/2,1]}\frac{1-\alpha}{2\alpha}\left(E-\widebar{H}_{\alpha}^{\uparrow}(A|B)_\rho\right)\,,
\end{align*}
for all rates $E < E_c$ below a critical rate $E_c$, while the strong converse exponent is given by~\cite{berta2026strong}
\begin{align*}
    sc_{\rm merg}(A|B,E)
    & =\sup_{\alpha\in[1,2]}\frac{2(\alpha-1)}{\alpha}\left(H_{\alpha,\alpha/2}^{1}(A|B)_\rho-E\right).
\end{align*}
\end{appbox}

\begin{table*}[t]
\begingroup
\small
\setlength{\tabcolsep}{2pt}
\renewcommand{\arraystretch}{1.2}
\begin{tabular*}{\textwidth}{@{\extracolsep{\fill}}llccc@{\extracolsep{28pt}}l@{\extracolsep{24pt}}l@{\extracolsep{24pt}}l@{\extracolsep{20pt}}l@{}}
 & & & & & \multicolumn{3}{l@{\extracolsep{\fill}}}{Dimension chain rules} & \\
Family & Symbol & \makebox[.17\textwidth][c]{\shortstack{DPI\\$\alpha<1$}}
& \makebox[.17\textwidth][c]{\shortstack{DPI\\$\alpha>1$}}
& \shortstack{Coarse-\\graining}
& ST & CL & QL & Dual \\
\hline
\rule{0pt}{3ex}%
Sandwiched, optimized
& $\widetilde H_\alpha^\uparrow$ & $[\tfrac12,1)$ & $(1,\infty]$
& $\checkmark$ & $\checkmark$ & $\checkmark$ & $\checkmark$
& $\widetilde H_{\alpha/(2\alpha-1)}^\uparrow$ \\
Sandwiched, marginal
& $\widetilde H_\alpha^\downarrow$ & $[\tfrac12,1)$ & $(1,\infty]$
& $\checkmark$ & $\times$ & $\times$ & $\times$
& $\widebar H_{1/\alpha}^\uparrow$ \\
Petz, optimized
& $\widebar H_\alpha^\uparrow$ & $[0,1)$ & $(1,2]$
& $\times$ & $\checkmark$ & $\times$ & $\times$
& $\widetilde H_{1/\alpha}^\downarrow$ \\
Petz, marginal
& $\widebar H_\alpha^\downarrow$ & $[0,1)$ & $(1,2]$
& $\times$ & $\times$ & $\times$ & $\times$
& $\widebar H_{2-\alpha}^\downarrow$ \\
Club-sandwiched
& $H^\lambda_{\alpha,\alpha}$
& \multicolumn{2}{c}{$|x_1|\geq1,\quad
1-|x_1|\leq\lambda\leq1$\textsuperscript{a}}
& $\checkmark$ & $\times^{\mathrm b}$ & $\times^{\mathrm b}$ & $\times^{\mathrm b}$
& $H^1_{\widehat\alpha,\widehat z}$\textsuperscript{c} \\
Three-parameter (RGT)
& $H^\lambda_{\alpha,z}$
& \multicolumn{2}{c}{$|x_1|\geq1,\quad
|x_2-x_1|+\bigl|x_3+2-|x_1|\bigr|\leq|x_1|
$\textsuperscript{a}}
& $\times^{\mathrm d}$
& $\times^{\mathrm b}$ & $\times^{\mathrm b}$ & $\times^{\mathrm b}$
& $H^{\widehat\lambda}_{\widehat\alpha,\widehat z}$\textsuperscript{c} \\
\end{tabular*}

\caption[Properties of quantum R\'enyi conditional entropies]{%
\textbf{Quantum R\'enyi conditional entropies.}
The dimension chain rules are system transfer (ST),
classical leakage (CL), and quantum leakage (QL). They read:
\begin{align*}
    \mathrm{ST}:\; \mathbb H(AC|B)
\leq \mathbb H(A|BC)+\log d_C; \quad
\mathrm{CL}:\; \mathbb H(A|BX)
\geq \mathbb H(A|B)-\log d_X; \quad
\mathrm{QL}:\; \mathbb H(A|BC)
\geq \mathbb H(A|B)-2\log d_C.
\end{align*}
\par
\textsuperscript{a} Use
$x_1=z/(1-\alpha)$,
$x_2=1/(1-\alpha)-\lambda$, and
$x_3=(z-1)/(1-\alpha)-\lambda$.
The DPI ranges apply to both $\alpha<1$ and $\alpha>1$.
For club-sandwiched, set $z=\alpha$.
\textsuperscript{b} Not valid throughout the family;
$\lambda=1$ satisfies ST, while
$\lambda=1$, $z=\alpha$ satisfies ST, CL, and QL.
\textsuperscript{c} Dual parameters:
$\widehat\alpha=(z+(\lambda-1)(\alpha-1))/(z+\lambda(\alpha-1))$,
$\widehat z=z/(z+\lambda(\alpha-1))$, and
$\widehat\lambda=(z-1)/(\alpha-1)$
(set $z=\alpha$ for club-sandwiched).
\textsuperscript{d} Coarse-graining holds only for $z=\alpha$.
}
\label{tab:renyi-conditional-entropy-properties}
\endgroup
\end{table*}

\subsection{R\'enyi mutual information}\label{subsec:renyi_mutual}

The standard mutual information can be written as
$I(A:B)=H(A)+H(B)-H(AB)$. This identity suggests defining a
R\'enyi mutual information by replacing each von Neumann entropy
with its R\'enyi counterpart. However, as in the conditional-entropy
setting discussed in Section~\ref{subsec:renyi_conditional}, this
prescription does not generally satisfy data processing under local
channels, one of the defining requirements for mutual information
in Section~\ref{sec:class/mutual-info}.
\begin{example}
Consider again the binary distribution from
Section~\ref{subsec:renyi_conditional}, with
$p(0,0)=0$, $p(0,1)=p(1,0)=1/4$, and $p(1,1)=1/2$.
The proposed definition gives
\begin{align}
H_2(A)+H_2(B)-H_2(AB)=\log\frac{24}{25}<0.
\end{align}
Discarding $B$ makes this expression zero, so local processing
increases it, contrary to the data-processing inequality.
\end{example}

The canonical constructions of Section~\ref{sec:class/canonical}
instead yield mutual informations satisfying the required properties
in appropriate parameter ranges. They compare the joint state with
a product reference state using a relative entropy, either fixing
the reference states to the corresponding marginals or optimizing
over one or both. These constructions coincide for the Umegaki
relative entropy but generally yield distinct R\'enyi mutual
informations.

\subsubsection{Max-mutual information}

The
max-relative entropy gives three canonical definitions of mutual information~\cite{ciganovic_2014}
\begin{align}
I_{\max}^{\,\uparrow\uparrow}(A:B)_\rho
&=D_{\max}(\rho_{AB}\|\rho_A\otimes\rho_B),\\
I_{\max}^{\,\uparrow\downarrow}(A:B)_\rho
&=\inf_{\sigma_B}D_{\max}(\rho_{AB}\|\rho_A\otimes\sigma_B),\\
I_{\max}^{\,\downarrow\downarrow}(A:B)_\rho
&=\inf_{\sigma_A,\sigma_B}D_{\max}
  (\rho_{AB}\|\sigma_A\otimes\sigma_B).
\end{align}
Here, the optimizations are over quantum states. In particular, it holds that $I_{\max}^{\,\downarrow\downarrow}
\leq I_{\max}^{\,\uparrow\downarrow}
\leq I_{\max}^{\,\uparrow\uparrow}$.

As discussed in Section~\ref{sec:class/mutual-info}, data processing
ensures that discarding a register cannot increase mutual information,
but does not bound the amount lost. The additional no-information-locking
property limits this loss to $\log d_X$ for a classical register $X$,
preventing a key of fixed size from revealing arbitrarily large
correlations as measured by that quantity. The singly optimized max-information $I_{\max}^{\,\uparrow\downarrow}$ satisfies the no-information-locking property~\cite{leditzky_2016}. The corresponding bound for the doubly optimized quantity follows analogously:
\begin{align}
I_{\max}^{\,\downarrow\downarrow}(A:BX)_\rho
&\leq I_{\max}^{\,\downarrow\downarrow}(A:B)_\rho+\log d_X\,.
\end{align}
For the singly optimized quantity, however, the property holds only when the classical register is included in the second argument, due to the asymmetry of its definition. By contrast, the quantities involving no optimization do not satisfy this property whatsoever, as illustrated by the following example.
\begin{example}
Let $X$ be uniformly distributed on $\{1,\ldots,d\}$, with $d\geq2$,
and let $Y$ be the binary indicator of the event $X=1$. Then
\begin{equation}
I_{\max}^{\,\uparrow\uparrow}(X:Y)=\log d,
\qquad
I_{\max}^{\,\uparrow\downarrow}(X:Y)=\log 2.
\end{equation}
Discarding $Y$ reduces both quantities to zero. Thus, a single
classical bit can unlock an arbitrarily large amount of
non-optimized max-information, whereas optimizing the reference
state on $Y$ limits the increase to $\log 2$.
\end{example}
When the additional register is quantum, the singly optimized bound holds with a factor of $2$~\cite{leditzky_2016}, and the doubly optimized bound again follows analogously:
\begin{align}
I_{\max}^{\,\downarrow\downarrow}(A:BC)_\rho
&\leq I_{\max}^{\,\downarrow\downarrow}(A:B)_\rho+2\log d_C\,.
\end{align}
As the next example shows, these bounds cannot be further improved in general.
\begin{example}
The classical bound is saturated by a uniform classical register
and its perfect copy. The quantum bound is saturated when a classical
register $X$ uniformly labels a maximally entangled basis of $BC$,
with $d_B=d_C$; the mutual information increases from zero for
$X:B$ to $2\log d_C$ for $X:BC$. These equalities hold for both
optimized max-informations, so neither
bound can be improved in general.
\end{example}

\subsubsection{R\'enyi mutual informations}
The max-mutual informations above are special cases of R\'enyi
mutual information. Classically, the singly minimized R\'enyi mutual
information was introduced by~\textcite{sibson1969radius}; the doubly
minimized version was introduced in~\cite{tomamichel2017operational}
and subsequently studied in~\cite{lapidoth2019two}. In the quantum
setting, the definition also depends on the chosen extension of the
classical R\'enyi relative entropy. The quantum Petz--R\'enyi and sandwiched R\'enyi constructions
play a prominent role owing to their connections with hypothesis
testing and operational bounds. The sandwiched variants are defined
as follows~\cite{beigi_2013,hayashi_2016-1}:
\begin{align}
\widetilde I_\alpha^{\,\uparrow\uparrow}(A:B)_\rho
&=\widetilde D_\alpha(\rho_{AB}\|\rho_A\otimes\rho_B),
\label{eq:i-quantum-0}\\
\widetilde I_\alpha^{\,\uparrow\downarrow}(A:B)_\rho
&=\inf_{\sigma_B}\widetilde D_\alpha
 (\rho_{AB}\|\rho_A\otimes\sigma_B),
\label{eq:i-quantum-1}\\
\widetilde I_\alpha^{\,\downarrow\downarrow}(A:B)_\rho
&=\inf_{\sigma_A,\sigma_B}\widetilde D_\alpha
 (\rho_{AB}\|\sigma_A\otimes\sigma_B).
\label{eq:i-quantum-2}
\end{align}
The Petz quantities, denoted by a bar, are defined analogously using
$\widebar D_\alpha$.
All three constructions are
monotonically increasing in $\alpha$ and converge to the standard
mutual information as $\alpha\to1$. They inherit local data processing
from the underlying divergence, for $\alpha\in[1/2,\infty]$ in the
sandwiched case and $\alpha\in[0,2]$ in the Petz--R\'enyi case. While additivity of the mutual information quantities involving no optimization follows immediately from that of their underlying divergences, the optimized cases are more subtle. The singly optimized variants are additive throughout the respective DPI ranges~\cite{hayashi_2016-1}. For the doubly optimized variants, however, additivity has been established for $\alpha\in[1/2,2]$ in the Petz case and $\alpha\in[2/3,\infty]$ in the sandwiched case~\cite{cheng2025tight,burri2024doubly}. Whether the additivity results extend to the full DPI range remains an open question.
In the range where these quantities are additive, they satisfy all the axioms required for mutual information discussed in
Section~\ref{sec: classification}.

At $\alpha=\infty$, these quantities satisfy $I_{\max}^{\uparrow\uparrow}=\widetilde I_\infty^{\uparrow\uparrow}$,
with analogous relations for the other optimization choices.

The singly optimized Petz mutual information admits both an explicit
optimizer and a closed-form expression~\cite{sharma2012strong,hayashi_2016-1},
whereas no analogous formula is known in general for its sandwiched
counterpart.

For a tripartite pure state $\rho_{ABC}$ and parameters satisfying $\alpha^{-1}+\beta^{-1}=2$, \textcite{hayashi_2016-1} showed that
\begin{align}
\widetilde I_\alpha^{\,\uparrow\downarrow}(A:B)_\rho
=-\inf_{\sigma_C}\widetilde D_\beta
(\rho_{AC}\|\rho_A^{-1}\otimes\sigma_C)\,.
\end{align}
This identity is commonly referred to as a duality relation for mutual information and can be viewed as a generalization of the duality relation~\eqref{relation 1} for the optimized sandwiched conditional entropy. More generally, replacing $\rho_A$ by a positive operator $\tau_A$ in the above relation yields a corresponding identity whose left-hand side is interpreted as a generalized R\'enyi mutual information. This generalized quantity need not, however, satisfy the usual requirements of a mutual information. Analogous duality relations hold for both the optimized Petz mutual information and the version defined using the marginal, with the same parameter relations as in~\eqref{relation 2} and~\eqref{relation 3}.

The no-information-locking bounds likewise extend to the sandwiched family for $\alpha\geq1/2$. For the singly optimized quantity, these bounds were established by~\textcite{leditzky_2016}; the corresponding bounds for the doubly optimized quantity follow analogously:
\begin{align}
\widetilde I_\alpha^{\,\downarrow\downarrow}(A:BX)_\rho
&\leq \widetilde I_\alpha^{\,\downarrow\downarrow}(A:B)_\rho+\log d_X,
\label{eq:renyi-mi-classical-leakage}\\
\widetilde I_\alpha^{\,\downarrow\downarrow}(A:BC)_\rho
&\leq \widetilde I_\alpha^{\,\downarrow\downarrow}(A:B)_\rho+2\log d_C,
\label{eq:renyi-mi-quantum-leakage}
\end{align}
where $X$ is classical and $C$ is quantum.

Chain rules have also been established for sandwiched R\'enyi mutual information, with parameter relations analogous to those for sandwiched conditional entropies. Choose parameters $\alpha,\beta,\gamma\in[\tfrac{1}{2},1)\cup(1,\infty]$ such that
\begin{align}
\frac{\alpha}{\alpha-1}
=\frac{\beta}{\beta-1}+\frac{\gamma}{\gamma-1}.
\end{align}
Then,~\textcite{mckinlay2020decomposition} showed that
\begin{align}
\widetilde I_\gamma^{\,\uparrow\downarrow}(A:B)_\rho
&\geq H_\beta(B)_\rho
-\widetilde H_\alpha^\downarrow(B|A)_\rho,\\
\widetilde I_\gamma^{\,\downarrow\downarrow}(A:B)_\rho
&\geq H_\beta(B)_\rho
-\widetilde H_\alpha^\uparrow(B|A)_\rho.
\end{align}
whenever $(\alpha-1)(\beta-1)(\gamma-1)>0$, with the inequality
reversed when $(\alpha-1)(\beta-1)(\gamma-1)<0$.

As in the conditional-entropy setting, further chain rules can be obtained by allowing the state to vary between the different entropic quantities, introducing so-called tilted states; see~\textcite{arqand2024mutual}.

R\'enyi mutual informations have found applications in the context of correlation detection discussed in
Application~\ref{app:HT-e}

\bigskip
\begin{appbox}{HT}{Product testing}{app:HT-e}
R\'enyi mutual information can be used to detect correlations of a bipartite state. Consider the task of distinguishing between a bipartite state $ \rho_{AB}^{\otimes n} $ and product states of the form $ \rho_A^{\otimes n} \otimes \sigma_{B^n} $, where $ \sigma_{B^n} $ is an arbitrary state on $ B^n $. It was shown in~\cite{hayashi_2016-1} that the Hoeffding and strong converse exponents are given by
\begin{align*}
&\sup_{\alpha \in (0,1)}\!\left\{\frac{\alpha-1}{\alpha}\Big(r-\widebar{I}^{\, \uparrow\downarrow}_\alpha(A:B)_\rho\Big)\right\} \,,\notag\\
&\hspace{6.04pt} \sup_{\alpha > 1} \hspace{5pt} \left\{ \frac{\alpha - 1}{\alpha} \Big( r - \widetilde{I}^{\ \uparrow\downarrow}_\alpha(A:B)_\rho \Big) \right\}.\notag
\end{align*}
These exponents are characterized by the singly-minimized mutual informations. Similarly, the non-optimized version characterizes the task of distinguishing between $ n $ copies of $ \rho_{AB} $ and the corresponding product of its marginals, $(\rho_A \otimes \rho_B)^{\otimes n}$.
Finally, the doubly optimized variant characterizes, under some additional assumptions, the discrimination between $\rho_{AB}^{\otimes n}$ and any product state $\sigma_{A^n} \otimes \sigma_{B^n}$, with $\sigma_{A^n}$ and $\sigma_{B^n}$ arbitrary states on $A^n$ and $B^n$~\cite{burri2024doublyP,burri2024doubly}.
\end{appbox}

R\'enyi mutual informations have also found applications in
channel-coding exponents~\cite{gupta_2015,mosonyi2017strong} and
quantitative bounds for decoupling, convex splitting, and soft
covering~\cite{li2024operational,li_2025,cheng2025tight,cheng2023error}.

\subsubsection{R\'enyi conditional mutual information}
R\'enyi conditional mutual information extends mutual information to settings in which a conditioning system is also present. Mirroring the properties of the standard conditional mutual information discussed in Section~\ref{sec:standard}, desirable properties include additivity under tensor products and data processing under local channels on both $A$ and $B$. Although not strictly required, it would also be desirable for these quantities to reflect some notion of conditional independence, for example through connections to recoverability or closeness to quantum Markov states. \textcite{berta14CMI} initiated the study of this problem from this perspective, introducing Petz and sandwiched generalizations that satisfy non-negativity, duality relations, and data processing under local channels on one of the two unconditioned subsystems. However, data processing on both subsystems, as well as general monotonicity in the R\'enyi parameter, was left open in that work. More recently,~\textcite{rubboli2026log} introduced the log-Euclidean R\'enyi conditional mutual information,
which satisfies the desired monotonicity and data-processing properties for $\alpha\in[0,1]$ and converges to the standard conditional mutual information as $\alpha\to1$. They also established R\'enyi analogues of several bounds for the CMI discussed in Section~\ref{sec:standard/mutual_cond}. Finding a fully satisfactory Petz or sandwiched R\'enyi conditional mutual information nevertheless remains an open problem.


\section{One-shot information measures}
\label{sec:one-shot}

\newcommand{\prismtikz}[1]{%
  \vcenter{\hbox{\tikz[scale=#1]{
    \draw[line width=0.6pt] (0,0) -- (1,0) -- (0.5,0.866) -- cycle;
    \fill (0.5, 0.289) circle (0.12);  %
  }}}%
}

\newcommand{\prism}{\mathchoice
  {\prismtikz{0.22}}   %
  {\prismtikz{0.18}}   %
  {\prismtikz{0.13}}   %
  {\prismtikz{0.10}}   %
}

Entropic quantities in information theory, just like thermodynamic entropies, typically emerge in the macroscopic behavior of independent, non-interacting physical systems.
One-shot information theory aims to give a refined understanding of operational properties beyond such average properties.
Despite its name, it is not merely concerned with what can be achieved with a single copy of a physical system or a single instance of a communication channel --- any one such system may, for instance, represent many subsystems with strong correlations that prevent the application of standard i.i.d.\ approaches. More broadly, this setting therefore studies information processing beyond i.i.d., which includes both non-asymptotic (finite-resource) results as well as asymptotic approaches to the properties of correlated systems.

Operational tasks are generally studied by characterizing the trade-offs between their performance --- governed by a figure of merit such as the size of a code, the number of extractable bits, etc.\,---\,and their error $\ve$, representing the probability of failure or the accuracy of the process.
Modern approaches to one-shot quantum information aim to tightly connect the figures of merit with various entropies and divergences.
The purpose of doing so is to cleanly separate information-theoretic arguments from statistical ones: one first reduces the study of a physical problem to the study of a more familiar entropic measure, to which known tools and techniques can then be applied.
Broadly speaking, the aim is then to approximate the best achievable performance of a task up to error $\ve$ in terms of some entropic quantity $\H^\ve$, or vice versa, to estimate the best achievable error given a demand on performance.
The accuracy of this approximation then governs which properties of the quantity $\H^\ve$ directly carry over to the operational problem at hand, and improving the entropic approximation can give refinements beyond average asymptotic behavior.

Two main one-shot frameworks, which emerged from seemingly different motivations but ultimately are very closely related, are the smooth entropy formalism and the information spectrum approach.


\subsection{Smooth entropic information measures}
\label{sec:one-shot/smooth}


\subsubsection{Operational origins of smoothing}

The use of smooth entropies can be traced to classical works on randomness extraction and privacy amplification~\cite{impagliazzo1989leftover,bennett_1995,hastad_1999}, which is a motivating application.\footnote{Note that early works such as~\textcite{hastad_1999} and~\textcite{cachin1997smooth} used the term `smoothing' to refer to the process of randomness extraction itself; this term took on a different meaning in \poscite{renner_2004,renner2005simple} formalization, which is now standard and is followed here.}
While the entropy $H(X)$ naturally measures the randomness of a physical system, all of this randomness may not be accessible in one-shot contexts, necessitating more conservative estimates.
Recalling from Application~\ref{app:guessing-a} that the min-entropy $H_{\min}(X)$ measures the probability that the value of a random variable $X$ can be guessed with only a single sample, it provides a natural one-shot measure of randomness.
The achievability result of~\textcite{impagliazzo1989leftover} showed that, indeed, $H_{\min}(X)$ can be used to bound the size of uniform randomness that can be extracted from a given source (see Application~\ref{app:privacy}).
However, understanding the min-entropy as a `worst-case' entropy variant, with more samples we would expect to recover the average properties given by the entropy $H(X)$ itself. This does not appear immediate: the additivity $H_{\min}(X^n)_{p^{\otimes n}} = n H_{\min}(X)_{p}$ seemingly means that even with many i.i.d.\ samples, the randomness of a source is still governed by the min-entropy.
The crux is that $H_{\min}$ can significantly underestimate the randomness that is accessible in practice, and to characterize the performance more precisely, one needs to apply the bounds to a `smoothing' of $p$ --- this can be thought of as choosing another distribution $p' \epsclose p$ which approximates $p$ up to total variation distance $T(p,p') \leq \ve$.

\begin{example}
Earlier we considered the distribution $p = \bigl(\frac{1}{2}, \frac{1}{8}, \frac{1}{8}, \frac{1}{8}, \frac{1}{16}, \frac{1}{16}\bigr)$, which is dominated by its largest element $1/2$ and hence $H_{\min}(X)_p = 1$. Allowing for smoothing tempers the peak in the distribution, increasing the min-entropy. For $\ve = 1/4$, redistributing weight from the largest entry to the two smallest entries gives $p' = \bigl(\frac{1}{4}, \frac{1}{8}, \frac{1}{8}, \frac{1}{8}, \frac{3}{16}, \frac{3}{16}\bigr)$ and $ H_{\min}(X)_{p'} = 2$, already close to $H(X)_p = \frac{17}{8}$. Further smoothing can even increase the min-entropy beyond the Shannon entropy of the original distribution: for $\ve = 1/3$, the uniform distribution on six outcomes lies in the trace-distance ball around $p$ and has min-entropy $\log_2 6 > H(X)_p$.
\end{example}

The randomness available in $p$ is thus more accurately reflected by the smooth min-entropy
\begin{equation}\begin{aligned}
  H^\smooth\ve T_{\min}(X)_{p} = \max_{p' \epsclose p} H_{\min}(X)_{p'}.
\end{aligned}\end{equation}
As can be deduced already from \poscite{shannon1948mathematical} asymptotic equipartition property, the entropy is then indeed recovered in the asymptotic i.i.d.\ limit:
\begin{equation}\begin{aligned}\label{eq:Hmin_AEP}
  \reg\!\left[H_{\min}^\smooth\ve T \right]\!(X)_{p}  = \lim_{n\to\infty} \frac1n H^\smooth\ve T_{\min}(X^n)_{p^{\otimes n}} = H(X)_p
\end{aligned}\end{equation}
for all $\ve \in (0,1)$.
The notion of smoothing thus forms a bridge between one-shot treatment and average asymptotic properties, at the same time reflecting the natural operational assumptions of most information-theoretic, statistical, or cryptographic tasks, in which errors may only vanish asymptotically.
However, as visualized in Figure~\ref{fig:aep}, this equipartition property may not be sufficient to give an accurate approximation to the value of $H^\smooth\ve T_{\min}(X^n)_{p^{\otimes n}}$ for small $n$, which motivates the study of various refinements.

\begin{inplacefigure}
    \centering
    \begingroup%
\providecommand{\EntropyPlotFont}{\footnotesize}%
\setlength{\unitlength}{0.01\linewidth}%
\begin{picture}(100,74.346607)%
  \put(4.426934,6.163261){\includegraphics[width=96\unitlength]{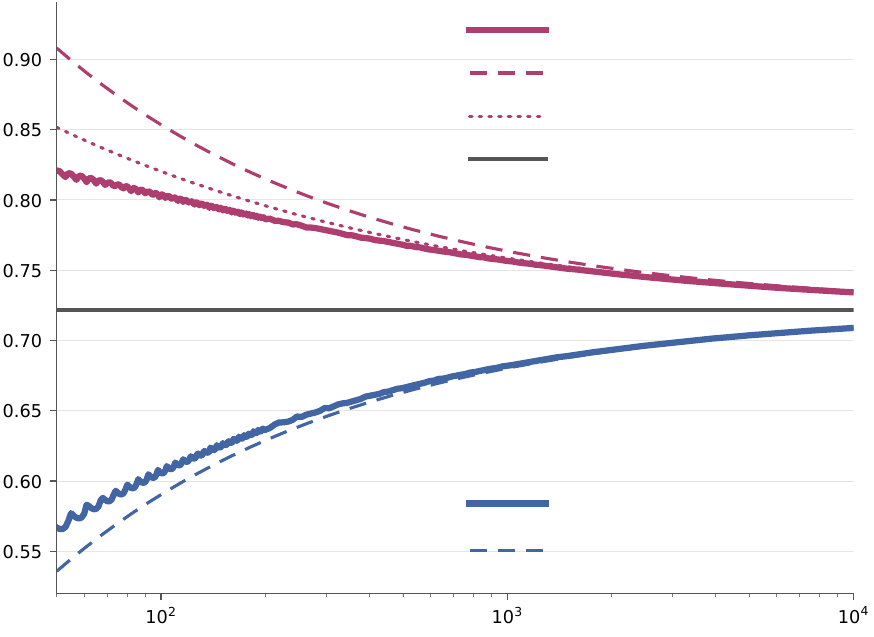}}%
  \put(55.207287,3.009701){\makebox(0,0){\EntropyPlotFont Number of samples $n$ (log scale)}}%
  \put(2.009701,42.370926){\makebox(0,0){\rotatebox{90}{\EntropyPlotFont Entropy per sample (bits)}}}%
  \put(66.525923,21){\makebox(0,0)[l]{\EntropyPlotFont $H_{\min}^{\ve,T}(X^n)/n$}}%
  \put(66.525923,15.6){\makebox(0,0)[l]{\EntropyPlotFont Second = third order}}%
  \put(66.525923,73){\makebox(0,0)[l]{\EntropyPlotFont $H_0^{\ve,T}(X^n)/n$}}%
  \put(66.525923,68.2){\makebox(0,0)[l]{\EntropyPlotFont Second order}}%
  \put(66.525923,63.4){\makebox(0,0)[l]{\EntropyPlotFont Third order}}%
  \put(66.525923,58.4){\makebox(0,0)[l]{\EntropyPlotFont $H(X)$}}%
\end{picture}%
\endgroup%
    \caption{\textbf{Asymptotic equipartition of smooth entropies.} Smooth R\'enyi entropies satisfy the equipartition $H_{\alpha}^{\ve,T}(X^n) = n H(X) + o(n)$, which we plot for the smooth min-entropy and smooth order-zero entropy of $n$ independent $\mathrm{Bernoulli}(0.2)$ trials with $\varepsilon=0.05$. The dashed and dotted curves represent the second- and third-order expansions, that is, refined statements of the equipartition that retain terms of order $O(\sqrt{n})$ and $O(\log n)$, respectively~\cite{hayashi_2008,hayashi_2019}.}
    \label{fig:aep}
\end{inplacefigure}

Complementing the achievability results, \textcite{renner2005simple} then showed that, up to asymptotically vanishing error terms, the one-shot estimate given by the smooth min-entropy $H^\smooth\ve T_{\min}(X)$ can be considered approximately optimal: precisely, the amount of randomness that can be extracted up to error $\ve$ through any physical means, denoted $\ell_\ve(X)_p$, is bounded as
\begin{equation}\begin{aligned}
  H_{\min}^{\ve,T} (X)_p \geq \ell_\ve(X)_p \geq H_{\min}^{\ve-\mu,T}(X)_p - \log \frac{1}{4 \mu^2}
\end{aligned}\end{equation}
for any $\mu \in (0, \ve)$.
The works of~\textcite{renner_2004,renner2005simple} also extended the smooth entropy framework to more general R\'enyi entropies  $H_\alpha^\smooth\ve T$, making the point that both the asymptotic equipartition in~\eqref{eq:Hmin_AEP} as well as the randomness estimates can equivalently use essentially any R\'enyi entropy of order $\alpha > 1$, with $H_{\min}^\smooth\ve T$ representing the minimal such choice.

The realization that the approach based on smooth entropies can be applied beyond randomness extraction traces to \poscite{renner2005simple} study of the complementary task of data compression in terms of R\'enyi entropies of order $\alpha < 1$.
\poscite{renner2008security} extension of smooth entropies to quantum states, followed by the development of a broad smooth entropic framework that allows the one-shot quantities to reflect many natural properties of conditional entropy and mutual information, led to it becoming a very general formalism through which a number of operational problems are analyzed.

\begin{appbox}{randomness}{Randomness extraction}{app:privacy}
Randomness extraction is concerned with turning a source of imperfect randomness $X$ into a uniformly random output $Z$.
When the extraction is performed in a way that guarantees security against an adversary $E$, potentially correlated with the source, this is termed \emph{privacy amplification}. %
The procedure is implemented by applying a random hash function $f: X \to Z$ with the help of an ancillary source of randomness in the form of a so-called seed.
\smallskip

A fundamental bound on extractable randomness is provided by the leftover hash lemma, whose classical statements in \poscite{impagliazzo1989leftover} and~\poscite{bennett_1995} foundational works were later extended to quantum adversaries by~\textcite{renner2005universally} and subsequently refined by~\textcite{tomamichel_2011} and~\textcite{regula_2026}.
In its basic form, it says that if the hash functions are chosen so that the probability of hash collisions is suitably constrained, then the collision entropy of the source can be used to estimate the accuracy of randomness extraction.
Precisely, randomness extraction from $\rho_{XE}$ yields an output state $\rho^f_{ZE}$ whose average distance to a uniformly random state $\pi_Z$, independent of the adversary's system, satisfies
\begin{equation*}\begin{aligned}
  \mathbb{E}_f  \frac12 \norm{ \rho^f_{ZE} - \pi_{Z} \otimes \rho_E}_1 \leq \sqrt{ \frac{|Z|}{4} \, \exp\!\left( - H_{2}^{\meas,\uparrow} (X|E)_\rho\right) };
\end{aligned}\end{equation*}
the collision entropy $H_{2}^{\meas}$ here is often relaxed to $H_{\min}(X|E)_\rho$.
However, this bound on its own provides an overly pessimistic estimate of the randomness accessible in the system $X$. The use of smooth entropies leads to the improved bounds
\begin{equation*}\begin{aligned}
  \mathbb{E}_f \frac12 \norm{ \rho^f_{ZE} - \pi_{Z} \otimes \rho_E}_1 &\leq \ve + \sqrt{ \frac{|Z|}{4} \, \exp\!\left( - H_{2}^{\smooth\ve\meas} (X|E)_\rho\right) }\\
  &\leq \ve +  \sqrt{ \frac{|Z|}{3} \, \exp\!\left( - H_{\min}^{\sqrt\ve} (X|E)_\rho\right) }.
\end{aligned}\end{equation*}
These estimates are approximately optimal in a number of ways: they tightly bound both the randomness that is extractable at the single shot level, as well as its i.i.d.\ asymptotic behavior.

\end{appbox}

\subsubsection{Smooth conditional min- and max-entropies}
\label{subsec:smooth_minmax}

In defining quantum smooth entropies, there is an inherent ambiguity regarding which quantum entropies to employ and how smoothing should be defined, leading to the emergence of several different variants. We begin by reviewing the framework based on two smooth conditional entropies that has by now become standard in many applications.

We recall that $\smash{H_{\min} = \widetilde{H}_{\infty}^\uparrow}$, while $\smash{H_{\max} = \widetilde{H}_{1/2}^\uparrow}$ stands here for the conditional entropy dual to $H_{\min}$ (see discussion in Sec.~\ref{subsec:renyi_conditional}).
The quantum smooth min-entropy $H_{\min}^\ve$ and smooth max-entropy $H_{\max}^\ve$ are consequently defined as
\begin{align}\label{eq:smooth_hmin}
  H_{\min}^\ve (A|B)_\rho &= \max_{\rho' \in \ball_\ve^P(\rho)} H_{\min}(A|B)_{\rho'},\\
  H_{\max}^\ve (A|B)_\rho &= \min_{\rho' \in \ball_\ve^P(\rho)} H_{\max}(A|B)_{\rho'},
\end{align}
where the set $\ball_\ve^P(\rho)$ denotes all sub-normalized states $\rho'$, i.e.\ positive operators with $\Tr \rho' \leq 1$, with purified distance $P(\rho,\rho') \leq \ve$; equivalently, their fidelity must satisfy $F(\rho,\rho') \geq 1-\ve^2$.
The appearance of sub-normalized states in the smoothing ball might appear counter-intuitive at first sight. In the classical case, this is often understood as discarding part of the probability mass of a distribution, resulting in an unnormalized sub-distribution --- this indeed features in many definitions of classical smooth entropies~\cite{renner2005simple}. Sub-normalized quantum states can be similarly understood as substates whose part of the mass is in an inaccessible incoherent subspace. The following example shows why this is needed to ensure isometric invariance of the resulting entropies.
\begin{example}
Suppose that smoothing were restricted to normalized states, and
let $\ve \in (0, 1/\sqrt{2}]$. On a one-dimensional system $A$,
the state $\rho_A=1$ is the only normalized state and, thus, smoothing would not affect its min-entropy, which is zero.
Now embed $A$ isometrically into a two-dimensional system $A'$
so that $\rho_{A'}=|0\rangle\langle0|$. The normalized state
$\tau_{A'}
  =(1-\varepsilon^2)|0\rangle\langle0|
   +\varepsilon^2|1\rangle\langle1|$
satisfies %
 \begin{align}
   P(\rho_{A'},\tau_{A'})=\varepsilon,
   \quad
   H_{\min}(A')_{\tau}
   =-\log(1-\varepsilon^2)>0.
\end{align}
Thus, smoothing over normalized states yields different
values before and after embedding, violating isometric
invariance.

Allowing sub-normalised states removes this discrepancy:
the state ${\rho}_A' = (1-\varepsilon^2)\rho_A$ is already
available in the original one-dimensional system and satisfies $P(\rho_A,{\rho}_A')=\varepsilon$ and $H_{\min}(A)_{{\rho}'} =-\log(1-\varepsilon^2)$.
The sub-normalized smoothing thus gives exactly %
 \begin{align}
   H_{\min}^{\varepsilon}(A)_{\rho}
   =H_{\min}^{\varepsilon}(A')_{\rho}
   =-\log(1-\varepsilon^2),
 \end{align}
independently of the embedding.
\end{example}

The asymptotic equipartition property, established in the quantum case by~\textcite{renner2008security} and then in the below form in~\textcite{tomamichel08_aep,tomamichel2016book}, tells us that
\begin{equation}\begin{aligned}
  \reg\!\left[H_{\max}^\ve\right]\! (A|B)_\rho = \reg\!\left[H_{\min}^\ve\right]\! (A|B)_\rho = H(A|B)_\rho
\end{aligned}\end{equation}
for all $\ve \in (0,1)$, meaning that the expected first-order asymptotic behavior is recovered.

The standard definitions in~\eqref{eq:smooth_hmin} differ from \poscite{renner2008security} earliest constructions of smooth entropies, with one notable difference being the use of purified distance. This choice can be motivated by many important properties that are enabled by Uhlmann's theorem for the fidelity, ensuring that smooth min- and max-entropy obey the properties expected of a conditional entropy~\cite{tomamichel2016book}. Both $H_{\min}^\ve$ and $H_{\max}^\ve$ satisfy isometric invariance~\cite{tomamichel2010duality}, coarse-graining monotonicity~\cite{tomamichel2013hierarchy}, and each of them inherits the standard conditional entropic chain rules (as in Table~\ref{tab:renyi-conditional-entropy-properties}) from its parent entropy, notably including the leakage chain rules~\cite{winkler_2011}
\begin{align}
H_{\min}^\ve(A|BX) &\geq H_{\min}^\ve(A|B) - \log d_X , \\
H_{\min}^\ve(A|BC) &\geq H_{\min}^\ve(A|B) - 2\log d_C ,
\end{align}
for classical registers $X$ and general quantum registers $C$.

Importantly, \textcite{tomamichel2010duality} showed that the entropic duality between $H_{\min}$ and $H_{\max}$~\cite{konig2009operational} extends to this definition of smooth conditional entropies:
\begin{equation}\begin{aligned}
  H_{\max}^\ve &= \dual\!\left[ H^{\ve}_{\min} \right]. \label{eq:smooth-duality}
\end{aligned}\end{equation}
This leads to many fundamental connections and trade-offs between the two --- a prominent example being how
the conditional entropies $H_{\min}^\ve$ and $H_{\max}^\ve$ complement each other in uncertainty relations~\cite{tomamichel2011uncertainty}.

\begin{appbox}{uncertainty}{Smooth entropic uncertainty relations}{app:uncertainty-smooth}
A useful feature of the smooth entropic framework is that it can mirror properties of the conditional entropy $H(A|B)$, including in particular uncertainty relations akin to the ones discussed in Applications~\ref{app:uncertainty-b}--\ref{app:uncertainty-c}.
Precisely, for any tripartite state $\rho_{ABC}$ and any two POVMs
on system A, $\{M_z\}_z$ and
$\{N_x\}_x$, the smooth entropy uncertainty relation of~\textcite{tomamichel2011uncertainty} states that
\begin{equation}
    H_{\min}^{\varepsilon}(Z|C) + H_{\max}^{\varepsilon}(X|B) \geq \log \frac{1}{c},
    \label{eq:qkd-eur}
\end{equation}
where $Z$ and $X$ denote the outcomes of $\{M_z\}_z$ and
$\{N_x\}_x$, respectively, and
$c=\max_{z,x}\|\sqrt{M_z}\sqrt{N_x}\|_\infty^2$
quantifies the overlap, and hence the compatibility, of the two
measurements.
\end{appbox}

Both $H_{\min}^\ve$ and $H_{\max}^\ve$ admit convenient variational forms that can be evaluated as semidefinite programs. Combining the SDP for the min-entropy~\eqref{eq:min-entropy-sdp-dual} with Uhlmann's theorem to simplify the optimization involved in purified distance smoothing, one has~\cite{tomamichel2016book}
\begin{align}
  &\exp(-H_{\min}^\ve(A|B)_\rho) \notag \\
  &= \min \Big\{ \Tr \sigma_B \,:\, \rho'_{ABC} \geq 0,\, \Tr \rho'_{ABC} \leq 1, \,\sigma_B \geq 0, \notag\\
   &\qquad\quad \Tr [ \rho'_{ABC} \rho_{ABC} ] \geq 1 -\ve^2, \, \id_A \otimes \sigma_B \geq \rho'_{AB} \Big\} \label{eq:hmin_sdp} \\
&= \sup \Big\{ (1-\ve^2)\nu-\mu \,:\,
       M_{AB}\geq0,\ \Tr_A M_{AB}\leq\id_B, \notag\\
  &\qquad  \mu,\nu\geq0,\
       M_{AB}\otimes\id_C+\mu\id_{ABC}
       \geq\nu\rho_{ABC}\Big\} , \label{eq:hmin_sdp_dual}
\end{align}
where $\rho_{ABC}$ is any purification of $\rho_{AB}$ and the supremum is attained for  $0<\ve<1$.
The corresponding program for $H_{\max}^\ve$ can be obtained as~\cite{vitanov_2013}
\begin{align}
  &\exp(H_{\max}^\ve(A|B)_\rho) \notag\\
&= \min \Big \{ \lambda \,:\,  \rho'_{ABC} \geq 0,\, \Tr \rho'_{ABC} \leq 1, \,Z_{AB} \geq 0,\notag\\
&\qquad\qquad\quad  \Tr [ \rho'_{ABC} \rho_{ABC} ] \geq 1 -\ve^2,\, \notag\\
&\qquad\qquad\quad Z_{AB} \otimes \id_C \geq \rho'_{ABC}, \, \lambda \id_B \geq Z_B \Big\} \label{eq:hmax_sdp} .
\end{align}

For normalized states $\rho_{AB}$, the $\ve$-smooth entropies take values in the ranges
\begin{align}
-\log m \leq H_{\min}^{\varepsilon}(A|B)_{\rho} + \log(1-\ve^2) &\leq \log d_A,\\
-\log m \leq H_{\max}^{\varepsilon}(A|B)_{\rho} - \log(1-\ve^2) &\leq \log d_A
\end{align}
where $m = \min\{ d_A, d_B\}$, with the lower bounds attained by a maximally entangled state of Schmidt rank $m$ and the upper bounds by a maximally mixed state.

The allowance for smoothing can make the two entropies exhibit very different behavior; there is no universal ordering between the two, meaning that $H_{\min}^\ve(A|B)_\rho \not\leq H_{\max}^\ve(A|B)_\rho$ in general, and it is possible to compare them in either
direction by varying the smoothing parameters.
For any normalized state $\rho_{AB}$ and angles
$\varphi,\vartheta\in[0,\pi/2)$ with
$\varphi+\vartheta<\pi/2$, one has
\cite{tomamichel2016book}
\begin{align}\label{eq:weakstrong1}
  H_{\min}^{\sin\varphi}(A|B)_\rho &\leq H_{\max}^{\sin\vartheta}(A|B)_\rho -2\log\cos(\varphi+\vartheta).
\end{align}
Writing the smoothing parameters as $\ve=\sin\varphi$ and
$\delta=\sin\vartheta$, the condition is equivalently
$\ve^2+\delta^2<1$.
When the parameters instead satisfy $\ve^2+\delta^2\geq1$,
the reverse comparison holds:
\begin{align}
  H_{\min}^{\ve}(A|B)_\rho \geq H_{\max}^{\delta}(A|B)_\rho \, . \label{eq:min-larger-than-max}
\end{align}
The reverse comparison follows by combining Eq.~\eqref{eq:Dmax_DH} below~\cite{regula_2025} with~\textcite{nuradha_2024-1}.

Together, the two smooth entropies can also be used to generalize the more general chain rules for sandwiched R\'enyi conditional entropies, albeit only in an approximate form: we have in particular the inequalities~\cite{vitanov_2013}
\begin{align}
  H_{\min}^\ve(AB|C)_\rho &\geq H_{\min}^{\delta-\mu}(A|BC)_\rho + H_{\min}^{\ve-2\delta}(B|C)_\rho - g(\mu),\\
  H_{\min}^{\ve-2\delta}(AB|C)_\rho &\leq H_{\min}^{\ve}(A|BC)_\rho + H_{\max}^{\delta - \mu}(B|C)_\rho + 2 g(\mu),\\
  H_{\min}^{\ve-2\delta}(AB|C)_\rho &\leq H_{\max}^{\delta-\mu}(A|BC)_\rho + H_{\min}^{\ve}(B|C)_\rho + 3 g(\mu/2).
\end{align}
for any parameters $0 < \mu \leq \delta < \ve/2$, where $g(\mu) = - \log (1 - \smash{\sqrt{1 - 4\mu^2}})$.

The original motivation for the introduction of the smooth entropic formalism by~\textcite{renner2008security} was to provide a rigorous framework for the study of quantum key distribution.
Subsequent improvements in the characterization of smooth entropies, including in particular their uncertainty relations and chain rules, allowed for a more flexible and realistic analysis of quantum cryptographic assumptions~\cite{tomamichel_2012,tomamichel2017largelyself}. Bounds involving smooth entropies continue to form a key element of proofs of cryptographic security~\cite{dupuis_2020,metger_2023,arqand2025generalized,tan_rehashed}, see Application~\ref{app:qkd}.

\begin{appbox}{qkdapp}{Security of quantum key distribution}{app:qkd}
One approach to rigorously proving the security of quantum key distribution (QKD) is based on entropic uncertainty relations~\cite{mayers_2001}.   Employing the smooth entropic framework allows for precise estimates of key rates achievable in practical implementations of QKD protocols such as BB84~\cite{bb84}.

\smallskip
The basic structure of a BB84 protocol begins with two parties (Alice and Bob) each randomly selecting one of two complementary bases and using it to measure their share of a joint quantum state $\rho_{AB}$; repeating this for $n$ rounds ends with Alice possessing a raw bitstring $X$ and Bob possessing a bitstring $X'$. The parties then coordinate to perform statistical tests and error correction procedures to increase the probability that the strings match and can be used as a cryptographic key.
However, since these steps must make use of a public communication channel, there is a possibility that some amount of information leaks to an adversary $E$.
The final step of the protocol is thus privacy amplification, meant to ensure security from any adversary.

\smallskip
To estimate the length $\ell$ of the key shared by Alice and Bob that can be guaranteed to be secure, one then needs to combine several key properties of smooth entropies.
Importantly, the privacy amplification step uses the leftover hash lemma (Application~\ref{app:privacy}) to ensure that
$\ell \gtrapprox H_{\min}^{\ve}(X|EC)$, where the smoothing parameter $\ve$ may be adjusted depending on the required security guarantees, and where $C$ is a classical transcript that possibly leaks to Eve. Assuming that $C$ takes at most $2^m$ different values, the leakage chain rule tells us that the min-entropy can decrease by at most $m$ bits, i.e.\ $H_{\min}^{\ve}(X|EC) \geq H_{\min}^{\ve}(X|E)- m$~\cite{renner2008security,winkler_2011}.
The achievable key length is thus directly constrained by the uncertainty that the eavesdropper has about the key, measured by the min-entropy. This can be tightly bounded using the smooth entropic uncertainty relation of~\textcite{tomamichel2011uncertainty} (Application~\ref{app:uncertainty-smooth}), giving
\begin{equation*}\begin{aligned}
  \ell \gtrapprox H_{\min}^{\ve}(X|EC)_\rho \geq n - H^{\ve}_{\max} (X|X')_\rho - m.
\end{aligned}\end{equation*}
Both the max-entropy $H^{\ve}_{\max} (X|X')$ and the amount of leaked information $m$ can now be efficiently estimated from the statistical tests that Alice and Bob used to verify the correlations between their bitstrings~\cite{tomamichel_2012}. This approach provides some of the most competitive and efficiently computable bounds on key rates achievable in practical quantum key distribution~\cite{tupkary_2026}.
\end{appbox}

More generally, the framework finds widespread use in the analysis of many results in one-shot quantum information. The first variants of the smooth min-entropy were used by \textcite{renner2005universally} and \textcite{renner2008security} to characterize the task of privacy amplification with quantum adversaries. The applications were later extended to more general tasks such as decoupling~\cite{dupuis_2014} and convex split~\cite{anshu_2017}, which form the fundamental building blocks of many achievability results in quantum information.
In a typical decoupling protocol, one applies a unitary to a system and discards part of it, aiming to make the retained subsystem approximately independent of a reference. Conditional entropy bounds quantify the size of the retained subsystem compatible with a prescribed error.
Decoupling also underlies information recovery in the Hayden--Preskill model of evaporating black holes, providing a physical application of conditional  smooth entropies~\cite{hayden07,dupuis_2014}.

The smooth max-entropy was first used to establish one-shot constraints on data compression and information reconciliation~\cite{renner2008security,renes_2012}.
The combination of the two entropies is often used to characterize quantum information tasks through an approach based on decoupling~\cite{dupuis_2014}, which finds use in quantum state merging~\cite{berta_2009}, distillation of secret key~\cite{renes_2012} and entanglement~\cite{berta_2009,buscemi_2010}, or the more general quantum state redistribution~\cite{berta_2016}.
The duality of the entropies in fact can be understood as reflecting the operational duality of the tasks they describe~\cite{renes_2011,renes_2018-1}.

A particular strength and appeal of the entropic framework outlined above is that a large number of operational tasks can be analyzed using just two quantities, $H_{\min}^\ve$ and $H_{\max}^\ve$.
As~\textcite{renner_2004} describe, this stems from the fact that smooth entropies of order $\alpha < 1$ all exhibit similar behavior, as do those of order $\alpha > 1$; it then suffices to pick one of each.
However, already the early works of~\textcite{hastad_1999} and~\textcite{bennett_1995} made the point that the collision entropy $H_2(X)$ gives a tighter estimate on randomness extraction than $H_{\min}(X)$.
Some problems such as classical data compression are also naturally connected to $H_0(X)$ rather than the max-entropy~\cite{renner_2004}.
The relevance of this in information-theoretic contexts was made explicit by~\textcite{hayashi2011exponential}, who showed how a more careful choice of a R\'enyi entropy $H_\alpha$ in the smoothed framework can refine bounds on the asymptotics of the given problem.
Other instances of smooth R\'enyi entropic quantities thus make appearances in various contexts, often giving tighter bounds on certain tasks or serving as convenient proof tools; this motivates a careful examination of other possible definitions.

\subsubsection{Defining smooth divergences}

Just as in the preceding sections, one can understand smooth entropies as emanating from underlying divergences.
Here we will present a consolidating overview of the smoothing notions for quantum states that have emerged in the literature. Although we stress that no single definition of smoothing suffices to encapsulate all the required properties and tightly characterize all operational problems of interest, in operational applications two definitions are distinguished: one based on purified distance smoothing, and one on the familiar notion of lifting classical divergences by measurements.

Smooth divergences are defined using R\'enyi relative entropies. Letting $\D_\alpha$ denote a quantum R\'enyi divergence and $\Delta$ a chosen measure of distance, a basic recipe for constructing a corresponding smooth R\'enyi relative entropy is~(\citealp{renner_2004}; \citealp{datta2009min})
\begin{equation}\begin{aligned}\label{eq:smooth_general_definition}
  \D^\smooth\ve\Delta_\alpha (\rho \| \sigma) = \begin{cases} \displaystyle \inf_{\rho' \in \ball_\ve^\Delta(\rho)} \D_\alpha(\rho' \| \sigma) & \alpha > 1 \\
  \displaystyle \sup_{\rho' \in \ball_\ve^\Delta(\rho)} \D_\alpha(\rho' \| \sigma) & \alpha < 1\,,\end{cases}
\end{aligned}\end{equation}
where $\ball_\ve^\Delta(\rho)$ denotes a set of operators that approximate the state $\rho$ up to distance $\ve$ as measured by $\Delta$.

Classical definitions rarely feature a choice of distance $\Delta$ other than total variation distance. Conventional quantum definitions employ either the trace distance $T$ or the purified distance $P$, with the choice often predetermined by the specifics of the operational task that one aims to characterize.
The set $\ball_\ve^\Delta(\rho)$ is sometimes understood as consisting of quantum states, while sometimes it considers $\rho'$ as sub-normalized operators with $\Tr \rho' \leq 1$; while for some divergences such as $D_{\max}$ the choice has little bearing on the properties of the resulting smooth  divergence~\cite{regula_2025}, for some other divergences such as $\wt{D}_\alpha$ with $\alpha < 1$ it may be necessary to allow sub-normalized states for the smooth quantity to obey data processing~\cite{rubboli_2022}, and we already saw the importance of sub-normalized smoothing for conditional entropies in Sec.~\ref{subsec:smooth_minmax}. %
We thus consider
\begin{align}
  \ball_\ve^T (\rho) = \left\{ \rho' : \rho' \geq 0,\; \Tr \rho' \leq 1, \; T(\rho',\rho) \leq \ve \right\},\\
    \ball_\ve^P (\rho) = \left\{ \rho' : \rho' \geq 0,\; \Tr \rho' \leq 1, \; P(\rho',\rho) \leq \ve \right\},
\end{align}
where $T$ and $P$ denote here the trace and purified distance extended to metrics for sub-normalized states~\cite{tomamichel2016book}. With this definition, we have for instance that $\wt{D}_\alpha^\smooth\ve P(\rho\|\sigma)$ satisfies data processing for $\alpha \in [1/2,\infty]$~\cite{rubboli_2022}.
Examples of commonly encountered divergences in this class are the smooth max-relative entropy $D_{\max}^\smooth\ve P$~\cite{datta2009min} and the smooth divergence~$\wt{D}_{1/2}^\smooth\ve P$~\cite{tomamichel13duality}, sometimes also called the smooth min-relative entropy.

Many properties and bounds of the smooth relative entropies rely on the choice of smoothing.
Depending on the desired level of precision, the distinction between trace and purified distance may not be critical --- the \customtextcite{fuchsvandegraaf1999}{Fuchs--van de Graaf} inequalities ensure, for example, that
\begin{equation}\begin{aligned}
  \D^\smooth\ve P_\alpha (\rho \| \sigma) \geq \D^\smooth\ve T_\alpha (\rho \| \sigma) \geq \D^\smooth{\sqrt{2\ve}}{P}_\alpha (\rho \| \sigma)
\end{aligned}\end{equation}
for all $\alpha > 1$, meaning that the first-order asymptotic behavior of the quantities is effectively the same. However, a more refined analysis of operational tasks often requires a precise understanding of the scaling in terms of error $\ve$, for which these crude bounds are insufficient.
At the same time, the properties themselves may be a motivation for choosing a different distance: for example, in the study of quantum key distribution the most meaningful measure of error probability is given by the trace distance~\cite{portmann_2022}, but as we saw, smooth entropies defined using purified distance satisfy a number of desirable properties that often make them a more useful choice~\cite{tomamichel2010duality}.

\begin{inplacefigure}
    \centering
\begingroup%
\definecolor{purified}{HTML}{245A81}
\definecolor{tracedistance}{HTML}{BBD7EA}
\definecolor{traceoutline}{HTML}{4F91BC}
\providecommand{\PurifiedKey}{%
  \tikz[baseline=.4ex]{%
    \fill[purified] (0,0) rectangle (1.5em,.85em);
    \draw[purified,line width=.5pt] (0,0) rectangle (1.5em,.85em);}}
\providecommand{\TraceKey}{%
  \tikz[baseline=.4ex]{%
    \fill[tracedistance] (0,0) rectangle (1.5em,.85em);
    \draw[traceoutline,line width=.5pt,dash pattern=on 2pt off 1.3pt]
      (0,0) rectangle (1.5em,.85em);}}

\begin{tikzpicture}
  \node[anchor=south west,inner sep=0] (plot) at (0,0)
    {\includegraphics[width=\columnwidth]{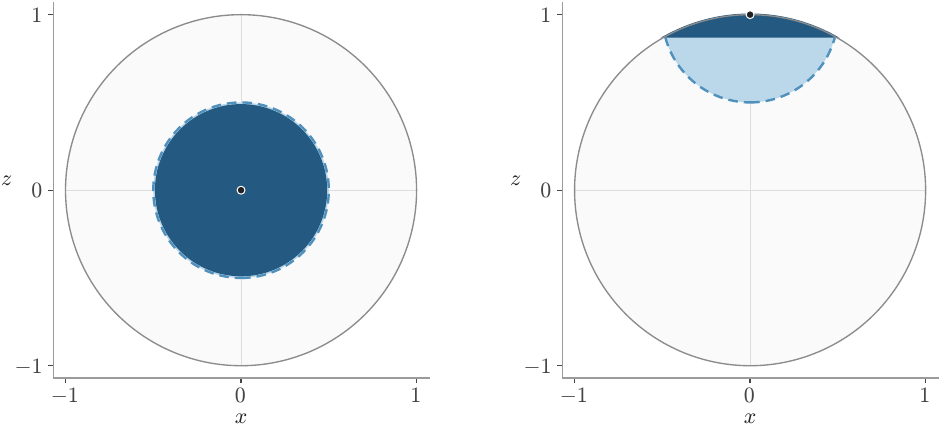}};
  \begin{scope}[x={(plot.south east)},y={(plot.north west)}]
    \node[anchor=south west,xshift=3pt,yshift=2pt,text=white]
      at (0.2592410714,0.47) {$\rho$};
    \node[anchor=south west,xshift=3pt,yshift=2pt]
      at (0.7857589286,0.9389953271) {$\rho$};

    \node[anchor=north] at (0.2792410714,-0.015)
      {$\rho=\id/2$};
    \node[anchor=north] at (0.7857589286,-0.015)
      {$\rho=|0\rangle\langle0|$};

    \node[anchor=north] at (0.59,-0.18)
      {\begin{tabular}{@{}l@{\hspace{2em}}l@{}}
         \PurifiedKey\quad States in $\ball_\ve^P(\rho) \cap \ball_\ve^T(\rho) $  \\[5pt]
         \TraceKey\quad States in $\ball_\ve^T(\rho)$ only
       \end{tabular}};
  \end{scope}
\end{tikzpicture}
\caption{\textbf{The different shape of trace and purified distance balls.} The cross-section of the Bloch sphere is visualized for $\ve = 1/4$.} %
\label{fig:distances}
\endgroup%
\end{inplacefigure}

In addition to these smoothing notions, natural generalizations of classical smooth divergences to the quantum context can be obtained by considering the minimal extensions of classical smooth divergences~\cite{regula_2025,regula_2026}.
As discussed in Sec.~\ref{sec:class/lifting}, the trace and purified distance are the minimal extensions of the corresponding classical metrics; however, instead of smoothing with these distances as in~\eqref{eq:smooth_general_definition}, one may consider extensions of classical smooth divergences themselves, leading to different quantum divergences whose properties often closely connect with their classical counterparts. We define the measured smooth R\'enyi relative entropies
\begin{equation}\begin{aligned}\label{eq:measured_smoothed_alpha}
  D^\smooth\ve\meas_\alpha (\rho \| \sigma) = \sup_{\cM} \, D^\smooth\ve T_\alpha \big(\cM(\rho)\big\| \cM(\sigma)\big),
\end{aligned}\end{equation}
with the maximization being over all measurement channels.
Smooth divergences defined in this way satisfy the data processing inequality for all $\alpha \in [0,\infty]$.
In relevant cases, the optimization over measurements in~\eqref{eq:measured_smoothed_alpha} often can be reduced to a less daunting form. Notably, the minimal divergence in this family is the smooth min-relative entropy $D_{\min}^\smooth\ve\meas = D_0^\smooth\ve\meas$, which often appears under the name of hypothesis testing relative entropy $D_H^\ve$~\cite{buscemi_2010,wang2012one},
as it exactly measures the best type II error probability in binary hypothesis testing subject to the type I error being at most $\ve$; that is,
\begin{equation}\begin{aligned}
  &D^\smooth\ve\meas_{\min}(\rho\|\sigma) \\ %
  &= - \!\log \min \!\big\{ \Tr T \sigma : 0 \leq T \leq I,\; \Tr T \rho \geq 1 - \ve \big\}.
\end{aligned}\end{equation}
This class of divergences includes also a variant of smooth max-relative entropy~$D_{\max}^\smooth\ve\meas$~\cite{datta2014second} that we will see to be related to the information spectrum method.
Several of the quantities $D^\smooth\ve\meas_\alpha$ can also be expressed as a smoothing of the measured R\'enyi relative entropies $D^\meas_\alpha$ in the conventional sense of Eq.~\eqref{eq:smooth_general_definition}, although doing so requires optimization over a set $\ball_\ve^\meas(\rho)$ that includes non-positive Hermitian operators~\cite{regula_2026}.

The quantum asymptotic equipartition property~\cite{tomamichel08_aep,tomamichel2016book} again shows that the variants of smoothing considered here all converge to the Umegaki relative entropy asymptotically, and in particular that
\begin{equation}\begin{aligned}
  \reg[\wt{D}_\alpha^\smooth\ve P] = \reg[\wt{D}_\alpha^\smooth\ve T] = \reg[D_\alpha^\smooth\ve\meas] = D
\end{aligned}\end{equation}
for all $\ve \in (0,1)$ and all $\alpha \in [0,1)\cup(1,\infty]$.

\begin{appbox}{thermo}{One-shot thermodynamic state conversion}{app:thermo-oneshot}
Smooth divergences provide sufficient conditions for converting $\rho$
into a state close to $\sigma$, while preserving the Gibbs state.
In particular, if
\begin{align*}
D_{\min}^{\smooth{\ve_1}\meas}(\rho\|\gamma)
\geq D_{\max}^{\smooth{\ve_2}T}(\sigma\|\gamma),
\end{align*}
then there exists a Gibbs-preserving map  $\cE$ such that
$T\bigl(\cE(\rho),\sigma\bigr)\leq\ve_1+\ve_2$~\cite{buscemi2019information, wang2019resource}.
On its own, $D_{\min}^{\smooth{\ve}\meas}(\rho\|\gamma)$ governs the amount of work that can be extracted from $\rho$ up to error $\ve$, while $D_{\max}^{\smooth{\ve}T}(\rho\|\gamma)$ the work cost of preparing $\rho$ \cite{horodecki2013fundamental}.
For many independent copies, the asymptotic equipartition property
of the smooth divergences recovers the free-energy criterion
of Application~\ref{app:free-energy}.

\smallskip
Generalizations of such bounds form the foundation of one-shot frameworks for the manipulation of broader types of quantum resources~\cite{liu_2019}, including axiomatic approaches to entanglement transformations~\cite{brandao_2010-1,brandao_2011}.
\end{appbox}

\subsubsection{Other smooth conditional R\'enyi entropies}

Given a smooth R\'enyi divergence $\D_\alpha^\smooth\ve\Delta$, optimized and unoptimized variants of a conditional entropy can be defined as
\begin{align}
  \H_\alpha^{\smooth\ve\Delta,\downarrow}(A|B)_\rho &= \log d_A - \D_\alpha^{\smooth\ve\Delta} (\rho_{AB} \| \pi_A \otimes \rho_B),\\
  \H_\alpha^{\smooth\ve\Delta,\uparrow}(A|B)_\rho &= \log d_A - \inf_{\sigma_B}  \label{eq:smooth_halpha} \D_\alpha^{\smooth\ve\Delta} (\rho_{AB} \| \pi_A \otimes \sigma_B).
\end{align}
The representative cases of smooth conditional min- and max-entropy that we encountered before are then $H_{\min}^\ve = \wt{H}_{\infty}^{\smooth\ve P,\uparrow}$, based on the smooth max-relative entropy $D_{\max}$, and $H_{\max}^\ve = \wt{H}_{1/2}^{\smooth\ve P,\uparrow}$, based on a smoothing of $\wt{D}_{1/2}$.\footnote{One may notice that the order of the two optimizations in Eq.~\eqref{eq:smooth_halpha} is interchanged compared to the definitions in~\eqref{eq:smooth_hmin} --- namely, the infimum over $\sigma_B$ is taken only after smoothing. We choose this convention to straightforwardly account for measured smooth divergences. However, the interchange does not affect the previously considered quantities $H_{\min}^\ve$ and $H_{\max}^\ve$: for $H_{\min}^\ve$, this is immediate since both of the optimizations are minimizations; for $H_{\max}^\ve$, it can be seen by using the SDP characterization in Eq.~\eqref{eq:hmax_sdp} and invoking a minimax theorem.}

In the classical applications of smooth entropies (e.g.~\citealp{renner_2004}) as well as in some quantum definitions including~\poscite{renner2008security} early formalism, smooth min-entropy is accompanied not by its entropic dual $H^\ve_{\max} = \wt{H}_{1/2}^{\smooth\ve P,\uparrow}$, but rather by variants built from the conditional R\'enyi entropy of order 0. For quantum states, this is typically the smooth min-relative entropy $D_{\min}^{\smooth\ve\meas}$ (hypothesis testing relative entropy), whose corresponding conditional entropy $H_0^{\smooth\ve\meas,\uparrow}(A|B)_\rho$
was first defined in the study of one-shot quantum capacity by~\textcite{buscemi2010quantum} and finds use e.g.\ in the analysis of source compression with quantum side information~\cite{tomamichel2013hierarchy,cheng2023simple}.

Furthermore, in tasks such as privacy amplification against quantum side information~\cite{tomamichel_2011} as well as its coherent quantum extension of decoupling~\cite{dupuis_2014}, achievability results are in fact naturally obtained in terms of a conditional R\'enyi entropy of order~2, and retaining entropic bounds at this order can improve over the min-entropy in approximating the operational performance~\cite{hayashi2011exponential}.  Specifically, if the accuracy is quantified in terms of trace distance, the choice of measured smooth collision entropy $H_{2}^{\smooth\ve\meas,\uparrow}(A|B)_\rho$ yields significantly improved estimates~\cite{regula_2026,tan_rehashed}; the resulting one-shot bound is sufficient to tightly constrain also the second-order asymptotics and error exponents of privacy amplification, which does not appear possible with other types of smoothing.
A different approach to yield improved bounds, applicable in particular to the purified distance, is to employ a modified notion of smoothing called partial smoothing~\cite{hayashi_2014,anshu_2020}. This construction makes use of the fact that many operational tasks are only concerned with operations applied to one marginal of a joint system. By suitably constraining the marginals of the states $\rho'_{AB}$ in the smoothing ball~$\ball_\ve^P(\rho_{AB})$ of $H_{\min}^\ve(A|B)_\rho$ to satisfy $\rho'_B \leq \rho_B$, the scaling of the resulting bounds can be improved~\cite{anshu_2020,abdelhadi_2020,berta_2026}.

\begin{appbox}{compression}{One-shot quantum source compression}{app:one-shot-compression}
Given a known pure-state ensemble with average state $\rho$, the task of (visible) compression concerns a setting where the encoder knows which state was emitted. The optimal performance, $q_{\rm comp}^{\ve}(\rho)$, is given by the minimal number of qubits needed to guarantee an ensemble-average reconstruction error of at most $\ve$. It evaluates exactly to~\cite{PhysRevA.66.032321}
\begin{align*}
q_{\rm comp}^{\ve}(\rho)
&=\left\lceil\min_{P:\,\tr[P\rho]\geq1-\ve}\log\operatorname{rank}P \right\rceil
=\left\lceil H_{0}^{\smooth\ve T}(\rho) \right\rceil,
\end{align*}
where $P$ is an orthogonal projector.
In this case, the trace distance smoothing is equivalent to $H_{0}^{\smooth\ve\meas}(\rho) = - D_{\min}^{\smooth\ve\meas}(\rho \| \id)$ up to integer rounding.
For i.i.d.\ sources, the asymptotic equipartition property immediately tells us that the asymptotic cost equals $H(\rho)$; however, by analyzing the asymptotic behavior of the smooth entropy one can also precisely determine the error and strong converse exponents of the task, which are governed by R\'enyi entropies~\cite{PhysRevA.66.032321}.

\smallskip
Quantum state merging (Application~\ref{app:state-merging}) can be understood as a generalization that allows the receiver to exploit quantum side information. With free classical communication, the net number of qubits $q_{\rm merg}^\ve(\rho_{AB})$ that Alice needs to communicate in order to ensure a purified distance error of at most $\eps$ can be estimated as~\cite{berta_2009}
\begin{align*}
H_{\max}^\ve(A|B)_\rho\leq q_{\rm merg}^\ve(\rho_{AB})\leq H_{\max}^{\ve-\delta}(A|B)_\rho+O\!\left(\log\tfrac1\delta\right).
\end{align*}
The sharpest one-shot bounds employ entropic duality $H_{\max}^{\varepsilon}(A|B)_\psi=-H_{\min}^{\varepsilon}(A|R)_\psi$ together with the partial smoothing approach of~\textcite{anshu_2020} to account for the unchanging,  inaccessible purifying system $R$.
\end{appbox}

\subsubsection{Smooth mutual information}

The smooth divergence $D^{\smooth\ve\meas}_{\min}$ is often used to define the one-shot mutual information quantity
\begin{align}
  I_{\min}^{\smooth\ve\meas,\uparrow\downarrow}(A\!:\!B)_\rho &= \inf_{\sigma_B} D^\smooth\ve\meas_{\min} (\rho_{AB} \| \rho_A \otimes \sigma_B)\,.
\end{align}
This is sometimes also denoted $I_{H}^{\ve}$; it also makes appearances in its unoptimized form $I_{\min}^{\smooth\ve\meas,\upuparrow}$.
Through connections with hypothesis testing, this quantity naturally finds applications in the study of classical and quantum communication~\cite{hayashi2003capacity}.
The first explicit appearance of $I_{\min}^{\smooth\ve\meas,\uparrow\downarrow}(A:B)_\rho$ was in the analysis of one-shot quantum channel coding by~\textcite{buscemi2010quantum} and~\textcite{wang2012one}.
Concurrent advances were made in classical information theory, where \textcite{polyanskiy_2010} established a `meta-converse' bound, stated precisely in terms of $I_{\min}^{\smooth\ve\meas,\uparrow\downarrow}(A:B)_\rho$, that led to a consolidation and strengthening of known converse bounds on one-shot channel coding; extensions of these ideas found use in quantum information~\cite{leung_2015,tomamichel_2016-1,wang2019converse}.
The development of the technique of position-based coding by \textcite{anshu_2019-1} then led to a widespread use of entropic quantities based on the hypothesis testing quantity $D_{\min}^\smooth\ve\meas$ in various achievability proofs for quantum communication, e.g.~\cite{qi_2018}.
The one-shot analysis of these results shares a general structure that employs near-optimal decoders based on the pretty-good measurement~\cite{beigi_2014}, which can be broadly applied to achievability results in one-shot quantum information~\cite{cheng2023simple}.

Other commonly encountered smooth mutual information quantities are built from the max-relative entropy. As discussed in Section~\ref{subsec:renyi_mutual}, the optimized unsmoothed max-information satisfies leakage chain rules that need not hold for its unoptimized counterpart. This motivates choosing its optimized smooth generalization, known as smooth max-information~\cite{ciganovic_2014}:
\begin{align}
  I_{\max}^{\smooth\ve P,\uparrow\downarrow}(A\!:\!B)_\rho &= \inf_{\sigma_B} D^\smooth\ve P_{\max} (\rho_{AB} \| \rho_A \otimes \sigma_B).
\end{align}
This one-shot quantity finds applications in bounds on the tasks of quantum state merging and quantum state splitting~\cite{berta_2009,berta_2011}, which are important primitives from which a broad number of coherent, fully quantum protocols in quantum information theory can be derived~\cite{abeyesinghe2009mother}.
Extensions of the smooth max-information are also used to describe the simulation cost of quantum channels~\cite{berta_2011,fang_2020}.
Modern achievability bounds in this setting exploit the convex split lemma~\cite{anshu_2017}. The lemma mixes over which of many auxiliary registers is correlated with a given system, while the remaining registers contain independent copies of a fixed state. With sufficiently many registers, this mixture is close to a product; max-relative entropy bounds control the required number. Similarly to decoupling-based proofs, the one-shot formulation of such results can be tightened by exploiting collision entropic quantities such as $I_2^{\smooth\ve P, \uparrow\downarrow}$~\cite{cheng2025tight,gour_2025} or partial smoothing~\cite{anshu_2020}.

\begin{appbox}{simulation}{Quantum channel simulation}{app:channel-simulation}
The task dual to channel coding is given by channel simulation, where one aims to use a noiseless channel to approximate the action of a fixed noisy channel. This task can be tightly characterized when the  parties are in a possession of shared entanglement that they can use to assist their communication. The minimal  communication cost $q^\ve_{\rm sim}(\mathcal{N})$ for error at most $\eps$ becomes \cite{berta_2011}
\begin{align*}
\frac12 I_{\max}^{\smooth{\ve}P,\uparrow\downarrow}(A':B)_{\omega}&\leq q^\ve_{\rm sim}(\mathcal{N})\\
    &\leq \frac12 I_{\max}^{\smooth{\ve-\delta}P,\uparrow\downarrow}(A':B)_{\omega}+O\!\left(\log\tfrac1\delta\right)
\end{align*}
maximized over input states $\psi_{A'A}$, where $\omega=(\idc_{A'}\otimes\cN)(\psi)$.
The same maximized mutual information, without the factor $1/2$, measures the classical communication cost~\cite{Bennett02}.
In the asymptotic i.i.d.\ limit, the quantum reverse Shannon theorem gives the cost per channel use as~\cite{berta_2011,6757002}
\begin{align*}
\lim_{n\to\infty} \frac1n q^\ve_{\rm sim}(\cN^{\otimes n})= \frac12 \max_{\psi_{A'A}} I(A':B)_\omega,
\end{align*}
which equals the entanglement-assisted channel capacity.
Due to this, the optimal rate at which one channel can simulate any other channel, assuming free entanglement assistance, is neatly quantified by the ratio of the corresponding channel mutual informations~\cite{Bennett02}.
\end{appbox}

\subsection{Information spectrum approach}
\label{sec:one-shot/infspec}

Many classical information-theoretic problems can be understood as the study of sequences of random variables, say $X^n = (X_1, \ldots, X_n)$, distributed according to the laws given by one of two probability distributions, $p_n$ or $q_n$.
We have seen throughout this review the important role played by the log-likelihood ratio $L^n = \log \frac{\Prob_{p_n}(X^n)}{\Prob_{q_n}(X^n)}$, which reduces the understanding of many statistical and information-theoretic problems to characterizing the asymptotic convergence of a single random variable.
Noting in particular that $\EE_{p_n}[L^n] = D(p_n\|q_n)$, for IID distributions $p_n = p^{\otimes n}$ and $q_n = q^{\otimes n}$, the weak law of large numbers tells us the sample average $\frac1n L^n$ converges to $D(p\|q)$ in probability under $p^{\otimes n}$. This asymptotic equipartition property is obeyed by more general classes of sequences~\cite{breiman_1957}, but in full generality, the sequence $L^n$ need not converge. The approach of information spectrum, introduced by~\textcite{verdue1994infospectrum} and developed by~\textcite{han02book}, aims to understand the whole spectrum of possible asymptotic behaviors of this random variable.
As a key benefit, this allows one to express operational quantities --- such as error exponents of hypothesis testing or capacities of channels --- in terms of asymptotic quantities known as the inf- and sup-spectral divergence rates~\cite{han02book}, even when an asymptotic equipartition property does not hold. Although not always leading to computable entropic quantities, such reformulation can unify and simplify the study of the asymptotic properties of information-theoretic problems.

On a technical level, the information spectrum approach is concerned with the behavior of the cumulative distribution function of $L^n$ given by $\Prob_{p_n}(L^n \leq x)$, or more precisely its inverse function
\begin{equation}\begin{aligned}
  D^\ve_s (p_n \| q_n) = \sup \left\{ \gamma \,:\, \Prob_{p_n}\!\left( L^n \leq \gamma \right) \leq \ve \right\},
\end{aligned}\end{equation}
which --- despite not actually being a divergence, in the sense that it is not monotone --- is sometimes called the information spectrum divergence.
The weak law of large numbers ensures the basic property that $D^\ve_s(p^{\otimes n} \| q^{\otimes n}) = n D(p\|q) + o(n)$; however, an important feature of $D^\ve_s$ is that it encodes much more information than just the expectation of the log-likelihood ratio.
In particular, the central limit theorem of Berry--Esseen can be used to easily evaluate terms that scale as $O(\sqrt{n})$ and even higher-order corrections, helping describe the speed of convergence of $\frac1n L^n$ (see Figure~\ref{fig:aep}).
Higher-order analysis of this type is used to refine the asymptotic study of information-theoretic tasks~\cite{strassen_1962}, and the connection with information spectrum allowed for broad extensions of such results~\cite{hayashi_2009-1}.
This is also often called small-deviation (when $\ve$ is a constant) or moderate-deviation analysis (when $\ve$ decays sub-exponentially), and it should be distinguished from the large-deviation regime where $\ve$ exhibits an exponential decay (such as in the study of error exponents).

The generality and adaptivity of the information spectrum approach led to it quickly finding use in quantum information theory~\cite{hayashi2003capacity,nagaoka2007infospectrum}, motivating its further developments. Here, given sequences of quantum states $(\rho_n)_n$ and $(\sigma_n)_n$, quantum generalisations of classical Neyman--Pearson tests are the \textcite{helstrom_1969}--\textcite{holevo_1973} tests of the form $\{ \rho_n > t \sigma_n \}$ and $\{ \rho_n \leq t \sigma_n \}$, where $t$ denotes a fixed threshold and the notation $\{ A > B \}$ (respectively, $\{A \leq B\}$) denotes the projection onto the subspace where the operator $A - B$ is positive definite (respectively, negative semidefinite). In this formulation, the information spectrum divergence is defined as~\cite{tomamichel2013hierarchy}
\begin{equation}\begin{aligned}
  D^\ve_s(\rho_n\|\sigma_n) = \sup \big\{ \gamma \,:\, \Tr\left[ \rho_n \!\left\{ \rho_n \leq \exp(\gamma) \, \sigma_n  \right\}\right] \leq \ve \big\}.
\end{aligned}\end{equation}
Its asymptotics are given by the quantum inf- and sup-spectral divergence rates~\cite{nagaoka2007infospectrum}, defined respectively as
\begin{align}
  &\infspec((\rho_n)_n\|(\sigma_n)_n) \\&= \sup \left\{ \gamma  \,:\, \liminf_{n\to\infty}\, \Tr\left[ \rho_n \left\{ \rho_n \geq \exp(n\gamma) \, \sigma_n \right\}\right] = 1 \right\}, \nonumber\\
  &\supspec((\rho_n)_n\|(\sigma_n)_n)\\
  &= \inf \left\{ \gamma \,:\,  \limsup_{n\to\infty} \,\Tr\left[ \rho_n \left\{ \rho_n \geq \exp(n\gamma) \, \sigma_n  \right\}\right] = 0 \right\}, \nonumber
\end{align}
which both reduce to the quantum relative entropy $D(\rho\|\sigma)$ if $\rho_n = \rho^{\otimes n}$ and $\sigma_n = \sigma^{\otimes n}$.

\textcite{tomamichel2013hierarchy} generalized the classical central limit theorem, giving the exact second-order expansion
\begin{equation}\begin{aligned}\label{eq:berry}
  D^\ve_s (\rho^{\otimes n} \| \sigma^{\otimes n}) =  n D(\rho\|\sigma) &+ \sqrt{n V(\rho\|\sigma)} \, \Phi^{-1}(\ve) + O(\log n),
\end{aligned}\end{equation}
where $\Phi^{-1}$ denotes the inverse function of the cumulative normal distribution $\Phi(\ve) = \frac{1}{\sqrt{2\pi}}\int_{-\infty}^\ve e^{-t^2/2} \mathrm{d}t$, and
\begin{equation}\begin{aligned} \label{eq:relative_entropy_variance}
V(\rho\|\sigma) = \Tr \left( \rho ( \log \rho - \log \sigma)^2 \right) - D(\rho\|\sigma)^2
\end{aligned}\end{equation}
is the relative entropy variance. The latter quantity can be understood as quantifying the fluctuations in the distinguishability of the two states. It also emerges from the local behavior of quantum R\'enyi divergences~\cite{lin2015investigating} as
\begin{equation}\begin{aligned}
\frac{V(\rho\|\sigma)}{2 \log e}  = \left. \frac{\mathrm{d}}{\mathrm{d}\alpha} \widetilde{D}_\alpha(\rho\|\sigma) \right|_{\alpha = 1} = \left. \frac{\mathrm{d}}{\mathrm{d}\alpha} \widebar{D}_\alpha(\rho\|\sigma) \right|_{\alpha = 1}.
\end{aligned}\end{equation}
Notation such as $V(A|B)_\rho$ is often used in analogy with entropic quantities, although $V(\rho\|\sigma)$ is not a divergence.

Some of the earliest systematic one-shot achievability results in quantum information were implicitly phrased precisely in terms of the quantity $D^\ve_s$~\cite{hayashi2003capacity}. The information spectrum approach was thus gradually reinterpreted as a one-shot approach that allows for a precise understanding of trade-offs between the errors of information-processing tasks and their asymptotic rates.
This was rigorously formalized by~\textcite{tomamichel2013hierarchy}, who showed a number of connections between $D^\ve_s$ and other operational and entropic quantities used in quantum information, and allowed for the introduction of second-order analysis into the broader quantum realm (independently done by~\textcite{li2014second}).
Such an approach found particular use in the analysis of quantum hypothesis testing~\cite{tomamichel2013hierarchy,li2014second}, randomness extraction~\cite{tomamichel2013hierarchy,shen_2024}, channel coding~\cite{tomamichel_2015,tomamichel_2016-1}, and entanglement manipulation~\cite{hayashi_2006,kumagai_2013,datta2014second} among others.
Despite not formally being a divergence, $D^\ve_s$ is also sometimes used to define entropy-like quantities, with early such appearances including the quantity $H_{\rm Robert}^\ve(A|B)$ used as a proof tool by~\textcite{vitanov_2013}.

\begin{appbox}{Axiomatic}{Second-order analysis of pure-state entanglement transformation}{app:axiom-c}
Leading-order analysis tells us that the conversion of $n$ copies of a pure entangled state $\psi_{AB}$ into $m(n)$ copies of $\phi_{AB}$ with an arbitrarily small error $\ve$ can be realized for a number of copies that scales as $m(n) = n \frac{H(\psi_A)}{H(\phi_A)} + o(n)$, allowing for a conversion rate of $H(\psi_A) / H(\phi_A)$~\cite{bennett_1996-2}.
This can be refined to a second-order approximation as~\cite{kumagai_2013}
\begin{equation*}\begin{aligned}
  m(n) = n\, \frac{H(\psi_A)}{H(\phi_A)} + \sqrt{n}\, \frac{\sqrt{V(\psi_A)}}{H(\phi_A)} Z^{-1}_{\nu(\psi,\phi)}(\ve) + o(\sqrt{n}),
\end{aligned}\end{equation*}
where $Z_{\nu(\psi,\phi)}$ is a Rayleigh-normal cumulative distribution that depends in particular on the ratio $\nu(\psi,\phi) = \frac{H(\psi_A) V(\phi_A)}{H(\phi_A) V(\psi_A)}$, measuring the fluctuations in the entanglement content.

\smallskip
This has several consequences: it tells us, for example, that even for two states such that $H(\psi_A) = H(\phi_A)$, it may not be possible to convert $n$ copies of $\psi_{AB}$ into exactly $n$ copies of $\phi_{AB}$ with vanishing error, since a sublinear number of copies could be lost in the conversion~\cite{kumagai_2013}. This loss is precisely quantified by the second-order corrections.
However, the phenomenon can be mitigated by carefully choosing the states involved in the transformation so that $\nu(\psi,\phi) = 1$, which leads to `resonance' between the fluctuations and allows the transformation to be realized without incurring any loss in the number of copies whatsoever~\cite{korzekwa_2019}. Such analysis has particular implications for practical regimes where the numbers of manipulated systems are heavily constrained and finite-size effects have a large impact.
\end{appbox}

\subsection{Connections and equivalences between one-shot approaches}

Although we have introduced smooth relative entropies and the information spectrum approach as different frameworks with distinct motivations and applications, it soon emerged that the two approaches are very closely related, in some cases making them completely interchangeable.
A basic example is their asymptotic equivalence, which extends to unstructured states beyond i.i.d.: \textcite{datta_2009-1} showed that for any sequences $(\rho_n)_n$ and $(\sigma_n)_n$, it holds that
\begin{align}
  \infspec((\rho_n)\|(\sigma_n)) &= \lim_{\ve\to0} \liminf_{n\to\infty} \frac1n D_{\min}^\smooth\ve\meas (\rho_n \| \sigma_n) \nonumber\\
  &= \lim_{\ve\to0} \liminf_{n\to\infty} \frac1n \wt{D}_{1/2}^\smooth\ve P (\rho_n \| \sigma_n),\\
  \supspec((\rho_n)\|(\sigma_n)) &= \lim_{\ve\to0} \limsup_{n\to\infty} \frac1n D^\smooth\ve P_{\max} (\rho_n \| \sigma_n).
\end{align}
This relation was later refined by \textcite{tomamichel2013hierarchy} into more precise one-shot bounds that led to a number of useful connections. Most importantly, the information spectrum approach is closely related to hypothesis testing, and thus to the measured smooth min-relative entropy $D_{\min}^\smooth\ve\meas$, as
\begin{equation}\begin{aligned}\label{eq:Ds_DH}
  D^\ve_s(\rho\|\sigma) + \log \frac{1}{1-\ve}
  &\leq  D_{\min}^\smooth\ve\meas(\rho\|\sigma) \\
  &\leq  D^{\ve+\delta}_s(\rho\|\sigma) + \log\frac1\delta
\end{aligned}\end{equation}
holds for any two states $\rho$ and $\sigma$, and $\ve \in (0,1)$, and any $\delta \in (0, 1- \ve)$.  This property often allows for an extension of properties of $D^\ve_s$, importantly including its second-order expansion, to the study of hypothesis testing and other operational tasks that are described by $D_{\min}^\smooth\ve\meas$.
The smooth divergence $D_{\min}^\smooth\ve\meas$ can also be closely connected to smooth variants of $\wt{D}_{1/2}$ \cite{dupuis_2012}, which can be understood as the divergence underlying the smooth max-entropy $H_{\max}^\ve$.

Another convenient bridge in the zoo of one-shot relative entropies is provided by the measured smooth max-relative entropy $D_{\max}^\smooth\ve\meas$.
It was first introduced as a modified variant of the information spectrum divergence by \textcite{datta2014second}, using the rather different definition
\begin{align}
  &D_{\max}^\smooth\ve\meas(\rho\|\sigma)\\
  &= \inf \left\{ \gamma \,:\, \Tr\left[ (\rho - \exp(\gamma) \, \sigma) \big\{ \rho \geq \exp(\gamma) \, \sigma  \big\} \right] \leq \ve \right\}. \nonumber
\end{align}
This quantity was shown by \textcite{datta2014second} to satisfy many of the same properties as $D^{1-\ve}_s$, but with the added benefit of obeying the data processing inequality. The equivalence of~\poscite{datta2014second} definition with $D_{\max}^\smooth\ve\meas$ as defined before, i.e.\ the minimal extension of smooth max-relative entropy, was shown by~\textcite{regula_2025}.
The same function independently appeared in the classical literature on differential privacy~\cite{dwork_2014}, leading to its subsequent applications in this context also in quantum information theory~\cite{hirche_2023,nuradha_2024}.

Employing the measured smooth divergences $D_{\max}^\smooth\ve\meas$ and $D_{\min}^\smooth\ve\meas$ in place of $D_s^\ve$ thus allows us to formally unify the information spectrum approach with the one-shot framework of smooth relative entropies.
The two quantities obey a form of complementary equivalence~\cite{regula_2025},
\begin{equation}\begin{aligned}
  D_{\min}^\smooth\ve\meas (\rho\|\sigma) = \inf_{\delta \in [0, 1-\ve)} \left[ D_{\max}^\smooth\delta\meas(\rho\|\sigma) - \log(1-\ve-\delta) \right],%
\end{aligned}\end{equation}
which leads to the bounds
\begin{align}
   D^\smooth\ve\meas_{\max} (\rho \| \sigma) + \log\frac{1}{\ve(1-\ve)} &
  \leq D^\smooth{1-\ve}\meas_{\min}(\rho\|\sigma) \\&\leq D^{\smooth{\ve-\mu}\meas}_{\max}(\rho\|\sigma) + \log\frac{1}{\mu} \nonumber
\end{align}
for all $\ve \in (0,1)$ and all $\mu \in (0,\ve)$.
This shows that the two divergences are connected in a dual manner: the behavior of one when the smoothing parameter $\ve$ is small is approximated by the behavior of the other when the smoothing parameter $1-\ve$ is large.
This explains also the common occurrences of $D_{\max}^\smooth\ve\meas$ as a bound in the study of the strong converse regime of quantum hypothesis testing~\cite{nagaoka2007infospectrum,mosonyi2015quantum,lami_2025-1}.
The more standard max-relative entropy smoothed with purified distance can be closely related to $D_{\max}^\smooth\ve\meas$~\cite{regula_2025}, yielding analogous dual bounds
\begin{equation}\begin{aligned}\label{eq:Dmax_DH}
  &D^\smooth{\sqrt\ve}{P}_{\max} (\rho \| \sigma) + \log\frac1\ve
  \leq D^\smooth{1-\ve}\meas_{\min}(\rho\|\sigma) \\ &\leq D^\smooth{\sqrt{\ve-\mu}}{P}_{\max}(\rho\|\sigma) + \log\frac{F_2(1-\ve,\ve-\mu)}{\mu^2}.
\end{aligned}\end{equation}
This mirrors the complementary behavior we saw between smooth min- and max-entropies in Sec.~\ref{subsec:smooth_minmax}.
Earlier variants of such bounds found use in strong converse~\cite{datta_2013-1} and second-order~\cite{tomamichel2013hierarchy} properties of operational tasks.

\begin{appbox}{capacity}{From one-shot to asymptotic classical--quantum channel coding}{app:one-shot-coding}
Consider a classical--quantum channel $x\mapsto\rho_B^x$
and the joint state
$\omega_{XB}=\sum_xp_x\ketbra{x}{x}\otimes\rho_B^x$.
The largest number $M^\ve$ of messages transmissible
in one use with average error at most $\ve$ obeys
the so-called meta-converse bound~\cite{matthews2014finite}
\begin{align*}
\log M^\ve\leq\max_{\{p_x\}}
I_{\min}^{\smooth\ve\meas,\uparrow\downarrow}(X:B)_\omega.
\end{align*}
The bound can be achieved exactly in the setting of activated non-signaling assistance, in which strong correlations between the communicating parties are exploited and a noiseless classical channel is borrowed to assist the communication~\cite{wang2019converse}.
More generally, achievability bounds without additional assistance can also be obtained in terms of the same smooth mutual information  $I_{\min}^{\smooth\ve\meas,\uparrow\downarrow}(X:B)$ up to adjusting smoothing parameters and
logarithmic corrections~\cite{wang2012one,cheng2023simple}.

\smallskip
Applied to $n$ independent uses, these bounds recover
the Holevo capacity $C$ in Application~\ref{app:CC}
and quantify the approach to it in different regimes.
For fixed error $\ve\in(0,1)$, second-order analysis
determines the $\sqrt n$ correction to the leading
term $nC$ in the logarithm of the message size,
governed by the channel
dispersion~\cite{tomamichel_2015}.
Moderate-deviation analysis instead describes the
trade-off between approaching capacity and requiring
the error to vanish~\cite{chubb2017moderate}.
At fixed rates separated from capacity, error and
strong-converse exponents describe exponential decay
of the error below capacity and of the success
probability above capacity. Their characterization requires further analysis through R\'enyi mutual informations~\cite{mosonyi2017strong,renes_2025} and optimal error exponents are only known in a high-rate regime above a critical rate.
\end{appbox}

Smooth divergences are often related to R\'enyi relative entropies in order to more easily analyze their asymptotic behavior and characterize the exponents of operational tasks. We have for example that, for all $\alpha \in (1,\infty)$ and $\beta \in (0,1)$, \cite{mosonyi2015quantum,audenaert_2008}
\begin{equation}\begin{aligned}
  &\widetilde{D}_\alpha (\rho\|\sigma) + \frac{\alpha}{\alpha-1} \log \frac{1}{1-\ve} \\
  \geq \;& D_{\min}^\smooth\ve\meas (\rho\|\sigma)
   \geq \widebar{D}_\beta(\rho\|\sigma) - \frac{\beta}{1-\beta} \log \frac{1}{\ve};
\end{aligned}\end{equation}
these give tight bounds on, respectively, the strong converse and error exponent of quantum hypothesis testing.
For relative entropies smoothed with purified distance, bounds of this form were first obtained in the study of the quantum asymptotic equipartition property~\cite{tomamichel08_aep,tomamichel2016book}.
For the max-relative entropy in particular, subsequent improvements established that~\cite{regula_2025,wang2019resource}
\begin{equation}\begin{aligned}
  &\widetilde{D}_\alpha (\rho\|\sigma) + \frac{1}{\alpha-1} \log \frac{1}{\ve} \\
  \geq\;& D_{\max}^\smooth{\sqrt\ve}{P} (\rho\|\sigma)
   \geq \widetilde{D}_\beta(\rho\|\sigma) - \frac{\beta}{1-\beta} \log \frac{1}{1-\ve}
\end{aligned}\end{equation}
for $\alpha \in (1,\infty)$ and $\beta \in (\frac12,1)$, which again tightly constrains the asymptotic behavior of the smooth divergence. The bounds also directly translate to inequalities between smooth entropic quantities like $H_{\min}^\ve$ or $I_{\min}^{\smooth\ve\meas,\uparrow\downarrow}$ and corresponding R\'enyi entropies like $\wt{H}_\alpha^\uparrow$ or $\widebar{I}_{\alpha}^{\uparrow\downarrow}$. The one-shot forms of the inequalities can be further tightened and extended to other orders of smooth R\'enyi divergences~\cite{regula_2025,gour_2026}.



\section{Further entropic measures}
\label{sec:further_divergences}

\subsection{Quantum generalisations of Csisz\'ar's \texorpdfstring{$f$}{f}-divergence}

A broad and unifying framework for classical distinguishability measures is provided by the family of $f$-divergences~\cite{morimoto1963markov,ali1966general,csiszar1967divergence}. Let $f:(0,\infty)\to \mathbb{R}$ be a convex function satisfying $f(1)=0$. For two probability distributions $p$ and $q$, the classical $f$-divergence is defined as
\begin{align}
    D_f(p\|q)
    =\sum_x q_x f\left(\frac{p_x}{q_x}\right) \,.
    \label{eq:classical-f-divergence}
\end{align}
The convexity of $f$ guarantees that the resulting divergences satisfy the DPI. The framework
encompasses several familiar information measures. The Kullback--Leibler
divergence is recovered by choosing $f(t)=t\log t$, for example.

\subsubsection{Petz \texorpdfstring{$f$}{f}-divergences}
Classical $f$-divergences admit several inequivalent extensions to the
quantum setting. To account for the noncommutativity of quantum states,
Petz~\cite{petz1985quasi,petz1986quasi} replaced the classical likelihood
ratio $p_x/q_x$ with the relative modular operator
$\Delta_{\rho\mid\sigma} =
\cL_\rho\cR_\sigma^{-1}$, where
$\cL_\rho(X) = \rho X$ and
$\cR_\sigma(X) = X\sigma$. The resulting \emph{Petz
$f$-divergence} is defined by
\begin{align}
 \widebar{D}_f(\rho\|\sigma)=
 \tr\!\left[
   \sigma^{1/2}
   f\!\left(\Delta_{\rho\mid\sigma}\right)
   \bigl(\sigma^{1/2}\bigr)
 \right].
 \label{eq:Petz-f-divergence}
\end{align}
In finite dimensions, this construction admits a transparent spectral
representation. Let
$\rho=\sum_x\lambda_x\ketbra{e_x}{e_x}$ and
$\sigma=\sum_y\mu_y\ketbra{f_y}{f_y}$. The operators
$\ketbra{e_x}{f_y}$ diagonalize the relative modular operator, since
$\Delta_{\rho\mid\sigma}(\ketbra{e_x}{f_y})
=(\lambda_x/\mu_y)\ketbra{e_x}{f_y}$. Consequently,
Eq.~\eqref{eq:Petz-f-divergence} yields
\begin{align}
 \widebar{D}_f(\rho\|\sigma)
 =
 \sum_{x,y}\mu_y
 \lvert\braket{e_x}{f_y}\rvert^2
 f\!\left(\frac{\lambda_x}{\mu_y}\right).
\end{align}
The right-hand side is precisely the classical $f$-divergence between
the Nussbaum--Szko{\l}a distributions
$p^{\rho,\sigma}(x,y)
=\lambda_x\lvert\braket{e_x}{f_y}\rvert^2$ and
$q^{\rho,\sigma}(x,y)
=\mu_y\lvert\braket{e_x}{f_y}\rvert^2$, introduced in
Section~\ref{sec: Petz Renyi relative entropy}. When $\rho$ and
$\sigma$ commute, they may be diagonalized in a common eigenbasis, and
the expression reduces to the classical $f$-divergence between their
corresponding eigenvalue distributions.

The choice $f(t)=t\log t$ recovers the Umegaki relative entropy. More
generally, for
$f_\alpha(t)=(t^\alpha-1)/(\alpha-1)$, the Petz Rényi relative entropy $\widebar{D}_\alpha$ introduced in Section~\ref{sec: Petz Renyi relative entropy} can be expressed as
\begin{align}
 \widebar{D}_\alpha(\rho\|\sigma)
 =
 \frac{1}{\alpha-1}
 \log\!\left[
  1+(\alpha-1) \widebar{D}_{f_\alpha}(\rho\|\sigma)
 \right]. \label{eq:renyi-from-f}
\end{align}
Unlike in the classical setting, ordinary convexity of $f$ does not
suffice to ensure the DPI; a sufficient quantum analogue is operator
convexity~\cite{petz1985quasi,petz1986quasi}. The function
$t\mapsto t^\alpha$ is operator concave for
$\alpha\in(0,1]$ and operator convex for
$\alpha\in[1,2]$. Consequently, the Petz R\'enyi relative entropy
satisfies the DPI in this entire range.

\subsubsection{Optimised \texorpdfstring{$f$}{f}-divergences}

Wilde~\cite{wilde_2018-1} introduced a construction based on
optimising a modular-operator expression over an auxiliary state.
For an operator-monotone decreasing function $g$, define
\begin{align}
 Q_g^{\mathrm{opt}}(\rho\|\sigma)
=\sup_{\substack{\tau>0\\ \tr[\tau]=1}}
 \tr\!\left[
   \rho^{1/2}
   g\!\left(\Delta_{\sigma\mid\tau}\right)
   \bigl(\rho^{1/2}\bigr)
 \right].
\end{align}
These quantities satisfy DPI.
For the power generators $f_\alpha(t)$,
choose
$g_\alpha(t)=\operatorname{sgn}(\alpha-1)
t^{(1-\alpha)/\alpha}$, with
$\alpha\in[1/2,1)\cup(1,\infty)$.
Normalising the resulting quantities gives
\begin{equation}
\begin{aligned}
 D_{f_\alpha}^{\mathrm{opt}}(\rho\|\sigma)
 &=\frac{
 \bigl|Q_{g_\alpha}^{\mathrm{opt}}(\rho\|\sigma)\bigr|^\alpha-1
 }{\alpha-1} \\
 &=\frac{1}{\alpha-1}
 \left\{
 \tr\!\left[
 \left(
 \sigma^{\frac{1-\alpha}{2\alpha}}
 \rho
 \sigma^{\frac{1-\alpha}{2\alpha}}
 \right)^\alpha
 \right]-1
 \right\}.
\end{aligned}
\end{equation}
These agree with the classical $f_\alpha$-divergences on commuting
states and inherit DPI. Applying the same logarithmic transformation
as in~\eqref{eq:renyi-from-f}
recovers the sandwiched R\'enyi relative entropy. For general convex $f$, it is not always possible to find a corresponding $g$ and construct an optimized $f$-divergence.

\subsubsection{Hockey-stick divergences}
The classical \emph{hockey-stick divergence} is the
$f$-divergence generated by $f_\gamma(u)=(u-\gamma)_+$, namely, for $\gamma\geq 0$, it is defined as $E_\gamma(p\|q)
 =
 \sum_x\max\!\left\{p(x)-\gamma q(x),0\right\}$.
A natural quantum extension is given by~\cite{sharma2012strong} \begin{align} E_\gamma(\rho\|\sigma) =\tr\!\left[(\rho-\gamma\sigma)_+\right] = \max_{0\leq\Lambda\leq\id} \tr\!\left[\Lambda(\rho-\gamma\sigma)\right], \end{align}
where $X_+$ denotes the positive part of a Hermitian operator $X$. The variational characterization makes the DPI immediate.
At $\gamma=1$,
the hockey-stick divergence reduces to the trace distance,
$E_1(\rho\|\sigma)=\frac{1}{2}\norm{\rho-\sigma}_1$.

\begin{appbox}{HT}{Binary state discrimination}{app:hockey-stick-guessing}
The hockey-stick divergence admits a direct interpretation in binary
state discrimination, owing to the relation
\begin{align*}
  E_\gamma(\rho\|\sigma) = \frac12 \| \rho - \gamma \sigma\|_1 + \frac12(1-\gamma) \,
\end{align*}
which relates it to the trace distance.

Suppose that a source prepares $\rho$ with
prior probability $0<t<1$ and $\sigma$ with probability $1-t$.
Interpreting the outcome associated with $0\leq\Lambda\leq\id$ as the
guess ``$\rho$'', the optimal success probability is
\begin{align*}
 P_{\mathrm{guess}}^{(t)}(\rho,\sigma)
 &=
 \max_{0\leq\Lambda\leq\id}
 \left\{
 t\tr[\Lambda\rho]
 +(1-t)\tr[(\id-\Lambda)\sigma]
 \right\}
 \notag\\
 &=
 1-t+tE_{\frac{1-t}{t}}(\rho\|\sigma).
\end{align*}
Thus, $tE_{(1-t)/t}(\rho\|\sigma)$ quantifies the improvement over the
strategy that always guesses $\sigma$. In particular, for equal priors,
$t=1/2$, the above expression reduces to the Helstrom formula.
Consequently, the family $\{E_\gamma\}_{\gamma>0}$ completely
characterizes the performance of the pair in binary guessing
games (see, e.g., \textcite{gour2026quantum}).
\end{appbox}

Hockey-stick divergences admit a geometric interpretation through
testing regions, which collect all pairs of acceptance probabilities
obtainable from two states by binary tests. The ordering
$E_\gamma(\rho\|\sigma)\geq E_\gamma(\rho'\|\sigma')$ for every
$\gamma\geq0$ is equivalent to inclusion of the corresponding testing
regions and defines the preorder known as {quantum Lorenz
majorization}~\cite{buscemi2017quantum}. Operationally, it means that
$(\rho,\sigma)$ is at least as distinguishable as
$(\rho',\sigma')$ in every binary test.

For qubit dichotomies, the Alberti--Uhlmann theorem states that this
hockey-stick ordering is equivalent to convertibility under a quantum
channel: it holds if and only if there exists a channel
$\mathcal{E}$ such that
$\mathcal{E}(\rho)=\rho'$ and
$\mathcal{E}(\sigma)=\sigma'$~\cite{alberti1980problem}. In higher
dimensions, the ordering remains necessary but is no longer sufficient
for channel convertibility~\cite{matsumoto2014example}.

Beyond their role in statistical comparison and state-pair
convertibility, hockey-stick divergences have found applications to
strong-converse bounds~\cite{sharma2012strong} and, more recently, to
quantum differential privacy~\cite{hirche_2023}.

\subsubsection{Hockey-stick (Hirche--Tomamichel) \texorpdfstring{$f$}{f}-divergences}
\label{sec:integral-f}
In the classical setting, \textcite{sason16} showed that $f$-divergences
generated by twice continuously differentiable convex functions admit a
representation as weighted integrals of hockey-stick divergences. Building on \textcite{frenkel2023integral} and \textcite{jencova24},
\textcite{hirche2024quantum} introduced integral quantum $f$-divergences by replacing the classical
hockey-stick divergences with their quantum counterparts. For a twice continuously differentiable convex function
$f:(0,\infty)\to\mathbb R$ with $f(1)=0$, this gives
\begin{align}
 &D_f^{\mathrm{HT}}(\rho\|\sigma) \nonumber\\
 &=\int_1^\infty
 \left(
   f''(\gamma)E_\gamma(\rho\|\sigma)
   +\frac{f''(\gamma^{-1})}{\gamma^3}
    E_\gamma(\sigma\|\rho)
 \right)
 \,\mathrm{d}\gamma .
 \label{eq:HT-f-divergence}
\end{align}
For an extension to convex $f$ without differentiability
assumptions, see \textcite{hiai2026hockey} and \textcite{liu_2025}.

As weighted integrals of hockey-stick divergences, these quantities
inherit several of their properties, including the data processing inequality. However, these divergences are generally not additive; in fact, under some regularity
assumptions, additivity uniquely singles out
the Umegaki relative entropy among them~\cite{gour2026quantum}.

\begin{example}
In the extension to general convex generators, the choice
$f(t)=\tfrac12|t-1|$ yields the trace distance,
\begin{equation}
 D_f^{\mathrm{HT}}(\rho\|\sigma)
 = E_1(\rho\|\sigma)
 = \frac12\|\rho-\sigma\|_1 .
\end{equation}
For $f(t)=t\log t$, this is precisely
the Frenkel representation~\cite{frenkel2023integral} of the Umegaki relative entropy in~\eqref{eq:Frenkel}.
\end{example}

The construction analogous to~\eqref{eq:renyi-from-f} gives a variant of R\'enyi divergences, $D_{\alpha}^{\rm HT}$, that in general differs from conventional R\'enyi relative entropies discussed in Section~\ref{sec:additive}.
However, their asymptotic
regularization recovers the Petz and
sandwiched R\'enyi relative entropies for $\alpha<1$ and
$\alpha>1$, respectively~\cite{hirche2024quantum}
\begin{align}
 \lim_{n\to\infty}\frac{1}{n}
 D_\alpha^{\mathrm{HT}}
 \left(\rho^{\otimes n}\middle\|\sigma^{\otimes n}\right)
 =
 \begin{cases}
  \widebar{D}_\alpha(\rho\|\sigma),
    & 0<\alpha<1,\\
  \widetilde{D}_\alpha(\rho\|\sigma),
    & \alpha>1.
 \end{cases}
 \label{eq:HT-renyi-regularization}
\end{align}
These divergences can also be expressed in a so-called layer cake form~\cite{liu_2025}
\begin{equation}\begin{aligned}
  D_\alpha^{\mathrm{HT}}(\rho\|\sigma) = \frac{1}{\alpha-1} \log \left( \alpha \int_0^\infty \gamma^{\alpha-1} \Tr \left[ \sigma \{ \rho > \gamma \sigma \} \right] \mathrm{d}\gamma \right),
\end{aligned}\end{equation}
where we recall that $\{A > B \}$ denotes the spectral projection onto the positive part of $A-B$. This formulation recently found applications in establishing tight bounds for the exponents of many quantum information tasks~\cite{cheng_2025,cheng_2025-1}.

\subsubsection{Minimal and maximal \texorpdfstring{$f$}{f}-divergences}
Given a classical $f$-divergence, one can construct a quantum
$f$-divergence by lifting the classical definition through the
measurement and preparation maps described in
Section~\ref{sec:class/lifting}. Any quantum extension that
agrees with the classical definition on commuting states is then bounded between the
measured and prepared extensions. When $f$ is
operator convex and $\supp(\rho)\subseteq\supp(\sigma)$, the prepared
extension admits the following closed-form expression~\cite{matsumoto2018maxdivergence}:
\begin{align}
    D_f^{\prep}(\rho\|\sigma)
    =
    \tr\!\left[
        \sigma
        f\!\left(\sigma^{-1/2}\rho\sigma^{-1/2}\right)
    \right].
\end{align}
The choice $f(t)=t\log t$ yields the Belavkin--Staszewski relative
entropy, whereas power functions yield, up to a suitable rewriting, the
geometric R\'enyi divergences introduced in
Section~\ref{sec:renyi/maximal}.

\subsection{Quantum \texorpdfstring{$\chi^2$}{chi-squared} divergences}
\label{sec:chi2}
The classical $\chi^2$ divergence is the $f$-divergence generated by
$f(t)=(t-1)^2$, namely,
\begin{align}
    \chi^2(p\|q)
    =
    \sum_x \frac{\bigl(p(x)-q(x)\bigr)^2}{q(x)}.
\end{align}
It is related to the order-two R\'enyi relative entropy through
$D_2(p\|q)=\log\bigl(1+\chi^2(p\|q)\bigr)$.
It describes the local behavior of smooth
$f$-divergences: for a faithful distribution $q$, a perturbation
$p_\epsilon=(1-\epsilon)q+\epsilon p$, and a twice
continuously differentiable generator $f$,
\begin{align}
    D_f(p_\epsilon\|q)
    =
    \frac{f''(1)}{2}\,\epsilon^2\chi^2(p\|q)
    +o(\epsilon^2).
\end{align}
Thus, when $f''(1)>0$, these divergences have the same sensitivity to
small perturbations up to normalization and induce the same local
(Fisher information) metric. In the quantum setting, different
extensions can instead induce distinct metrics, revealing differences
that persist even for nearby states
\cite{petz1996metrics, lesniewski1999metrics}.

This connection motivates the metric-based family of quantum
$\chi^2$ divergences. A Riemannian metric on the manifold of faithful
density operators is a smoothly varying family of positive-definite
symmetric bilinear forms on its tangent spaces.
\poscite{petz1996metrics} classification shows that every such metric
that contracts under quantum channels, appropriately normalized, is parametrized by an
operator-monotone-decreasing function
$k:(0,\infty)\to(0,\infty)$ satisfying
$k(1)=1$ and $k(t^{-1})=t k(t)$.
Evaluating the corresponding metric on the displacement $\rho-\sigma$
gives the quantum $\chi^2$ divergence
\begin{align}
    \chi_k^2(\rho\|\sigma)
    =
    \Tr\!\left[
        (\rho-\sigma)\,
        \cR_\sigma^{-1}
        k(\cL_\sigma\cR_\sigma^{-1})(\rho-\sigma)
    \right].
    \label{eq:quantum-chi2}
\end{align}
These quantities recover the classical $\chi^2$ divergence on commuting
states and satisfy DPI by monotonicity of the underlying
metric~\cite{petz1996metrics}.

A useful subfamily is obtained from
\smash{$k_s(t)=\tfrac12(t^{-s}+t^{s-1})$}, with $s\in[0,\tfrac12]$.
The corresponding divergences have the explicit form $\chi_{k_s}^2(\rho\|\sigma)
    =
    \Tr\!\left[\rho\sigma^{-s}\rho\sigma^{s-1}\right]-1$.
Their endpoints recover the order-two Petz, geometric, and sandwiched
R\'enyi divergences:
\begin{align}
    \widebar D_2(\rho\|\sigma)
    =
    D_2^{\prep}(\rho\|\sigma)
    &=
    \log\bigl(1+\chi_{k_0}^2(\rho\|\sigma)\bigr),\\
    \widetilde D_2(\rho\|\sigma)
    &=
    \log\bigl(1+\chi_{k_{1/2}}^2(\rho\|\sigma)\bigr).
\end{align}
The Bures kernel $k_{\rm B}(t)=2/(1+t)$ similarly yields the
measured R\'enyi divergence~\cite{temme2015goodness}:
\begin{align}
    D_2^{\meas}(\rho\|\sigma)
    =
    \log\bigl(1+\chi_{\rm B}^2(\rho\|\sigma)\bigr).
\end{align}
The BKM kernel $k_{\rm BKM}(t)=\ln t/(t-1)$ gives the divergence
$\chi^2$ introduced in~\eqref{eq: chi square Umegaki}, which governs
the local behavior of Umegaki relative entropy.
Writing $\sigma=\sum_i p_i\ketbra{i}{i}$, it admits the equivalent
representations~\cite{lesniewski1999metrics}
\begin{align}
    \chi^2(\rho\|\sigma)
    &=
    \sum_{i,j}
    \bigl|\bra{i}(\rho-\sigma)\ket{j}\bigr|^2
    \begin{cases}
        \frac{\ln p_i-\ln p_j}{p_i-p_j},
        & p_i\ne p_j,\\[6pt]
        \frac{1}{p_i},
        & p_i=p_j.
    \end{cases}
\end{align}
Although DPI and classical agreement do not force every quantum
extension of $\chi^2$ to have the quadratic form
\eqref{eq:quantum-chi2}, this family contains the minimal
and maximal extensions.

Returning to local behavior, a smooth, positive-definite Hessian of a
divergence satisfying DPI defines a monotone Riemannian metric.
Petz's classification therefore identifies its normalized quadratic
term with a member of the quantum $\chi^2$ family
\cite{lesniewski1999metrics}.
For the R\'enyi divergences, writing
$\rho_\epsilon=(1-\epsilon)\sigma+\epsilon\rho$, this gives
\begin{align}
    \D_\alpha(\rho_\epsilon\|\sigma)
    =
    \frac{\alpha\log e}{2}\,\epsilon^2
    \chi_{\mathbb{k}_\alpha}^2(\rho\|\sigma)
    +o(\epsilon^2).
    \label{eq:renyi-local-chi2}
\end{align}
Table~\ref{tab:renyi-local-chi2} lists pairs of divergences $\D_{\alpha}$
and kernels $\mathbb{k}_{\alpha}$.
\begin{table}[t]
    \centering
    \small
    \setlength{\tabcolsep}{3pt}
    \begin{tabular}{@{}c|@{\hspace{4pt}}lc@{}}
        $\D_\alpha$ & $\mathbb{k}_\alpha(t)$
        & Range of $\alpha$ \\
        \hline
        $\widebar D_\alpha$
        & $\displaystyle
        \widebar{k}_\alpha(t)=
        \frac{t^\alpha+t^{1-\alpha}-t-1}
             {\alpha(\alpha-1)(t-1)^2}$
        & $0<\alpha\le2$\\[10pt]
        $\widetilde D_\alpha$
        & $\displaystyle
        \widetilde{k}_\alpha(t)=
        \frac{1-t^{(1-\alpha)/\alpha}}
             {(\alpha-1)(t^{1/\alpha}-1)}$
        & $\tfrac12\le\alpha<\infty$\\[10pt]
        $\widehat D_\alpha$
        & $\displaystyle
        \widehat{k}_\alpha(t)=\frac{1+t}{2t}$
        & $0<\alpha\le2$\\[10pt]
        $D_\alpha^{\rm HT}$
        & $\displaystyle
        k_\alpha^{\rm HT}(t)=\frac{\ln t}{t-1}$
        & $0<\alpha<\infty$\\[6pt]
    \end{tabular}
    \caption{Local behaviour of quantum R\'enyi divergences~\cite{takahashi2017information,hirche2024quantum,tan2026tight}.}
    \label{tab:renyi-local-chi2}
\end{table}
The Petz and sandwiched kernels depend on $\alpha$ and both converge
at $\alpha=1$ to the BKM kernel
$k_{\rm BKM}(t)=\ln t/(t-1)$, recovering the local behavior of
Umegaki relative entropy in~\eqref{eq: chi square Umegaki}.
The geometric and Hirche--Tomamichel families instead retain,
respectively, the maximal and BKM metrics at every order.

\begin{appbox}{Contraction}{Contraction coefficients from local divergences}{app:contraction-coefficient}
The modified log-Sobolev constant $\kappa$ in Application~\ref{app:relaxation} controls global relative-entropy contraction. For a channel $\Phi$ and reference state $\sigma$, recall the contraction coefficient
\begin{align*}
\eta_{\D_\alpha}(\Phi,\sigma)
=\sup_{\rho\ne\sigma}
\frac{\D_\alpha(\Phi(\rho)\|\Phi(\sigma))}
     {\D_\alpha(\rho\|\sigma)}
\leq1.
\end{align*}
The local expansion~\eqref{eq:renyi-local-chi2} gives
\begin{align*}
\eta_{\chi_{\mathbb{k}_\alpha}^2}(\Phi,\sigma)
\leq\eta_{\D_\alpha}(\Phi,\sigma)
\end{align*}
for the pairs in Table~\ref{tab:renyi-local-chi2}. The quadratic structure makes the $\chi^2$ coefficient accessible through a generalized eigenvalue problem~\cite{hiai15}. For Umegaki relative entropy, the local metric is BKM. Applied to $\Phi=\mathcal T_t$ fixing $\sigma$, its coefficient lower-bounds $\eta(\mathcal T_t,\sigma)$ and therefore gives upper bounds on $\kappa$. Establishing the modified log-Sobolev bound requires control over all states and all times. Under suitable regularity and detailed-balance assumptions, local coefficients also determine asymptotic contraction rates~\cite{hirche2024quantum,tan2026tight}.
\end{appbox}

\subsection{Bregman pseudo-divergences}

Let $F$ be a differentiable convex function defined on a convex region of a finite-dimensional Euclidean space. The~\textcite{bregman1967divergence} (pseudo-)divergence generated by $F$ is
defined as
\begin{align}
    B_F(x\|y)
    =
    F(x)-F(y)-\langle\nabla F(y),x-y\rangle.
\end{align}
Geometrically, $B_F(x\|y)$ is the difference between $F(x)$ and the
first-order affine approximation of $F$ at $y$. Bregman divergences are
widely used in optimization problems whose natural geometry is not
Euclidean. In particular, mirror-descent methods use them in place of
the squared Euclidean distance, allowing each update to reflect the
geometry of the feasible set~\cite{nemirovski1983problem,beck2003mirror}.

In the quantum case, they are defined as
\begin{align}
    B_F(\rho\|\sigma)
    =
    F(\rho)-F(\sigma)
    -
    \tr\!\left[
        \nabla F(\sigma)(\rho-\sigma)
    \right].
    \label{eq:Bregman-quantum}
\end{align}
A prominent example is obtained from the negative von Neumann entropy
$F(\rho)=\tr[\rho\log\rho]$, for which $B_F(\rho\|\sigma)$ reduces to
the Umegaki relative entropy.

Although Umegaki relative entropy satisfies the DPI, Bregman pseudo-divergences generally do not, as convexity guarantees only their
non-negativity. A prominent counterexample is the squared
Hilbert--Schmidt distance, generated by
$F(\rho)=\|\rho\|_2^2$.
Under standard regularity assumptions, in dimension $d\geq3$, the
only Bregman pseudo-divergences that are contractive under all positive
trace-preserving maps are the non-negative multiples of Umegaki
relative entropy~\cite{harremoes2017divergence}. This is the quantum
counterpart of the corresponding characterization of the
Kullback--Leibler divergence~\cite{jiao2014curious}.

\subsection{Tsallis pseudo-entropy}

The~\textcite{tsallis1988possible} entropies provide another
parametrization of the R\'enyi entropies~\cite{harvda1967quantification}. For $\alpha>0$, $\alpha\neq1$, they are defined as
\begin{align}
 T_\alpha(\rho)
 &=\frac{\log e}{1-\alpha}
   \left(\tr[\rho^\alpha]-1\right) \,.
 \label{eq:Tsallis-entropy}
\end{align}
 They converge
to the von Neumann entropy as $\alpha\to1$. Crucially, they are not additive for tensor products, which is why we only refer to them as pseudo-entropies; they do, however, obey the
pseudo-additivity law
\begin{align}
 T_\alpha(\rho\otimes\sigma)
 =T_\alpha(\rho)+T_\alpha(\sigma)
 +\frac{1-\alpha}{\log e}
 T_\alpha(\rho)T_\alpha(\sigma) \,.
\end{align}

\subsection{Observational entropy and restricted measurements}

Already~\textcite{vonneumann1932} made the point that in order to measure the entropy $H(\rho)$ of a quantum system precisely, one must be able to make perfect measurements in its eigenbasis, and demanding that this be possible for an arbitrary quantum state may not realistically reflect the capabilities of a physical observer. As an alternative, he proposed what is now known as the observational entropy
\begin{equation}\begin{aligned}
	H_{O}(\rho) = - \sum_i \Tr[M_i \rho] \big( \log \Tr[M_i \rho] - \log \Tr M_i \big),
\end{aligned}\end{equation}
where $O = (M_i)_i$ is a POVM representing a given observable.
More recently, it has appeared in the study of quantum measurements and their role in applications such as thermodynamics~\cite{safranek_2021,buscemi_2023}.
We note, however, that it is not a bona fide entropy in the sense used in this review.

This can be understood more generally as lifting classical divergences by restricted classes of measurements: one can define
\begin{equation}\begin{aligned}
	\D^{\meas,\mathfrak{R}} (\rho \| \sigma) = \sup_{\cM \in \mathfrak{R}} \D(\cM(\rho)\|\cM(\sigma))
\end{aligned}\end{equation}
where, in contrast with Sec.~\ref{sec:class/lifting}, the optimization is not over all measurement channels $\cM$ but rather a subset $\mathfrak{R}$ thereof.
Such extensions were first formalized for distinguishability norms by~\textcite{matthews_2009} in the study of data hiding phenomena, and for the quantum relative entropy in~\poscite{piani_2009-1} work on entanglement theory. They have later found use in the study of quantum correlations~\cite{brandao11,li14} (see Application~\ref{app:SSSAI-b}).
Although the quantities $\D^{\meas,\mathfrak{R}}$ are not necessarily divergences in the sense that they might not obey the data processing inequality, their restricted notions of monotonicity suffice in specific applications.

\subsection{Computational entropy}

A prominent framework of entropies with restrictions on measurements emerged already in classical cryptographic applications.
\textcite{hastad_1999} considered what is now known as the HILL pseudo-entropy: it is defined analogously to smooth min-entropy discussed in Sec.~\ref{sec:one-shot}, but instead of conventional notions of distance it employs the notion of computational distinguishability, where measurements are restricted to those that can be implemented by an efficient (computationally bounded) algorithm.
These and other computational entropy notions were studied systematically by \textcite{barak_2003} and form an important part of the theory of computational pseudorandomness~\cite{vadhan_2012}.
Compared with standard randomness extractors, this allows for extraction of strings that appear random only to constrained observers, applying even to sources from which extraction of perfect randomness is impossible.
Once again, however, care needs to be taken, as such extensions may not possess properties expected of entropies such as chain rules~\cite{krenn_2016}.

Quantum analogues of HILL and related pseudo-entropies were first studied by \textcite{chen_2017}, with further extensions and generalizations developed more recently~\cite{avidan_2026,avidan_2026-1}.
Their definitions connect to the operational interpretations of min-entropy as guessing probability or maximal singlet fraction (Applications~\ref{app:guessing-a}--\ref{app:singletfraction}), expressing the computational restrictions through bounds on quantum circuit size; the fully quantum extension of~\textcite{avidan_2026-1} then defines a corresponding computational max-entropy through duality.

A complementary line of work connects computational constraints to thermodynamics and quantum hypothesis testing. Building on a resource theory of quantum uncomplexity~\cite{yungerhalpern_2022}, \textcite{munson_2025} developed complexity-dependent entropy measures that quantify the trade-off between circuit complexity and the work cost of information erasure. More recently, \textcite{meyer_2025} introduced a computational relative entropy through hypothesis testing with polynomially many copies and quantum gates, deriving also a pseudo-entropy that characterizes efficient quantum compression.


\section{Information measures for Gaussian states}
\label{sec:gaussian}

So far, we have considered only finite-dimensional quantum systems. However, many systems that are of interest in physics are intrinsically infinite dimensional. Important examples of these are continuous-variable systems, which are composed of finite collections of harmonic oscillators, or modes. In this section, we explain how to extend some of the above notions to any quantum system with a separable Hilbert space $A$.

\subsection{Infinite-dimensional entropies and relative entropies}

An operator $T$ acting on a separable Hilbert space $A$ is bounded if $\|T\|_\infty = \sup_{\ket{x}\in A:\, \|\ket{x}\|=1} \left\| T\ket{x}\right\| < \infty$.
A bounded operator $T$ is of trace class if $\|T\|_1$ is finite, and is a Hilbert--Schmidt operator if $\|T\|_2$ is finite. For a bounded self-adjoint operator, we define its range as $\ran(T) = \{T\ket{x}:\, \ket{x}\in A\}$, and its support $\supp(T)$ as the (norm) closure of the range. We will also need to deal with operators $T$ that are unbounded, and thus only defined on a subspace $\dom(T)\subseteq A$, called the domain.

\subsubsection{Von Neumann and R\'enyi entropies}

The analogy with continuous-variable probability density functions may suggest that notions such as differential entropy become relevant in the study of entropies of infinite-dimensional quantum states. However, this is not necessarily the case: density operators on infinite-dimensional separable Hilbert spaces still have a discrete, rather than continuous, spectrum. Indeed, our original definition~\eqref{eq:von_Neumann} of von Neumann entropy can be extended without difficulty to the infinite-dimensional (separable) case: for a positive semidefinite trace class operator $\rho$ on $A$ with spectral decomposition given by the series $\rho=\sum_i p_i \proj{e_i}$, we can set
\begin{equation}
H(\rho) = \sum_i (-p_i \log p_i)\, ,
\label{eq:von_Neumann_infinite}
\end{equation}
where the sum is well defined because %
only finitely many of its terms can be negative\,---\,in fact, for a state, all terms are non-negative. Note that it is possible to have $H(\rho) = +\infty$, for example by considering the \textcite{wehrl1978review} state
\begin{equation}
\rho_W = \sum_{i=1}^\infty \frac{1}{c\, i \log^2(i+1)}\, \proj{i}\, ,
\label{eq:Wehrl_state}
\end{equation}
where $c
< \infty$ is a normalization constant.

Remarkably, the entropy of an infinite-dimensional state $\rho\in \cS(A)$ can be obtained by taking an appropriate limit of the entropies of its finite-dimensional truncations:
\begin{align}
H(\rho) &= \sup_P \left\{ H(P\rho P) + \log \tr[P\rho] \right\} \label{eq:infinite_dim_entropy_as_sup} \\
&= \lim_{n\to\infty} H\big(P_n\rho P_n\big) , \label{eq:infinite_dim_entropy_as_lim}
\end{align}
where the supremum is over all finite-dimensional projectors $P$, and on the second line we considered an arbitrary sequence of finite-dimensional orthogonal projectors
that converges strongly to the identity (i.e., such that $\lim_{n\to\infty} \| P_n\ket{\psi} - \ket{\psi}\| = 0$ for all fixed $\ket{\psi}\in A$). The variational formula~\eqref{eq:infinite_dim_entropy_as_sup} is due to \textcite{lindblad1974expectations}, while the approximation formula~\eqref{eq:infinite_dim_entropy_as_lim} is due to B.\ Simon \cite{lieb1973proof}.

Also R\'enyi entropies do not pose any substantial problem for infinite-dimensional systems: for $\alpha \in (0,1) \cup (1,\infty)$ and a state with spectral decomposition $\rho=\sum_i p_i \ketbra{e_i}{e_i}$, we can we can follow~\eqref{eq:def_Renyi_entropy} and set
\begin{align}
    H_\alpha(\rho) = \frac{1}{1-\alpha} \log\left(\sum\nolimits_i p_i^\alpha\right) ,
\end{align}
where the sum of the series is well defined, although possibly infinite for $\alpha\in (0,1)$; in that case, we set $H_\alpha(\rho)=+\infty$.

\subsubsection{Relative entropy}

Given two positive semidefinite trace class operators $\rho,\sigma$, how is their Umegaki relative entropy $D(\rho\|\sigma)$ defined?  Without loss of generality, we assume that the support condition $\supp(\rho)\subseteq \supp(\sigma)$ is obeyed; otherwise, we simply set $D(\rho\|\sigma) = +\infty$. Looking at~\eqref{eq:Umegaki}, we are immediately faced with the problem of assigning a meaning to the trace of the possibly unbounded operator $\rho \left(\log \rho - \log \sigma\right)$. Fortunately, there is an easy fix, which involves writing out the spectral decompositions $\rho = \sum_i p_i \proj{e_i}$ (with $(\ket{e_i})_i$ a complete orthonormal basis) and $\sigma = \sum_j q_j \proj{f_j}$; up to restricting the second sum, we can assume that $q_j>0$ for all $j$. %
We can now extend~\eqref{eq:Umegaki} by setting~\cite{umegaki1962conditional, lindblad1974expectations}\footnote{One can verify that the value of the right-hand side of~\eqref{eq:Umegaki_infinite} does not depend on which spectral decompositions of $\rho$ and $\sigma$ we choose.}
\begin{align}
D(\rho\|\sigma) &=\! \sum_{i,j} \left|\braket{e_i}{f_j}\right|^2\! f(p_i,q_j) + (\log e) \Tr[\rho \!-\! \sigma],
\label{eq:Umegaki_infinite}
\end{align}
with $f(x,y) = x \log (x/y) + (\log e) (y-x)$.
Note also that the additive term outside the summation is designed to cancel out the contribution of the linear terms $q_j-p_i$ in $f(p_i,q_j)$, producing an expression that is essentially equivalent to the right-hand side of~\eqref{eq:Umegaki}. However, a key advantage of the expression in~\eqref{eq:Umegaki_infinite} is that the value of the sum is now well defined, as all addends are non-negative: in fact, from the convexity of the function $x\mapsto x \log x$ one can deduce that $f(x,y)\geq 0$ for all $x,y>0$.

Like the entropy, also the relative entropy can be recovered by considering finite-dimensional approximations of the states: namely, \textcite{lindblad1974expectations} has shown that
\begin{align}
& D(\rho\|\sigma) \nonumber \\
&\ = \sup_P \left\{ D(P\rho P\| P\sigma P) + (\log e) \tr[(\id\!-\!P)(\rho\!-\!\sigma)] \right\} \label{eq:infinite_dim_relent_as_sup} \\
&\ = \lim_{n\to\infty} D(P_n\rho P_n\| P_n\sigma P_n) , \label{eq:infinite_dim_relent_as_lim}
\end{align}
where, as in~\eqref{eq:infinite_dim_entropy_as_sup}--\eqref{eq:infinite_dim_entropy_as_lim}, the supremum in~\eqref{eq:infinite_dim_relent_as_sup} is over all finite-dimensional orthogonal projectors $P$, and in~\eqref{eq:infinite_dim_relent_as_lim} we considered an arbitrary sequence of finite-dimensional orthogonal projectors converging strongly to the identity.

If $H(\rho)<\infty$ and $\supp(\rho)\subseteq \supp(\sigma)$, then~\eqref{eq:Umegaki_infinite} can be recast into the more familiar form
\begin{equation}
D(\rho\|\sigma) = -H(\rho) - \tr\left[ \rho \log \sigma\right] ;
\label{eq:Umegaki_splitting_infinite}
\end{equation}
the right-hand side is well defined, as the only term that can possibly diverge is the second, which we agree to interpret as the series $-\tr\left[ \rho\log \sigma\right] = -\sum_{i,j} \left|\braket{e_i}{f_j}\right|^2 p_i \log q_j$.

Essentially all properties of $H(\rho)$ and $D(\rho\|\sigma)$ that do not directly involve the Hilbert space dimension carry over to the infinite-dimensional separable case. In particular, $D(\rho\|\sigma)$ is a relative entropy, in the sense that it is monotone and additive, for states it obeys the variational principle~\eqref{eq:qreltentropy-variational}, as already established by \textcite{petz88}, and it admits the integral representation~\eqref{eq:Frenkel}, as argued by \textcite{jencova24}. In the variational principle, one should assume that $0< c \id \leq W \leq C \id$ for some $c,C>0$, so that $\log W$ is well defined; moreover, $\exp(\log \sigma + \log W)$ has to be interpreted as $\Pi^\dag\! \exp\big(\Pi(\log \sigma + \log W)\Pi^\dag\big)\, \Pi$, where $\Pi:A\to \supp(\sigma)$ is the orthogonal projector onto $\supp(\sigma)$. The variational formula~\eqref{eq:variational_H} holds, and one should restrict the infimum to self-adjoint operators $X$ that are bounded from above. The semi-continuity inequality~\eqref{eq:qrelentropy-cont} also carries over, as it does not involve the dimension, but only the max-relative entropy, whose definition does not pose any technical problem in infinite dimension.

\subsubsection{R\'enyi relative entropies}

To define R\'enyi relative entropies in infinite-dimensional systems, it is useful to follow the approach of \textcite{mosonyi2023strong}, which rests on the unifying perspective offered by $\alpha$-$z$ relative entropies. These are defined by~\eqref{def: alpha-z relative entropies} in the finite-dimensional case, and we can readily extend the definition by setting, for all $(\alpha,z)\in \big((0,1)\cup(1,\infty)\big)\times (0,\infty)$ and all pairs of positive semi-definite trace class operators $\rho,\sigma$,
\begin{equation}\begin{aligned} \label{eq:alpha-z-infinite-dimension}
D_{\alpha,z}(\rho\|\sigma) = \frac{1}{\alpha-1} \log \big\| \sigma^{\frac{1-\alpha}{2z}} \rho^{\frac{\alpha}{2z}} \big\|^{2z}_{2z}\, ,
\end{aligned}\end{equation}
where $\|T\|_{2z} = \big(\tr (TT^\dag)^{z}\big)^{\frac{1}{2z}}$ is the Schatten $2z$-norm. In the case where $\alpha\in (0,1)$, the above expression is always well defined, although possibly infinite\,---\,this happens precisely if $\supp(\rho) \perp \supp(\sigma)$. Indeed, $\sigma^{\frac{1-\alpha}{2z}} \rho^{\frac{\alpha}{2z}}$, which is the product of two bounded operators, always has finite Schatten $2z$-norm, as a simple application of the H\"older inequality reveals. For $\alpha\in (1,\infty)$, the situation is more complicated. We first restrict to the case where $\supp(\rho)\subseteq \supp(\sigma)$, setting $D_{\alpha,z}(\rho\|\sigma) = +\infty$ otherwise. In order to make sense of the product that appears in~\eqref{eq:alpha-z-infinite-dimension}, we need to make sure that the range of the operator that acts first is contained in the domain of the operator that acts second. This amounts to requiring that $\ran\big(\rho^{\frac{\alpha}{2z}}\big) \subseteq \dom\big(\sigma^{\frac{1-\alpha}{2z}}\big) = \ran\big(\sigma^{\frac{\alpha-1}{2z}}\big)$. Due to a lemma by \textcite{Douglas1966}, this is equivalent to the existence of some $\lambda\geq 0$ such that $\rho^{\frac{\alpha}{z}} \leq \lambda\,\sigma^{\frac{\alpha-1}{z}}$. If these conditions are violated we set immediately $D_{\alpha,z}(\rho\|\sigma) = +\infty$. If they are obeyed, then $\sigma^{\frac{1-\alpha}{2z}} \rho^{\frac{\alpha}{2z}}$ is well defined and bounded; we can therefore compute its Schatten $2z$-norm, which may be finite or infinite\,---\,in this latter case, we again set $D_{\alpha,z}(\rho\|\sigma) = +\infty$.

The Petz--R\'enyi and the sandwiched R\'enyi relative entropies are obtained by setting $z=1$ and $z=\alpha$, respectively:
\begin{align}
\widebar{D}_{\alpha}(\rho\|\sigma) &= \frac{1}{\alpha-1} \log \big\| \sigma^{\frac{1-\alpha}{2}} \rho^{\frac{\alpha}{2}} \big\|^{2}_{2}\, , \label{eq:Petz_def_infinite_dim} \\
\widetilde{D}_{\alpha}(\rho\|\sigma) &= \frac{1}{\alpha-1} \log \big\| \sigma^{\frac{1-\alpha}{2\alpha}} \rho^{\frac12} \big\|^{2\alpha}_{2\alpha}\, , \label{eq:sandwiched_def_infinite_dim}
\end{align}
For $\alpha>1$, a specialization of the above consistency requirements yields the conditions $\rho^\alpha \leq \lambda \sigma^{\alpha-1}$ for Petz and $\rho \leq \lambda \sigma^{\frac{\alpha-1}{\alpha}}$ for sandwiched. Finally, the definition of max-relative entropy via the optimization on the first line of~\eqref{eq:Dmax} poses no problem.

\subsubsection{(Dis)continuity properties}

An identity that does not carry over is the relation $H(\rho) = \log d - D(\rho\|\pi)$ in~\eqref{eq:entropy_from_relative_entropy}, as it involves the maximally mixed state $\pi$, which does not exist in infinite-dimensional systems. The same holds for the continuity relation~\eqref{eq:Fannes_Audenaert}, which trivializes if one sets $d=+\infty$. This is no coincidence, as it can be shown that, if the underlying system is infinite dimensional, the entropy is %
discontinuous at any state where it is finite: given any state $\rho$ and any $\ve>0$, one can find some state $\tilde{\rho}$ such that $\frac12 \|\rho - \tilde{\rho}\|_1\leq \ve$ and $H(\tilde{\rho}) = +\infty$~\cite{wehrl1978review}.\footnote{To construct $\tilde{\rho}$, one truncates off a tail with sufficiently small weight from the spectral expansion of $\rho$, and adds a rescaled multiple of the Wehrl state from~\eqref{eq:Wehrl_state}.}

While general continuity properties break down, entropies and relative entropies in infinite-dimensional systems are still typically lower semi-continuous. Starting from the von Neumann entropy, we have that, for a generic sequence $(\rho_n)_n$ of states on the same system,
\begin{equation}
\lim_{n\to\infty} \left\|\rho - \rho_n\right\|_1 = 0\quad \Rightarrow\quad H(\rho) \leq \liminf_{n\to\infty} H(\rho_n) ,
\end{equation}
where $\rho$ is the limit of the sequence, and convergence in trace norm guarantees that it is also a state. Similarly, for the relative entropy we have
\begin{equation}
\left. \begin{array}{l} \lim\limits_{n\to\infty} \left\|\rho - \rho_n\right\|_1 = 0 \\[.5ex] \lim\limits_{n\to\infty} \hspace{-1.3pt} \left\|\sigma - \sigma_n\right\|_1 = 0 \end{array} \right\} \ \, \Rightarrow\ \, D(\rho\|\sigma) \leq \liminf_{n\to\infty} D(\rho_n\|\sigma_n) .
\end{equation}
Note that strict inequality can hold also in the finite-dimensional case: for two sequences of distinct pure states converging to the same pure state, the left-hand side vanishes, and the right-hand side is $+\infty$.

\subsection{Continuous-variable systems}

A set of $n$ harmonic oscillators, also called ``modes'', is represented by the Hilbert space $A^n = L^2(\bbR^n) \simeq L^2(\bbR)^{\otimes n}$ of square-integrable complex-valued functions on $\bbR^n$. Denoting the annihilation and creation operators on the $j^\text{th}$ mode by $a_j$ and $a_j^\dag$, respectively, a handier representation of $A^n$ can be obtained by introducing the orthonormal basis of Fock states $\{ \ket{\boldsymbol{k}}\}_{\boldsymbol{k}\in \bbN^n}$, where, for $k = (k_1,\ldots,k_n)^\intercal$, we set $\ket{k} = \ket{k_1}\otimes \ldots \otimes \ket{k_n}$, and, on each mode, $\ket{h} = (a^\dag)^h\ket{0}/\sqrt{h!}$, where $\ket{0}$ denotes the vacuum state.
The canonical Hamiltonian, which counts the total excitation number, can be written as $N_n = \sum_{j=1}^n {a_j}^\dag {a_j}^{\vphantom{\dag}}$; its eigenvectors are precisely the Fock states $\ket{k}$, and the associated eigenvalues are $N_n\ket{k} = \big(\sum_{j=1}^n k_j \big)\ket{k}$. The annihilation and creation operators and the associated quadratures $x_j = (a_j+{a_j}^\dag)/\sqrt2$ and $p_j = -i (a_j-{a_j}^\dag)/\sqrt2$, as well as any %
polynomials thereof, can be thought of as acting on the dense subspace of finite linear combinations of Fock states.

We will now briefly describe the phase space formalism of continuous-variable quantum systems, mostly following the presentations by~\textcite{HOLEVO} and~\textcite{BUCCO}. To this end, it will be convenient to group the quadratures into the vector $R= (x_1,p_1,\ldots,x_n,p_n)^\intercal$, so that the canonical commutation relations $[x_j,p_k]=i\delta_{jk}$ take the more compact form
\begin{equation}
\left[R , R^\intercal\right] = i\Omega = i \left(\begin{matrix} 0 & 1 \\ -1 & 0 \end{matrix}\right)^{\oplus n} ;
\end{equation}
here, the object on the left-hand side is a matrix of operators on $A^n$ with entries $\left[R , R^\intercal\right]_{ab} = \left[R_a , R_b\right]$, and on the right-hand side we identified scalars with multiples of the identity operators on $A^n$. A key role in the theory is played by the phase space, i.e., the Euclidean space $\bbR^{2n}$ that emerges naturally from this picture. An arbitrary phase space vector $s\in \bbR^{2n}$ can be decomposed as $s = s_1\oplus \ldots \oplus s_n$, where each $s_j\in \bbR^2$ ($j=1,\ldots,n$) is obtained by grouping together the two components of $s$ corresponding to the $j^\text{th}$ mode.

Given $s\in \bbR^{2n}$, the associated displacement operator is the unitary operator~\cite{Weyl1927}
\begin{equation}
D(s) = e^{-i s^\intercal \Omega R} = D(s_1)\otimes \ldots\otimes D(s_n)\, ,
\label{eq:displacement}
\end{equation}
where $s = s_1\oplus \ldots \oplus s_n$. Note that $D(s)^\dag = D(-s) = D(s)^{-1}$. One can also show that $D(-s) R_j D(s) = R_j + s_j$, where scalars represent multiples of the identity operators.

Hamiltonians that are polynomials of degree at most two in the quadrature operators, i.e., of the form $H = \frac12 R^\intercal K R + w^\intercal R$, where $w\in \bbR^{2n}$ and $K$ is a $2n\times 2n$ real symmetric matrix, induce unitaries of the form $e^{-iH}$. These operators generate the group of so-called Gaussian unitaries. In the special case where $w=0$, a unitary defined by a single exponential acts as $e^{iH} R_j e^{-iH} = \sum_k S_{jk} R_k$, where $S = e^{\Omega K}$ is a matrix belonging to the symplectic group, defined by the identity $S\Omega S^\intercal = \Omega$. As it turns out, any symplectic matrix $S$ can be associated with a Gaussian unitary $U_S$ satisfying $U_S^\dag R_j U_S^{\vphantom{\dag}} = \sum_k S_{jk} R_k$, which is unique up to a global phase; the map $S\mapsto U_S$ is often called the projective metaplectic representation. Moreover, every Gaussian unitary can be written, up to a global phase, as a product $D(t)U_S$ of a displacement operator $D(t)$ and a unitary $U_S$ associated with a symplectic matrix.

Given any trace class operator $T$ on $n$ modes, we can construct the associated \textcite{Wigner} function $W_T:\bbR^{2n}\to \bbC$, defined by \cite{Grossmann1976, Hillery1984}
\begin{align}
W_T(u) &= \frac{1}{\pi^n} \Tr \big[ T D(u) \Pi D(-u) \big] \label{eq:Wigner_function_1} \\
&= \int \frac{\dd^{2n}s}{(2\pi)^{2n}}\ \Tr[T D(s)]\, e^{-i u^\intercal \Omega s} , \label{eq:Wigner_function_2}
\end{align}
where $\Pi = (-1)^{N_n} = \bigotimes_{j=1}^n (-1)^{a_j^\dag a_j^{\vphantom{\dag}}}$ is the parity operator, whose eigenvectors are the Fock states $\{\ket{\boldsymbol{k}}\}_{\boldsymbol{k}\in \bbN^n}$, with associated eigenvalue given by $(-1)^{\sum_{j=1}^n k_j}$. In~\eqref{eq:Wigner_function_2}, we expressed $W_T$ as the Fourier transform of the so-called characteristic function $s\mapsto \Tr[T D(s)]$.

The correspondence
$T\mapsto W_T$ can be extended to a fully fledged isomorphism between the space of Hilbert--Schmidt operators on $A^n$ and the space of square-integrable complex-valued functions on the phase space $\bbR^{2n}$. In particular, we can fully describe an operator by specifying %
its Wigner function.
One verifies that $\int \dd^{2n}u\ W_T(u) = \tr[T]$ whenever $W_T$ is integrable, which suggests an interpretation of $W_\rho$ as the ``quasi-probability distribution'' induced by a state $\rho$ on the phase space. However, $W_\rho$ can take, in general, negative values.

We can now define a Gaussian state as a state whose Wigner function is a Gaussian probability distribution. We will call $m\in \bbR^{2n}$ the associated vector of first moments, and $V/2 \in \bbR^{2n\times 2n}$ the corresponding covariance matrix. Note that $V$ is positive definite. Using normalization, we can write the associated Gaussian state as the state $\rho_\G[m,V]$ with Wigner function
\begin{equation}
W_{\rho_\G[m,V]}(u) = \frac{1}{\pi^n\sqrt{\det V}}\, e^{-(u-m)^\intercal V^{-1} (u-m)} .
\label{eq:Wigner_Gaussian}
\end{equation}
We refer to $m$ as the mean of the state, and to $V$ as its quantum covariance matrix. These names are justified by the fact that
\begin{align}
\tr\big[\rho_\G[m,V] R \big] &= m , \\
\tr\big[\rho_\G[m,V] \{R-m,\,(R-m)^\intercal\} \big] &= V ,
\end{align}
where $\{X,Y\} = XY+YX$ is the anti-commutator; these are to be interpreted as entry-wise equalities. While any $m\in \bbR^{2n}$ can be the mean of a Gaussian state, only the real symmetric matrices $V$ that satisfy the bona fide condition
\begin{equation}
V + i\Omega \geq 0
\label{eq:bona_fide}
\end{equation}
can be quantum covariance matrices of valid quantum states. While any $V$ that satisfies~\eqref{eq:bona_fide} is automatically positive definite, the converse is not true. If $V$ is positive definite but does not satisfy~\eqref{eq:bona_fide}, the operator $\rho_\G[m,V]$ defined by~\eqref{eq:Wigner_Gaussian} is still trace class but no longer positive semidefinite. To verify~\eqref{eq:bona_fide}, we can multiply both sides from the left and from the right by $V^{-1/2}$. The matrix $V^{1/2}i\Omega V^{1/2}$ is Hermitian and skew-symmetric, and thus has a spectrum of the form
\begin{align}
\spec\left(V^{1/2} i\Omega V^{1/2}\right) &= \spec\left(Vi\Omega\right) \\
&= \left\{ \nu_1, -\nu_1,\ldots, \nu_n, -\nu_n\right\} .
\end{align}
The numbers $\nu_1,\ldots,\nu_n$ are called the symplectic eigenvalues of $V$. Using this observation, it is straightforward to verify that %
\eqref{eq:bona_fide} can be simply rephrased as the set of inequalities
\begin{align}
\nu_j \geq 1\, ,\qquad \forall\ j=1,\ldots, n\, .
\end{align}

Some notable examples of single-mode Gaussian states are: coherent states~\cite{Schroedinger1926-coherent, Glauber1963, Sudarshan1963}, which are pure states of the form $\rho_\G[m,\one] = \ketbra{m}{m}$, where $m\in \bbR^{2}$ is arbitrary, and can be obtained by applying the displacement operator $D(m)$ to the vacuum state; and thermal states, which are of the form $\rho_\G[0,(2k+1)\one] = \tau_k$, where $k \geq 0$ is called their mean photon number.

A fundamental observation, sometimes known as William\-son's theorem, is that any Gaussian state can be rotated, via a displacement followed by a symplectic unitary, into a product of thermal states (one per mode) of possibly different temperatures. Once the Gaussian state is in that form, its support can be readily identified as the tensor product of all the local Hilbert spaces with non-zero temperature, padded with vacuum states on the zero-temperature modes. Leveraging this fact, one can show that
\begin{align} \label{eq:support_inclusion_Gaussian}
&\supp\big(\rho_\G[m_1,V_1]\big) \subseteq \supp\big(\rho_\G[m_2,V_2]\big) \iff \nonumber \\
&\ker(V_2\!+\!i\Omega) \subseteq \ker(V_1\!+\!i\Omega)\ \text{and}\ m_1\!-\!m_2\perp \ker(V_2\!+\!i\Omega) .
\end{align}

\subsection{Entropies of Gaussian states}

Throughout this subsection, $V$ denotes the quantum covariance matrix in the convention of~\eqref{eq:Wigner_Gaussian}; in particular, the vacuum covariance is $\id$.
Let $\nu_1(V),\ldots,\nu_n(V)$ be the symplectic eigenvalues of $V$. The von Neumann entropy of a Gaussian state depends only on these symplectic eigenvalues~\cite{Agarwal1971, holevo1999}:
\begin{align}
H\left( \rho_\G[m,V] \right) &= \sum_{j=1}^n g\!\left( \frac{\nu_j(V)-1}{2}\right) , \label{eq:entropy_Gaussian_state} \\
g(x) &= (x+1)\log (x+1) - x\log x\, . \label{eq:g}
\end{align}
The function $g$ is sometimes called the bosonic entropy function. The absence of $m$ from the formula reflects the fact that first moments are changed by displacement unitaries, and unitaries cannot affect the entropy. Since $g(0)=0$, pure Gaussian modes, corresponding to symplectic eigenvalues equal to $1$, do not contribute. Note that a thermal state with mean photon number $N\geq 0$ has entropy $g(N)$.

\begin{appbox}{capacity}{Communication through lossy bosonic channels}{app:gaussian-capacity}
For a single-mode pure-loss channel $\mathcal L_\eta$ of transmissivity $\eta\in[0,1]$ and average input photon number at most $E$, the unassisted classical capacity is~\cite{giovannetti2004classical}
\begin{align*}
C(\mathcal L_\eta,E) = g(\eta E)\, .
\end{align*}
With binary logarithms, $C$ is in bits per channel use. This formula can be extended to the case where the environment is at non-zero temperature, in which case the resulting channel, called a thermal attenuator and denoted $\mathcal{L}_{\eta,N}$, is characterized by two parameters, the transmissivity $\eta\in[0,1]$ and the mean photon number $N\geq 0$ of the environment. Following the solution of the Gaussian optimizer conjecture, its unassisted classical capacity has been calculated to be~\cite{Giovadd, Giovadd-CMP}
\begin{align*}
C(\mathcal L_{\eta,N},E) = g\left(\eta E + (1-\eta) N\right) - g\left( (1-\eta) N \right) .
\end{align*}
In both cases, a Gaussian ensemble of coherent states achieves this rate with collective decoding. Each conditional output is a displaced thermal state with mean photon number $(1-\eta)N$, while their average is a thermal state with mean photon number $\eta E + (1-\eta)N$. Since displacements do not change entropy, the Holevo information associated with this ensemble equals the difference between the entropies of these two thermal states. The formula for the capacity follows because this ensemble maximizes the Holevo information under the energy constraint, and this maximum is additive over multiple uses. In fact, such additivity has recently been established for all single-mode Gaussian channels~\cite{Adesso2026}.
\end{appbox}

We note in passing the following integral representation of the von Neumann entropy of a Gaussian state:
\begin{align}
H\left(\rho_\G[m,V]\right) &= n\log\frac{e}{2} + \frac12\int_0^1 \dd t\ \log\det(V+it\Omega)\, .
\label{gaussian_entropy_integral}
\end{align}

The R\'enyi entropies are obtained by applying the same normal-mode decomposition found in~\cite{holevo1999}. For $\alpha>0$, $\alpha\neq 1$, we obtain
\begin{align}
H_\alpha \left( \rho_\G[m,V] \right) &= \sum_{j=1}^n g_\alpha\!\left( \frac{\nu_j(V)-1}{2}\right) , \label{eq:Renyi_entropy_Gaussian_state} \\
g_\alpha(x) &= \frac{1}{\alpha-1} \log\left((x+1)^\alpha - x^\alpha\right) .
\label{eq:g_alpha}
\end{align}
As expected, $g_\alpha(x)\to g(x)$ as $\alpha\to1$. R\'enyi entropies with $0<\alpha\leq 2$ admit an integral representation that generalizes~\eqref{gaussian_entropy_integral}:
\begin{align}
H_\alpha\left(\rho_\G[m,V]\right) &= n\left(\frac{\log\alpha}{\alpha-1}-\log2\right) \nonumber \\
&\quad + \frac12\int_0^1 \dd t\ p_\alpha(t)\log\det(V+it\Omega)\, ,
\label{gaussian_renyi_integral}
\end{align}
where $p_\alpha$ is the probability density on $(0,1)$ given by
\begin{align}
p_\alpha(t) &= -\frac{4\alpha\sin(\pi\alpha)}{\pi(\alpha\!-\!1)} \nonumber \\
&\quad \cdot \frac{(1-t^2)^{\alpha-1}}{(1+t)^{2\alpha}+(1-t)^{2\alpha}-2(1-t^2)^\alpha\cos(\pi\alpha)}\, .
\label{gaussian_renyi_density}
\end{align}
In the limit $\alpha\to 1$, the density becomes $p_1(t)=1$, and~\eqref{gaussian_renyi_integral} reduces to~\eqref{gaussian_entropy_integral}. At $\alpha=2$, the probability measure is instead a point mass at $t=0$, giving $H_2(\rho_\G[m,V])=\frac12\log\det V$.

A notable consequence of the above integral representations is in the context of the entropy-cone problem for quantum Gaussian states (cf.\ Section~\ref{sec:standard/inequalities}). Consider an $n$-partite continuous-variable quantum system $A_1\ldots A_n$, where each $A_j$ is composed of arbitrarily many modes. For a fixed $\alpha>0$, we are then interested in the inequalities of the form
\begin{align}
\label{eq: Renyi entropy cone}
    \sum_{S\subseteq[n]} c_S H_\alpha(\rho_S) \geq 0\, ,
\end{align}
with real coefficients $\{c_S\}_{S\subseteq[n]}$, that are obeyed by all Gaussian states $\rho_{A_1\ldots A_n}$. A simple consequence of~\eqref{gaussian_entropy_integral} and~\eqref{gaussian_renyi_integral} is that, \emph{for every fixed $0 < \alpha \le 2$, a balanced homogeneous linear entropy inequality holds for the R\'enyi-$\alpha$ entropies of all bosonic Gaussian states if and only if it holds for all classical vector Gaussians, with $H_\alpha$ replaced by the differential (Shannon) entropy.}\footnote{The classical-to-quantum implication follows from~\eqref{gaussian_entropy_integral} and~\eqref{gaussian_renyi_integral} by noting that, up to constants that cancel by balance, $\frac12 \log \det V$ equals the entropy of a classical Gaussian vector with covariance matrix $V$. Conversely, we can encode a classical Gaussian vector $X$ into coherent states as $\rho_t = \int\dd x\,p_X(x) \ketbra{tx}{tx}$, where $t>0$ is a scale factor. As $t\to\infty$, its marginal R\'enyi entropies approximate the corresponding differential entropies of $X$ up to terms linear in the number of modes, which again cancel out.} Recall that an entropy inequality is balanced when $\sum_{S:\,j\in S} c_S=0$ for every party $j=1,\ldots,n$. In particular, the case $\alpha=1$ answers in the affirmative the question raised on p.~2 of~\cite{Gross2013}. In the case $\alpha=2$, some of the resulting R\'enyi entropy inequalities (including strong subadditivity), which are naturally phrased in terms of logarithms of minors of a quantum covariance matrix, can be strengthened to operator inequalities involving Schur complements \cite{Lami16, LL-log-det}.

\subsection{Relative entropy and fidelity between two Gaussian states}

We now turn to formulas involving two Gaussian states. For non-faithful inputs, the expressions below are understood as limits with $V\to V+\ve\id$ and $V'\to V'+\ve\id$, as $\ve\to 0^+$, within the stated parameter and finiteness regimes.
For the relative entropy, it is convenient to introduce the Gibbs-matrix function
\begin{align}
G(V) = 2i\Omega\, \operatorname{arccoth}(Vi\Omega) ,
\end{align}
where $\operatorname{arccoth}(x)=\frac12\log\frac{x+1}{x-1}$, and its application to matrices is understood through functional calculus. With this notation, the Umegaki relative entropy can be expressed as~\cite{PLOB}
\begin{align}
&D\big(\rho_\G[m,V] \,\big\|\, \rho_\G[m',V']\big) \nonumber \\
&= \frac12 \log\frac{\det(V'\!+\!i\Omega)}{\det(V\!+\!i\Omega)} + \frac14 \Tr\!\left[ V\big(G(V')-G(V)\big) \right] \nonumber \\
&\quad + \frac12 (m-m')^\intercal G(V') (m-m')\, .
\end{align}
For Gaussian states, this relative entropy is finite if and only if $\supp(\rho_\G[m,V]) \subseteq \supp(\rho_\G[m',V'])$, which can be checked using~\eqref{eq:support_inclusion_Gaussian}. Indeed, after removing the common pure modes, the second state is faithful and its logarithm is a quadratic polynomial in the quadratures, whose expectation in the first state is necessarily finite.

For the fidelity and for the R\'enyi relative entropies below, we use the functions
\begin{align}
f_t(x) = \frac{\left(1+\frac1x\right)^t + \left(1-\frac1x\right)^t}{\left(1+\frac1x\right)^t - \left(1-\frac1x\right)^t}\, , \qquad t>0,
\end{align}
which are monotonic and bijective on $x \in (-\infty,-1] \cup [1,\infty)$ for all $t>0$. The fidelity between two Gaussian states can then be written as~\cite{twamley1996, paraoanu1998, scutaru1998, paraoanu2000, wang_2000, marian2012, banchi2015}
\begin{align}
&F\big(\rho_\G[m,V],\rho_\G[m',V']\big) \nonumber \\
&= \sqrt{\frac{\det Z_{1/2}}{\det\!\left(\frac{V+V'}{2}\right)}}\, e^{- (m-m')^\intercal (V+V')^{-1} (m-m')} ,
\end{align}
where
\begin{align}
Z_{1/2} &= f_{1/2}\big(W_{1/2} i\Omega\big)i\Omega , \\
W_{1/2} &= V - \sqrt{\id + (V\Omega)^{-2}}\, V (V+V')^{-1} V\, \sqrt{\id + (\Omega V)^{-2}}\, .
\end{align}
Although $Z_{1/2}$ is not manifestly symmetric under $V\leftrightarrow V'$, its determinant can be shown to be. Indeed, the relevant part of the spectrum of the matrix $W_{1/2} i\Omega$ can also be shown to be invariant under the exchange $V\leftrightarrow V'$.

\subsection{Other R\'enyi divergences}

Closed formulas are also available for the Petz and the sandwiched R\'enyi relative entropies of Gaussian states~\cite{calsamiglia2008, pirandola2008, lami2018}. We state here the formulas corresponding to the two parameter ranges most often associated with the direct and strong-converse regimes of quantum hypothesis testing, referring the reader to the original works for more details.
These formulas can be derived systematically by a repeated application of the ``golden rule'' derived by \textcite{LDW2018}. This technique, which is in essence a straightforward application of the Baker--Campbell--Hausdorff formula, allows us to calculate products of powers of Gaussian operators by mapping the problem into one involving matrices only.

For two $n$-mode quantum covariance matrices $V$ and $V'$, and some $\alpha \in (0,1)$, define
\begin{align}
V_\alpha &= f_\alpha(Vi\Omega)i\Omega\, , \\
V'_{1-\alpha} &= f_{1-\alpha}(V'i\Omega)i\Omega\, .
\end{align}
Then the Petz--R\'enyi relative entropy between the two corresponding Gaussian states is \cite{calsamiglia2008, pirandola2008}
\begin{equation}
\begin{aligned}
\label{gaussian_petz_relative_entropy}
&\widebar{D}_\alpha\big(\rho_\G[m,V] \,\big\|\, \rho_\G[m',V']\big) \\
&= \frac{1}{1\!-\!\alpha} \bigg\{ (\log e) (m-m')^\intercal (V_\alpha+V'_{1-\alpha})^{-1} (m-m') \\
&\; + \frac{1}{2} \log \frac{\big(\!\det(V\!+\!i\Omega)\big)^\alpha \big(\!\det(V' \!+\! i\Omega) \big)^{1-\alpha}\det\!\big(V_\alpha \!+\! V'_{1-\alpha}\big)}{\det(V_\alpha\!+\!i\Omega)\det\big(V'_{1-\alpha}\!+\! i\Omega\big)} \bigg\}\, .
\end{aligned}
\end{equation}
The formula depends on the first moments exclusively through their difference $m-m'$, as is dictated by unitary invariance. Note that, for $0<\alpha<1$, the Petz--R\'enyi relative entropy (and, hence, also the sandwiched one) is finite for every pair of Gaussian states.

The Chernoff exponent between $\rho_\G[m,V]$ and $\rho_\G[m',V']$ can be calculated, as in the finite-dimensional case, as the supremum of $(1-\alpha) \widebar{D}_\alpha\big(\rho_\G[m,V] \,\big\|\, \rho_\G[m',V']\big)$ over $\alpha\in (0,1)$ (Application~\ref{app:HT-b}). Due to the formula~\eqref{gaussian_petz_relative_entropy}, this optimization can be solved numerically (and sometimes analytically) for every pair of Gaussian states~\cite{calsamiglia2008, pirandola2008}.

\begin{appbox}{HT}{Quantum illumination}{app:quantum-illumination}
In the task of quantum illumination~\cite{tan2008illumination}, one sends one mode of a two-mode squeezed vacuum state $\ket{\psi_{N_S}} = \sum_{n=0}^\infty \frac{N_S^{n/2}}{(N_S+1)^{(n+1)/2}}\,\ket{n}\ket{n}$ towards a target region, while retaining the other. It is possible to show that $\ket{\psi_{N_S}}$, which is a purification of a thermal state with mean photon number $N_S$, is a Gaussian state. The goal is to determine whether the target region contains a partially reflective object or not, assuming that the returned background has the same mean photon number $N_B$ under both hypotheses. The overall state of the two-mode return--idler system is called $\rho_0$ if no object is present and $\rho_1$ if an object is present. A little thought reveals that $\rho_0$ is simply a product of two thermal states, one with mean photon number $N_B$ (the return) and one with mean photon number $N_S$ (the idler). Modeling the reflection as a thermal attenuator $\mathcal{L}_{\eta,\,N_B/(1-\eta)}$, where $\eta\in (0,1)$ encodes the object reflectivity, we see that $\rho_1 = (\mathcal{L}_{\eta,\,N_B/(1-\eta)}\otimes \idc)(\ketbra{\psi_{N_S}}{\psi_{N_S}})$. Therefore, both $\rho_0$ and $\rho_1$ are Gaussian states.

For equally likely hypotheses, independent repetitions and optimal collective detection, the average error probability is governed by the Chernoff error exponent (Application~\ref{app:HT-b}), which can be calculated as $C(\rho_0,\rho_1)=\sup_{0<\alpha<1}(1-\alpha)\,\widebar{D}_\alpha(\rho_0\|\rho_1)$. Specializing~\eqref{gaussian_petz_relative_entropy} to this case allows for a numerical evaluation of this formula~\cite{tan2008illumination}. In the limit of low signal brightness ($N_S\ll 1$), high noise ($N_B\gg 1$), and weak reflectivity ($\eta\ll 1$), already the choice $\alpha=1/2$ yields an achievable error exponent asymptotically four times the optimal exponent obtained by sending coherent states at the same transmitted signal energy. Quantum illumination, therefore, significantly outperforms coherent-state illumination.
\end{appbox}

In the case where the second state is faithful, i.e., when all symplectic eigenvalues of $V'$ are strictly larger than $1$, the Petz--R\'enyi relative entropy $\widebar{D}_\alpha\big(\rho_\G[m,V] \,\big\|\, \rho_\G[m',V']\big)$ for $\alpha>1$ is finite if and only if\footnote{Its necessity was left open by \textcite{lami2018}, and follows from a simple argument: assuming, up to applying a Gaussian unitary, that $\sigma$ is a product of thermal states, we can write $\sigma^{1-\alpha}$ as a positive integral of coherent-state projectors, understood as a quadratic form; the squared Hilbert--Schmidt norm in~\eqref{eq:Petz_def_infinite_dim} can then be evaluated in terms of the matrix elements $\bra{s}\rho^\alpha\ket{s}$; the resulting Gaussian integral converges exactly when the covariance difference in~\eqref{gaussian_petz_finiteness} is strictly positive.}
\begin{equation}
f_{\alpha-1}(V'i\Omega)i\Omega > f_\alpha(Vi\Omega)i\Omega\, .
\label{gaussian_petz_finiteness}
\end{equation}
Here, the matrices on the two sides are the covariance matrices of the normalised powers $\sigma^{\alpha-1}$ and $\rho^\alpha$, respectively. If the second state is not faithful, i.e., if some of the symplectic eigenvalues of $V'$ are equal to $1$, then we first need to make sure that $\supp(\rho)\subseteq \supp(\sigma)$, which can be done by checking~\eqref{eq:support_inclusion_Gaussian}, and then we need to apply~\eqref{gaussian_petz_finiteness} to the two Gaussian states obtained by removing the common pure modes. This support reduction also applies to the sandwiched relative entropy below, and needs to be performed before the regularization $V\to V+\eps\id$, $V'\to V'+\eps\id$, $\eps\to 0^+$.

In the case where $\widebar{D}_\alpha\big(\rho_\G[m,V] \,\big\|\, \rho_\G[m',V']\big) < \infty$, the expression in~\eqref{gaussian_petz_relative_entropy} also holds for $\alpha>1$, upon extending $f_{-t}=-f_t$ for $t>0$.

For the sandwiched R\'enyi relative entropy with $\alpha\in(1,\infty)$, we need the auxiliary matrices
\begin{align}
U_\alpha &= f_{1-1/\alpha}\big(V'i\Omega\big)i\Omega , \\
W_\alpha &= V + \sqrt{\id \!+\! (V\Omega)^{-2}}\, V \big(U_\alpha \!-\! V\big)^{-1} V \sqrt{\id \!+\! (\Omega V)^{-2}}\, , \\
Z_\alpha &= f_\alpha\big(W_\alpha i\Omega\big)i\Omega .
\end{align}
When the second state is faithful, the sandwiched R\'enyi relative entropy between $\rho_\G[m,V]$ and $\rho_\G[m',V']$ is finite if and only if $U_\alpha > V$,\footnote{The necessity of this condition was also left open in \cite{lami2018}. However, finiteness implies that the operator $T=\sigma^{-\frac{\alpha-1}{2\alpha}}\rho^{\frac12}$ is bounded and $\tr[(T^\dag T)^\alpha]<\infty$. Applying Jensen's inequality to the coherent-state expectations of $T^\dag T$ forces the corresponding Gaussian integral to converge, which is equivalent to $U_\alpha>V$.} and in that case it is given by~\cite{lami2018}
\begin{align}
&\widetilde{D}_\alpha\big(\rho_\G[m,V] \,\big\|\, \rho_\G[m',V']\big) \nonumber \\
&= \frac{\alpha}{\alpha\!-\!1} (\log e) (m-m')^\intercal \big(U_\alpha \!-\! V\big)^{-1} (m-m') \nonumber \\
&\quad + \frac{1}{2(\alpha\!-\!1)} \log \frac{\big(\det(V'\!+\!i\Omega)\big)^{\alpha-1}\det(Z_\alpha\!+\!i\Omega)}{\big(\det(V\!+\!i\Omega)\big)^\alpha}\, .
\end{align}
In the limit $\alpha\to 1$, the above expressions reduce to the Umegaki relative entropy displayed in the previous subsection. For faithful Gaussian states, the derivatives at $\alpha=1$ of both the Petz and the sandwiched relative entropies equal $\frac{1}{2\log e}\, V(\rho\|\sigma)$, where $V(\rho\|\sigma)$ is the relative entropy variance of~\eqref{eq:relative_entropy_variance}; for Gaussian states, \textcite{wilde2017} have computed this variance explicitly.

For the max-relative entropy, the condition $V'>V$ is sufficient for finiteness, and $V'\geq V$ is necessary. A general formula is known if the former condition is obeyed; in this case, define
\begin{align}
W_\infty &= V + \sqrt{\id\!+\!(V\Omega)^{-2}}\, V (V'\!-\!V)^{-1} V \sqrt{\id\!+\!(\Omega V)^{-2}}\, .
\label{gaussian_max_covariance}
\end{align}
Writing $\nu_j(W_\infty)$ for its symplectic eigenvalues, \textcite{lami2018} have obtained
\begin{align}
&D_{\max}\big(\rho_\G[m,V] \,\big\|\, \rho_\G[m',V']\big) \nonumber \\
&\quad = (\log e)(m-m')^\intercal (V'-V)^{-1}(m-m') \nonumber \\
&\quad \quad + \log\frac{\det(V'+i\Omega)}{\sqrt{\det(V'-V)}\prod_{j=1}^n\big(\nu_j(W_\infty)+1\big)}\, .
\label{gaussian_max_relative_entropy}
\end{align}
Incidentally, the non-strict inequality $V'\geq V$ alone does not guarantee finiteness of the max-relative entropy.\footnote{For example, the centred one-mode pair with $V=\id$ and $V'=\operatorname{diag}(1,2)$ has infinite max-relative entropy.} To our knowledge, a complete characterization of the conditions under which $D_{\max}\big(\rho_\G[m,V] \,\big\|\, \rho_\G[m',V']\big)<\infty$ on the boundary where $V'-V \geq 0$ has non-trivial kernel remains open.


\section{Information measures for states on von Neumann algebras}
\label{sec:algebraic}

Entropic information measures can equally well be defined for general infinite-dimensional systems, which cannot necessarily be described by separable Hilbert spaces. This is, e.g., of importance for applications in quantum field theory, quantum gravity, and for studying the thermodynamic limit of quantum many body systems. One might also look at information-theoretic tasks in general systems for the sake of gaining insight into operator algebra theory. Now, the mathematical difficulty for this extension is that in general no trace functional exists, whereas all entropic measures defined so far crucially featured a trace. Luckily, it turns out that relative entropies and mutual information stay well-defined concepts with all the usual mathematical properties, whereas entropy and conditional entropy are only meaningful in restricted contexts \cite{petz93-book}.

The algebraic definition of quantum relative entropy for general quantum systems was given by \textcite{araki75,araki77}, the definition for Petz Rényi relative entropies by \textcite{petz1985quasi}, the definition for sandwiched Rényi relative entropies by \textcite{berta18} and independently by \textcite{jencova18,jencova21}. Further, the special case of quantum fidelity for $\alpha=1/2$ was already introduced by \textcite{uhlmann_1976}. Some variants of smooth entropies \cite{furrer14,berta15,fawzi25}, $\alpha$--$z$ Rényi divergences \cite{hiai2022}, and \poscite{belavkin1982c} relative entropy were also studied for general quantum systems.

In the following, we briefly outline how entropic information measures can be formulated in terms of standard objects from operator algebra theory. We neither discuss in any detail specific applications of quantum entropies in this section, nor do we attempt a comprehensive discussion of their mathematical properties, as these largely carry over from the finite-dimensional setting (provided the necessary technical care is taken).

The motivation of the algebraic approach to quantum theory is to treat all quantum systems on one mathematical footing, independent of the number of degrees of freedom. For this, the central mathematical object is no longer an underlying Hilbert space $\mathcal{H}$ of (pure) states, but rather the algebra $\mathcal{M}$ generated by the possible observables of the quantum system. When thought of in a concrete representation, this corresponds to a sub-algebra $\mathcal{M}\subseteq\mathcal{B}(\mathcal{H})$\,---\,called a von Neumann algebra\,---\,of the linear, bounded operators $\mathcal{B}(\mathcal{H})$ on some Hilbert space $\mathcal{H}$, in the appropriate topology. Equivalently, one might think of the corresponding abstract object, termed $W^*$-algebra. States are then no longer represented as density operators, but rather by positive, linear, normalized, normal (continuous in the ultra-weak topology) functionals on $\mathcal{M}$.

For states $\sigma$ that are faithful (giving strictly positive weight to every non-zero positive operator), the extension of the quantum relative entropy to general algebraic quantum systems can then be understood by rewriting the finite-dimensional definition from Eq.~\eqref{eq:Umegaki} via the $\ket{\rho}=\sum_i\rho^{1/2}\ket{i}\otimes\ket{i}$ purification\,---\,termed pretty good by \textcite{winter04}\,---\,as
\begin{align}\label{eq:umegaki-algebraic}
    D(\rho\|\sigma)=\bra{\rho}\log\Delta(\rho/\sigma)\ket{\rho}
\end{align}
where the so-called relative modular operator is defined as
\begin{align}
    \Delta(\rho/\sigma)=L(\sigma^{-1})R(\rho)\,,
\end{align}
and $L(a)$ and $R(a)$ denote the left and right multiplication by an element $a\in\mathcal{B}(\mathcal{H})$, respectively. Now, the crucial point is that for general algebraic systems, the relative modular operator remains a well-defined object. We refer the interested reader to the textbook by \textcite{petz93-book} for technical details.

The quantum relative entropy defined in this way is then indeed monotone under quantum channels \cite{araki75,araki77}, while other properties such as continuity have to be replaced with the weaker lower semi-continuity. For the generalization of standard and more advanced quantum entropy inequalities as discussed in Section \ref{sec:standard}, we refer to \textcite{petz1986sufficiency,petz1988sufficiency,faulkner22,faulkner22-1,junge22,hollands23}. As a general rule, relative entropy and mutual information remain well-behaved distance and correlation measures, whereas entropy and conditional entropy easily diverge for general quantum systems \cite{petz93-book}.

\begin{appbox}{HT}{Discriminating states on von Neumann algebras}{app:HT-f}
A notable information-theoretic application understood for general quantum systems is hypothesis testing, as pioneered by \textcite{jaksic12} and later refined by \textcite{hiai2021} as well as \textcite{junge25,lin2026exactexponentssmoothingmaxrelative}. Here, exactly the quantum relative entropy in Eq.~\eqref{eq:umegaki-algebraic}, the Petz--R\'enyi relative entropy in Eq.~\eqref{eq:petz-algebraic}, and the algebraic sandwiched R\'enyi relative entropies have operational interpretations that exactly mirror the finite-dimensional cases presented in Applications \ref{app:HT-d} and \ref{app:HT-b}.
\end{appbox}

More general relative entropies are again defined via the spatial derivative, in particular \poscite{petz1985quasi} R\'enyi relative entropies as
\begin{align}\label{eq:petz-algebraic}
    \widebar{D}_\alpha(\rho\|\sigma)=\frac{1}{\alpha-1}\log{\bra{\rho}\Delta^{\alpha-1}(\rho/\sigma)\ket{\rho}}\,,
\end{align}
under appropriate technical assumptions. The extension of the sandwiched R\'enyi relative entropies is less direct and works on an abstract level either via \poscite{araki82} weighted non-commutative vector valued $L_p$-spaces \cite{berta18} or alternatively via \poscite{kosaki84} version \cite{jencova18}\,---\,leading to equivalent expressions \cite{jencova21}. Recalling the finite-dimensional definition from Eq.~\eqref{eq:sandwiched-renyi} for states $\rho$ and $\sigma$, the crucial step is to rewrite the corresponding trace functional for $\alpha>1$ as \cite{lennert13renyi}
\begin{align}
    \tr \left[ \left( \sigma^{\frac{1-\alpha}{2\alpha}} \rho \sigma^{\frac{1-\alpha}{2\alpha}} \right)^{\alpha} \right]&=\tr \left[ \left( \rho^{\frac{1}{2}} \sigma^{\frac{1-\alpha}{\alpha}}\rho^{\frac{1}{2}} \right)^{\alpha} \right]\\
    &=\left(\sup_{\omega}\tr \left[\omega^{\frac{\alpha-1}{\alpha}}\rho^{\frac{1}{2}}\sigma^{\frac{1-\alpha}{\alpha}}\rho^{\frac{1}{2}} \right]\right)^\alpha\\
    &=\left(\sup_{\omega}\;\bra{\rho}\Delta^{\frac{\alpha-1}{\alpha}}(\omega/\sigma)\ket{\rho}\right)^\alpha\,,
\end{align}
where the optimization is over $\omega>0$ with $\tr[\omega]=1$. This then allows one to define the algebraic sandwiched R\'enyi relative entropies in this variational form \cite{berta18}.

The algebraic entropy formalism is particularly relevant to quantum field theory, where relative entropy remains well defined even when individual entanglement entropies diverge~\cite{Longo2018,Hollands2021,Hollands2023}.


\bigskip

\section{Outlook and conclusions}
\label{sec: outlook}
\label{sec: conclusion}

In this journey through the landscape of quantum entropies, we aimed to systematically classify the broad variety of entropic measures that find use in characterizing quantum information, while at the same time demonstrating how the rich structure of quantum systems makes this diverse landscape a necessity.
Our axiomatic approach in Section~\ref{sec: classification} showed that many properties of divergences and conditional entropies can be completely characterized by the natural physical principles that they are meant to describe.
Indeed, in Section~\ref{sec:standard} we saw how standard entropic quantities naturally emerge as measures of different aspects of quantum information.
However, the range of applications discussed throughout our review illustrated how, depending on the problem at hand, different properties of the information measures may be required, which leads to the plethora of information measures discussed in Sections~\ref{sec:additive} and~\ref{sec:one-shot}.
Due to this complex structure, many questions remain open\,---\,both within the axiomatic framework, as well as in the practical applications.
For example, quantum generalizations of R\'enyi conditional mutual information are still not well understood.
Recent progress on log-Euclidean constructions
motivates the search for Petz and sandwiched generalizations that combine data processing, additivity, and duality with quantitative recovery bounds and operational interpretations.
More generally, determining the entropic constraints on multipartite quantum correlations remains a fundamental challenge. Progress on these questions would strengthen the connection between the axiomatic and operational approaches developed in this review, clarifying which properties select the information measures relevant to a given task, and which features of quantum correlations those measures reveal.
Many operational problems such as the communication of quantum information are also prohibitively difficult to characterize in full generality, due in large part to phenomena such as non-additivity; precisely understanding the structure of quantum channels and the entropic quantities best suited to characterize them presents many exciting opportunities for further developments at the intersection of information theory, quantum physics, and mathematics.


\smallskip

At the heart of our review was the notion of entropy.
Since its inception, interpretations of entropy have gained much attention even beyond physics, in Western culture as a whole. The somewhat anti-teleological nature of the concept has been, to many thinkers, especially appalling: the world seems to gain a purpose, but that purpose, rather than a universal ordering principle, is the ``heat death''\,---\, the state of maximum disorder, i.e., of maximum entropy. \textcite{Clausius1867} himself seems to have been the first to speculate in this direction, but the poetic idea of entropy as a symbol of dissolution revealed itself to be exceptionally appealing. Written in the aftermath of the First World War, \poscite%
{SPENGLER}
philosophy of history treatise, aptly titled ``The Decline of the West'', draws an intrepid parallelism with the Twilight of the Gods of Norse mythology:
\begin{quote}
What the myth of G\"otterd\"ammerung signified of old, the irreligious form of it, the theory of Entropy, signifies to-day\,---\,world's end as completion of an inwardly necessary evolution.
\end{quote}
Our aims in this review were more modest.
We merely intended to illustrate the multifaceted beauty of the notion of entropy in quantum physics and information theory, all in the hope that a structured presentation of the topic might help reduce the entropy in the Reader's mind. If, instead, we have only succeeded in increasing it, rest assured that this was not done on purpose.

\medskip
\noindent\textit{Acknowledgments.} We thank Andreas Winter for many discussions and his patient advice throughout this project. R.R.\ acknowledges financial support from the ERC grant GIFNEQ 101163938. B.R.\ acknowledges the support of the Japan Science and Technology Agency (JST) PRESTO grant no.\ JPMJPR25FB and the RIKEN Pioneering Project `Mathematical Foundation of Quantum Information'. M.B.\ acknowledges support from the European Research Council (ERC Grant Agreement No.\ 948139) and the Excellence Cluster Matter and Light for Quantum Computing (ML4Q-2). L.L.\ acknowledges financial support from the European Union under the European Research Council (ERC Grant Agreement No.~101165230). M.T.\ is supported by the NRF Investigatorship award (NRF-NRFI10-2024-0006).

\appendix


\section{Universal properties of relative entropies}
\label{app:universal}
In Section~\ref{sec: classification}, we introduced relative entropies as normalized functionals satisfying monotonicity under quantum channels and additivity under tensor products.
These information-theoretic requirements impose strong structural constraints.
We collect several consequences below, following the systematic treatment by~\textcite{gour2025book}.

The following properties hold for every relative entropy $\D$ (see, e.g., \textcite{gour2021axiomatic,gour2020extensions}).
Unless otherwise stated, the arguments are normalized states.

\begin{enumerate}[itemsep=0pt,label=\Roman*.]
    \item $\D$ is non-negative:
    \begin{align}
        \D(\rho\|\sigma) \geq 0 \,.
    \end{align}
    This follows from data processing and $\D(\rho\|\rho)=0$, by applying a channel that replaces its input with a fixed state.

    \item $\D$ is jointly continuous on pairs of states with full support.

    \item $\D$ is either faithful or has discontinuities at rank-deficient states.
    Here, faithfulness means that $\D(\rho\|\sigma)=0$ implies $\rho=\sigma$.\footnote{The min-relative entropy $D_{\min}$ is not faithful and can be discontinuous when the support of its first argument changes.}

    \item $\D$ satisfies a triangle inequality of the form
    \begin{align}
        \label{eq:Dtriangle}
        \D(\rho\|\sigma)
        \leq \D(\rho\|\tau) + D_{\max}(\tau\|\sigma) \,.
    \end{align}
\end{enumerate}

Data processing also implies invariance under isometries: an isometric channel can be reversed on its image, so applying data processing in both directions gives equality.
Relative entropies satisfy the following stronger statement, using their extension to a subnormalized second argument~\cite{gour2025book}.

\begin{enumerate}[itemsep=0pt,resume,label=\Roman*.]
    \item $\D$ is strongly invariant under isometries.
    For any isometry $U$ and states $\rho$ and $\sigma$ on its input and output Hilbert spaces, respectively,
    \begin{align}
        \D\big(U\rho U^{\dag}\big\|\sigma\big)
        \geq \D\big(\rho\big\|U^{\dag}\sigma U\big) \,,
    \end{align}
    with equality if $\sigma$ commutes with the projector $UU^{\dag}$ onto the image of $U$.
\end{enumerate}

The operator $U^{\dag}\sigma U$ need not have unit trace, since $\sigma$ can have weight outside the image of $U$.
The equality statement can equivalently be expressed using normalized states as
\begin{align}
    \label{eq:Dsupport}
    \D\big(\rho\oplus 0\big\|
        \lambda\sigma\oplus(1-\lambda)\tau\big)
    = \D(\rho\|\sigma)-\log\lambda
\end{align}
for all compatible triples of states $(\rho,\sigma,\tau)$ and $0<\lambda\leq 1$.

Thus, when the second argument is block diagonal with respect to the support of the first, its block outside that support can be discarded, provided the retained block is left subnormalized.
Renormalizing introduces the correction $-\log\lambda$ in~\eqref{eq:Dsupport}.

A further useful consequence is operator anti-monotonicity in the second argument.
\textcite{colbeck2012uncertainty} derive this property from data processing and~\eqref{eq:Dsupport}.

\begin{enumerate}[itemsep=0pt,resume,label=\Roman*.]
    \item $\D$ is operator anti-monotone in the second argument:
    if $S\leq T$, then
    \begin{align}
        \D(R\|S)\geq\D(R\|T)
    \end{align}
    for all positive semidefinite operators $R$, $S$, and $T$ in the domain of the extension.
\end{enumerate}

\section{A zoo of divergences}
\label{app: all entropies}

In Figure~\ref{fig:important_entropies}, we illustrated the R\'enyi entropic information measures of greatest operational relevance. Below we present a more comprehensive diagram encompassing (almost) all the divergences considered in this review.

\begin{inplacefigure}
\centering
\begin{tikzpicture}

    \node at (0,0) {\includegraphics[width=0.45\textwidth]{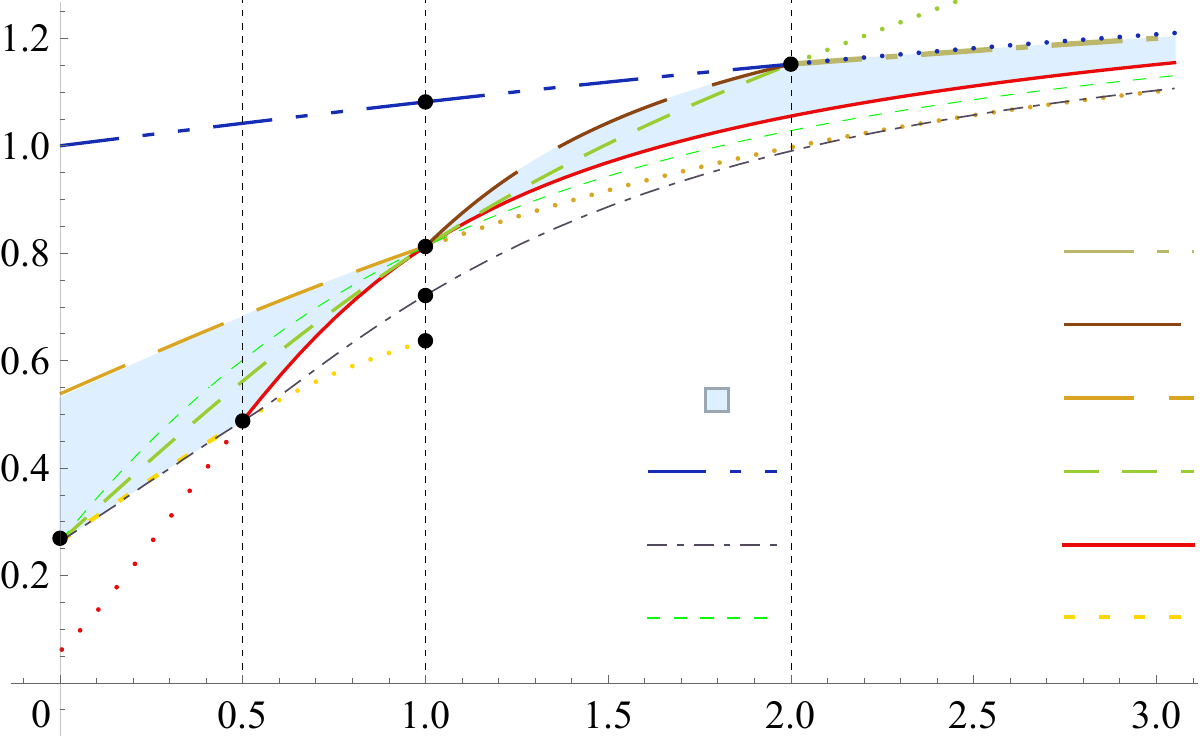}};

    \node at (-3.2,-1.28) {\small$D_{\min}$};
    \node at (-2,-0.5) {\small$-\log{F}$};
    \node at (-1.34,0.94) {\small$D$};
    \node at (-0.83,0.47) {\small $D^{\mathbb{M}}$};
     \node at (-0.88,0.12) {\small $D^\star$};
    \node at (-0.98,2.05) {\small$\widehat{D}$};
    \node at (1.07,2.25) {\small$\widebar{D}_2$};
    \node at (2.8,2.38) {\small $D_{\max}$};
    \draw[->] (3.2,2.38) -- (3.5,2.40)
    node[midway, right, sloped] {\tiny $\;+\infty$};
    \node at (2.3,0.82) {\small$D_{\alpha,\alpha-1}$};
    \node at (2.3,0.3) {\small$D_{\alpha,\alpha/2}$};
    \node at (2.6,-0.1) {\small$D^\flat_\alpha$};
    \node at (2.6,-0.6) {\small$\widebar{D}_\alpha$};
    \node at (2.6,-1.13) {\small $\widetilde{D}_\alpha$};\node at (2.3,-1.63) {\small $D_{\alpha,1-\alpha}$};

     \node at (-0.1,-1.1) {\small $D^{\mathbb{M}}_\alpha$};
    \node at (-0.1,-1.6) {\small $D^{\mathrm{HT}}_{\alpha}$};
    \node at (-0.1,-0.65) {\small$\widehat{D}_\alpha$};
    \node at (0.2,-0.19) {\small$D_{\alpha,z}$};
    \node at (4.2,-2.2) {\large$\alpha$};


\end{tikzpicture}

\caption{R\'enyi divergences in the literature evaluated for the states
\vskip-5pt
\begin{equation*}
 \rho = \tfrac{1}{6}
 \left(\begin{smallmatrix}
 1 & 1 & 0 \\[1pt]
 1 & 3 & 2  \\[1pt]
 0 & 2 & 2
 \end{smallmatrix}\right), \quad
 \sigma = \tfrac{1}{8}
 \left(\begin{smallmatrix}
 5 & 1 & 0 \\[1pt]
 1 & 2 & 1 \\[1pt]
 0 & 1 & 1
 \end{smallmatrix}\right).
\end{equation*}%
The dotted lines indicate parameter regimes in which the corresponding relative entropies fail to satisfy DPI. Across all $\alpha$, the smallest and largest quantities are the min- and max-relative entropies, $D_{\min}$ and $D_{\max}$, respectively. For fixed $\alpha$, the smallest (additive) relative entropy is the reverse sandwiched R\'enyi relative entropy for $\alpha \in [0,1/2]$ and the sandwiched R\'enyi relative entropy for $\alpha \geq 1/2$. In their respective DPI-violating regimes, both families lie below the measured relative entropy, which provides a lower bound for all divergences satisfying DPI. For $\alpha \in [0,2]$, the geometric relative entropy is the largest relative entropy, while for $\alpha > 2$, the largest is $D_{\alpha,\alpha-1}$.}

\label{fig:all_entropies}
\end{inplacefigure}


\makeatletter
\def\@bibstyle{apsrmp4-2-fixed}
\makeatother
\bibliography{library_new.bib}

\end{document}